%% file: Hoyle_Thesis.tex
\documentclass[a4paper,12pt,twoside]{report}
\usepackage{a4wide,exscale}
\usepackage{fancyhdr}
\usepackage{amsmath}
\usepackage{amsfonts}
\usepackage{amssymb}
\usepackage{amsthm}
\usepackage{graphicx}
\usepackage{dsfont}
\usepackage[numbers]{natbib}
\usepackage{verbatim}
\usepackage{ifthen}
\usepackage{subfigure}
\newtheorem{thm}{Theorem}[section]

\newtheorem{coro}[thm]{Corollary}

\newtheorem{defn}[thm]{Definition}
\newtheorem{rem}[thm]{Remark}
\newtheorem{prop}[thm]{Proposition}
\newtheorem*{prop*}{Proposition}

\newcommand{\p}[1]{\ensuremath{\mathbb{#1}}}
\newcommand{\gb}[1]{\ensuremath{\gamma_{#1 T}}}
\newcommand{\gbb}[1]{\ensuremath{\bar{\gamma}_{#1 T}}}
\newcommand{\levy}{{L\'evy }}
\newcommand{\D}[1]{\ensuremath{\Delta_{#1}}}
\newcommand{\tD}[1]{\ensuremath{\Delta^{\pi}_{#1}}}
\newcommand{\mymark}{%
	\ifthenelse{\equal{\rightmark}{}}
		{\leftmark}
		{\rightmark}}
\def\1{\mathds{1}}
\def\R{\mathbb{R}}

\def\Q{\mathbb{Q}}
\def\E{\mathbb{E}}
\def\N{\mathbb{N}}
\def\d{\,\mathrm{d}}
\def\dd{\mathrm{d}}
\def\half{\frac{1}{2}}
\def\var{\mathrm{Var}}
\def\cov{\mathrm{Cov}}
\def\th{\theta}
\def\e{\mathrm{e}}
\def\b{\beta}
\def\g{\gamma}
\def\G{\Gamma}
\def\a{\alpha}

\def\s{\sigma}
\def\l{\lambda}
\def\law{\overset{\textrm{law}}{=}}
\def\tp{\mathrm{T}}
\def\F{\mathcal{F}}
\def\lrb{\mathit{LRB}}
\def\lrbc{\mathit{LRB}_{\mathcal{C}}}
\def\lrbd{\mathit{LRB}_{\mathcal{D}}}
\def\i{\mathrm{i}}

\def\SHS{stable-1/2 subordinator }
\def\SH{stable-1/2 }
\def\k{\kappa}
\def\t{\tau}

\begin{document}

\bibliographystyle{plainnat_ed}

\input{fancy}
\input{FrontPage}

\newpage
\input{Abstract}

\newpage

\input{Acknowledgements}

\newpage
\input{Declaration}

\clearpage
\fancyhead{}
\fancyfoot{}
\pagestyle{fancy} 
\renewcommand{\chaptermark}[1]{\markboth{\thechapter \ #1}{}}
\fancyhead[RO,LE]{\sffamily\small \thepage}
\fancyhead[LO,RE]{\sffamily\small \nouppercase{\rightmark}}
\pagestyle{fancy}
\renewcommand{\headrulewidth}{0.4pt}
\renewcommand{\footrulewidth}{0.0pt}
\setcounter{tocdepth}{1}
\tableofcontents
\listoffigures


\clearpage

\input{fancy}
\input{chap01}

\clearpage
\input{chap02}
\input{chap03}
\input{chap04}

\input{chap05}
\input{chap06}
\input{chap07}
\input{chap08}
\input{chap09}

\appendix
\input{chap10}

\clearpage
\input{Literature}

\end{document}

%% file: fancy.tex
\fancyhead{}
\fancyfoot{}
\pagestyle{fancy} 
\renewcommand{\chaptermark}[1]{\markboth{\thechapter \ #1}{}}
\renewcommand{\sectionmark}[1]{\markright{\thesection\ #1}}
\fancyhead[RO,LE]{\sffamily\small \thepage}
\fancyhead[LO,RE]{\sffamily\small \mymark}
\pagestyle{fancy}
\renewcommand{\headrulewidth}{0.4pt}
\renewcommand{\footrulewidth}{0.0pt}

%% file: FrontPage.tex
\thispagestyle{empty}%
\null\vskip0.2in%
\begin{center}
\LARGE{{\bf Information-Based Models \\ for Finance and Insurance\\}}
\end{center}
\vfill
\begin{center}
{\Large {\bf by}}\\
\mbox{} \\
{\Large {\bf Anthony Edward Vickerstaff Hoyle}}
\end{center}
\vfill
\begin{center}
\large{\bf{Department of Mathematics \\ Imperial College London \\
London SW7 2AZ \\ United Kingdom}}
\end{center}
\vfill
\begin{center}
\large{\bf{Submitted to Imperial College London \\
for the degree of \\
Doctor of Philosophy}}
\end{center}
\vfill
\begin{center}
{\large\bf{2010}}
\end{center}

%% file: Abstract.tex
\mbox{}\newline \vspace{5mm} \mbox{}\newline \LARGE
{\bf Abstract} 

\normalsize \vspace{5mm}

\noindent
In financial markets, the information that traders have about an asset is reflected in its price.
The arrival of new information then leads to price changes.
The `information-based framework' of Brody, Hughston and Macrina (BHM) isolates the emergence of information, and examines its role as a driver of price dynamics.
This approach has led to the development of new models that capture a broad range of price behaviour.
This thesis extends the work of BHM by introducing a wider class of processes for the generation of the market filtration.
In the BHM framework, each asset is associated with a collection of random cash flows.
The asset price is the sum of the discounted expectations of the cash flows.
Expectations are taken with respect (i) an appropriate measure, and (ii) the filtration generated by a set of so-called 
information processes that carry noisy or imperfect market information about the cash flows.
To model the flow of information, we introduce a class of processes termed \levy random bridges (LRBs), generalising the Brownian and gamma information processes of BHM.
Conditioned on its terminal value, an LRB is identical in law to a \levy bridge.
We consider in detail the case where the asset generates a single cash flow $X_T$ at a fixed date $T$.
The flow of information about $X_T$ is modelled by an LRB with random terminal value $X_T$.
An explicit expression for the price process is found by working out the discounted conditional expectation of $X_T$ with respect to the natural filtration of the LRB.
New models are constructed using information processes related to the Poisson process, the Cauchy process, the stable-1/2 subordinator, the variance-gamma process, and the normal inverse-Gaussian process.
These are applied to the valuation of credit-risky bonds, vanilla and exotic options, and non-life insurance liabilities.

%% file: Acknowledgements.tex
\mbox{}\newline \vspace{5mm} \mbox{}\newline \LARGE
{\bf Acknowledgements} \normalsize \vspace{5mm}

\noindent 
I am very grateful to Lane P.~Hughston, my supervisor, for his help and support.
With his breadth of knowledge, Lane's teachings have stretched beyond mathematical finance to such eclectic subjects as quantum mechanics, Italian opera, and Indian literature.
Many thanks go my co-author Andrea Macrina whose enthusiasm for this work often pushed me forward.
I would like to express my gratitude to members of the mathematical finance groups at Imperial College London and King's College London, and members of the London Graduate School in Mathematical Finance, with particular thanks going to my teachers M.~Davis, H.~Geman, W.~Shaw, and M.~Zervos; and also to P.~Barrieu, D.~Brody, M.~Pistorius, and R.~Norberg.
I am grateful to F.~Delbaen, and M.~Schweizer, and other members of the mathematical finance group at ETH Z\"urich for their hospitality and support.
I also wish to thank D.~Taylor for his hospitality at the University of the Witwatersrand.
This work was supported by an EPSRC Doctoral Training Grant and a European Science Foundation research visit grant under the Advanced Mathematical Methods in Finance programme (AMaMeF).
Part of this work has undertaken when I was student at King's College London.
Finally, for their unfailing support throughout my studies, I must thank my family---Mum, Dad, Charlotte and Thomas---and Christine.

%% file: Declaration.tex
\begin{center}
\phantom{} \vspace{8cm} \noindent The work presented in this
thesis is my own.\\ \vspace{4cm} A.E.V.~Hoyle
\end{center}

%% file: chap01.tex
\chapter{Introduction and summary}

The formation of prices in financial markets has long been a concern for economists.
Macroeconomic factors, market microstructure, and investor preferences constitute an inexhaustive list of constituents that play a part.
It is a daunting task to formulate a parsimonious pricing model that incorporates even this short list of effects, and is capable of delivering useful and timely results.
It is not surprising then that analysis of price formation tends to focus on a single or small number of relevant elements at any one time.

The information that traders and investors have about an asset is reflected in its price.
Information about an asset might include information about any of the effects that influence the formation of its price.
The arrival of new information leads to changes in the price of the asset.
Qualitatively speaking, if information about an asset arrives infrequently and in large lots, then its price process will exhibit large jumps.
Conversely, if information arrives smoothly and steadily, then the impact of information arrival over short time-scales will be modest.
Thus, the emergence of information as driver of price dynamics presents itself as interesting avenue of investigation.

The objective of this thesis is to provide a framework for the derivation of price dynamics of assets (or, indeed, the valuation dynamics of liabilities) through the modelling of information flow.
It has become commonplace in the mathematical finance literature to develop pricing models under the risk-neutral measure.
Two situations when this may be appropriate are (a) when the market is incomplete and there exists a multitude of equivalent martingale measures, and the selection of any one for pricing is made subjectively; and (b) when pricing models are calibrated to market prices (of options), since these prices are (theoretically) expectations under the risk-neutral measure.
We adopt a similar approach.
Throughout this work we build models under a `suitable' measure $\Q$.
At time $t$, we define the price of a contingent claim $X_T$ due at time $T$ as $P_{tT} \, \E_{\Q}[X_T \,|\, \F_t]$, where $P_{tT}$ is a discount factor, and $\{\F_t\}$ is the market filtration.
For this price to agree with the finance-theoretic arbitrage-free price, then the measure $\Q$ needs to be interpreted as the $T$-forward measure in the case of stochastic interest rates, or the risk-neutral measure in the case of deterministic interest rates.
If $X_T$ is the size of a future liability that needs to be accounted for (`booked'), then it may not be appropriate to use the market (i.e.~arbitrage-free) price for valuation. 
This is particularly true if the market for such a claim is illiquid.
In order to remain as general as possible we refrain from being precise in the interpretation of $\Q$, and leave that to the judgement of the implementer.

In a stochastic model, it is the filtration $\{\F_t\}$ that encodes the emergence of information.
Choosing an asset price process to be geometric Brownian motion, for example, implies that the process is adapted to a Brownian filtration.
Although this approach of implicitly choosing a filtration by specifying the law of a price process is common in mathematical finance, we wish to avoid it and to specify $\{\F_t\}$ directly.
In particular, we postulate the existence of a market information process $\{\xi_{tT}\}$ which generates $\{\F_t\}$.
Then prices are derived by taking conditional expectations of cash flows with respect to this filtration.

Earlier, we were careful to include the valuation of liabilities within our remit.
We will consider in detail how the methods we develop can be applied to the calculation of reserves that an insurance company should set aside to meet future claims.
A non-life insurance company may underwrite various risks for a particular year in return for premiums.
The company `incurs' claims over the one-year period, which means that losses are triggered that the company is liable to cover. 
However, there may be a delay between the date a loss is triggered and the day it is reported to the company, the total size of a claim may not be known when the claim is reported, and a claim may not be paid by a single cash-flow on a single date. 
The insurer may be paying for these losses for many years to come.
The problem is: How much money should the insurer reserve at a given time to cover all future claim payments?
This has implications for the company's accounting, tax liability, solvency, capital adequacy, and investment	strategy.

Chapter \ref{chap:levy_processes} begins with a brief introduction to \levy processes, strictly stable processes, and \levy bridges.
A \levy bridge is a stochastic process defined over a finite time horizon, and is a \levy process whose terminal value is known from the outset.
We provide a proof of the Markov property for \levy bridges.
For the remainder of the chapter we examine particular examples of these processes.
First are the well-known Brownian motion and the Brownian bridge;
these are Gaussian processes with continuous sample paths.
Brownian motion without drift is a stable process.
Then are the gamma process and gamma bridge, which are increasing processes.
The variance-gamma (VG) process is closely related to Brownian motion and the gamma process.
In particular, a VG process is constructed by subordinating a Brownian motion by an independent gamma process, or by taking the difference of two independent gamma processes.
We list some properties of the VG process and then examine VG bridges.
We are able to utilise the relationship between the VG process and Brownian motion and the gamma process to derive two constructions of the VG bridge.
These constructions write the VG bridge in terms of Brownian bridges, gamma bridges, and random volatility terms.
The second stable process we examine is the stable-1/2 subordinator.
This process can be constructed as the hitting times of a Brownian motion.
We choose to examine this subordinator over other stable subordinators because its density is known in closed form.
It will become apparent that this is a desirable feature in this work.
The stable-1/2 subordinator has heavy tails; the expected value of the process at any future date is infinite.
However, all positive moments of the stable-1/2 bridge exist.
By subordinating a Brownian motion with an independent stable-1/2 subordinator we can construct a Cauchy process.
The Cauchy process is the third stable process we consider.
It too is heavy tailed---its first moment does not exist.
We find that the first two moments of the Cauchy bridge do exist and are finite.
One of the effects of pinning the end point of the Cauchy process is to temper its wild behaviour.
The last process we consider that has a continuous state space is the normal inverse-Gaussian (NIG) process.
This process is constructed by subordinating a Brownian motion by an inverse-Gaussian (IG) process.
It is a very similar process to the VG process.
It is not surprising then that NIG bridges are similar to VG bridges.
Finally, we come to the Poisson process and the Poisson bridge.
These processes have a discrete state spaces.
The state space of the Poisson process is $\N_0$, and Possion bridges are restricted to a subset of $\N_0$.
We show how the Poisson bridge can be written as a Poisson process under a time change that is not independent of the process.

In Chapter \ref{chap:LRB} we define \levy random bridges (LRBs).
An LRB is a process defined over a finite time horizon whose bridge laws are the bridge laws of a \levy process.
It can be interpreted as being a \levy process conditioned to have a fixed marginal law at a fixed future date.
The motivation for LRBs is the desire for Markov processes, defined over a fixed time interval, over which we have a distributional degree of freedom for the terminal value.
We shall see that such processes are useful for asset pricing in the information-based framework.
We derive various properties of LRBs including: 
	that LRBs are Markov processes;
	that the law of an LRB is equivalent to the law of a \levy process (at least for all times up to the LRB's termination time);
	that LRBs have stationary increments;
	that the joint distribution of the increments of an LRB have a generalized multivariate Liouville distribution;
	and, if the path of an LRB is split into non-overlapping portions, then each portion is itself an LRB.
The marginal characteristic function often proves convenient in the analysis of a \levy process.
This is not the case for an LRB.
However, we are able to provide an expression for the transition law of a general LRB.

The information-based framework of Brody, Hughston \& Macrina is described in Chapter \ref{chap:info_based}.
In this framework, cash flows are functions of independent $X$-factors.
`Information processes' generate the market filtration, and reveal the value of the $X$-factors---thus, they reveal the value of the cash flows.
Cash flows are priced by discounting their expected value given the market information.
A general multi-factor set-up is described that allows for a rich dependency structure between cash flows.
Much of the analysis is presented for a single $X$-factor market where the only information process is an LRB.
The final value of the LRB is set to be the value of the $X$-factor.
Using a single factor may sound restrictive, but this includes some well-known one-dimensional exponential \levy models as special cases (the Black-Scholes model was recovered from an information-based model by \citet{BHM2}).
Using Bayesian methods, we are able to derive the dynamics of the price of a cash flow.
From this, we can price European call options on the cash flow price.
An expression for the call price is given in a general LRB-information model.
The material in Chapters \ref{chap:LRB} and \ref{chap:info_based} appears in \citep{HHM1}.

In Chapter \ref{chap:VG} we examine a one-parameter VG random bridge and apply it to information-based pricing.
We derive two terminal value decompositions of the VG random bridge using the decomposition of the VG bridge.
We show that a three parameter VG process can be recovered by scaling a standard VG random bridge which has an asymmetric VG terminal law.
This allows the VG equity model of \citet{MCC1998} to be derived as a special case of a single $X$-factor information-based model, when information is provided by a standard VG random bridge.
We price a binary bond in a VG information model and include a rate parameter $\s$ which acts like a volatility parameter.
We provide two algorithms for the simulation of sample paths of the VG random bridge, and provide plots of simulated binary bond prices.

We develop a non-life reserving model in Chapter \ref{chap:insurance} using a stable-1/2 random bridge to simulate the accumulation of paid claims, allowing for an arbitrary choice of \emph{a priori} distribution for the ultimate loss.
Taking a Bayesian approach to the reserving problem, we derive the process of the conditional distribution of the ultimate loss.
The `best-estimate ultimate loss process' is given by the conditional expectation of the ultimate loss.
We derive explicit expressions for the best-estimate ultimate loss process, and for expected recoveries arising from aggregate excess-of-loss reinsurance treaties.
Use of a deterministic time change allows for the matching of any initial (increasing) development pattern for the paid claims.
We show that these methods are well-suited to the modelling of claims where there is a non-trivial probability of catastrophic loss.
The generalized inverse-Gaussian (GIG) distribution is shown to be a natural choice for the \emph{a priori} ultimate loss distribution.
For particular GIG parameter choices, the best-estimate ultimate loss process can be written as a rational function of the paid-claims process.
We extend the model to include a second paid-claims process, and allow the two processes to be dependent.
The results obtained can be applied to the modelling of multiple lines of business or multiple origin years.
The multidimensional model has the attractive property that the dimensionality of calculations remains low, regardless of the number of paid-claims processes.
An algorithm is provided for the simulation of the paid-claims processes.
The material in this chapter appears in \citep{HHM2}.

In Chapter \ref{chap:Cauchy} we derive properties of Cauchy random bridges, and describe a method for simulating sample paths.
Modelling the information process as a Cauchy random bridge, we examine the pricing of a binary bond in an information-based model.
We derive an explicit expression for the price of a call option on the bond price.

Chapter \ref{chap:NIG} closely follows Chapter \ref{chap:VG}.
Only, in this case, we examine the NIG random bridge, and use it to model an information process, as opposed to using the VG random bridge.
Since the NIG and VG processes are similar, the information-based models their respective random bridges produce are similar.

In Chapter \ref{chap:Poisson} we provide an example of an LRB with a discrete state-space, the Poisson random bridge.
Poisson random bridges are counting processes, and we derive expressions for the waiting times between jumps in terms of the probability generating function of the terminal distribution.
We show that a Poisson random bridge can be written as a Poisson process with a state-dependent intensity, and we derive an explicit expression for the intensity process.
When the terminal distribution of a Poisson random bridge is a negative binomial distribution, we show that all of the increment distributions of the process are negative binomial.
We then generalise this result to show that a Poisson random bridge with a mixed Poisson terminal distribution is a mixed Poisson process.
That is, the distribution of any increment of the process is a Poisson distribution with a mixed mean.
By making the jump sizes of the PRB random we construct the compound Possion random bridge.
We derive an expression for the characteristic function of compound Poisson random bridge.
Finally, we price an $n$th-to-default credit swap in a model where defaults occur at the jump times of a Poisson random bridge.
In this credit swap the buyer pays a premium in return for a lump-sum payment on the event of the $n$th default from a basket of credit risks.

%% file: chap02.tex
%
%

\chapter{\levy processes and \levy bridges} \label{chap:levy_processes}

We fix a probability space $(\Omega,\Q,\F)$, and assume that all processes and filtrations under consideration are c\`adl\`ag.
Unless otherwise stated, when discussing a stochastic process we assume that the process takes values in $\R$, begins at time 0, and the filtration is that generated by the process itself.
We work with a finite time horizon $[0,T]$.

\section{\levy processes} \label{sec:levy_processes}
This section and the next summarise a few well-known results about one-dimensional \levy processes and stable processes, further details of which can be found in \citet{Bert1996}, \citet{Kyp2006}, and \citet{Sato1999}.
A \levy process is a stochastically-continuous process that starts from the value 0, and has stationary, independent increments.
An increasing \levy process is called a \emph{subordinator}.
For $\{L_t\}$ a \levy process, its \emph{characteristic exponent} $\Psi:\R\rightarrow\mathbb{C}$ is defined by
\begin{equation}
	\E[\e^{\i\l L_t}]= \exp(-t\Psi(\l)), \qquad \l\in\R.
\end{equation}
The characteristic exponent of a \levy process characterises its law,
and its form is prescribed by the L\'evy-Khintchine formula:
\begin{equation}
	\label{eq:LK}
	\Psi(\lambda)=\i a \lambda+\half \s^2 \lambda^2+\int_{-\infty}^{\infty}(1-\e^{\i x \lambda}+\i x\lambda\1_{\{|x|<1\}})\Pi(\dd x),
\end{equation}
where $a\in\R$, $\s>0$, and $\Pi$ is a measure (the \emph{\levy measure}) on $\R\backslash\{0\}$ such that 
\begin{equation}
	\int_{-\infty}^{\infty}(1 \wedge |x|^2)\,\Pi(\dd x)<\infty.
\end{equation}

There are particular subclasses of \levy processes that we shall consider, defined as follows:
\begin{defn}
	Let $\{L_t\}_{0\leq t \leq T}$ and $\{M_t\}_{0\leq t \leq T}$ be \levy processes.
	Then we write
	\begin{enumerate}
		\item
			$\{L_t\}\in\mathcal{C}[0,T]$ if the density of $L_t$ exists for every $t\in (0,T]$,
		\item
			$\{M_t\}\in\mathcal{D}$ if the marginal law of $M_t$ is discrete for some $t>0$.
	\end{enumerate}
\end{defn}
\noindent

\begin{rem}
	If the marginal law of $M_t$ is discrete for some $t>0$, then the marginal law of $M_t$ is discrete for all $t>0$.
	The density of $L_t$ exists if and only if its law is absolutely continuous with respect to the Lebesgue measure.
	In general, the absolute continuity of $L_t$ depends on $t$; thus 
	$\mathcal{C}[0,T_2]\subseteq \mathcal{C}[0,T_1]$ for $T_1\leq T_2$.
	See \emph{\citet[chap.~5]{Sato1999}} for further details on the time dependence of distributional properties of \levy processes.
\end{rem}

We reserve the notation $f_t(x)$ to represent the density of $L_t$ for some $\{L_t\}\in\mathcal{C}[0,T]$.
Hence $f_t:\R\rightarrow \R_+$ and $\Q[L_t\in \dd x]=f_t(x)\d x$.
We reserve $Q_t(a)$ to represent the probability mass function of $M_t$ for some $\{M_t\}\in\mathcal{D}$.
We denote the state-space of $\{M_t\}$ by $\{a_i\}\subset\R$.
Hence $Q_t:\{a_i\}\rightarrow [0,1]$ and $\Q[M_t=a_i]=Q_t(a_i)$.
We assume that the sequence $\{a_i\}$ is strictly increasing.

The transition probabilities of \levy processes satisfy the convolution identities
\begin{align}
		f_t(x)&=\int_{-\infty}^{\infty} f_{t-s}(x-y)f_s(y)\d y &&\text{for $\{L_t\}\in\mathcal{C}[0,T]$},
		\\\intertext{and}
		Q_t(a_n)&=\sum_{m=-\infty}^{\infty}Q_{t-s}(a_n-a_m)Q_s(a_m) &&\text{for $\{M_t\}\in\mathcal{D}$},
\end{align}
for $0\leq s<t\leq T$.
These are the Chapman-Kolmogorov equations for the processes $\{L_t\}$ and $\{M_t\}$.

The law of any c\`adl\`ag stochastic process is characterised by its finite-dimensional distributions.
The finite-dimensional densities of $\{L_t\}_{0\leq t\leq T}$ exist and, with the understanding that $x_0=t_0=0$, they are given by
\begin{equation} 
	\Q[L_{t_1} \in\dd x_1, \ldots,L_{t_n}\in\dd x_n]=
	\prod_{i=1}^n \left[f_{t_i-t_{i-1}}(x_i-x_{i-1}) \d x_i\right],
\end{equation}
for every $n\in\mathbb{N}_+$, every $0<t_1<\cdots<t_n\leq T$, and every $(x_1,\ldots,x_n)\in\R^n$.
With the understanding that $a_{k_0}=t_0=0$, the finite-dimensional probabilities of $\{M_t\}$ are
\begin{equation} 
	\Q[M_{t_1} = a_{k_1}, \ldots,M_{t_n} = a_{k_n}]=\prod_{i=1}^n Q_{t_i-t_{i-1}}(a_{k_i}-a_{k_{i-1}}),
\end{equation}
for every $n\in\mathbb{N}_+$, every $0<t_1<\cdots<t_n$, and every $(k_1,\ldots,k_n)\in\mathbb{Z}^n$.

\section{Stable processes} \label{sec:stable}
The (strictly) stable processes form a subclass of the \levy processes.
We say that a \levy process $\{S^{\a}_t\}$ is a \emph{stable process with index $\a$} (or \emph{stable-$\alpha$ process}) if its characteristic exponent satisfies 
\begin{equation}
	\label{eq:stableDef}
	\Psi(k\l)=k^{\a}\Psi(\l),
\end{equation} 
for every $k>0$ and every $\l\in\R$; $\a$ is restricted to values in $(0,2]$.
The process $\{S^{\a}_t\}$ satisfies the scaling property 
\begin{equation}
	\{k^{-1/\a}S_{kt}^{\a}\}_{t\geq 0}\law\{S_t^{\a}\}_{t\geq 0} \qquad \text{for $k>0$}.
\end{equation}
Equation (\ref{eq:stableDef}) and the L\'evy-Khintchine formula restrict the characteristic exponent and the \levy measure of $\{S^{\a}_t\}$ to take an explicit form which depends on $\a$.
When $\a\in(0,1)\cup(1,2)$,
\begin{equation}
	\Psi(\l)=\k|\l|^{\a}(1-\i\b\mathrm{sign}(\l)\tan(\pi\a/2)), 	\label{eq:cess}
\end{equation}
where $\k>0$, and $\b\in[-1,1]$.
In this case the \levy measure can be written
\begin{equation}
	\Pi(\dd x)=	\left\{\begin{aligned}
								&\k^+x^{-\a-1} \d x && \text{for $x>0$,}
								\\ &\k^-|x|^{-\a-1} \d x && \text{for $x<0$,}
							\end{aligned}\right.
\end{equation}
where $\k^+$ and $\k^-$ are non-negative numbers satisfying
\begin{equation}
	\b=\frac{\k^+-\k^-}{\k^++\k^-}.
\end{equation}
If $\b=1$ then the process exhibits only positive jumps, if $\b=0$ the process is symmetric, and if $\b=-1$ the process exhibits only negative jumps.
If $\a=2$ then $\Psi(\l)=\half \s^2\l^2$, and $\Pi(\dd x)=0$.
In this case $\{S^{\a}_t\}$ is a Brownian motion (without drift).
If $\a=1$ then $\Psi(\l)=\i a \l+c|\l|$, for $c>0$, and $\Pi(\dd x)=cx^{-2}\d x$.
In this case $\{S^{\a}_t\}$ is a Cauchy process with drift.

Excluding the case when $\a=2$, stable processes are heavy-tailed processes.
The fractional moments of the stable random variable $S^{\a}_t$ ($\a<2$) satisfy
\begin{align}
	\E[|S^{\a}_t|^p]&<\infty && \text{if $p<\a$},
	\\ \E[|S^{\a}_t|^p]&=\infty && \text{if $p\geq \a$}.
\end{align}
Hence the second moment of $S^{\a}_t$ is infinite for $\a<2$, and the expected value of $S^{\a}_t$ does not exist or is infinite for $\a\leq 1$.

The density of $S^{\a}_t$ exists and is continuous for any $\a$, and so $\{S^{\a}_t\}\in\mathcal{C}[0,T]$ for any $T>0$.
However, this density can be expressed in terms of elementary functions only when $\{S^{\a}_t\}$ is a Brownian motion, a Cauchy process, or a stable subordinator with index $\a=1/2$.
We will examine each of these special cases later in this chapter.
(See \citet[XVII.6]{Feller2} for examples of series representations for the density of a stable random variable with arbitrary index $\a$.)

If $\{S^{\a}_t\}$ is a stable subordinator then the Laplace transform of $S^{\a}_t$ exists and is given by
\begin{equation}
	 \E[\exp\left(-\l S^{\a}_t\right)]=\exp\left(-\k t \l^{\a}\right) \quad \text{for $\l\geq 0$,}
\end{equation}
where $\k>0$, and $\a$ must be further restricted to $0<\a<1$.

\section{\levy bridges}
A bridge is a stochastic process that is pinned to some fixed point at a fixed future time.
Bridges of Markov processes were constructed and analysed by \citet{FPY1993} in a general setting.
In this section we focus on the bridges of \levy processes in the classes $\mathcal{C}[0,T]$ and $\mathcal{D}$.
In particular we have the following:
\begin{prop}
	\label{prop:LB_Markov}
	The bridges of processes in $\mathcal{C}[0,T]$ and $\mathcal{D}$ are Markov processes.
\end{prop}
\begin{proof}
	We need to the show that the process $\{L_t\}\in\mathcal{C}[0,T]$ is a Markov process when we know that $L_T=x$, for some constant $x$
	such that $0<f_T(x)<\infty$.
	(It will be explained later why the condition that $0<f_T(x)<\infty$ is required to ensure that the law of the bridge process is well defined.)
	In other words, we need to show that
	\begin{equation} 
		\Q\left[L_{t} \leq y \,|\, L_{t_1}=x_1,\ldots,L_{t_m}=x_m,L_T=x\right]
					= \Q\left[L_{t} \leq y \,|\, L_{t_m}=x_m,L_T=x\right],
	\end{equation}
	for all $m \in \mathbb{N}_+$, all $(x_1,\ldots,x_m,y)\in\R^{m+1}$, and all $0\leq t_1<\cdots<t_m<t\leq T$.
	It is crucial to the proof that $\{L_t\}$ has independent increments.
	Let us write
	\begin{align}
		\Delta_i&=L_{t_i}-L_{t_{i-1}},
		\\\delta_i&=x_i-x_{i-1},
	\end{align}
	for $1\leq i\leq m$, where $t_0=0$ and $x_0=0$.
	Then we have:
	\begin{align}
		&\Q\left[\left. L_{t} \leq y \,\right| L_{t_1}=x_1,\ldots,L_{t_m}=x_m,L_T=x\right] \nonumber
		\\&\quad=\Q\left[\left. L_{t}-L_{t_m} \leq y-x_m \,\right| \Delta_1=\delta_1,\ldots,\Delta_m=\delta_m,L_T-L_{t_m}=x-x_m\right] \nonumber
		\\&\quad=\Q\left[\left. L_{t}-L_{t_m} \leq y-x_m \,\right| L_T-L_{t_m}=x-x_m\right] \nonumber
		\\&\quad=\Q\left[\left. L_{t}-L_{t_m} \leq y-x_m \,\right| L_T-L_{t_m}=x-x_m,L_{t_m}=x_m\right] \nonumber
		\\&\quad=\Q\left[\left. L_{t} \leq y \,\right| L_T=x,L_{t_m}=x_m\right].
	\end{align}
	The proof for processes in class $\mathcal{D}$ is similar.
\end{proof}

Let $\{L_t\}\in\mathcal{C}[0,T]$, and let $\{L^{(z)}_{tT}\}_{0 \leq t\leq T}$ be an $\{L_t\}$-bridge to the value $z\in\R$ at time $T$.
For the transition probabilities of the bridge process to be well defined, we require that $0<f_T(z)<\infty$.
By the Bayes theorem we have
\begin{align}
	\Q\left[L^{(z)}_{tT}\in \dd y \left|\, L^{(z)}_{sT}=x \right.\right]&=
				\Q\left[L_{t}\in \dd y \left|\, L_{s}=x, L_T=z \right.\right] \nonumber
	\\ &=\frac{\Q\left[L_{t}\in \dd y, L_T\in \dd z \left|\, L_{s}=x \right.\right]}{\Q\left[L_T\in\dd z \left|\, L_{s}=x \right.\right]} \nonumber
	\\ &=\frac{f_{t-s}(y-x)f_{T-t}(z-y)}{f_{T-s}(z-x)} \d y, \label{eq:br_den}
\end{align}
for $0\leq s<t<T$.
We define the marginal bridge density $f_{tT}(y;z)$ by
\begin{equation}
	\label{eq:def_ftT}
	f_{tT}(y;z)=\frac{f_t(y)f_{T-t}(z-y)}{f_T(z)}.
\end{equation}
In this way
\begin{equation}
	\label{eq:LB_trans}
	\Q\left[L^{(z)}_{tT}\in \dd y \left|\, L^{(z)}_{sT}=x \right.\right]=f_{t-s,T-s}(y-x;z-x) \d y.
\end{equation}
The condition $0<f_T(z)<\infty$ is enough to ensure that
\begin{equation}
	y\mapsto f_{t-s,T-s}(y-L^{(z)}_{sT};z-L^{(z)}_{sT})
\end{equation}
is a well-defined density for almost every value of $L^{(z)}_{sT}$.
To see this, note that
\begin{align}
	& \int_{-\infty}^{\infty}\int_{-\infty}^{\infty} f_{t-s,T-s}(y-x;z-x) \, \Q\left[L^{(z)}_{sT}\in\dd x\right]  \dd y \nonumber
	\\ &\qquad=\int_{-\infty}^{\infty}\int_{-\infty}^{\infty} f_{t-s,T-s}(y-x;z-x) f_{s,T}(x;z) \d x  \d y \nonumber
	\\ &\qquad=\int_{-\infty}^{\infty}\frac{f_{T-t}(z-y)}{f_T(z)}\int_{-\infty}^{\infty} f_{t-s}(y-x) f_{s}(x) \d x  \d y \nonumber
	\\ &\qquad=\frac{1}{f_T(z)} \int_{-\infty}^{\infty} f_{T-t}(z-y) f_t(y) \d y=1. \label{eq:ImADensity}
\end{align}
From (\ref{eq:ImADensity}) it follows that
\begin{equation}
	\label{eq:check_den}
	\Q\left[ \int_{-\infty}^{\infty} f_{t-s,T-s}(y-L^{(z)}_{sT};z-L^{(z)}_{sT}) \d y=1\right]=1.
\end{equation}

Let $\{M_t\}\in\mathcal{D}$, and let $\{M^{(k)}_{tT}\}_{0 \leq t\leq T}$ be an $\{M_t\}$-bridge to the value $a_k$ at time $T$,
so $\Q[M^{(k)}_{TT}=a_k]=1$.
For the transition probabilities of the bridge to be well defined, we require that $\Q[M_T=a_k]=Q_T(a_k)>0$.
Then the Bayes theorem gives
\begin{align}
	\Q\left[M^{(k)}_{tT}=a_j \left|\, M^{(k)}_{sT}=a_i \right.\right]&=
				\Q\left[M_{t}=a_j \left|\, M_{s}=a_i, M_T=a_k \right.\right] \nonumber
	\\ &=\frac{\Q\left[M_{t}=a_j, M_T=a_k \left|\, M_{s}=a_i \right.\right]}{\Q\left[M_T=a_k \left|\, M_{s}=a_i \right.\right]} \nonumber
	\\ &=\frac{Q_{t-s}(a_j-a_i)Q_{T-t}(a_k-a_j)}{Q_{T-s}(a_k-a_i)}, \label{eq:ratio}
\end{align}
where $s,t$ satisfy $0\leq s<t<T$.
Note that if $Q_T(a_k)=0$, then the ratio (\ref{eq:ratio}) is not well defined when $s=0$.

We provide sufficient conditions for the integrability of \levy bridges:
\begin{prop}
	If there exists a constant $C<\infty$ such that $|x|^{1+\b} f_t(x)$ is bounded for $|x|\geq C$ and all $t\in(0,T]$, then
	\begin{equation}
		\int_{-\infty}^{\infty} |x|^{1+2\a} \, f_{tT}(x;z) \d x<\infty,
	\end{equation}
	for every $\a\in(0,\b)$.
	In other words, 
	\begin{equation}
			\E\left[\left|L_{tT}^{(z)}\right|^{1+2\a}\right]<\infty \qquad (0\leq t \leq T).
	\end{equation}
	Similarly, if there exists a constant $C$ such that $|a_i|^{1+\b} Q_t(a_i)$ is bounded for $|i|\geq C$ and all $t\in(0,T]$, then
	\begin{equation}
			\E\left[\left|M_{tT}^{(z)}\right|^{1+2\a}\right]<\infty \qquad (0\leq t \leq T).
	\end{equation}
\end{prop}
\begin{proof}
	We prove the proposition in the continuous case. 
	The discrete case is similar.
	The cases $t=0$ and $t=T$ are trivial, so we will assume that $t\in(0,T)$.
	
	Fix $\a\in(0,\b)$ and assume that
	\begin{equation}
		\label{eq:limit}
		|x|^{1+\b} f_t(x)<K<\infty \qquad \text{for $|x|\geq C$ and all $t\in(0,T]$.}
	\end{equation}
	First we prove that $\int |x|^{\a} f_t(x) \d x<\infty$ for $t\in(0,T)$.
	We have
	\begin{align}
		\int_{-\infty}^{\infty} |x|^{\a} f_t(x) \d x &= \int_{-C}^{C} |x|^{\a} f_t(x) \d x+\int_{(-\infty,-C]\cup [C,\infty)} |x|^{\a} f_t(x) \d x \nonumber
		\\ &< 2 C^{\a}+2K \int_C^{\infty} \frac{x^{\a}}{x^{1+\b}} \d x \nonumber
		\\ &=2 C^{\a}+\frac{2K}{\b-\a}C^{\a-\b} <\infty.
	\end{align}
	For $y\in\R$, we can generalise this to
	\begin{align}
		\int_{-\infty}^{\infty} |x+y|^{\a} f_t(x) \d x &\leq \int_{-\infty}^{\infty} (|x|+|y|)^{\a} f_t(x) \d x  \nonumber
		\\ &=\int_{-y}^{y} (|x|+|y|)^{\a} f_t(x) \d x +\int_{(-\infty,-y]\cup[y,\infty)} (|x|+|y|)^{\a} f_t(x) \d x  \nonumber
		\\ &\leq 2^{\a}|y|^{\a}+2^{\a}\int_{-\infty}^{\infty} |x|^{\a} f_t(x) \d x <\infty.
	\end{align}
	Finally, we have
	\begin{align}
		\int_{-\infty}^{\infty} |x|^{1+2\a} f_{tT}(x;z) \d x&=\int_{-C}^{C} |x|^{1+2\a} f_{tT}(x;z) \d x
				+\int_{(-\infty,-C]\cup [C,\infty)} |x|^{1+2\a} f_{tT}(x;z) \d x \nonumber
		\\ &< 2C^{1+2\a}+\frac{K}{f_T(z)} \int_{-\infty}^{\infty} |x|^{\a} f_{T-t}(z-x) \d x \nonumber
		\\ &= 2C^{1+2\a}+\frac{K}{f_T(z)} \int_{-\infty}^{\infty} |y-z|^{\a} f_{T-t}(y) \d y<\infty.
	\end{align}
\end{proof}

\section{Stable bridges}
Bridges of stable processes inherit a scaling property:
\begin{prop}
	\label{prop:StableBridge}
	Let $\{S_t\}$ be a stable process with index $\a$, and let $k>0$ be a constant. 
	If $\{S_{tT}^{(z)}\}$ is a bridge of $\{S_t\}$ to the value $z$ at time $T$, and $\{S_{t,kT}^{(\zeta)}\}$ is a bridge of $\{S_t\}$ to the value $\zeta=k^{1/\a}z$ at time $kT$, then
	\begin{equation}
		\{S^{(z)}_{tT}\} \law \{S^{(\zeta)}_{kt,kT}\}. 
	\end{equation}
\end{prop}
\begin{proof}
	Denote the density of $S_t$ by $f_t(x)$.
	From the scaling property of stable processes, we have
	\begin{equation}
		S_t \law k^{-1/\a} S_{kt}.
	\end{equation}
	It follows that
	\begin{equation}
		f_t(x)=k^{1/\a} f_{kt}\left( k^{1/\a} x\right).
	\end{equation}
	From (\ref{eq:br_den}) we have
	\begin{align}
		\Q\left[S^{(z)}_{tT}\in \dd y \left|\, S^{(z)}_{sT}=x \right.  \right]
					&=\frac{f_{t-s}(y-x)f_{T-t}(z-y)}{f_{T-s}(z-x)}\d y \nonumber
			\\ 	&=k^{1/\a} \frac{f_{kt-ks}\left(k^{1/\a}y-k^{1/\a}x\right)f_{kT-kt}\left(\zeta-k^{1/\a}y\right)}
									{f_{kT-ks}\left(\zeta-k^{1/\a}x\right)}\d y \nonumber
			\\	&=\Q\left[k^{-1/\a} S^{(\zeta)}_{kt,kT}\in \dd y \left|\, k^{-1/\a} S^{(\zeta)}_{ks,kT}=x \right. \right].
	\end{align}
\end{proof}

\section{Brownian motion and Brownian bridge}
\subsection{Brownian motion}
Brownian motion is a \levy process, and is a Gaussian process (i.e.~all of its finite-dimensional distributions are multivariate normal).
Gaussian processes are characterised by their mean and covariance functions.
In the case of a one-dimensional Brownian motion $\{B_t\}$, these are
\begin{align}
	&\E[B_{t}]=\th t, &\cov[B_s,B_t]=\s^2 \min(s,t),
\end{align}
where $\th\in\R$ is the drift of $\{B_t\}$, and $\s>0$ is the diffusion coefficient.
The characteristic exponent of $\{B_t\}$ is
\begin{equation}
	\Psi(\l)=-\i \th \l +\half \s^2\l^2.
\end{equation}
The density of $B_t$ is
\begin{equation}
	f_t(x)=\frac{1}{\s \sqrt{2\pi t}} \exp\left(-\half \frac{(x-\th t)^2}{\s^2 t}  \right).
\end{equation}
The sample paths of Brownian motion are continuous but nowhere differentiable.
When $\th=0$, $\{B_t\}$ is a stable process with index $\a=2$, and satisfies the scaling identity
\begin{equation}
	\{k^{-1/2} B_{kt}\} \law \{B_t\} \qquad \text{for $k>0$.}
\end{equation}
When $\th=0$ and $\s=1$ we say that $\{B_t\}$ is a \emph{standard} Brownian motion (or Weiner process).

\subsection{Brownian bridge}
A Brownian bridge is also a Gaussian process.
Let $\{\b^{(z)}_{tT}\}_{0\leq t\leq T}$ be a standard Brownian bridge to the point $z\in\R$.
The mean and covariance functions of $\{\b^{(z)}_{tT}\}$ are
\begin{align}
	&\E[\b^{(z)}_{tT}]=\frac{t}{T}z, &\cov[\b^{(z)}_{sT},\b^{(z)}_{tT}]=\min(s,t)-\frac{st}{T}.
\end{align}
It follows that
\begin{equation}
	\left\{ k^{-1/2} \b_{kt,kT}^{(k^{1/2}z)} \right\}\law \{\b^{(z)}_{tT}\} \qquad \text{for $k>0$.}
\end{equation}
Let $\{W_t\}$ be a standard Brownian motion, and define the process $\{W^{(z)}_{tT}\}$ by
\begin{equation}
	W^{(z)}_{tT}= W_t+\frac{t}{T}(z-W_T) \qquad(0\leq t\leq T). \label{eq:BBBM}
\end{equation}
Calculating the mean and covariance functions for this process verifies that it is a standard Brownian bridge to the value $z$ at time $T$.
It is also notable and easily verified that $\{W^{(z)}_{tT}\}$ is independent of $W_T$.

\section{Gamma process and gamma bridge} \label{sec:gamma}
\subsection{Gamma process} \label{subsec:gamma_process}
A gamma process is a subordinator with gamma-distributed increments.
The law of a gamma process is uniquely determined by its mean and variance at time 1.
Both of these quantities are positive.
Let $\{\g_t\}$ be a gamma process with mean 1 and variance $m^{-1}>0$ at time 1; then
\begin{align}
	\label{eq:gamma_choice}
	&\E[\g_t]=t, &\var[\g_t]=t/m.
\end{align}
The density of $\g_t$ is
\begin{equation}
	\label{eq:gamma_m_den}
	g_t^{(m)}(x)=  \1_{\{x>0\}} \frac{ m^{mt}}{\G[mt]} x^{mt-1} \e^{-mx},
\end{equation}
where $\G[z]$ is the gamma function, defined as usual for $x>0$ by
\begin{equation}
	\G[x]=\int_0^\infty u^{x-1} \e^{-u} \d u.
\end{equation}
Hence $\{\g_t\}\in\mathcal{C}[0,T]$ for any $T>0$.
Due to the scaling property of the gamma distribution, if $\kappa>0$ then the process $\{\kappa \g_t\}$ is a gamma process with mean $\kappa$, and variance $m^{-1}\kappa^2$ at $t=1$.
The characteristic exponent of $\{\g_t\}$ is
\begin{equation}
	\Psi(\l)=m \log[1-\i\l /m],
\end{equation}
and so the characteristic function of $\g_t$ is
\begin{equation}
	\E[\e^{\i\l\g_t}]=(1-\i\l/m)^{-mt}.
\end{equation}
In the limit $m\rightarrow \infty$ this characteristic function is $\e^{\i\l t}$, which is the characteristic function of the Dirac measure centred at $t$.
It follows that $\{\g_t\}\overset{\text{law}}{\longrightarrow} \{t\}$ as $m\rightarrow \infty$.

It should be noted that we find it convenient here to use a somewhat different parametrisation scheme for the family of gamma processes from that presented in \citet{BHM3}.
In their scheme the `basic' gamma process $\{\g_t\}$ is characterised by a single parameter $m$, with units of inverse time,  such that $\E[\g_t]=mt$ and $\var[\g_t]=mt$.
The `general' gamma process is then obtained by considering a `scaled' process $\{\k \g_t\}$ where $\k>0$ is a constant.
Clearly $\E[\k \g_t]=\k m t$ and $\var[\k \g_t]=\k^2 m t$.
Thus, if the mean rate $\mu$ and variance rate $\s^2$ of a gamma process is specified, then we have $m=\mu^2/\s^2$ and $\k=\s^2/\mu$.
The scheme introduced in (\ref{eq:gamma_choice}) above is equivalent to the choice $\k=m^{-1}$ in the BHM scheme.
The advantage of the choice (\ref{eq:gamma_choice}) for present purposes as the basic process is that it gives $\{\g_t\}$ the dimensionality of time, and hence makes it suitable as a basis for a time change.

\subsection{Gamma bridge}
Gamma bridges exhibit a number of remarkable similarities to Brownian bridges, some of which have been presented by \citet{EY2004}.
Let $\{\gb{t}\}_{0\leq t\leq T}$ be a gamma bridge with final value 1 associated with the gamma process $\{\g_t\}$.
The transition law of $\{\gb{t}\}$ is given by
\begin{align}
	\Q\left[\gb{t}\in \dd y \left|\, \gb{s}=x\right.\right]&=\Q\left[\g_t\in \dd y \left|\, \g_s=x, \g_T=1 \right.\right] \nonumber
		\\ &=\frac{g^{(m)}_{t-s}(y-x)g^{(m)}_{T-t}(1-y)}{g^{(m)}_{T-s}(1-x)} \nonumber
		\\ &=\1_{\{x<y< 1\}}\frac{\left(\frac{y-x}{1-x}\right)^{m(t-s)-1} \left(\frac{1-y}{1-x}\right)^{m(T-t)-1}}
					{(1-x)\mathrm{B}[m(t-s),m(T-t)]} \d y, \label{eq:gambrtrans}
\end{align}
for $0\leq s <t\leq T$ and $x\geq 0$.
Here $\mathrm{B}[\a,\b]$ is the beta function, defined for $\a,\b>0$ by
\begin{equation}
	\mathrm{B}[\a,\b]=\int_0^1 x^{\a-1}(1-x)^{\b-1}\d x=\frac{\G[\a]\G[\b]}{\G[\a+\b]}.
\end{equation}
If the gamma bridge $\{\g_{tT}\}$ has reached the value $x$ at time $s$, then it must yet travel a distance $1-x$ over the time period $(s,T]$.
Equation (\ref{eq:gambrtrans}) shows that the proportion of this distance that the gamma bridge will cover over $(s,t]$ is a random variable with a beta distribution
(with parameters $\a=m(t-s)$ and $\b=m(T-t)$).
The conditional characteristic function of $\g_{tT}$ is
\begin{equation}
	\label{eq:gambrcf}
	\E\left[\left. \e^{\i\l\gb{t}} \,\right| \gb{s}=x \right]=M[m(t-s),m(T-s),\i(1-x)\l],
\end{equation}
where $M[\a,\b,z]$ is Kummer's confluent hypergeometric function of the first kind, which can be expanded as the power series \citep[13.1.2]{AS1964}
\begin{equation}
\label{eq:Kummer}
	M[\a,\b,z]=1+\frac{\a}{\b}z+\frac{\a(\a+1)}{\b(\b+1)}\frac{z}{2!}+\frac{\a(\a+1)(\a+2)}{\b(\b+1)(\b+2)}\frac{z}{3!}+\cdots.
\end{equation}
For brevity, we will later refer to (\ref{eq:Kummer}) as `Kummer's function $M[\a,\b,z]$'.
Taking the limit as $m\rightarrow\infty$ in (\ref{eq:gambrcf}), we have
\begin{align}
	\E\left[\left. \e^{\i\l\gb{t}} \,\right| \gb{s} =x\right]&\rightarrow \sum_{k=0}^{\infty}\left(\frac{t-s}{T-s} \right)^k \frac{(\i(1-x)\l)^k}{k!} \nonumber
		\\&=\exp\left( \i\frac{t-s}{T-s}(1-x)\l \right),
\end{align}
which is the characteristic function of the Dirac measure centred at $(1-x)(t-s)/(T-s)$.
It then follows from the Markov property of gamma bridges that $\{\gb{t}\}\overset{\text{law}}{\longrightarrow}\{t/T\}$ as $m\rightarrow \infty$.
It is a property of gamma processes that the renormalised process $\{\g_t/\g_T\}_{0\leq t \leq T}$ is independent of $\g_T$
(indeed, this independence property characterises the gamma process among \levy processes).
This leads to the remarkable identity
\begin{equation}
	\label{eq:gamma_ratio}
	\left\{ \frac{\g_t}{\g_T} \right\}\law \{\gb{t}\}.
\end{equation}
The identity (\ref{eq:gamma_ratio}) can be proved by showing that the process on the left-hand side is Markov, and then verifying that its transition law is the same as (\ref{eq:gambrtrans}), which can be done using the results in \citet{BHM3}.
Two further properties of gamma bridges follow immediately from (\ref{eq:gamma_ratio}).
The first is that the bridge of the scaled gamma process $\{\kappa \g_t\}$ is, for any $\k>0$, identical in law to the bridge of the unscaled process.
Hence, for fixed $m$ and fixed terminal value, the bridge of the gamma process defined by (\ref{eq:gamma_choice}) is identical to the bridge of the basic BHM gamma process.
The second property is that a $\{\g_t\}$-bridge to the value $z>0$ at time $T$ is identical in law to the process $\{z \gb{t}\}$. 

\section{Variance-gamma process and variance-gamma bridge} \label{sec:VG}

\subsection{Variance-gamma process}
A variance-gamma (VG) process is a Brownian motion with drift, subordinated by an independent gamma process.
Letting $\{W_t\}$ denote a standard Brownian motion, we define the process $\{V_t\}$ by
\begin{equation}
	V_t=\s W_{\g_t} + \th \g_t \qquad \text{$\s>0$ and $\th \in\R$.}
\end{equation}
Then $\{V_t\}$ is a VG process in its most general form (on the real line).
The mean of the gamma process $\{\g_t\}$ at $t=1$ was fixed as unity, but (in terms of the law of the VG process) varying this is equivalent to an appropriate change of the parameters $\s$ and $\th$.
When $\th=0$ we say that $\{V_t\}$ is a symmetric VG process; if, in addition, $\s=1$ then we say that $\{V_t\}$ is a standard VG process.
The characteristic exponent of $\{V_t\}$ is given by
\begin{equation}
	\Psi(\l)=m \log\left[1-\frac{\i\th\l}{m}+\frac{\s^2\l^2}{2m}\right],
\end{equation}
where $m^{-1}$ is the variance of $\g_1$.
This can be decomposed as
\begin{equation}
	\label{eq:decomp_VG_CF}
	\Psi(\l)=m \log\left[1-\frac{\i\mu_+\l}{m}\right]+m \log\left[1+\frac{\i\mu_-\l}{m}\right],
\end{equation}
where
\begin{align}
	\mu_+&=\half\left(\sqrt{\th^2+2m\s^2}+\th\right),
	\\\intertext{and} \mu_-&=\half\left(\sqrt{\th^2+2m\s^2}-\th\right).
\end{align}
The right-hand side of (\ref{eq:decomp_VG_CF}) is the sum of two characteristic exponents.
The first corresponds to a gamma process, and the second corresponds to a decreasing \levy process whose absolute value is a gamma process.
It follows that, for $\{\g_t^{(+)}\}$ and $\{\g_t^{(-)}\}$ independent copies of $\{\g_t\}$, we have
\begin{equation}
	\{V_t\} \law \left\{\mu_+ \g_t^{(+)}-\mu_- \g_t^{(-)}\right\}.
	\label{eq:VGis2Gamma}
\end{equation}
The characteristic function of $V_t$ is
\begin{equation}
	\label{eq:VGCF}
	\E[\e^{\i\l V_t}]=\left(1-\frac{\i\th\l}{m}+\frac{\s^2\l^2}{2m}\right)^{-mt}.
\end{equation}
In the limit as $m\rightarrow\infty$, this characteristic function tends to
\begin{equation}
	\exp\left(\i\th \l t-\tfrac{1}{2} \s^2\l^2 t\right).
\end{equation}
It follows that
\begin{align}
	\{V_t\}\overset{\text{law}}{\longrightarrow}\{\s W_t+\th t\} &&\text{as $m\rightarrow \infty$.}
\end{align}

The distribution of $V_t$ is normal with a gamma-mixed mean and variance.
The density of $V_t$ can be shown to be \citep{MCC1998}
\begin{equation}
	f^{(m,\th,\s)}_t(x)=\sqrt{\frac{2}{\pi}}\frac{m^{mt}\e^{\th x/\s^2}}{\s \G[mt]}\left(\frac{x^2}{2m\s^2+\th^2} \right)^{\frac{mt}{2}-\frac{1}{4}} K_{mt-\half}\left[ \s^{-2}\sqrt{x^2(2m\s^2+\th^2)} \right],
\end{equation}
where $K_{\nu}[z]$ is the modified Bessel function of the third kind.
We see that $\{V_t\}\in\mathcal{C}[0,T]$ for all $T>0$.
Each of the following facts about $K_{\nu}[z]$ can be found in \citet[9.6]{AS1964}:
\begin{align}
	&\text{1.} &&0<K_{\nu}(z)<\infty && \text{for $\nu\in\R$ and $z>0$,}
	\\&\text{2.} &&K_{-\nu}[z]=K_{\nu}[z],
	\\&\text{3.} &&\lim_{z\rightarrow 0} \frac{K_0[z]}{-\log z}=1,
	\\&\text{4.} &&\lim_{z\rightarrow 0} \frac{K_{\nu}[z]}{\tfrac{1}{2} \G[\nu](\tfrac{1}{2} z)^{-\nu}}=1 && \text{for $\nu>0$,}
	\\&\text{5.} &&\lim_{z\rightarrow \infty} \frac{K_{\nu}[z]}{\sqrt{\frac{\pi}{2z}}\e^{-z}}=1. \label{eq:K_inf}
\end{align}
In the case when $\nu$ is half an odd number we have following expression for $K_{\nu}[z]$ from \citep[10.2.15]{AS1964}:
\begin{equation}
	\label{eq:BesselHalfInt}
	K_{ n+\frac{1}{2}}[z]=\sqrt{\frac{\pi}{2z}}\e^{-z} \sum_{j=0}^{n} (n+\tfrac{1}{2},j) (2z)^{-j}, \quad\text{for $n\in\mathbb{N}$,}
\end{equation}
where $(m,n)$ is Hankel's symbol,
\begin{equation}
	(m,n)=\frac{\G[m+\tfrac{1}{2}+n]}{n!\, \G[m+\tfrac{1}{2}-n]}.
\end{equation}
Writing
\begin{align}
	f_t^{(m)}(x)
	&=f^{(m,0,1)}_t(x) \nonumber
	\\ &=\sqrt{\frac{2}{\pi}}\frac{m^{mt}}{\G[mt]}\left(\frac{x^2}{2m} \right)^{\frac{mt}{2}-\frac{1}{4}} 
			K_{mt-\half}\left[ \sqrt{2mx^2} \right], \label{eq:VGden}
\end{align}
we have
\begin{equation}
	f^{(m,\th,\s)}_t(x)=\frac{1}{\s}\e^{\th x/\s^2}\left(k_{(m,\th,\s)}\right)^{1-2mt} f_t^{(m)}\left( k_{(m,\th,\s)}x /\s\right),
	\label{eq:AVGden}
\end{equation}
where
\begin{equation}
	\label{eq:VG_k}
	k_{(m,\th,\s)}=\sqrt{1+\frac{\th^2}{2m\s^2}}.
\end{equation}
Here $f_t^{(m)}(x)$ is the density of the standard VG random variable $W(\g_t)$.
Since $K_{\nu}[z]$ is finite for $z>0$, $f^{(m,\th,\s)}_t(x)$ is finite away from zero.
In other words, $0<f^{(m,\th,\s)}_t(x)<\infty$ for all $t>0$ and all $x\in\R \backslash \{0\}$.
When $x=0$, we have
\begin{equation}
	f^{(m,\th,\s)}_t(0)=\left\{
	\begin{aligned}
		&\frac{1}{\s}\sqrt{\frac{m}{2\pi}}\left(1+\frac{\th^2}{2m\s^2} \right)^{\half-mt} \frac{\G[mt-1/2]}{\G[mt]} &&\text{for $t>(2m)^{-1}$,}
		\\ &+\infty &&\text{for $0<t\leq (2m)^{-1}$.}
	\end{aligned}
	\right. \label{eq:VGdenat0}
\end{equation}

\subsection{Variance gamma bridge}

Let $\{V_t\}$ be a VG process with parameter set $\{m,\th,\s\}$;
and let $\{V_{tT}^{(z)}\}$ be a $\{V_t\}$-bridge to the value $z\in\R$ at time $T$.
Since we must have that the density of $V_T$ is positive and finite at $z$, it follows from (\ref{eq:VGdenat0}) that
either $z \neq 0$, or $T>(2m)^{-1}$.
The transition law of $\{V^{(z)}_{tT}\}$ is given by
\begin{align}
	\Q\left[\left. V^{(z)}_{tT}\in\dd y \,\right| V^{(z)}_{sT}=x \right]	
	 &=\frac{f^{(m,\th,\s)}_{t-s}(y-x)f^{(m,\th,\s)}_{T-t}(z-y)}{f^{(m,\th,\s)}_{T-s}(z-x)} \d y \nonumber
	\\ &=\frac{k}{\s}\frac{f^{(m)}_{t-s}\left(\frac{k}{\s}(y-x)\right)f^{(m)}_{T-t}\left(\frac{k}{\s}(z-y)\right)}
			{f^{(m)}_{T-s}\left(\frac{k}{\s}(z-x)\right)} \d y \nonumber
	\\ &= \Q\left[\left. \frac{\s}{k} U^{(kz/\s)}_{tT}\in\dd y \,\right| \frac{\s}{k} U^{(kz/\s)}_{sT}=x \right],
\end{align}
for $0\leq s<t<T$, $k=k_{(m,\th,\s)}$, and $\{U^{(kz/\s)}_{tT}\}$ is a $\{W(\g_t)\}$-bridge to $kz/\s$ at time $T$.
This shows that a bridge corresponding to any VG process is identical in law to a scaled bridge of a standard VG process.
We focus now on bridges of the standard VG process.
It follows from (\ref{eq:BBBM}) that
\begin{equation}
	\{W_t\}_{0\leq t \leq 1}\law \{\b_{t}+t W_1\}_{0\leq t \leq 1}, \label{eq:BBBM2}
\end{equation}
where $\{\b_t\}$ is a Brownian bridge, independent of $W_1$, ending at the value 0 at time $t=1$.
Also, from the time-scaling property of Brownian motion,
\begin{equation}
	\{W(Xt)\}\law \{\sqrt{X} \, W(t)\}, \label{eq:scaling}
\end{equation}
for $X$ a positive random variable independent of $\{W_t\}$.
We can use these identities, with some properties of gamma bridges, to derive a terminal-value decomposition of a standard VG process:
For $t\in[0,T]$, we have
\begin{align}
	\{W(\g_t)\} &=\left\{W\left(\g_T \frac{\g_t}{\g_T}  \right) \right\}  \nonumber
	\\ &\law \left\{W(\g_T \gb{t} ) \right\} \label{eq:decomp2}
	\\ &\law \left\{\sqrt{\g_T} \, W(\gb{t}) \right\} \label{eq:decomp3}
	\\ &\law \left\{\gb{t}\sqrt{\g_T} \, W(1)+\sqrt{\g_T} \, \b(\gb{t}) \right\} \label{eq:decomp4}
	\\ &\law \left\{\gb{t} W(\g_T) + \sqrt{\g_T} \, \b(\gb{t}) \right\}. \label{eq:decomp5}
\end{align}
In the above, (\ref{eq:decomp2}) holds since $\{W(t)\}$, $\{\g_t/\g_T\}$, and $\g_T$ are independent;
(\ref{eq:decomp3}) follows from (\ref{eq:scaling});
(\ref{eq:decomp4}) follows from (\ref{eq:BBBM2});
(\ref{eq:decomp5}) also follows from (\ref{eq:scaling}).
Note that in (\ref{eq:decomp5}) the Brownian bridge $\{\b(t)\}_{0\leq t\leq 1}$, the gamma bridge $\{\gb{t}\}_{0\leq t \leq T}$,
and the random vector $(\g_T,W(\g_T))$ are independent.
The joint density of $(\g_T,W(\g_T))$ is
\begin{equation}
	(y,z)\mapsto\frac{1}{\sqrt{2\pi y}}\exp\left(-\half \frac{z^2}{y}\right) g^{(m)}_T(y).
\end{equation}
Hence, given $W(\g_T)=z$, the conditional density of $\g_T$ is
\begin{equation}
	y\mapsto \frac{1}{f^{(m)}_T(z) \sqrt{2\pi y}}\exp\left(-\half \frac{z^2}{y}\right) g^{(m)}_T(y).
\end{equation}
Some simplification shows that this is the generalized inverse-Gaussian (GIG) density $f_{\textit{GIG}}(y;mT-1/2,|z|,\sqrt{2m})$, where
\begin{equation}
	 f_{\textit{GIG}}(x;\l,\delta,\g)= 
	 \1_{\{x>0\}} \left( \frac{\g}{\delta}\right)^{\l} \frac{1}{2\,K_{\l}[\g\delta]}x^{\l-1}\exp\left(-\tfrac{1}{2}(\delta^2x^{-1}+\g^2x) \right).
\end{equation}
It follows from (\ref{eq:decomp5}) that the standard VG bridge satisfies
\begin{equation}
	\label{eq:VGdecompI}
	\{U^{(z)}_{tT}\} \law \left\{ z \gb{t} +\sqrt{\Sigma_z} \, \b(\gb{t}) \right\},
\end{equation}
where $\Sigma_z$ has a GIG distribution with parameter set $\{mT-1/2,|z|,\sqrt{2m}\}$.

From (\ref{eq:VGis2Gamma}) we have that 
\begin{equation}
	\{W(\g_t)\} \law \left\{ \mu \left( \bar{\g}_{t}-\g_{t} \right) \right\},
\end{equation}
where $\{\g_t\}$ and $\{\bar{\g}_t\}$ are independent, identical gamma processes with parameter $m$, and $\mu=(m/2)^{-1/2}$.
We can use this to derive an alternative representation of the standard VG bridge.
Define the process $\{G_t\}$ by setting $G_t=\bar{\g}_{t}-\g_{t}$. 
Then we have
\begin{align}
	\{G_t\}&=\left\{ \bar{\g}_{t}-\g_{t} \right\}  \nonumber
	\\ &\law \left\{  \bar{\g}_T \gbb{t}-\g_{T}\gb{t} \right\} \nonumber
	\\ &= \left\{ G_T \gbb{t}+\g_T\left(\gbb{t}-\gb{t}  \right) \right\},
\end{align}
where $\{\gb{t}\}$ and $\{\gbb{t}\}$ are identical gamma bridges, independent of each other, and independent of the random vector $(\g_T,G_T)$.
The joint density of $(\g_T,G_T)$ is given by
\begin{equation}
	(y,z)\mapsto g^{(m)}_T(y)g^{(m)}_T(y+z).
\end{equation}
Given that $G_T=z$, the conditional density of $\g_T$ is
\begin{equation}
	y\mapsto \1_{\{y>(-z)^+\}}\frac{m^{2mt}}{\G[mT]^2 f_T^{(m)}(z)} (yz+y^2)^{mT-1} \e^{-m(z+2y)},
	\label{eq:condGam}
\end{equation}
where $x^+$ denotes the positive part of $x$, so $(-z)^+=\max(0,-z)$.
Then we have
\begin{equation}
	\{U^{(z)}_{tT}\}\law \left\{ z\gbb{t} +  Y_z \,\mu\left(\gbb{t}-\gb{t} \right) \right\},  \label{eq:VGdecompII}
\end{equation}
where $Y_z>0$ is a random variable with density (\ref{eq:condGam}).

\section{Stable-1/2 subordinator and stable-1/2 bridge} \label{sec:stable_half}

\subsection{Stable-1/2 subordinator}

Let $\{S_t\}$ be a \SHS.
The characteristic exponent of $\{S_t\}$ is
\begin{equation}
	\Psi(\l)=\frac{c}{\sqrt 2}|\l|^{1/2}(1-\i\, \mathrm{sign}(\l)),
\end{equation}
where $c>0$ is a parameter related to $\k$ in (\ref{eq:cess}) by $c=\k\sqrt{2}$.
Then $\{S_t\}$ satisfies the scaling property $\{k^{-2}S_{kt}\}\law \{S_t\}$, for $k>0$.
The random variable $S_t$ has a `\levy distribution' with density
\begin{equation} 
	\label{eq:StableDen}
	f_t(x)=\1_{\{x>0\}} \,\frac{ct}{\sqrt{2\pi}\,x^{3/2}} \, \exp\left(-\half \frac{c^2t^2}{x}  \right).
\end{equation}
We call $c$ the `activity parameter' of $\{S_t\}$.
The density (\ref{eq:StableDen}) is bounded for all $t>0$ and is strictly positive for all $x>0$, hence $\{S_t\}\in\mathcal{C}[0,T]$ for all $T>0$.
Integrating (\ref{eq:StableDen}) yields the distribution function
\begin{equation}
	\int_{0}^{x}f_t(y) \d y= 2\,\Phi\!\left[-ctx^{-1/2} \right],
\end{equation}
where $\Phi[x]$ is the standard normal distribution function.
The random variable $S_t$ has infinite mean, indeed $\E[S_t^p]<\infty$ if and only if $p<1/2$.
The density of $1/S_t$ is
\begin{equation} 
	x\mapsto \1_{\{x>0\}} \frac{ct}{\sqrt 2 \, \G[1/2]}x^{-1/2} \, \exp\left(-\half c^2t^2x  \right).
\end{equation}
Thus the increments of $\{S_t\}$ are distributed as reciprocals of gamma random variables.
For $\{W_t\}$ a standard Brownian motion, define the exceedence times $\{\tau_t\}_{t\geq 0}$ by 
\begin{equation}
	\tau_t=\inf\{s: W_s> ct\}.
\end{equation}
From \citet[X.7]{Feller2}, we then have
\begin{equation}
	\{S_t\}\law \{\tau_t\}.
\end{equation}

\subsection{Stable-1/2 bridge}
Fix $z>0$ and let $\{S_{tT}^{(z)}\}_{0\leq t\leq T}$ be a bridge of the process $\{S_t\}$ to the value $z$ and time $T$.
We call $\{S^{(z)}_{tT}\}$ a stable-1/2 bridge.
The density function of the random variable $S^{(z)}_{tT}$ is
\begin{align}
	f_{tT}(y;z)&=\frac{f_t(y)f_{T-t}(z-y)}{f_T(z)} \nonumber
		\\&=\1_{\{ 0<y\leq z\}}\frac{1}{\sqrt{{2\pi}}} \frac{ct(T-t)}{T}\frac{\exp\left( -\frac{1}{2}  \frac{c^2(Ty-tz)^2}{yz(z-y)}\right)}{\left(y-y^2/z\right)^{3/2}}.
		\label{eq:kernel}
\end{align}		
This density is bounded, and has bounded support, so 
\begin{equation}
	\E\left[\left(S^{(z)}_{tT}\right)^p\right]<\infty \qquad \text{for $p>0$.}
\end{equation}

\begin{rem}
	From Proposition \ref{prop:StableBridge}, stable-1/2 bridges satisfy the following scaling property: 
	For $k>0$ a constant, and $\{ S^{(z)}_{tT}\}$ a \SH bridge, we have
	\begin{equation}
			\left\{ S^{(z)}_{tT} \right\}_{0\leq t \leq T}\law\left\{k^{-2} S^{(k^2z)}_{kt,kT} \right\}_{0\leq t \leq T}.
	\end{equation}
\end{rem}

Integrating the density (\ref{eq:kernel}) yields the following (details of the proof can be found in Appendix \ref{apdx:SBDF}):
\begin{prop}
	\label{prop:DFIGLRB}
	For $y\in[0,z]$, the distribution function of the random variable $S^{(z)}_{tT}$ is given by
	\begin{equation}
		F_{tT}(y;z)=
			\Phi\left[\frac{c(Ty-tz)}{\sqrt{yz(z-y)}}\right]+\left(1-\frac{2t}{T} \right)\e^{2c^2t(T-t)/z}\,
				\Phi\left[\frac{c((2t-T)y-tz)}{\sqrt{yz(z-y)}}\right] \label{eq:DFIGLRB}.
	\end{equation}
\end{prop}

\begin{rem}
When $t=T/2$, the second term in the distribution function (\ref{eq:DFIGLRB}) vanishes.
The distribution function is then analytically invertible, and we obtain the identity
\begin{equation}
	\label{eq:GT2}
	S^{(z)}_{T/2,T}\law \half z\left(1+\frac{Z}{\sqrt{c^2T^2/z+Z^2}} \right),
\end{equation}
where $Z$ is a standard normal random variable.
\end{rem}

\begin{coro}
	\begin{equation}
			\{S^{(z)}_{tT}\} \overset{\text{law}}{\longrightarrow}\{\tfrac{t}{T} z\} \qquad \text{as $c\rightarrow \infty$}.
	\end{equation}
\end{coro}
\begin{proof}
	Fix $z>0$.
	It is sufficient to show that
	\begin{equation}
		\lim_{c\rightarrow\infty} F_{tT}(y;z)=\1_{\{Ty\geq tz\}} \qquad \text{for Lebesgue-a.e.~$y\in (0,z)$},
	\end{equation}
	since this is equivalent to
	\begin{equation}
		\lim_{c\rightarrow\infty} \Q\left[\left|S^{(z)}_{tT}-\tfrac{t}{T} z \right|<\varepsilon \right]=1 
																																			\qquad \text{for all $t\in [0,T]$ and any $\varepsilon>0$.}
	\end{equation}
	Define $\a$ by
	\begin{equation}
		\a=-\frac{(2t-T)y-tz}{\sqrt{yz(z-y)}},
	\end{equation}
	and note that $\a>0$ for $y\in(0,z)$.
	The inequality \citep[7.1.13]{AS1964} states
	\begin{equation}
		\e^{x^2} \int_{x}^{\infty} \e^{-t^2} \d t\leq \frac{1}{x+\sqrt{x^2+4/\pi}} \qquad (x>0),
	\end{equation}
	from which we deduce
	\begin{align}
		\label{eq:ineq}
		\e^{2c^2t(T-t)/z}\Phi[-\a c]&\leq \e^{2c^2t(T-t)/z}\sqrt{\frac{2}{\pi}} \frac{\e^{-\a^2c^2/2}}{\a c+\sqrt{\a^2c^2+2/\pi}} \nonumber
		\\ &=\sqrt{\frac{2}{\pi}} \frac{\exp\left(-c^2\frac{(Ty-tz)^2}{2y(z-y)z}\right)}{\a c+\sqrt{\a^2c^2+2/\pi}}.
	\end{align}
	Since the left-hand side of (\ref{eq:ineq}) is positive, we see that
	\begin{equation}
		\lim_{c\rightarrow\infty} \e^{2c^2t(T-t)/z}\Phi[-\a c]=0.
	\end{equation}
	Then we have
	\begin{align}
		\lim_{c\rightarrow\infty} F_{tT}(y;z)&=\lim_{c\rightarrow\infty} \Phi\left[\frac{c(Ty-tz)}{\sqrt{yz(z-y)}}\right] 
				+\left(1-\frac{2t}{T} \right)\lim_{c\rightarrow\infty} \e^{2c^2t(T-t)/z}\Phi[-\a c] \nonumber
				\\&= \1_{\{ Ty-tz\geq 0 \}}-\tfrac{1}{2} \1_{\{ Ty=tz \}},
	\end{align}
	which completes the proof.
\end{proof}

The proof of the following proposition can be found in Appendix \ref{apdx:SBmom}:
\begin{prop}
	\label{prop:mom}
	Define the incomplete first moment of $S^{(z)}_{tT}$ by
	\begin{equation}
		M_{tT}(y;z)=\int_{0}^y u \, f_{tT}(u;z) \d u \qquad (0\leq y\leq z).
	\end{equation}
	Then we have
	\begin{equation}
		M_{tT}(y;z)=\frac{t}{T}z \left\{ \Phi\left[\frac{c(Ty-tz)}{\sqrt{yz(z-y)}}\right]-\e^{2c^2t(T-t)/z}\,
					\Phi\left[\frac{c((2t-T)y-tz)}{\sqrt{yz(z-y)}}\right] \right\},	
	\end{equation}
	and the second moment of $S^{(z)}_{tT}$ is given by
	\begin{equation}
		\E\left[ \left(S^{(z)}_{tT}\right)^2 \right]=
		\frac{t}{T}z^2\left\{1-c(T-t)\e^{\frac{c^2T^2}{2z}}\sqrt{\frac{2\pi}{z}}\,\Phi\left[-cTz^{-1/2} \right]  \right\}.
	\end{equation}
\end{prop}

\begin{coro}
	\label{coro:mom}
	\begin{align}
		\E[S^{(z)}_{tT}]&=\frac{t}{T}z,
		\\ \var[S^{(z)}_{tT}] &=\frac{t(T-t)}{T^2}z^2 \left\{1-
			\e^{\frac{c^2T^2}{2z}}\sqrt{\frac{2\pi c^2T^2}{z}}\,\Phi\left[-\sqrt{\frac{c^2T^2}{z}} \right] \right\}.
	\end{align}
\end{coro}

\begin{rem}
In general, we have
	\begin{equation}
		\E\left[S_{tT}^{(z)}\left|\, S_{sT}^{(z)}=x \right.\right]=\frac{T-t}{T-s} x+\frac{t-s}{T-s}z,
	\end{equation}
	and
	\begin{multline}
		\E\left[\left.\left(S_{tT}^{(z)}\right)^2 \,\right| S_{sT}^{(z)}=x \right] 
		\\=\frac{t-s}{T-s}(z-x)^2
			\left\{1-c(T-t)\e^{\frac{c^2(T-s)^2}{2(z-x)}}\sqrt{\frac{2\pi}{(z-x)}}\,\Phi\left[-c\frac{T-s}{\sqrt{z-x}} \right]  \right\},
	\end{multline}
for $0\leq s <t <T$.
\end{rem}

\begin{figure}[ht]
	\begin{center}
		\subfigure[$c=1$.]{\includegraphics[scale=.9]{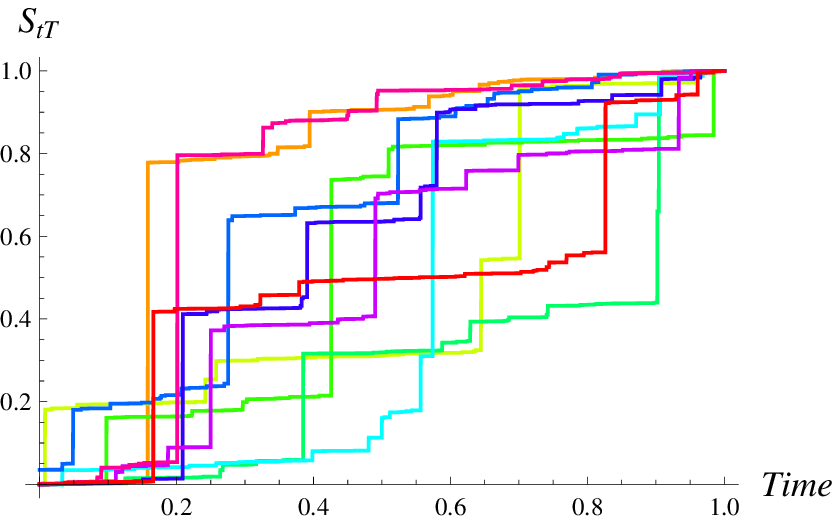}}
		\subfigure[$c=5$.]{\includegraphics[scale=.9]{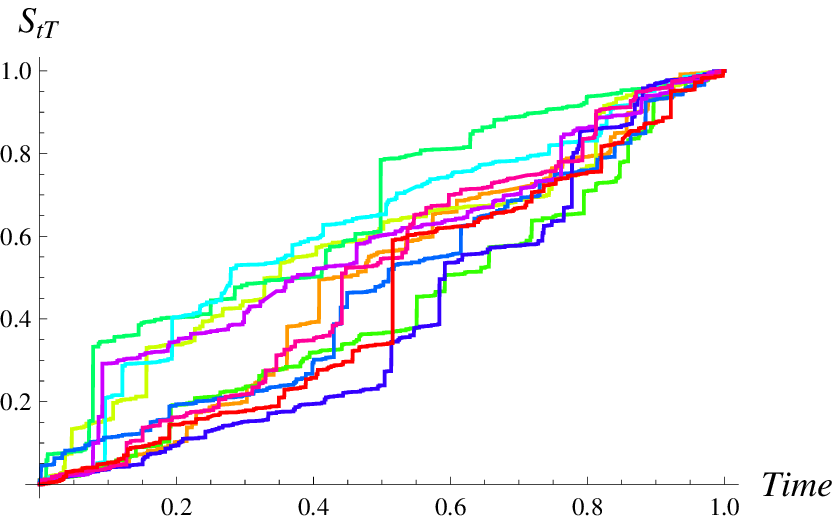}}
	\end{center}
	\caption[Stable-1/2 bridge simulations]{%
			Simulations of the stable-1/2 bridge demonstrating the influence of the activity parameter $c$.
			Qualitatively speaking, increasing the value of $c$ decreases the frequency of large jumps, and increases the frequency of small jumps.
			}
\end{figure}

\section{Cauchy process and Cauchy bridge} \label{sec:Cauchy}

\subsection{Cauchy process}

The Cauchy process is a stable process with index $\a=1$.
Let $\{Z_t\}$ be a driftless Cauchy process.
The characteristic exponent of $\{Z_t\}$ is
\begin{equation}
	\label{eq:CauchyCF}
	\Psi(\l)=c|\l| \qquad \text{for $\l\in\R$},
\end{equation}
where $c>0$ is a scale parameter.
The scaling property of $\{Z_t\}$ can be written
\begin{equation}
	\{k^{-1} Z_{kt}\} \law \{Z_t\} \qquad \text{for $k>0$.}
\end{equation}
The density function of $Z_t$ is
\begin{equation}
	\label{eq:CauchyDensity}
	f_t(y)=\frac{ct}{\pi(y^2+c^2t^2)},
\end{equation}
which is symmetric about $y=0$,
and the distribution function is
\begin{equation}
	\int_{-\infty}^yf_t(x) \d x=\half+\frac{1}{\pi} \arctan\left[ \frac{y}{ct}\right].
\end{equation}
We have $\E[|Z_t|]=\infty$, but $\E[|Z_t|^p]<\infty$ for $0<p<1$.
Let $\{S_t\}$ be a stable subordinator with index $\a=1/2$ (as described in the previous section).
From \citet[X.9]{Feller2}, if $\{W(t)\}$ is a standard Brownian motion then we have
\begin{equation}
	\{Z_t\}\law \{W(S_t)\}.
\end{equation}

\subsection{Cauchy bridge}

Fix $z\in\R$ and let $\{Z_{tT}^{(z)}\}$ be a bridge of the Cauchy process $\{Z_t\}$ terminating at the value $z$ at time $T$.
The density of the random variable $Z_{tT}^{(z)}$ is
\begin{align}
	f_{tT}(y;z)&=\frac{f_t(y)f_{T-t}(z-y)}{f_T(z)} \nonumber
	\\ &= \frac{ct(T-t)}{\pi T} \frac{z^2+c^2T^2}{(y^2+c^2t^2)((z-y)^2+c^2(T-t)^2)}. \label{eq:CBden}
\end{align}
Integrating (\ref{eq:CBden}) yields the following (details of the proof can be found in Appendix \ref{apdx:CBDF}):
\begin{prop}
	\label{prop:CBDF}
	The distribution function of $Z_{tT}^{(z)}$ is
	\begin{align}
		F_{tT}(y;z)=\half&+\frac{(T-t)(c^2T(T-2t)+z^2)}{\pi T(c^2(T-2t)^2+z^2)} \arctan\left[ \frac{y}{ct} \right]\nonumber
		\\ &+\frac{t(c^2T(T-2t)-z^2)}{\pi T(c^2(T-2t)^2+z^2)} \arctan \left[\frac{z-y}{c(T-t)} \right]\nonumber
		\\ &+\frac{ct(T-t)z}{\pi T(c^2(T-2t)^2+z^2)} \log\left[\frac{y^2+c^2t^2}{(z-y)^2+c^2(T-t)^2}  \right],
	\end{align}
	where $0<t<T$, $c>0$.
\end{prop}

\begin{rem}
	It follows from Proposition \ref{prop:StableBridge} that, for fixed $k>0$, the Cauchy bridge $\{Z_{tT}^{(z)}\}$ exhibits the scaling property
	\begin{equation}
		\label{eq:CauchyScaling}
		\left\{k^{-1} Z_{kt,kT}^{(kz)} \right\}_{0\leq t\leq T} \law \left\{Z_{tT}^{(z)} \right\}_{0\leq t\leq T}.
	\end{equation}
\end{rem}

The first moment of a Cauchy process is not well defined, and the second is infinite.
However, the first two moments of the Cauchy bridge exist and are finite.
Hence conditioning a Cauchy process on its final value can temper its behaviour.
The proof of the following proposition can be found in Appendix \ref{apdx:CBmom}:
\begin{prop}
	\label{prop:CBmom}
	The first two moments of $Z_{tT}^{(z)}$ exist, and are given by
	\begin{align}
		\E\left[Z_{tT}^{(z)} \right]&=\frac{t}{T}z,
		\\ \E\left[\left(Z_{tT}^{(z)}\right)^2 \right] &=\frac{t}{T}(z^2+c^2T(T-t)).
	\end{align}
\end{prop}

\begin{rem}
In general we have
	\begin{align}
		\E\left[Z_{tT}^{(z)}\left|\, Z_{sT}^{(z)}=x \right.\right]&=\frac{T-t}{T-s} x+\frac{t-s}{T-s}z, \label{eq:CBexp}
		\\ \E\left[\left.\left(Z_{tT}^{(z)}\right)^2 \,\right| Z_{sT}^{(z)}=x \right]&=
						\frac{T-t}{T-s} x^2 +\frac{t-s}{T-s}(z^2+c^2(T-s)(T-t)), \label{eq:CBexp2}
	\end{align}
for $0\leq s <t <T$.
\end{rem}

\begin{rem}
The third moment of $Z^{(z)}_{tT}$ does not exist; however
\begin{equation}
	\E\left[\left|Z_{tT}^{(z)}\right|^p \right]<\infty \qquad \text{for $p\in(0,3)$.}
\end{equation}
\end{rem}

\section{Normal inverse-Gaussian process and normal inverse-Gaussian bridge}

\subsection{Inverse-Gaussian process} \label{subsec:IG}
The inverse-Gaussian (IG) process is a subordinator with \levy exponent
\begin{equation}
	\Psi(\l)=c(\sqrt{\g^2-2\i\l}-\g),
\end{equation}
for $c>0$ and $\g>0$ constants.
Let $\{X_t\}$ be an inverse-Gaussian process.
The density function of $X_t$ is
\begin{align}
	q_t(x)&=\1_{\{x>0\}} \frac{c t\e^{\g c t}}{\sqrt{2\pi}}\frac{1}{x^{3/2}}\exp\left(-\tfrac{1}{2}(c^2t^2x^{-1}+\g^2x) \right) \nonumber
	\\ &=\1_{\{x>0\}} \frac{c t}{\sqrt{2\pi}}\frac{1}{x^{3/2}}\exp\left(-\frac{\g^2}{2 x}\left(x-\frac{c}{\g}t\right)^2 \right).
	\label{eq:IGDen}
\end{align}
For $k\in\R$, the moment $m_t^{(k)}=\E[X_t^k]$ can be written
\begin{align}
	m_t^{(k)}&=\left(\frac{ct}{\g} \right)^{k} \frac{K_{k-1/2}[\g c t]}{K_{1/2}[\g c t]} \nonumber
	\\&=\sqrt{\frac{2}{\pi}}\,\g\e^{\g c t}\left(\frac{c t}{\g} \right)^{k+\frac{1}{2}} K_{k-1/2}[\g c t].
	\label{eq:IGmom}
\end{align}
Using the series representation of $K_{n+1/2}[z]$ given in (\ref{eq:BesselHalfInt}), the first four integer moments simplify to
\begin{align}
	m_t^{(1)}&=\frac{c t}{\g},
	\\ m_t^{(2)}&=\frac{c t}{\g^3}(1+\g c t),
	\\ m_t^{(3)}&=\frac{c t}{\g^5}(3+3\g c t+\g^2 c^2 t^2),
	\\ m_t^{(4)}&=\frac{c t}{\g^7}(15+15\g c t+6 \g^2 c^2 t^2+\g^3 c^3t^3).
\end{align}
The variance of $X_t$ is $ct/\g^3$. Hence, similar to the gamma process, the law of an IG process is characterised by its mean and variance at $t=1$.

For $\{W_t\}$ a Brownian motion, define the exceedence times $\{\tau_t\}$ by
\begin{equation}
	\tau_t=\inf \{t\geq 0: c^{-1}W_t+c^{-1}\g t>t\}.
\end{equation}
Then we have
\begin{equation}
	\{X_t\} \law \{\tau_t\}.
\end{equation}

\begin{prop}
	A bridge of the process $\{X_t\}$ to a fixed point $z>0$ at time $T$ is identical in law to a bridge process of the stable-1/2 subordinator $\{S_t\}$.
\end{prop}
\begin{proof}
	The (one-dimensional) marginal density of the IG bridge is
	\begin{align}
		f_{tT}(y;z)&=\frac{q_t(y)q_{T-t}(z-y)}{q_T(z)} \nonumber
		\\&=\1_{\{ 0<y\leq z\}}\frac{1}{\sqrt{{2\pi}}} 
					\frac{ct(T-t)}{T}\frac{\exp\left( -\frac{1}{2} 	\frac{c^2(Ty-tz)^2}{yz(z-y)}\right)}{\left(y-y^2/z\right)^{3/2}}.
	\end{align}
	This is identical to the marginal density of the stable-1/2 bridge given in (\ref{eq:kernel}).
	It follows from (\ref{eq:LB_trans}) that the transition law of the IG bridge is identical to the transition law of the stable-1/2 bridge.
\end{proof}

\subsection{Normal inverse-Gaussian process}
We construct the normal inverse-Gaussian (NIG) process by subordinating a Brownian motion by an independent IG process.
Again let $\{X_t\}$ be an IG process, but this time set $c=\g$.
This ensures that $\E[X_t]=t$, so $\{X_t\}$ has units of time.
Then we define the NIG process $\{Y_t\}$ by setting
\begin{equation}
	Y_t=\s W(X_t) + \th X_t,
\end{equation}
for $\s>0$ and $\th\in\R$ constants.
In terms of the law of the NIG process, allowing the mean rate of $\{X_t\}$ to differ from unity is equivalent to an appropriate change in the parameters $\s$ and $\th$.
We say that $\{Y_t\}$ is a standard NIG process if $\th=0$ and $\s=1$.

The characteristic exponent of $\{Y_t\}$ is
\begin{equation}
	\Psi(\l)=c\sqrt{c^2+\s^2\l^2-2\i\th\l}-c^2.
\end{equation}
The density of $Y_t$ is
\begin{equation}
	\label{eq:VG_Full_den}
	f^{(c,\th,\s)}_t(x)=\frac{ct}{\s\pi}\e^{c^2 t+\th x/\s^2} \sqrt{\frac{c^2\s^2+\th^2}{c^2\s^2t^2+x^2}}
		K_1\left[ \s^{-2}\sqrt{(\th^2+c^2\s^2)(c^2\s^2 t^2+x^2)} \right].
\end{equation}
Note that $f^{(c,\th,\s)}_t(x)$ is bounded and continuous everywhere.
Hence $\{Y_t\}\in\mathcal{C}[0,T]$ for all $T>0$.
The density of the standard NIG variable $W(X_t)$ is
\begin{align}
	f^{(c)}_t(x)&=f^{(c,0,1)}_t(x) \nonumber
	\\&=\frac{c^2t\,\e^{c^2 t}}{\pi\sqrt{c^2t^2+x^2}} K_1\left[ c\sqrt{c^2t^2+x^2} \right]. \label{eq:NIG_den}
\end{align}
After some rearrangement, we find
\begin{equation}
	\label{eq:NIG_den_simple}
	f^{(c,\th,\s)}_t(x)=\frac{k}{\s} \e^{(c^2-\a^2)t+\th x/\s^2}  f^{(\a)}_t\left(k x/\s  \right),
\end{equation}
where $k\geq 1$ and $\a>0$ are given by
\begin{equation}
	\label{eq:NIG_k_a}
	k^2=c^{-1}\sqrt{\th^2+c^2}, \qquad \a^2=c\sqrt{\th^2+c^2}.
\end{equation}

\subsection{Normal inverse-Gaussian bridge}
Let $\{Y_t\}$ be an NIG process with parameter set $\{c,\th,\s\}$,
and let $\{Y^{(z)}_{tT}\}$ be a $\{Y_t\}$-bridge to the value $z\in\R$ at time $T$.
The transition law of $\{Y^{(z)}_{tT}\}$ is
\begin{align}
	\Q\left[\left. Y^{(z)}_{tT}\in\dd y \,\right| Y^{(z)}_{sT}=x \right]	
	 &=\frac{f^{(c,\th,\s)}_{t-s}(y-x)f^{(c,\th,\s)}_{T-t}(z-y)}{f^{(c,\th,\s)}_{T-s}(z-x)} \d y \nonumber
	\\ &=\frac{k}{\s}\frac{f^{(\a)}_{t-s}\left(\frac{k}{\s}(y-x)\right)f^{(\a)}_{T-t}\left(\frac{k}{\s}(z-y)\right)}
			{f^{(\a)}_{T-s}\left(\frac{k}{\s}(z-x)\right)} \d y \nonumber
	\\ &= \Q\left[\left. \frac{\s}{k} U^{(kz/\s)}_{tT}\in\dd y \,\right| \frac{\s}{k} U^{(kz/\s)}_{sT}=x \right],
\end{align}
where (i) $0\leq s <t<T$, (ii) $k$ and $\a$ are given by (\ref{eq:NIG_k_a}), and (iii) $\{U^{(kz/\s)}_{sT}\}$ is a standard NIG bridge to $kz/\s$ at time $T$ with parameter $\a$.
Thus, the three parameter NIG bridge is identical in law to a scaled standard NIG bridge.

\section{Poisson process and Poisson bridge} \label{sec:Poisson}

\subsection{Poisson process}
Let $\{N_t\}$ be a Poisson process with intensity $\l>0$.
Then $\{N_t\}$ is a \levy process and, for each $n\in\N_0$, $\Q[N_t=n]=Q_t(n)$ where
\begin{equation}
	\label{eq:Poisson}
	Q_t(n)=\1_{\{n\geq 0\}}\frac{\e^{-\l t} (\l t)^n}{n!}.
\end{equation}
The characteristic exponent of $\{N_t\}$ is
\begin{equation}
	\Psi(u)=\l(1-\e^{\i u}).
\end{equation}
Since $\{N_t\}$ has state space $\mathbb{N}_0$ and is increasing in increments of 1, it is a so-called counting process.
The moment generating function of $N_t$ exists, and is given by
\begin{equation}
	\label{eq:cfpo}
	\E\left[ \e^{u N_t} \right]=\exp\left( \l t(\e^{ u}-1) \right) \qquad \text{for $u\in\R$}.
\end{equation}
We have
\begin{equation}
	\E[N_t]=\var[N_t]=\l t.
\end{equation}
All higher moments of $N_t$ are finite, and can be found by differentiating (\ref{eq:cfpo}).
Define the sequence $\{T_i\}_{i=0}^{\infty}$ of jump times of $\{N_t\}$ by
\begin{equation}
	T_i=\inf \left\{ t\geq 0: N_t\geq i \right\},
\end{equation}
and the let $\{W_i\}$ be the sequence of waiting times $W_i=T_i-T_{i-1}$.
The $W_i$'s are independent, identically distributed random variables with
\begin{equation}
	\Q[W_i>t]=\Q[N_t=0]=\e^{-\lambda t},
\end{equation}
i.e.~the waiting times of Poisson processes are exponentially distributed.

\subsection{Poisson bridge}

Let $\{N^{(k)}_{tT}\}_{0\leq t \leq T}$ be a Poisson bridge to the value $k\in\mathbb{N}_+$ at time $T$
(we are excluding the case where $N^{(k)}_{TT}=0$).
The distribution of $N_{tT}^{(k)}$ is
\begin{align}
	\Q\left[N_{tT}^{(k)}=j\right] &=\frac{\Q[N_t=j, N_T=k]}{\Q[N_T=k]} \nonumber
	\\  &= \frac{Q_t(j)Q_{T-t}(k-j)}{Q_T(k)} \nonumber
	\\	&= \1_{\{0\leq j \leq k\}} \binom{k}{j} \left(\frac{t}{T} \right)^j \left(1-\frac{t}{T}\right)^{k-j},
\end{align}
for $0<t<T$.
Hence $N_{tT}^{(k)}$ has a $\mathrm{Binomial}(k,t/T)$ distribution (note that this holds for any choice of $\l$).
The probability generating function of $N_{tT}^{(k)}$ is
\begin{equation}
	\E\left[ z^{N_{tT}^{(k)}} \right]=\left( 1-\frac{t}{T}+\frac{t}{T}z \right)^k \qquad (z>0).
\end{equation}
All moments of $N_{tT}^{(k)}$ exist, and the first two are
\begin{align}
	\E\left[N_{tT}^{(k)}\right]&=k\frac{t}{T},
	\\ \E\left[\left(N_{tT}^{(k)}\right)^2\right]&=k\frac{t}{T}\left(1-\frac{t}{T}\right)+k^2\left(\frac{t}{T}\right)^2. \label{eq:PBmom2}
\end{align}
We denote the jump times of $\{N^{(k)}_{tT}\}$ by $\{T_i^{(k)}\}_{i=1}^k$, and the waiting times by $\{W_i^{(k)}\}_{i=1}^k$.
It is well known that the jump times are distributed as ordered, independent uniform random variables (see, for example, \citet{Sato1999} or \citet{TK2004}).
In particular, for $U_{(1)}<\cdots<U_{(k)}$ an ordered sequence of independent standard uniform random variables, we have
\begin{align}
	(T_1^{(k)},\ldots,T_k^{(k)})&\law \frac{T}{T_{k+1}}(T_1,\ldots,T_k)  \label{eq:timevec}
	\\ &\law T(U_{(1)},\ldots,U_{(k)}),
\end{align}
where $\{T_i\}$ are the jump times of the Poisson process $\{N_t\}$.
It is also well known that the jump time $T_{k+1}$ is independent of the vector
\begin{equation}
	\frac{T}{T_{k+1}}(T_1,\ldots,T_k).
\end{equation}
The distributional identity (\ref{eq:timevec}) is equivalent to the following:
\begin{equation}
	(W_1^{(k)},\ldots,W_k^{(k)}) \law \frac{T}{\sum_{i=1}^{k+1} W_i}(W_1,\ldots,W_k),
\end{equation}
where $\{W_i\}$ are the waiting times of $\{N_t\}$ (i.e.~they are independent, identically-distributed exponential random variables).
The ordered uniform random variable $U_{(i)}$ has a beta distribution with parameters $\a=i$ and $\b=k-i+1$.
Hence we have
\begin{equation}
	\label{eq:PB_JumpTime}
	\Q[T_i^{(k)} \leq t ]=I_{t/T}[i,k-i+1].
\end{equation}
Here $I_z[\a,\b]$ is the regularized incomplete beta function which is defined by
\begin{equation}
	I_z[\a,\b]=\frac{\int_0^zy^{\a-1}(1-y)^{\b-1} \d y}{\int_0^1y^{\a-1}(1-y)^{\b-1} \d y}, \quad \text{for $\a,\b>0$.}
\end{equation}
Define a process $\{\bar{N}_{tT}^{(k)}\}$ for  $0\leq t<T$ by
\begin{equation}
			\bar{N}_{tT}^{(k)}=N\!\left( \tfrac{t}{T} T_{k+1} \right),
			\qquad \bar{N}_{TT}^{(k)}=k.
\end{equation}
Then $\{\bar{N}_{tT}^{(k)}\}_{0\leq t \leq T}$ is a counting process that jumps at the times $\frac{T}{T_{k+1}}(T_1,\ldots,T_k)$,
and is independent of $T_{k+1}$.
It follows that
\begin{equation}
	\{\bar{N}_{tT}^{(k)}\}_{0\leq t \leq T} \law \{N_{tT}^{(k)}\}_{0\leq t \leq T}.
\end{equation}

%% file: chap03.tex
%
%

\chapter{\levy random bridges} \label{chap:LRB}
The idea of information-based asset pricing is to model the flow of information in financial markets and hence to construct the market filtration explicitly.
Let $X_T$ be a random variable (a market factor), with a given \emph{a priori} distribution.
The value of $X_T$ will be revealed to the market at time $T$.
We wish to construct an information process $\{\xi_{tT}\}$ such that $\xi_{TT}=X_T$.
We can then use the filtration generated by $\{\xi_{tT}\}$ to model the information that market participants have about $X_T$.
One problem to overcome is how to ensure that the marginal law of $\xi_{TT}$ is the \emph{a priori} law of $X_T$.

Two explicit forms for the information process have been considered in the literature.
The first is
\begin{equation}
	\label{eq:BB}
	\xi_{tT}=\frac{t}{T}X_T+\beta_{tT} \qquad (0\leq t \leq T),
\end{equation}
where $\{\beta_{tT}\}_{0\leq t\leq T}$ is a Brownian bridge starting and ending at the value 0 (see \citep{BHM1,BHM2,HM2008,MPhD2006,BDFH2008,YR2007}).
The second is 
\begin{equation}
	\xi_{tT}=X_T \gb{t} \qquad (0\leq t \leq T),
\end{equation}
where $X_T>0$ and $\{\gb{t}\}_{0 \leq t\leq T}$ is a gamma bridge starting at the value 0 and ending at the value 1 (see \citep{BHM3}).
These forms share the property that each is identical in law to a \levy process conditioned to have the \emph{a priori} law of $X_T$ at time $T$.
The Brownian bridge information process is identical in law to a conditioned Brownian motion, and the gamma bridge information process is identical in law to a conditioned gamma process.

With this as motivation, in this chapter we define a class of processes that we call \levy random bridges (LRBs).
An LRB is identical in law to a \levy process conditioned to have a prespecified marginal law at $T$.
Later we shall use LRBs as information processes in information-based models.

\section{Defining LRBs}
An LRB can be described as a process whose bridge laws are \levy bridge laws.
In the definitions below we define LRBs by reference to their finite-dimensional distributions rather than as conditioned \levy processes.
This proves convenient in future calculations.

\begin{defn}
	\label{def:LRB}
	We say that the process $\{L_{tT}\}_{0\leq t\leq T}$ has the law $\lrb_{\mathcal{C}}([0,T],\{f_t\},\nu)$ if the following are satisfied:
	\begin{enumerate}
		\item $L_{TT}$ has marginal law $\nu$.
		\item There exists a \levy process $\{L_t\}\in\mathcal{C}[0,T]$ such that $L_t$ has density $f_t(x)$ for all $t\in(0,T]$.
		\item \label{item:measurecond} $\nu$ concentrates mass where $f_T(z)$ is positive and finite, i.e.~$0<f_T(z)<\infty$ for $\nu$-a.e.~$z$.
		\item \label{item:condLevy} For every $n\in\mathbb{N}_+$, every $0<t_1<\cdots<t_n<T$, every $(x_1,\ldots,x_n)\in\R^n$, and $\nu$-a.e.~$z$, we have
						\begin{equation*}
											\Q\left[L_{t_1,T}\leq x_1,\ldots,L_{t_n,T} \leq x_n \left|\, L_{TT} = z \right.\right]
											 =\Q\left[L_{t_1}\leq x_1,\ldots,L_{t_n} \leq x_n \left|\, L_{T} = z \right.\right].
						\end{equation*}	 
	\end{enumerate}
\end{defn}

\begin{rem}
	Conditions 2 and 3 in Definition \ref{def:LRB} are sufficient conditions for the right-hand side of the equation in condition 4 to make sense.
\end{rem}

\begin{defn}
	We say that the process $\{M_{tT}\}_{0\leq t\leq T}$ has the law$\lrb_{\mathcal{D}}([0,T],\{Q_t\},P)$ if the following are satisfied: 
	\begin{enumerate}
		\item $M_{TT}$ has probability mass function $P$.
		\item There exists a \levy process $\{M_t\}\in\mathcal{D}$ such that $M_t$ has marginal probability mass function $Q_t(a)$ for all $t\in(0,T]$.
		\item The law of $M_{TT}$ is absolutely continuous with respect to the law of $M_T$, i.e.~%
		\begin{equation*}
			\text{if } P(a)>0 \text{ then } Q_T(a)>0.
		\end{equation*}
		\item For every $n\in\mathbb{N}_+$, every $0<t_1<\cdots<t_n<T$, every $(k_1,\ldots,k_n)\in\mathbb{Z}^n$, and every $b$ such that $P(b)>0$, we have
						\begin{equation*}
								\Q\left[M_{t_1,T}= a_{k_1},\ldots,M_{t_n,T} = a_{k_n} \left|\, M_{TT} = b \right.\right]
											=\Q\left[M_{t_1}= a_{k_1},\ldots,M_{t_n} = a_{k_n} \left|\, M_{T} = b \right.\right].
						\end{equation*}
	\end{enumerate}
\end{defn}

\begin{defn}
For a fixed time $s<T$, if the law of the process $\{\eta_{s+t}\}_{0\leq t \leq T-s}$ is of the type \mbox{$\lrb_{\mathcal{C}}([0,T-s],\,\cdot\,,\,\cdot\,)$}, resp.~\mbox{$\lrb_{\mathcal{D}}([0,T-s],\,\cdot\,,\,\cdot\,)$},
then we say that $\{\eta_{t}\}_{s\leq t \leq T}$ has law $\lrb_{\mathcal{C}}([s,T],\,\cdot\,,\,\cdot\,)$, resp.~$\lrb_{\mathcal{D}}([s,T],\,\cdot\,,\,\cdot\,)$.
\end{defn}
If the law of a process is one of the $\lrb$-types defined above, then we say that it is a \levy random bridge (LRB).

\section{Finite-dimensional distributions} \label{sec:LRB_FDD}
For the rest of this chapter we assume that $\{L_{tT}\}$ and $\{M_{tT}\}$ are LRBs with laws $\lrbc([0,T],\{f_t\},\nu)$
and $\lrbd([0,T],\{Q_t\},P)$, respectively.
We also assume that $\{L_t\}$ is a \levy process such that $L_t$ has density $f_t(x)$ for $t\leq T$, and
$\{M_t\}$ is a \levy process such that $M_t$ has probability mass function $Q_t(a_i)$ for $t\leq T$.

The finite-dimensional distributions of $\{L_{tT}\}$ are given by
\begin{equation}
	\label{eq:LRBlaw}
	 \Q\left[ L_{t_1,T}\in \dd x_1,\ldots, L_{t_n,T}\in\dd x_n, L_{TT}\in\dd z \right]= 
 	\prod_{i=1}^{n} \left[f_{t_i-t_{i-1}}(x_i-x_{i-1}) \d x_i\right] \psi_{t_{n}}(\dd z;x_{n}),
\end{equation}
where the (un-normalised) measure $\psi_t(\dd z;\xi)$ is given by
\begin{align}
	\psi_0(\dd z;\xi)&=\nu(\dd z),
	\\\psi_t(\dd z;\xi)&=\frac{f_{T-t}(z-\xi)}{f_T(z)}\nu(\dd z), \label{eq:psit}
\end{align}
for $0<t<T$.
Given (\ref{eq:LRBlaw}), Kolmogorov's extension theorem ensures the existence of LRBs (up to a modification).
It follows from the definition of $\lrbc([0,T],\{f_t\},\nu)$ and equation (\ref{eq:check_den}) that
\begin{equation}
	f_{tT}(x;z)=\frac{f_t(x)f_{T-t}(z-x)}{f_T(z)}
\end{equation}
is a well-defined density (as a function of $x$) for $t<T$ and $\nu$-a.e.~$z$.
Then from (\ref{eq:LRBlaw}) the marginal law of $L_{tT}$ is given by
\begin{equation}
	\Q[L_{tT}\in \dd x]=f_t(x) \psi_t(\R;x) \d x=\int_{z=-\infty}^{\infty} f_{tT}(x;z) \,\nu(\dd z) \d x.
\end{equation}
Hence the density of $L_{tT}$ exists for $t<T$, and 
\begin{equation}
	0\leq \psi_t(\R;x)<\infty \qquad \text{for Lebesgue-a.e.~$x\in\mathrm{Support}(f_t)$.}
\end{equation}
In particular, we have
\begin{align}
	\label{eq:density_conditions}
	&0< \psi_t(\R;L_{tT})<\infty &&\text{and} &&0< f_{T-t}(x-L_{tT})<\infty
\end{align}
for a.e.~value of $L_{tT}$.
If $\nu(\{z\})=1$ for some point $z\in\R$, i.e.~$\Q[L_{TT}=z]=1$, then $\{L_{tT}\}$ is a \levy bridge.
If $\nu(\dd z)=f_T(z)\d z$, then $\{L_{tT}\}\law\{L_t\}$ for $t\in[0,T]$.

In the discrete case, the finite-dimensional probabilities of $\{M_{tT}\}$ are
\begin{equation}
	 \Q\left[ M_{t_1,T}= a_{k_1},\ldots, M_{t_n,T}= a_{k_n}, M_{TT}= z \right]= 
 	\\ \prod_{i=1}^{n} \left[Q_{t_i-t_{i-1}}(a_{k_i}-a_{k_{1-1}}) \right] \phi_{t_{n}}(z;a_{k_n}),
\end{equation}
where the function $\phi_t(z;\xi)$ is given by
\begin{align}
	\phi_0(z;\xi)&=P(z),
	\\\phi_t(z;\xi)&=\frac{Q_{T-t}(z-\xi)}{Q_T(z)}P(z),
\end{align}
for $0<t<T$.
If $P$ is identical to $Q_T$, then $\{M_{tT}\}\law\{M_t\}$ for $t\in[0,T]$.

The existing literature on information-based asset pricing exploits special properties Brownian and gamma bridges.
See \citet{EY2004} for an insight into how remarkable these bridges are.
The methods we use do not require special properties of particular \levy bridges.
However, we will often use the Brownian and gamma cases as examples, and the results we obtain agree with previous work.

Many of the results that follow are proved for the LRB $\{L_{tT}\}$, which has a continuous state-space.
Analogous results are provided for the discrete state-space process $\{M_{tT}\}$; details of proofs are omitted since they are similar to the continuous case.

\section{LRBs as conditioned \levy processes}
It is useful to interpret an LRB as a \levy process conditioned to have a specified marginal law $\nu$ at time $T$.
Suppose that the random variable $Z$ has law $\nu$, then:
\begin{align}
 &\Q\left[L_{t_1} \in\dd x_1, \ldots,L_{t_{n}}\in\dd x_{n}, L_{T}\in\dd z \left|\, L_T=Z \right.\right] \nonumber
 \\ &\qquad=	 \Q\left[L_{t_1} \in\dd x_1, \ldots,L_{t_{n}}\in\dd x_{n} \left|\, L_T=z\right.\right]  \nu( \dd z) \nonumber
 \\ &\qquad=	\frac{f_{T-t_{n-1}}(z-x_{n-1})}{f_T(z)} \, \prod_{i=1}^{n} \left[f_{t_i-t_{i-1}}(x_i-x_{i-1}) \d x_i\right]\nu( \dd z).
\end{align}
Hence the conditioned \levy process has law $\lrbc([0,T],\{f_t\},\nu)$.

\section{The Markov property}
In this section we show that LRBs are Markov processes.
The Markov property is a key tool in the application of LRBs to information-based asset pricing.
As will be seen below, the Markov property of an LRB follows from the Markov property of the associated \levy bridge processes.
Note that if we stop the LRB $\{L_{tT}\}$ at time $s>0$, and then restart it, the terminal law will no longer be the \emph{a priori} law $\nu$, but instead an updated law conditional on $L_{sT}$.

\subsection{Continuous state-space}
\begin{prop}
	\label{prop:markov}
	The process $\{L_{tT}\}_{0\leq t \leq T}$ is a Markov process with transition law
	\begin{equation}
		\label{eq:LRBtranslaw}
			\begin{aligned}
				\Q[L_{tT} \in \dd y \,|\, L_{sT}=x]&=\frac{\psi_t(\R;y)}{\psi_s(\R;x)} f_{t-s}(y-x)\d y,
				\\\Q[L_{TT} \in \dd y \,|\, L_{sT}=x]&=\frac{\psi_s(\dd y;x)}{\psi_s(\R;x)},
			\end{aligned}
	\end{equation}
	where $0\leq s<t<T$.
\end{prop}
\begin{proof}
	To show that $\{L_{tT}\}$ is Markov, it is sufficient to show that
	\begin{equation} 
		\Q\left[L_{tT} \leq y \,|\, L_{t_1,T}=x_1,\ldots,L_{t_m,T}=x_m\right]
					= \Q\left[L_{tT} \leq y \,|\, L_{t_m,T}=x_m\right],
	\end{equation}
	for all $m \in \mathbb{N}_+$, all $(x_1,\ldots,x_m,y)\in\R^{m+1}$, and all $0\leq t_1<\cdots<t_m<t\leq T$.
	When $t=T$ we apply the Bayes theorem to (\ref{eq:LRBlaw}) and obtain
	\begin{equation}
		 \Q\left[\left. L_{TT}\in\dd y  \,\right|  L_{t_1,T}=x_1,\ldots,L_{t_m,T}=x_m\right]=\frac{\psi_{t_m}(\dd y;x_m)}{\psi_{t_m}(\R;x_m)}.
	\end{equation}	 
	We need now only consider the case $t<T$.
	Proposition \ref{prop:LB_Markov} shows that \levy bridges are Markov processes; therefore,
	\begin{equation}
		\label{eq:LBMarkov}
		 \Q\left[L_{t}\leq y \,|\, L_{t_1}=x_1,\ldots,L_{t_m}=x_m,L_T=x\right]=
					 \Q\left[L_{t}\leq y \,|\, L_{t_m}=x_m,L_T=x\right].
	\end{equation}
	It is straightforward by Definition \ref{def:LRB} part \ref{item:condLevy} to show that LRBs are Markov processes.
	Indeed we have:
	\begin{align}
		&\Q\left[\left. L_{tT} \leq y \,\right| L_{t_1,T}=x_1,\ldots,L_{t_m,T}=x_m\right] \nonumber
		\\	&\qquad=\int_{-\infty}^{\infty} \Q\left[\left. L_{tT} \leq y \,\right| L_{t_1,T}=x_1,\ldots,L_{t_m,T}=x_m,L_{T,T}=x\right]\nu(\dd x) \nonumber
		\\	&\qquad=\int_{-\infty}^{\infty} \Q\left[\left. L_{t} \leq y \,\right| L_{t_1}=x_1,\ldots,L_{t_m}=x_m,L_{T}=x\right]\nu(\dd x) \nonumber
		\\	&\qquad=\int_{-\infty}^{\infty} \Q\left[\left. L_{t} \leq y \,\right| L_{t_m}=x_m,L_{T}=x\right]\nu(\dd x) \nonumber
		\\	&\qquad=\int_{-\infty}^{\infty} \Q\left[\left. L_{tT} \leq y \,\right| L_{t_m,T}=x_m,L_{T,T}=x\right]\nu(\dd x) \nonumber
		\\	&\qquad= \Q\left[\left. L_{tT} \leq y \,\right| L_{t_m,T}=x_m\right].
	\end{align}
	The form of the transition law of $\{L_{tT}\}$ appearing in (\ref{eq:LRBtranslaw}) follows from (\ref{eq:LRBlaw}).
\end{proof}

\bigskip

\noindent{\bf Example.}
In the Brownian case we set
\begin{equation}
	\label{eq:NormalDensity}
	f_t(z)=\frac{1}{\sqrt{2 \pi t}} \exp\left[-\frac{z^2}{2t} \right],
\end{equation}
for $t>0$.
Thus $f_t(x)$ is the marginal density of a standard Brownian motion at time $t$.
Then we have
\begin{equation}
	\Q[L_{tT} \in \dd y \,|\, L_{sT}=x]=
	\sqrt{\frac{T-s}{T-t}}\frac{\int_{-\infty}^{\infty}\e^{-\half\left[\frac{(z-y)^2}{T-t}-\frac{z^2}{T}\right]}\,\nu(\dd z)}
				{\int_{-\infty}^{\infty}\e^{-\half\left[\frac{(z-x)^2}{T-s}-\frac{z^2}{T}\right]}\,\nu(\dd z)} \frac{\e^{-\half \frac{(y-x)^2}{t-s}}}{\sqrt{2\pi (t-s)}}\d y,
\end{equation}
and
\begin{equation}
	\Q[L_{TT} \in \dd y \,|\, L_{sT}=x] =\frac{\e^{-\half\left[\frac{(y-x)^2}{T-s}-\frac{y^2}{T}\right]}\,\nu(\dd y)}
				{\int_{-\infty}^{\infty}\e^{-\half\left[\frac{(z-x)^2}{T-s}-\frac{z^2}{T}\right]}\,\nu(\dd z)} 
	=\frac{\e^{\frac{1}{T-s}\left[xy-\half \frac{s}{T}y^2\right]}\,\nu(\dd y)}
				{\int_{-\infty}^{\infty}\e^{\frac{1}{T-s}\left[xz-\half \frac{s}{T}z^2\right]}\,\nu(\dd z)}.
\end{equation}

\bigskip

\noindent{\bf Example.}
In the gamma case we consider the one-parameter family of processes indexed by $m>0$ described in Section \ref{subsec:gamma_process}. 
We set
\begin{equation}
	f_t(z)=  \1_{\{x>0\}} \frac{ m^{mt}}{\G[mt]} x^{mt-1} \e^{-mx};
\end{equation}
so $f_t(z)=g_t^{(m)}(x)$ for $g_t^{(m)}(x)$ given by (\ref{eq:gamma_m_den}).
These densities are the increment densities of the gamma process with mean $1$ and variance $m^{-1}$ at $t=1$.
Then
\begin{multline}
	\Q[L_{tT} \in \dd y \,|\, L_{sT}=x] \\
		=\frac{\1_{\{y>x\}}}{\mathrm{B}[m(T-t),m(t-s)]}\frac{\int_{y}^{\infty}(z-y)^{m(T-t)-1}z^{1-mT}\,\nu(\dd z)}{\int_{x}^{\infty}(z-x)^{m(T-s)-1}z^{1-mT}\,\nu(\dd z)}
			(y-x)^{m(t-s)-1} \d y,
\end{multline}
and
\begin{equation}
	\Q[L_{TT} \in \dd y \,|\, L_{sT}=x]
		=\frac{\1_{\{y>x\}}(y-x)^{m(T-s)-1}y^{1-mT}\,\nu(\dd y)}{\int_{x}^{\infty}(z-x)^{m(T-s)-1}z^{1-mT}\,\nu(\dd z)},
\end{equation}
where $\mathrm{B}[\a,\b]$ is the Beta function.

\subsection{Discrete state-space}
The analogous result to Proposition \ref{prop:markov} for the discrete case is provided below---the proof is similar.
\begin{prop}
	The process $\{M_{tT}\}_{0\leq t\leq T}$ has the Markov property, with transition probabilities given by
	\begin{equation}
		\label{eq:discreteProb}
		\begin{aligned}
		\Q\left[M_{tT} = a_j \left|\, M_{sT}=a_i\right.\right]
			&=\frac{\sum_{k=-\infty}^{\infty} \phi_t(a_k;a_j)}{\sum_{k=-\infty}^{\infty} \phi_s(a_k;a_i)} Q_{t-s}(a_j-a_i),
		\\\Q\left[M_{TT} =a_j \left|\, M_{sT}=a_i\right.\right]&=\frac{\phi_s(a_j;a_i)}{\sum_{k=-\infty}^{\infty} \phi_s(a_k;a_i)},
		\end{aligned}
	\end{equation}
	where $0\leq s<t<T$.
\end{prop}
\hfill{$\square$}

\section{Conditional terminal distributions}
Let $\{\F^L_t\}$ and $\{\F^{M}_t\}$ be the filtrations generated by $\{L_{tT}\}$ and $\{M_{tT}\}$, respectively.
\begin{defn}
Let $\nu_s$ be the $\F^L_s$-conditional law of the terminal value $L_{TT}$, 
and let $P_s$ be the $\F^M_s$-conditional probability mass function of the terminal value $M_{TT}$.
\end{defn}
We have $\nu_0(B)=\nu(B)$, and $P_0(a)=P(a)$.
Furthermore, when $s>0$, it follows from the results of the previous section that
\begin{align}
	\label{eq:C_terminal}
	\nu_s(B)&=\Q\left[L_{TT}\in B \left|\,\F_s^{L} \right.\right]=\frac{\psi_s(B; L_{sT})}{\psi_s(\R;L_{sT})},
	\\ \intertext{and}
	\label{eq:D_terminal}
	P_s(a_k)&=\Q\left[M_{TT}=a_k \left|\,\F_s^{M} \right.\right]=\frac{\phi_s(a_k;M_{sT})}{\sum_{j=-\infty}^{\infty} \phi_s(a_j;M_{sT})}.
\end{align}
When the \emph{a priori} $q$th moment of $L_{TT}$ is finite, the $\F^L_s$-conditional $q$th moment is finite and given by
\begin{equation}
	\label{eq:LpMart}
	\int_{-\infty}^{\infty} |z|^q \, \nu_s(\dd z).
\end{equation}
Similarly, when the \emph{a priori} $q$th moment of $M_{TT}$ is finite, the $\F^M_s$-conditional $q$th moment is finite and given by
\begin{equation}
	\label{eq:MpMart}
	\sum_{k=-\infty}^{\infty} |a_k|^q \, P_s(a_k).
\end{equation}
When they are finite, the quantities in (\ref{eq:LpMart}) and (\ref{eq:MpMart}) are martingales with respect to $\{\F^L_t\}$ and $\{\F^M_t\}$, respectively.
If $q\in\p{Z}$ then $\int |z|^q \, \nu(\dd z)<\infty$ ensures that $\int z^q \, \nu(\dd z)$ is a martingale,
and $\sum |a_k|^q P(a_k)<\infty$ ensures that $\sum a_k^q \,P(a_k)$ is a martingale.

When the terminal law $\nu$ admits a density, we denote it by $p(z)$, i.e.~$\nu(\dd z)=p(z)\d z$.
In this case the $L_{tT}$-conditional density of $L_{TT}$ exists, and we denote it by
\begin{equation}
	p_t(z)=\frac{\nu_t(\dd z)}{\dd z}=\frac{f_{T-t}(z-L_{tT})p(z)}{\psi_t(\R;L_{tT})f_T(z)}.
\end{equation}

\section{Measure changes} \label{sec:measure}
This section leads to Proposition \ref{prop:measure} which states that there exists a measure $\p{L}$ equivalent to $\Q$ under which the LRB $\{L_{tT}\}$ is a \levy process.
To begin the analysis it proves convenient to assume the existence of such a measure $\p{L}$, which we do, and
we further assume that under $\p{L}$ the density of $L_{tT}$ is $f_t(x)$.

Writing $\psi_t=\psi_t(\R;L_{tT})$, we can show that $\{\psi_t\}_{0\leq t<T}$ is an $\p{L}$-martingale (with respect to the filtration generated by $\{L_{tT}\}$).
In particular, for times $0\leq s<t$ we have
\begin{align}
	\E_{\p{L}}\left[\psi_t \left|\, \F^{L}_s \right.\right]
	&=\E_{\p{L}}\left[\left.\int_{-\infty}^{\infty}\frac{f_{T-t}(z-L_{tT})}{f_T(z)}\, \nu(\dd z)\,\right|\F^{L}_s  \right] \nonumber
	\\&=\E_{\p{L}}\left[\left.\int_{-\infty}^{\infty}\frac{f_{T-t}(z-L_{sT}-(L_{tT}-L_{sT}))}{f_T(z)}\, \nu(\dd z)\,\right|L_{sT} \right] \nonumber
	\\&=\int_{y=-\infty}^{\infty}\int_{z=-\infty}^{\infty}\frac{f_{T-t}(z-L_{sT}-y)}{f_T(z)}\, \nu(\dd z)\,f_{t-s}(y) \d y \nonumber
	\\&=\int_{z=-\infty}^{\infty}\frac{1}{f_T(z)}\int_{y=-\infty}^{\infty}f_{T-t}(z-L_{sT}-y)f_{t-s}(y)\d y \, \nu(\dd z) \nonumber
	\\&=\int_{z=-\infty}^{\infty}\frac{f_{T-s}(z-L_{sT})}{f_T(z)} \, \nu(\dd z) \nonumber
	\\ &=\psi_s.
\end{align}
Since $\psi_0=1$, we can define a probability measure $\p{L}^{\mathrm{rb}}$ by the Radon-Nikod\'ym derivative
\begin{equation} 
	\left.\frac{\dd \p{L}^{\mathrm{rb}}}{\dd \p{L}}\right|_{\F^{L}_t}=\psi_t \qquad \text{for $0\leq t <T$}. 
\end{equation}
It was noted in Section \ref{sec:LRB_FDD} that $0<\psi_t<\infty$, so $\p{L}^{\mathrm{rb}}$ is equivalent to $\p{L}$ for $t<T$.
For $s,t$ satisfying $0\leq s<t<T$, the transition law of $\{L_{tT}\}$ under $\p{L}^{\mathrm{rb}}$ is
\begin{align}
	\p{L}^{\mathrm{rb}}\left[L_{tT}\in\dd y \left|\, \F^{L}_{s}\right.\right]
		 &=\E_{\p{L}^{\mathrm{rb}}}\left[\1_{\{L_{tT}\in\dd y\}}  \left|\, \F^{L}_{s}\right.\right] \nonumber
	\\ &=\psi_s^{-1} \,\E_{\p{L}} \left[\psi_t \1_{\{L_{tT}\in\dd y\}}  \left|\, L_{sT}\right.\right] \nonumber
	\\ &=\psi_s^{-1} \int_{-\infty}^{\infty} \frac{f_{T-t}(z-y)}{f_T(z)} \, \nu(\dd z) \, f_{t-s}(y-L_{sT})\d y \nonumber
	\\ &= \frac{\psi_t(\R;y)}{\psi_s(\R;L_{sT})} f_{t-s}(y-L_{sT})\d y.
\end{align}
We see that $\{L_{tT}\}_{0\leq t<T}$ is a Markov process under the measure $\p{L}^{\mathrm{rb}}$.
Furthermore, by virtue of Proposition \ref{prop:markov}, $\{L_{tT}\}$ is an LRB with law $\lrbc([0,T],\{f_t\},\nu)$.

We can restate this result with reference to the measure $\Q$ as the following:
\begin{prop}
	\label{prop:measure}
	Let $\p{L}$ be defined by
	\begin{equation}
		\left. \frac{\dd \p{L}}{\dd \Q} \right|_{\F^L_t}= \psi_t(\R;L_{tT})^{-1}
	\end{equation}
	for $t\in[0,T)$.
	Then $\p{L}$ is a probability measure.
	Under $\p{L}$, $\{L_{tT}\}_{0\leq t <T}$ is a \levy process, and $L_{tT}$ has density $f_t(x)$.
\end{prop}
\hfill{$\square$}

In the case of a discrete state space a similar result is obtained:
\begin{prop}
	Let $\p{L}$ be defined by
	\begin{equation}
		\left. \frac{\dd \p{L}}{\dd \Q} \right|_{\F^M_t}= \left[ \sum_{k=-\infty}^{\infty} \phi_t(a_k;M_{tT}) \right]^{-1}
	\end{equation}
	for $t\in[0,T)$.
	Then $\p{L}$ is a probability measure.
	Under $\p{L}$, $\{M_{tT}\}_{0\leq t <T}$ is a \levy process, and $M_{tT}$ has mass function $Q_t(a)$.
\end{prop}
\hfill{$\square$}

\section{Dynamic consistency}
In this section we show that LRBs possess the so-called dynamic consistency property.
For $\{L_{tT}\}$, this property means the process $\{\eta_{t}\}$ defined by setting
\begin{equation}
	\label{eq:define_eta}
	\eta_{t}=L_{tT}-L_{sT} \qquad (s\leq t\leq T)
\end{equation}
is an LRB for fixed $s$ and $L_{sT}$ given.
Defining the filtration $\{\F^{\eta}_t\}$ by
\begin{equation}
	\F^{\eta}_t=\s\left(L_{sT}, \{\eta_u\}_{s\leq u\leq t}  \right),
\end{equation}
we see that
\begin{equation}
	\Q\left[ F\left(\{L_{uT}\}_{s\leq u \leq T}\right) \left|\, \F^{\eta}_t \right.\right]
		=	\Q\left[ F\left(\{L_{uT}\}_{s\leq u \leq T}\right) \left|\, \F^{L}_t \right.\right],
\end{equation}
for $0\leq s<t<T$ and $F$ an arbitrary measurable functional.
Suppose two market participants, trader A and trader B, watch the evolution of $\{L_{tT}\}$; trader A watching from $t=0$ and trader B watching from $t=s$.
The filtration of trader A, $\{\F^{L}_t\}$, is larger than the filtration of trader B, $\{\F^{\eta}_t\}$, but they have a common view of the future evolution of $\{L_{tT}\}$.
This is the Markov property.
The dynamic consistency property is stronger.
It states that the filtration of trader B can be regarded as being generated by an LRB, in this case $\{\eta_t\}$, plus some information about the current state of the world, in this case $L_{sT}$.
The \emph{a priori} terminal law of $\{\eta_t\}$ will not be $\nu$, but instead it will be an updated measure that depends on the current state $L_{sT}$.

Later we shall model the market filtration as being generated by a set of LRBs.
Through the dynamic consistency property, we can consider each market participant's filtration to be generated by a set of LRBs, regardless of the time in which they enter the market, and without their views being inconsistent with other participants.

The dynamic consistency property was introduced in \citet{BHM1} with regard to Brownian random bridges, and was shown by the same authors to hold for gamma random bridges in \citep{BHM3}.

Fix a time $s<T$.
Given $L_{sT}$, we define a process $\{\eta_{t}\}$ by (\ref{eq:define_eta}).
We shall show that $\{\eta_t\}$ is an LRB.
At time $s$, the law of $\eta_{T}$ is
\begin{equation} 
	\nu^*(A)=\nu_{s}(A+L_{sT}) \qquad \text{for all $A\in\mathcal{B}(\R)$,}
\end{equation}
where $A+y$ denotes the shifted set
\begin{equation}
	A+y= \left\{x: x-y\in A\right\}. 
\end{equation}
Given the terminal value $\eta_{T}$, the finite-dimensional distributions of $\{\eta_{t}\}$ are given by
\begin{align}
	&\Q\left[\left.\eta_{s+t_1}\in\dd x_1,\ldots,\eta_{s+t_n}\in\dd x_n  \,\right| L_{sT}, \eta_{T}=z\right] \nonumber
	\\ &\qquad=\Q\left[\left.L_{s+t_1,T}-L_{sT}\in\dd x_1,\ldots,L_{s+t_n,T}-L_{sT}\in\dd x_n  \,\right| L_{sT}, L_{TT}-L_{sT}=z\right] \nonumber
	\\ &\qquad=\Q\left[\left.L_{s+t_1}-L_{s}\in\dd x_1,\ldots,L_{s+t_n}-L_{s}\in\dd x_n  \,\right| L_{s}, L_{T}-L_{s}=z\right] \nonumber
	\\ &\qquad=\Q\left[\left.L_{t_1}\in\dd x_1,\ldots,L_{t_n}\in\dd x_n  \,\right| L_{T-s}=z\right]		 \nonumber
	\\ &\qquad=\frac{f_{T-s-t_n}\left(z-x_n\right)}{f_{T-s}\left(z\right)} \prod_{i=1}^{n} f_{t_i-t_{i-1}}\left(x_i-x_{i-1} \right),
\end{align}
for every $n\in\N_+$, every $0=t_0<t_1<\cdots<t_n<T-s$, and every $(x_1,\ldots,x_n)\in\R^n$, where $x_0=0$.
Then we have
\begin{multline}
\Q\left[\left.\eta_{s+t_1}\in\dd x_1,\ldots,\eta_{s+t_n}\in\dd x_n, \eta_{T}\in \dd z  \,\right| L_{sT}\right]
\\=\frac{f_{T-s-t_n}\left(z-x_n\right)}{f_{T-s}\left(z\right)} \prod_{i=1}^{n} f_{t_i-t_{i-1}}\left(x_i-x_{i-1} \right) \, \nu^*(\dd z).
\end{multline}
Comparison of this expression to (\ref{eq:LRBlaw}) shows that the process $\{\eta_{s+t}\}_{0\leq t\leq T-s}$ has the law $\lrbc([0,T-s],\{f_t\},\nu^*)$,
and so the law of $\{\eta_t\}_{s\leq t \leq T}$ is  $\lrbc([s,T],\{f_t\},\nu^*)$.

In the discrete case, we define $\{\eta_t\}$ by setting
\begin{align}
	\eta_t=M_{tT}-M_{sT} \qquad (s\leq t \leq T).
\end{align}
Then, given $M_{sT}$, $\{\eta_t\}$ has the law $\lrbd([s,T],\{Q_t\},P^*)$, where $P^*$ is defined by
\begin{equation}
	P^*(a)=P_s(a+M_{sT}).
\end{equation}

\section{Increments of LRBs}
The form of the transition law in Proposition \ref{prop:markov} shows that in general the increments of an LRB are not independent.
The special cases of LRBs with independent increments are discussed later.
A result that holds for all LRBs is that they have stationary increments:
\begin{prop}
	\label{prop:stat}
	For $s,t,u$ satisfying $0\leq s<u<T$ and $0<t\leq T-u$,	we have
	\begin{align}
		\Q\left[L_{u+t, T}-L_{uT} \leq z \left|\, L_{sT} \right.\right]&=\Q[L_{s+t,T}-L_{sT}\leq z\left|\, L_{sT} \right.],
		\\ \intertext{and}
		\Q\left[M_{u+t, T}-M_{uT} \leq z \left|\, M_{sT} \right.\right]&=\Q[M_{s+t,T}-M_{sT}\leq z\left|\, M_{sT} \right.].
	\end{align}
\end{prop}
\begin{proof}
	We provide the proof for $\{L_{tT}\}$.
	The proof for $\{M_{tT}\}$ is similar.
	Throughout the proof we assume that $t<T-u$.
	The case $t=T-u$ follows from the stochastic continuity of $\{L_{tT}\}$.
	First we consider the case $s=0$.
	It follows from (\ref{eq:LRBtranslaw}) that
	\begin{equation}
		\Q[L_{u+t, T}\in\dd y,L_{uT}\in\dd x]=\psi_{u+t}(\R;y)f_t(y-x)f_u(x) \d x \d y.
	\end{equation}
	Then we have
	\begin{align}
		\Q[L_{u+t, T}-L_{uT}\in\dd z,L_{uT}\in\dd x]
		 &=\psi_{u+t}(\R;z+x)f_t(z)f_u(x) \d x \d z \nonumber
		\\ &=\int_{w=-\infty}^{\infty}\frac{f_{T-(u+t)}(w-z-x)}{f_T(w)} \d w \, f_t(z) f_u(x) \d x \d z.
	\end{align}
	Integrating over $x$, and changing the order of integration yields
	\begin{align}
		\Q[L_{u+t, T}-L_{uT}\in\dd z]
			&=\int_{w=-\infty}^{\infty}\int_{x=-\infty}^{\infty}f_{T-(u+t)}(w-z-x)f_u(x) \d x \, \frac{\dd w}{f_T(w)} \, f_t(z)  \d z. \nonumber
		\\ &=\int_{w=-\infty}^{\infty}\frac{f_{T-t}(w-z)}{f_T(w)}\d w \, f_t(z)  \d z \nonumber
		\\ &=\psi_t(\R,z) f_t(z)  \d z \nonumber
		\\ &=\Q[L_{tT}\in\dd z].
	\end{align}
	
	For the case $s>0$, we use the dynamic consistency property.
	For $s$ fixed and $L_{sT}$ given, the process $\{\eta_{uT}\}_{s\leq u\leq T}=\{L_{uT}-L_{sT}\}_{s\leq u\leq T}$ is an LRB with the law $\lrbc([s,T],\{f_t\},\nu^*)$, where $\nu^*(A)=\nu_s(A+L_{sT})$.
	We have
	\begin{align}
		\Q\left[ L_{u+t , T} -L_{uT}\in \dd z \left|\, L_{sT} \right.\right]
		 &=\Q\left[ \eta_{u+t , T} -\eta_{uT}\in \dd z \left|\, L_{sT} \right.\right] \nonumber
		\\ &=\Q\left[ \eta_{t T} \in \dd z \left|\, L_{sT} \right.\right] \nonumber
		\\ &=\int_{-\infty}^{\infty} \frac{f_{T-t}(w-z)}{f_{T-s}(w)}\nu^*(\dd w) \, f_{t-s}(z) \d z \nonumber
		\\ &=\int_{-\infty}^{\infty} \frac{f_{T-t}(w-z+L_{sT})}{f_{T-s}(w-L_{sT})}\nu_s(\dd w) \, f_{t-s}(z) \d z \nonumber
		\\ &=\frac{1}{\psi_s(\R;L_{sT})}
				\int_{-\infty}^{\infty} \frac{f_{T-t}(w-z+L_{sT})}{f_{T}(w)}\nu(\dd w) \, f_{t-s}(z) \d z \nonumber
		\\ &=\frac{\psi_t(\R;z+L_{sT})}{\psi_s(\R;L_{sT})}\, f_{t-s}(z) \d z \nonumber
		\\ &=\Q[L_{tT}-L_{sT} \in \dd z\left|\, L_{sT} \right. ].
	\end{align}
\end{proof}
When $\{L_{tT}\}$ is integrable, the stationary increments property offers enough structure to allow the calculation of the expected value of $L_{tT}$:
\begin{coro}
\label{coro:ExpectLRB}
If $\E[|L_{tT}|]<\infty$ for all $t\in(0,T]$ then
\begin{align}
		\E\left[L_{tT} \left|\, L_{sT} \right.\right]=\frac{T-t}{T-s}L_{sT} + \frac{t-s}{T-s} \E\left[L_{TT} \left|\, L_{sT} \right.\right]
		&& (s<t),
		\\\intertext{and if $\E[|M_{tT}|]<\infty$ for all $t\in(0,T]$ then}
		\E\left[M_{tT} \left|\, M_{sT} \right.\right]=\frac{T-t}{T-s}M_{sT} + \frac{t-s}{T-s} \E\left[M_{TT} \left|\, M_{sT} \right.\right]
		&& (s<t).
\end{align}
\end{coro}
\begin{proof}
	We only provide the proof for $\{L_{tT}\}$ since the proof for $\{M_{tT}\}$ is similar.
	The case $t=T$ is immediate, so we assume that $t<T$.
	First we consider the case $s=0$.
	Suppose that $t=\frac{m}{n}T$, where $m,n \in \mathbb{N}_+$ and $m<n$.
	We wish to show that
	\begin{equation}
		\E[L_{tT}]=\frac{m}{n} \E[L_{TT}].
	\end{equation}
	Writing $L(t,T)=L_{tT}$ for clarity, define the random variables $\{\D{i}\}$ by
	\begin{equation}
		\D{i}=L\left(\tfrac{i}{n}T,T\right)-L\left(\tfrac{(i-1)}{n}T,T\right).
	\end{equation}
	It follows from Proposition \ref{prop:stat} that the $\D{i}$'s are identically distributed, and by assumption they are integrable.
	Hence we have
	\begin{equation}
			\E[\D{i}]=\frac{1}{n} \, \E\left[ \sum_{i=1}^n \D{i} \right]=\frac{1}{n}\,\E[L_{TT}].
	\end{equation}
	Then, as required, we have
	\begin{equation}
		\label{eq:rat_case}
		\E\left[L\left(\tfrac{m}{n}T,T\right)\right]=\E\left[\sum_{i=1}^m \D{i}\right]=\frac{m}{n}\,\E[L_{TT}].
	\end{equation}
	
	For general $t$, choose an increasing sequence of positive rational numbers $\{q_i\}$ such that $\lim_{i\rightarrow\infty} q_i=\frac{t}{T}$.
	From (\ref{eq:rat_case}) we have
	\begin{align}
		\E[L(t,T)]&=\E[L(q_i T,T)] +\E[L(t,T)-L(q_i T,T)] \nonumber
		\\ &= q_i \, \E[L(T,T)] +\E[L(t,T)-L(q_i T,T)].
	\end{align}
	By stochastic continuity, in the limit $i\rightarrow\infty$ the law of $L(t,T)-L(q_i T,T)$ tends to a Dirac measure centred at 0.
	Hence
	\begin{align}
		\E[L(t,T)]&= \lim_{i\rightarrow\infty} q_i \, \E[L(T,T)] +\E[L(t,T)-L(q_i T,T)] \nonumber
		\\ &=\frac{t}{T} \, \E[L(T,T)].
	\end{align}
	
	For the case where $s>0$, we use the dynamic consistency property.
	For $s$ fixed and $L_{sT}$ given, the process 
	\begin{equation}
			\eta_{tT}=L_{tT}-L_{sT} \qquad (s\leq t\leq T)
	\end{equation}
	is an LRB with law $\lrbc([s,T],\{f_t\},\nu^*)$, where $\nu^*(A)=\nu_s(A+L_{sT})$.
	Then we have
	\begin{align}
		\E\left[L_{tT} \left|\, L_{sT} \right.\right]&=L_{sT}+\E[\eta_{tT}\left|\, L_{sT} \right.] \nonumber
		\\ &=L_{sT}+\frac{t-s}{T-s} \int_{-\infty}^{\infty} z \, \nu^*(\dd z) \nonumber
		\\ &=L_{sT}+\frac{t-s}{T-s} \int_{-\infty}^{\infty} (z-L_{sT}) \, \nu_s(\dd z) \nonumber
		\\ &=\frac{T-s}{T-s} L_{sT}+\frac{t-s}{T-s} \E\left[L_{TT} \left|\, L_{sT} \right.\right].
	\end{align}
\end{proof}

We have shown that the increments of LRBs are stationary, and so it is then natural to ask when the increments are independent,
i.e.~when is an LRB a \levy process?
The answer lies in the functional form of $\psi_t(\R;y)$.

For $0\leq s<t<T$, the likelihood that $L_{tT}=y$ given that $L_{sT}=x$ is
\begin{equation}
	\label{eq:qden}
	q(t,y;s,x)=\frac{\psi_t(\R;y)}{\psi_s(\R;x)}f_{t-s}(y-x).
\end{equation}
If $\{L_{tT}\}$ has stationary, independent increments then 
\begin{equation}
	q(t,y;s,x)= q(t-s,y-x;0,0).
\end{equation}
Therefore the ratio
\begin{equation}
	\frac{\psi_t(\R;y)}{\psi_s(\R;x)}
\end{equation}
is a function of the differences $t-s$ and $y-x$.
Thus, if we have that
\begin{equation}
	\label{eq:psiind}
	\psi_t(\R;y)=a\exp(by+ct),
\end{equation}
for constants $a$, $b$ and $c$, then $\{L_{tT}\}$ is a \levy process.
There are constraints on $a$, $b$ and $c$ since (\ref{eq:qden}) is a probability density.
When $b=c=0$ we have $\nu(\dd z)=f_T(z)\d z$ which is the case where $\{L_{tT}\}\law\{L_t\}$.
Note that the subclass of LRBs that are \levy processes is small.

\bigskip

\noindent{\bf Example.}
In the Brownian case we consider a process $\{W_{tT}\}$ with law
	\[\lrbc([0,T],\{f_t\},f_T(z-\th T) \d z),\]
where $f_t(x)$ is the normal density with zero mean and variance $t$ given by (\ref{eq:NormalDensity}).
In other words, $\{W_{tT}\}$ is a standard Brownian motion conditioned so that $W_{TT}$ is a normal random variable with mean $\th T$ and variance $T$.
In this case, we have
\begin{align}
	\psi_t(\R;y)&=\int_{-\infty}^{\infty}\frac{f_{T-t}(z-y)}{f_{T}(z)} f_T(z-\th T) \d z \nonumber
	\\ &=\exp\left(\th y-\frac{\th}{2}t\right).
\end{align}
Simplifying the expression for the transition densities of the process $\{W_{tT}\}$ allows one to verify that $\{W_{tT}\}$ is a Brownian motion with drift $\th$.
It is notable, by Girsanov's theorem, that the process $\{\psi_t(\R;W_t)\}$ is the Radon-Nikod\'ym density process that transforms a standard Brownian motion into a Brownian motion with drift $\th$.
Hence we can alternatively deduce that $\{W_{tT}\}$ is a Brownian motion with drift $\th$ from the analysis in Section \ref{sec:measure}.

\bigskip

\noindent{\bf Example.}
In the gamma case, we consider a process $\{\G_{tT}\}$ with law 
\[\lrbc([0,T],\{f_t\},\kappa^{-1} f_T(z/\kappa) \d z),\]
where $f_t(x)=g^{(m)}_t(x)$ is the gamma density with mean $t$ and variance $t/m$ defined by (\ref{eq:gamma_m_den}), and $\kappa>0$ is constant.
Then $\{\G_{tT}\}$ is a gamma process with mean unity and variance $m^{-1}$ at $t=1$, 
conditioned so that $\G_{TT}$ has a gamma distribution with mean $\kappa T$ and variance $\kappa^2 T/m$.
We have:
\begin{align}
	\psi_t(\R;y)&=\int_{-\infty}^{\infty}\frac{f_{T-t}(z-y)}{f_{T}(z)} \frac{f_T(z/\kappa)}{\kappa} \d z \nonumber
	\\ &=\kappa^{-mt}\exp\left(m(1-\kappa^{-1}) y\right).
\end{align}
The transition density of $\{\G_{tT}\}$ is then
\begin{equation}
	\Q[\G_{tT}\in\dd y \,|\, \G_{sT}=x]=\1_{\{y>x\}}\frac{(y-x)^{m(t-s)-1} \e^{-m(y-x)/\kappa}}{(\kappa/m)^{m(t-s)} \G[m(t-s)]}\d y.
\end{equation}
Hence $\{\G_{tT}\}$ is a gamma process with mean $\kappa$ and variance $\kappa^2/m$ at $t=1$.

\subsection{Increment distributions}
Partition the time interval $[0,T]$ by $0=t_0<t_1<t_2<\cdots<t_n=T$.
Then define the increments $\{\D{i}\}_{i=1}^n$ and $\{\a_i\}_{i=1}^n$ by
\begin{align}
	\label{eq:deltadef}
	\D{i}&=L_{t_i,T}-L_{t_{i-1},T}
	\\ \a_i&=t_i-t_{i-1}.
\end{align}
Assume that $\nu$ has no \emph{continuous singular part} \citep{Sato1999}.
Denoting the Dirac delta function centred at $z$ by $\delta_z(x)$, $x\in\R$, we can then write
\begin{equation}
	\label{eq:nu_nosing}
	\nu(\dd z)=\sum_{i=-\infty}^{\infty} v_i \delta_{z_i}(z)\d z + p(z) \d z,
\end{equation}
for some $\{a_i\}\subset \R$, $\{z_i\}\subset \R_+$, and $p:\R\rightarrow\R_+$.
Here $p(z)$ is the density of the continuous part of $\nu$, and $v_i$ is a point mass of $\nu$ located at $z_i$.
From (\ref{eq:LRBlaw}), the joint law of the random vector $(\D{1},\ldots , \D{n})^{\tp}$ is given by
\begin{equation}
	\label{eq:incden}
	\Q[\D{1}\in\dd y_1\ldots,\D{n}\in\dd y_n]= \widetilde{f}\left( \sum_{i=1}^ny_i \right)	
						\prod_{i=1}^n f_{\a_i}(y_i) \d y_i,
\end{equation}
where
\begin{equation}
	\widetilde{f}(z)=\frac{p(z)+\sum_{i=-\infty}^{\infty} v_i \delta_{z_i}(z)}{f_T(z)}.
\end{equation}
Equation (\ref{eq:incden}) shows that $(\D{1},\ldots , \D{n})^{\tp}$ has a generalized multivariate Liouville distribution as defined by \citet{GRIV}.
The classical multivariate Liouville distribution is obtained when $f_t(x)$ is the density of a gamma distribution (see \citep{GRI,GRII,GRIII,FKN1990}).
A survey of Liouville distributions can be found in \citet{GR2001}.
\citet{BNJ1991} construct a generalized Liouville distribution by conditioning a vector of independent inverse-Gaussian random variables on their sum.

In the discrete case, the joint distribution of increments also has a generalized Liouville distribution.
Define the increments $\{D_i\}$ by
\begin{equation}
	D_{i}=M_{t_i,T}-M_{t_{i-1},T}.
\end{equation}
Then we can write
\begin{equation}
	\Q[D_{1}\in\dd y_1\ldots,D_{n}\in\dd y_n]= \widetilde{Q}\left( \sum_{i=1}^ny_i \right)	
						\prod_{i=1}^n \dd Q_{\a_i}(y_i),
\end{equation}
where
\begin{equation}
	\widetilde{Q}(z)=\frac{\sum_{i=-\infty}^{\infty} P(a_i) \delta_{a_i}(z)}{Q_T(z)}.
\end{equation}

\subsection{The reordering of increments}
We are able to extend the Markov property of LRBs.
If we partition the path of an LRB into increments, then the Markov property means that future increments depend on the past only through the \emph{sum} of past increments.
We will show that for LRBs the ordering of the increments does not matter for this to hold---%
given the values of any set of increments of an LRB (past or future), the other increments depend on this subset only through the sum of its elements.

Let $\pi$ be a permutation of $\{1,2,\ldots,n\}$.
We define the partial sum $S^{\pi}_m$ by
\begin{equation}
	S^{\pi}_m=\sum_{i=1}^m \D{\pi(i)} \qquad \text{for $m=1,2,\ldots,n$,}
\end{equation}
where the $\{\D{i}\}$ are defined as in (\ref{eq:deltadef}),
and we define the partition $0=t^{\pi}_0<t^{\pi}_1<\cdots<t^{\pi}_n=T$ by
\begin{equation}
	t^{\pi}_{j+1}=\sum_{i=1}^{j}\a_{\pi(i)} \qquad \text{for $j=1,2,\ldots,n-1$.}
\end{equation}

\begin{prop} \label{prop:inc}
	We may extend the Markov property of $\{L_{tT}\}$ to the following:
	\begin{multline}
		\label{eq:increment_prop}
		\Q\left[\D{\pi(m+1)}\leq y_{m+1},\ldots,\D{\pi(n)}\leq y_n \left|\, \D{\pi(1)},\ldots,\D{\pi(m)}\right.\right]=
			\\ \Q\left[\left.\D{\pi(m+1)}\leq y_{m+1},\ldots,\D{\pi(n)}\leq y_{n} \,\right| S^{\pi}_m \right].
	\end{multline}
	If $\nu$ has no singular continuous part, then
	\begin{multline}
		\Q\left[\left.\D{\pi(m+1)}\in\dd y_{m+1},\ldots,\D{\pi(n)}\in\dd y_n \,\right| S^{\pi}_m \right]=
		 \\ \frac{\widetilde{f}\left(S^{\pi}_m + \sum_{i=m+1}^n y_i\right)}{\psi_{t^{\pi}_m}(\R;S^{\pi}_m)} 
		\prod_{i=m+1}^n f_{\a_{\pi(i)}}(y_i)\d y_i.
	\end{multline}
\end{prop}
\begin{proof}
	Define the increments $\{\tD{i}\}$ by
	\begin{equation}
		\tD{i}=L_{t^{\pi}_{n},T}-L_{t^{\pi}_{n-1},T}.
	\end{equation}
	The law of the random vector $(\tD{1},\ldots , \tD{n-1},\sum_{1}^n\tD{i})^{\tp}$ is given by
	\begin{multline}
		\Q\left[\tD{1}\in\dd y_{1},\ldots,\tD{n-1}\in \dd y_{n-1},\sum_{i=1}^n\tD{i}\in\dd z\right]=
			\\	\frac{\nu(\dd z)}{f_T(z)}f_{\a_{\pi(n)}}\left(z-\sum_{i=1}^{n-1}y_i \right)\prod_{i=1}^{n-1} f_{\a_{\pi(i)}}(y_i) \d y_i.
	\end{multline}
	This is also the law of $(\D{\pi(1)},\ldots,\D{\pi(n-1)},\sum_1^n\D{\pi(i)})^{\tp}$, hence
	\begin{equation}
		(\D{\pi(1)},\ldots,\D{\pi(n)})\law (\tD{1},\ldots , \tD{n}).
	\end{equation}
	The Markov property of LRBs gives
	\begin{multline}
		\Q\left[\tD{m+1}\leq y_{m+1},\ldots,\tD{n}\leq y_n \left|\, \tD{1},\ldots,\tD{m}\right.\right]=
			\\ \Q\left[\tD{m+1}\leq y_{m+1},\ldots,\tD{n}\leq y_{n} \left|\, \sum_{i=1}^m\tD{i} \right.\right],
	\end{multline}
	and so we have
	\begin{multline}
		\Q\left[\D{\pi(m+1)}\leq y_{m+1},\ldots,\D{\pi(n)}\leq y_n \left|\, \D{\pi(1)},\ldots,\D{\pi(m)}\right.\right]=
			\\ \Q\left[\left.\D{\pi(m+1)}\leq y_{m+1},\ldots,\D{\pi(n)}\leq y_{n} \,\right| S^{\pi}_m \right].
	\end{multline}
	This proves the first part of the proposition.
	
	For the second part of the proof we assume that $\nu$ takes the form (\ref{eq:nu_nosing}).
	Note that $L_{t^{\pi}_m,T}=\sum_{i=1}^m \tD{i}$, and that the density of $L_{t^{\pi}_m,T}$ is	
	\begin{equation}
		x \mapsto f_{t^{\pi}_m}(x)\psi_{t^{\pi}_m}(\R;x)
		=\int_{z=-\infty}^{\infty} \frac{f_{t^{\pi}_m}(x)f_{T-t^{\pi}_m}(z-x)}{f_T(z)}\,\nu(\dd z).
	\end{equation}
  The elements of the vector $(L_{t^{\pi}_m,T},\tD{m+1},\ldots,\tD{n})^{\tp}$ are non-overlapping increments of $\{L_{tT}\}$, and the law of the vector is given by
	\begin{multline}
		\Q\left[L_{t^{\pi}_m,T}\in \dd x, \tD{m+1}\in\dd y_{m+1},\ldots,\tD{n}\in \dd y_{n}\right]=
		\\\widetilde{f}\left(x+ \sum_{i=m+1}^ny_i \right) f_{t^{\pi}_m}(x) \d x \,
					\prod_{i=m+1}^n f_{\a_{\pi(i)}}(y_i) \d y_i.
	\end{multline}
	Thus we have
	\begin{align}
		&\Q\left[\left.\tD{m+1}\in \dd y_{m+1},\ldots,\tD{n}\in \dd y_n \, \right| L_{t^{\pi}_m,T}=x\right] \nonumber
		\\ &\qquad\qquad\qquad=\frac{\Q\left[\tD{m+1}\in \dd y_{m+1},\ldots,\tD{n}\in \dd y_n , L_{t^{\pi}_m,T}\in \dd x \right]}
								{\Q\left[L_{t^{\pi}_m,T}\in \dd x \right]} \nonumber
		\\ &\qquad\qquad\qquad=\frac{\widetilde{f}\left(x+ \sum_{i=m+1}^ny_i \right)\prod_{i=m+1}^n f_{\a_{\pi(i)}}(y_i)}
					{\psi_{t^{\pi}_m}(\R;S^{\pi}_m)}.
	\end{align}
\end{proof}

We note that \citet{GRIV} prove that if $(\D{1},\D{2},\ldots,\D{n})^{\tp}$ has a generalized Liouville distribution then equation (\ref{eq:increment_prop}) holds.

\bigskip
We can use Proposition \ref{prop:inc} to extend the dynamic consistency property.
In particular we have the following:
\begin{coro}\label{coro:inc}
	\begin{enumerate}
		\item[a.]
			Fix times $s_1,T_1$ satisfying $0<T_1\leq T-s_1$.
			The time-shifted, space-shifted partial process
			\begin{equation}
				\eta_{t,T_1}^{(1)}=L_{s_1+t,T}-L_{s_1,T}, \qquad (0\leq t \leq T_1),
			\end{equation}
			is an LRB with the law $\lrbc([0,T_1],\{f_t\},\nu^{(1)})$, where $\nu^{(1)}$ is a probability law on $\R$ with density $f_{T_1}(x)\psi_{T_1}(\R;x)$.
		\item[b.]
			Construct partial processes $\{\eta^{(i)}_{t,T_i}\}$, $i=1,\ldots,n$, from non-overlapping portions of $\{L_{tT}\}$, in a similar way to that above.
			The intervals $[s_i,s_i+T_i]$, $i=1,\ldots,n$, are non-overlapping except possibly at the endpoints.
			Set $\eta^{(i)}_{t,T_i}=\eta^{(i)}_{T_i,T_i}$ when $t>T_i$.
			If $u>t$, then
			\begin{multline}
				\Q\left[\left. \eta^{(1)}_{u,T_1}-\eta^{(1)}_{t,T_1}\leq x_1,\ldots,\eta^{(n)}_{u,T_n}-\eta^{(n)}_{t,T_n}\leq x_n \,\right| \F^{\eta}_t \right]=
				\\\Q\left[ \eta^{(1)}_{u,T_1}-\eta^{(1)}_{t,T_1}\leq x_1,\ldots,\eta^{(n)}_{u,T_n}-\eta^{(n)}_{t,T_n}\leq x_n \left|\, \sum_{i=1}^n \eta^{(i)}_{t,T_i} \right.\right],
			\end{multline}
			where
			\begin{equation}
				\F^{\eta}_{t}=\s\left(\left\{ \eta^{(i)}_{s,T_i} \right\}_{0\leq s \leq t}, i=1,2,\ldots,n  \right).
			\end{equation}
	\end{enumerate}
\end{coro}

\begin{rem}
The partial processes of Corollary \ref{coro:inc} are dependent, and we have
\begin{equation}
	\Q\left[\left. \eta^{(i)}_{tT} \in \dd x \,\right| \F^{\eta}_s \right]=
	\Q\left[\eta^{(i)}_{tT} \in \dd x \left|\, \eta^{(i)}_{sT}, \sum_{j=1}^{n} \eta^{(j)}_{sT} \right.\right],
\end{equation}
for $0\leq s<t \leq T$.
\end{rem}

We state but do not prove a discrete analogue of Proposition \ref{prop:inc}, which is as follows:
\begin{prop}
	We may extend the Markov property of $\{M_{tT}\}$ to the following:
	\begin{multline}
		\Q\left[D_{\pi(m+1)}\leq y_{m+1},\ldots,D_{\pi(n)}\leq y_n \left|\, D_{\pi(1)},\ldots,D_{\pi(m)}\right.\right]=
			\\ \Q\left[\left.D_{\pi(m+1)}\leq y_{m+1},\ldots,D_{\pi(n)}\leq y_{n} \,\right| R^{\pi}_m \right],
	\end{multline}
	where $R^{\pi}_m=\sum_{i=1}^m D_{\pi(i)}$.
	Furthermore,
	\begin{equation}
		\Q\left[\left.D_{\pi(m+1)}= y_{m+1},\ldots,D_{\pi(n)}= y_n \,\right| D^{\pi}_m \right]=
		  \frac{\widetilde{Q}\left(R^{\pi}_m + \sum_{i=m+1}^n y_i\right)}{\sum_{k=-\infty}^{\infty}\phi_{t^{\pi}_m}(a_k;R^{\pi}_m)} 
		\prod_{i=m+1}^n Q_{\a_{\pi(i)}}(y_i).
	\end{equation}
\end{prop}

Corollary \ref{coro:inc} can be extended to include LRBs with discrete state-spaces.

%% file: chap04.tex
%
%

\chapter{Information-based asset pricing} \label{chap:info_based}
We apply LRBs to the modelling of information flow within the information-based framework of Brody, Hughston \& Macrina (BHM).
The approach was applied to credit risk in \citet{BHM1}, and this was extended to include stochastic interest rates in \citet{YR2007}.
A general asset pricing framework was proposed in \citet{BHM2} (see also \citet{MPhD2006}), and there have also been applications to inflation modelling (\citet{HM2008}), insider trading (\citet{BDFH2008}), insurance (\citet{BHM3}), and interest rate theory (\citet{HM2008}).

We model a financial market as a collection of cash flows occurring on fixed dates.
We assume that each cash flow can be expressed as a function of independent market factors, which we call $X$-factors.
Under the assumptions that interest rates are deterministic and that the pricing kernel is given, the no-arbitrage price of a cash flow is its discounted expected value under the risk-neutral measure.

Each $X$-factor is revealed to the market through a factor information process.
We model a factor information process as an LRB whose terminal value is the value of an $X$-factor.
The market information process is then a collection of all factor information processes, and this process generates the market filtration.
Hence, we are modelling the information flow in the market by our choice of $X$-factors and factor information processes.
The dynamics of cash flow prices are derived by taking conditional expectations with respect to the market filtration.

Stylistic path properties of price processes can be incorporated into the model by an appropriate choice of factor information processes.
For example, if a cash flow depends on an $X$-factor whose information process is a pure-jump LRB, then the cash flow's price process will be a pure-jump process.
Dependence between two cash flows can be modelled by expressing them as functions of a common set of $X$-factors.

To simplify matters, we examine the case of a cash flow that pays an amount equal to the value of a single $X$-factor.
We derive the price process for the cash flow, and we derive an expression for a call option on this price.
Although this model is simple in structure, because the class of information (LRB) processes is large, the results are quite general.
In the special case that the $X$-factor can take only two values, we recover a generalisation of the binary bond model of \citet{BHM1}.
All the results of this chapter are for a general LRB information process.
We consider specific examples in the subsequent chapters.

\section{BHM framework}
We fix a finite time horizon $[0,T]$ and a probability space $(\Omega,\F,\Q)$.
We assume that the risk-free rate of interest $\{r_t\}$ is deterministic, and that 
$r_t>0$ and $\int_t^{\infty}r_u \d u=\infty$, for all $t>0$.
Then the time-$s$ (no-arbitrage) price of a risk-free, zero-coupon bond maturing at time $t$ (paying a nominal amount of unity) is
\begin{equation}
	P_{st}=\exp\left( -\int_s^t r_u \d u \right) \qquad (s\leq t).
\end{equation}
For $t<T$, the time-$t$ price of an integrable contingent cash flow $H_T$, due at time $T$, is given by an expression of the form
\begin{equation}
	\label{eq:price_formula}
	H_{tT}=P_{tT} \, \E[H_T \,|\, \F_t],
\end{equation}
where $\{\F_t\}$ is the \emph{market filtration}.
The sigma-algebra $\F_t$ represents all the information available to market participants at time $t$.
In order for equation (\ref{eq:price_formula}) to be consistent with the theory of no-arbitrage pricing, we must interpret $\Q$ to be the risk-neutral measure.

In such a set-up, the dynamics of the price process $\{H_{tT}\}$ are implicitly determined by the evolution of the market filtration $\{\F_t\}$.
We assume the existence of a (possibly multi-dimensional) \emph{information process} $\{\xi_{tT}\}_{0\leq t \leq T}$ such that
\begin{equation}
	\F_t=\s \left(\{\xi_{sT}\}_{0\leq s \leq t} \right).
\end{equation}
So $\{\xi_{tT}\}$ is responsible for the delivery of all information to the market participants.
The task of modelling the emergence of information in the market is reduced to that of specifying the law of the information process $\{\xi_{tT}\}$.

\subsection{Single $X$-factor market}
We assume that the cash flow $H_T$ can be written in the form
\begin{equation}
	H_T=h(X_T),
\end{equation}
for some function $h(x)$, and some market factor $X_T$.
We call $X_T$ an $X$-factor.
We assume that $\{\xi_{tT}\}$ is a one-dimensional process such that $\xi_{TT}=X_T$.
Then we have
\begin{equation}
	H_{tT}=P_{tT}\, \E[h(X_T) \,|\, \F_t]=P_{tT}\, \E[h(\xi_{TT}) \,|\, \F_t],
\end{equation}
which ensures that $H_{TT}=H_T$.
In the case where $\{\xi_{tT}\}$ is a Markov process, we have
\begin{equation}
	H_{tT}= P_{tT} \,\E[h(\xi_{TT}) \,|\, \xi_{tT}].
\end{equation}

\subsection{Multiple $X$-factor market}
In the more general framework, we model a financial asset that generates the $N$ cash flows $H_{T_1},H_{T_2},\ldots,H_{T_N}$, which are to be received on the dates $T_1\leq T_2\leq \cdots \leq T_N$, respectively.
At time $T_k$, we assume that the vector of $X$-factors $X_{T_k}\in\R^{n_k}$ ($n_k\in\p{N}_+$) is revealed to the market, and we write
\begin{equation}
	X_{T_k}=\left(X^{(1)}_{T_k},X^{(2)}_{T_k},\ldots,X^{(n_k)}_{T_k}\right)^{\tp}.
\end{equation}
We assume the $X$-factors are mutually independent, and that
\begin{equation}
	H_{T_k}=h_k(X_{T_1},X_{T_2},\ldots,X_{T_k}),
\end{equation}
for some $h_k:\R^{n_1}\times\R^{n_2}\times\cdots\times\R^{n_k}\rightarrow\R$ which we call a cash-flow function.
For each $X$-factor $X^{(i)}_{T_j}$, there is a factor information process $\{\xi_t^{(i,j)}\}$ such that $\xi_t^{(i,j)}=X^{(i)}_{T_j}$ for $t\geq T_j$, and the factor information processes are mutually independent.
Setting $T=T_N$, we define the market information process $\{\xi_{tT}\}$ to be an $\R^{n_1+n_2+\cdots+n_N}$-valued process with each of its elements being a factor information process.
The market filtration $\{\F_t\}$ is generated by $\{\xi_{tT}\}$.
By construction, $H_{T_k}$ is $\F_t$-measurable for $t\geq T_k$.
The time-$t$ price of the cash flow $H_{T_k}$ is
\begin{equation}
	H_{t}^{(k)}=\left\{
		\begin{aligned}
			&P_{t,T_k} \, \E\left[\left. h_k(X_{T_1},X_{T_2},\ldots,X_{T_k}) \,\right| \F_t\right] && \text{for $t< T_k$,}
			\\ &0 &&\text{for $t\geq T_k$.}
		\end{aligned}
		\right.
\end{equation}
Here we adopt the the convention that cash flows have nil value at the time that they are due.
In other words, prices are quoted on an ex-dividend basis.
In this way the price process $\{H_t^{(k)}\}$ is right-continuous at $t=T_k$.
The asset price process is then
\begin{equation}
	H_{tT}= \sum_{k=1}^n  H^{(k)}_{t} \qquad (0\leq t\leq T).
\end{equation}

\section{\levy bridge information}
We consider a market with a single factor, which we denote $X_T$.
This $X$-factor is the size of a contingent cash flow to be received at time $T>0$, so we take $h(x)=x$.
For example, $X_T$ could be the redemption amount of a credit risky bond.
$X_T$ is assumed to be integrable and to have the \emph{a priori} probability law $\nu$ (we also exclude the case where $X_T$ is constant).
Information is supplied to the market by an information process $\{\xi_{tT}\}$.
The law of $\{\xi_{tT}\}$ is $\lrbc([0,T],\{f_t\},\nu)$, and we set $\xi_{TT}=X_T$.
We assume throughout this chapter that the information process has a continuous state-space; the results can be extended to include LRB information processes with discrete state-spaces.
Indeed, in Chapter \ref{chap:Poisson} we give a detailed example of a discrete LRB, and apply it to the pricing of credit derivatives.

Since the information process has the Markov property, the price of the cash flow $X_T$ is given by
\begin{equation} 
	X_{tT}=P_{tT}\, \E\left[X_T\left|\, \xi_{tT} \right. \right] \qquad (0\leq t \leq T). 
\end{equation}
We note that $X_T$ is $\F_T$-measurable and $X_{TT}=X_T$, but $X_T$ is not $\F_t$-measurable for $t<T$ since we have excluded the case where $X_T$ is constant.
For $t\in(0,T)$, the $\F_t$-conditional law of $X_T$ as given by equation (\ref{eq:C_terminal}) is
\begin{equation} 
	\nu_{t}(\dd z)=\frac{\psi_t(\dd z;\xi_{tT})}{\psi_t(\R;\xi_{tT})}, 
\end{equation}
where 
\begin{equation}
	\psi_t(\dd z;\xi)=\frac{f_{T-t}(z-\xi)}{f_T(z)} \d z.
\end{equation}
Then we have
\begin{equation}
	X_{tT}=P_{tT} \int_{-\infty}^{\infty} z \, \nu_t(\dd z).
\end{equation}
When $\nu$ admits a density $p(z)$, the $\F_t$-conditional density of $X_T$ exists and is given by
\begin{equation} 
	p_t(z)= \frac{f_{T-t}(z-\xi_{tT})p(z)}{\psi_t(\R;\xi_{tT}) f_T(z)}.
\end{equation}

\bigskip

\noindent{\bf Example.}
In the Brownian case the price is
\begin{equation}
	X_{tT}=P_{tT} \frac{\int_{-\infty}^{\infty} z \, \e^{\frac{1}{T-t}\left[\xi_{tT}z-\half \frac{t}{T}z^2\right]}\,\nu(\dd z)}
				{\int_{-\infty}^{\infty} \e^{\frac{1}{T-t}\left[\xi_{tT}z-\half \frac{t}{T}z^2\right]}\,\nu(\dd z)}.
\end{equation}
The following SDE can be derived for $\{X_{tT}\}$ (see \citep{BHM1,BHM2,MPhD2006,YR2007}):
\begin{equation}
	\label{eq:SDE}
	\dd X_{tT}=r_t X_{tT} \d t+ \frac{P_{tT} \var[X_T \,|\,\xi_{tT}]}{T-t} \d W_t,
\end{equation}
where $\{W_t\}$ is an $\{\F_t\}$-Brownian motion.

\bigskip

\noindent{\bf Example.}
In the gamma case we have
\begin{equation}
	X_{tT}=P_{tT}\frac{\int_{\xi_{tT}}^{\infty}(z-\xi_{tT})^{m(T-t)-1}z^{2-mT}\,\nu(\dd z)}{\int_{\xi_{tT}}^{\infty}(z-\xi_{tT})^{m(T-t)-1}z^{1-mT}\,\nu(\dd z)}.
\end{equation}

\section{European option pricing} \label{sec:Call_Option}
We consider the problem of pricing a European option on the price $X_{tT}$ at time $t$.
For a strike price $K$ and $0\leq s<t<T$, the time-$s$ price of a $t$-maturity call option on $X_{tT}$ is
\begin{equation} 
	C_{st}=P_{st} \, \E\left[\left.(X_{tT}-K)^+\,\right| \xi_{sT} \right].
\end{equation}
The expectation can be expanded in the form
\begin{align}
	\E_{\Q}\left[\left.(X_{tT}-K)^+\,\right| \xi_{sT}\right]
	 &=\E_{\Q}\left[\left.\left(P_{tT}\,\E[X_T\,|\,\xi_{tT}]-K  \right)^+\,\right| \xi_{sT}\right] \nonumber
	\\ &=\E_{\Q}\left[\left.\left(\int_{-\infty}^{\infty}(P_{tT}z-K) \, \nu_t(\dd z) \right)^+\,\right| \xi_{sT}\right] \nonumber
	\\ &=\E_{\Q}\left[\left.\frac{1}{\psi_t(\R;\xi_{tT})} \left(\int_{-\infty}^{\infty}(P_{tT}z-K) \, \psi_t(\dd z;\xi_{tT}) \right)^+\,\right| \xi_{sT}\right].
\end{align}
Recall that the Radon-Nikodym density process
\begin{equation}
	\left.\frac{\dd \p{L}}{\dd \Q}\right|_{\F_t}=\psi_t(\R;\xi_{tT})^{-1} 
\end{equation}
defines a measure $\p{L}$ under which $\{\xi_{tT}\}_{0\leq t<T}$ is a \levy process.
By changing measure, we find that the expectation is
\begin{multline}
	\frac{1}{\psi_s(\R;\xi_{sT})}\,
	\E_{\p{L}}\left[\left.\left(\int_{-\infty}^{\infty}(P_{tT}z-K) \, \psi_t(\dd z;\xi_{tT}) \right)^+\,\right| \xi_{sT}\right]=
	\\ \frac{1}{\psi_s(\R;\xi_{sT})}\,
	\E_{\p{L}}\left[\left.\left(\int_{-\infty}^{\infty}(P_{tT}z-K)\frac{f_{T-t}(z-\xi_{tT})}{f_T(z)} \, \nu(\dd z) \right)^+\,\right| \xi_{sT}\right].
\end{multline}
Equation (\ref{eq:density_conditions}) states that $0<f_{T-s}(z-\xi_{sT})<\infty$, and we have
\begin{equation}
	\label{eq:nu_s}
	\nu_s(\dd z)=\psi_s(\R;\xi_{sT})^{-1} \frac{f_{T-s}(z-\xi_{sT})}{f_T(z)} \, \nu(\dd z).
\end{equation}
Thus we can write the expectation in terms of the $\xi_{sT}$-conditional terminal law $\nu_s$ in the form
\begin{multline}
	\E_{\p{L}}\left[\left.\left(\int_{-\infty}^{\infty}
			(P_{tT}z-K)\frac{f_{T-t}(z-\xi_{tT})}{f_{T-s}(z-\xi_{sT})} \, \nu_s(\dd z) \right)^+\,\right| \xi_{sT}\right]
	=\\ \int_{-\infty}^{\infty} \left(\int_{-\infty}^{\infty}
			(P_{tT}z-K)\frac{f_{T-t}(z-x)}{f_{T-s}(z-\xi_{sT})} \, \nu_s(\dd z) \right)^+f_{t-s}(x-\xi_{sT}) \d x.
\end{multline}
We defined the (marginal) \levy bridge density $f_{tT}(x;z)$ by
\begin{equation} 
	f_{tT}(x;z)=\frac{f_{T-t}(z-x)f_{t}(x)}{f_{T}(z)}.
\end{equation}
From this we can define the $\xi_{sT}$-dependent law $\mu_{st}(\dd x;z)$ by
\begin{equation} 
	\mu_{st}(\dd x;z)=f_{t-s,T-s}(x-\xi_{sT},z-\xi_{sT}) \d x. 
\end{equation}
Thus $\mu_{st}(\dd x;z)$ is the time-$t$ marginal law of a \levy bridge starting at the value $\xi_{sT}$ at time $s$, and terminating at the value $z$ at time $T$.
Defining the set $B_t$ by
\begin{align} 
	B_{t}&=\left\{ x\in\R: \int_{-\infty}^{\infty}(P_{tT}z-K)\frac{f_{T-t}(z-x)}{f_{T-s}(z-\xi_{sT})} \, \nu_s(\dd z)>0 \right\}\nonumber
	\\ &=\left\{ x\in\R: \int_{-\infty}^{\infty}(P_{tT}z-K)\frac{f_{T-t}(z-x)}{f_{T}(z)} \, \nu(\dd z)>0 \right\},
\end{align}
where the second equality follows from (\ref{eq:nu_s}), the expectation reduces to
\begin{equation}
	\int_{-\infty}^{\infty}(P_{tT}z-K)\mu_{st}(B_{t};z) \, \nu_s(\dd z).
\end{equation}
Then the option price is given by
\begin{equation} 
	C_{st}=P_{st} \int_{-\infty}^{\infty}(P_{tT}z-K)\mu_{st}(B_{t};z) \, \nu_s(\dd z).
\end{equation}

We can write $X_{tT}=\Lambda(t,\xi_{tT})$, for $\Lambda$ a deterministic function.
The set $B_t$ can then be written
\begin{equation}
	B_t=\left\{ \xi\in\R: \Lambda(t,\xi)>K \right\}.
\end{equation}
We see that if $\Lambda$ is increasing in its second argument then $B_t=(\xi^*_t,\infty)$ for some critical value $\xi^*_t$ of the information process.
$\Lambda$ is monotonic if the $\{\xi_{tT}\}$ is a \levy process.

\bigskip

\noindent{\bf Example.}
In the Brownian case we have
\begin{equation}
	\Lambda(t,x)=P_{tT} \frac{\int_{-\infty}^{\infty} z \, \e^{\frac{1}{T-t}\left[x z-\half \frac{t}{T}z^2\right]}\,\nu(\dd z)}
				{\int_{-\infty}^{\infty} \e^{\frac{1}{T-t}\left[x z-\half \frac{t}{T}z^2\right]}\,\nu(\dd z)}.
\end{equation}
It can be shown that the function $\Lambda$ is increasing in its second argument (see \citep{BHM2,YR2007}); hence $B_t=(\xi_t^*,\infty)$ for the unique $\xi_t^*$ satisfying $\Lambda(t,\xi_t^*)=K$.
A short calculation verifies that $\mu_{st}(\dd x;z)$ is the normal law with mean $M(z)$ and variance $V$ given by
\begin{align}
	&M(z)=\frac{T-t}{T-s} \xi_{sT} + \frac{t-s}{T-s} z,
	&&V= \frac{t-s}{T-s}(T-t).
\end{align}
This is the time-$t$ marginal law of a Brownian bridge starting from the value $\xi_{sT}$ at time $s$, and finishing at the value $z$ at time $T$.
We have
\begin{equation}
	\mu_{st}(B_t;z)=1-\Phi\left[\frac{\xi_t^*-M(z)}{\sqrt{V}}\right]=\Phi\left[\frac{M(z)-\xi_t^*}{\sqrt{V}}  \right],
\end{equation}
where $\Phi[x]$ is the standard normal distribution function.
The option price is then
\begin{equation} 
	C_{st}=P_{sT} \int_{-\infty}^{\infty} z\, \Phi\left[\frac{M(z)-\xi_t^*}{\sqrt{V}}\right] \nu_s(\dd z)
			+P_{st}K \int_{-\infty}^{\infty}\Phi\left[\frac{M(z)-\xi_t^*}{\sqrt{V}}  \right] \nu_s(\dd z).
\end{equation}

\bigskip

\noindent{\bf Example.}
In the gamma case we have 
\begin{equation}
	\Lambda(t,x)=P_{tT}\frac{\int_{x}^{\infty}(z-x)^{m(T-t)-1}z^{2-mT}\,\nu(\dd z)}
				{\int_{x}^{\infty}(z-x)^{m(T-t)-1}z^{1-mT}\,\nu(\dd z)}.
\end{equation}
The monotonicity of $\Lambda(t,x)$ in $x$ is proved for $m(T-t)>1$ by \citet{BHM3}.
The authors also give a numerical example where $\Lambda(t,x)$ was not monotonic in $x$ for $m(T-t)<1$.
For all $t\in(0,T)$, we have
\begin{equation}
	\mu_{st}(\dd x;z)=\1_{\{\xi_{sT}<x< z\}}\, k(z)^{-1}
		\left(\frac{x-\xi_{st}}{z-\xi_{sT}}\right)^{m(t-s)-1} \left(\frac{z-y}{z-\xi_{sT}}\right)^{m(T-t)-1}
					 \d x,
\end{equation}
where $k(z)$ is the normalising constant
\begin{equation}
	k(z)={(z-\xi_{sT}) \, \mathrm{B}[m(t-s),m(T-t)]}.
\end{equation}So $\mu_{st}(\dd x;z)$ is an $(z-\xi_{sT})$-scaled, $\xi_{sT}$-shifted beta-law with parameters $\a=m(t-s)$ and $\b=m(T-t)$.
This is the time-$t$ marginal law of a gamma bridge starting at the value $\xi_{sT}$ at time $s$, and terminating at the value $x$ at time $T$.
When $m(T-t)>1$, a critical $\xi^*_t$ exists such that $\Lambda(t,\xi_t^*)=K$.
Then $B_t=(\xi^*_t,\infty)$, and
\begin{align}
	\mu_{st}(B_t;z)&=1-I\left[\frac{\xi^*_t-\xi_{sT}}{z-\xi_{sT}};m(t-s),m(T-t)\right]\nonumber
	\\ &=I\left[\frac{z-\xi^*_t}{z-\xi_{sT}};m(T-t),m(t-s)\right],
\end{align}
where $I[z;\a,\b]=I_z[\a,\b]$ is the regularized incomplete beta function.
The option price is then given by
\begin{multline} 
	C_{st}=P_{sT} \int_{\xi_{sT}}^{\infty} z\, I\left[\frac{z-\xi^*_t}{z-\xi_{sT}};m(T-t),m(t-s)\right] \nu_s(\dd z)
			\\+P_{st}K \int_{\xi_{sT}}^{\infty}I\left[\frac{z-\xi^*_t}{z-\xi_{sT}};m(T-t),m(t-s)\right] \nu_s(\dd z).
\end{multline}

\section{Binary bond} \label{sec:Binary_Bond}
The simplest non-trivial contingent cash flow is $X_T\in\{k_0,k_1\}$, for $k_0< k_1$.
This is the pay-off from a zero-coupon, credit-risky bond that has principal $k_1$, and a fixed recovery rate $k_0/k_1$ on default.
Assume that, \emph{a priori}, $\Q[X_T=k_0]=p>0$ and $\Q[X_T=k_1]=1-p$.
Then
\begin{align}
	\Q[X_{T}=k_0 \,|\, \xi_{tT} ]&=\left(1+{\frac{f_T(k_0)}{f_T(k_1)}\frac{f_{T-t}(k_1-\xi_{tT})}{f_{T-t}(k_0-\xi_{tT})}\frac{1-p}{p}}  \right)^{-1},
	\\\intertext{and} \Q[X_{T}=k_1 \,|\, \xi_{tT} ]&=\left(1+{\frac{f_T(k_1)}{f_T(k_0)}\frac{f_{T-t}(k_0-\xi_{tT})}{f_{T-t}(k_1-\xi_{tT})}\frac{p}{1-p}}  \right)^{-1}.
\end{align}
The bond price process $\{X_{tT}\}$ associated with the given terminal cash flow is given by
\begin{equation}
	X_{tT}=P_{tT} \left(k_0\,\Q[X_{T}=k_0 \,|\, \xi_{tT} ]+k_1\,\Q[X_{T}=k_1 \,|\, \xi_{tT} ]  \right) \qquad(0\leq t \leq T).
\end{equation}

\bigskip
\noindent{\bf Example.}
In the Brownian case we have
\begin{align}
	\Q[X_{T}=k_0 \,|\, \xi_{tT} ]&=
		\left(1+\exp\left[-\half \frac{k_1-k_0}{T-t}(\tfrac{t}{T}(k_0+k_1)-2\xi_{tT})\right] \frac{1-p}{p}  \right)^{-1},
	\\\intertext{and} \Q[X_{T}=k_1 \,|\, \xi_{tT} ]&=
	 \left(1+\exp\left[\half \frac{k_1-k_0}{T-t}(\tfrac{t}{T}(k_0+k_1)-2\xi_{tT})\right] \frac{p}{1-p}  \right)^{-1}.
\end{align}
Writing $\rho_i=\Q[X_{T}=k_i \,|\, \xi_{tT} ]$, note that
\begin{align}
	\var[X_T\,|\,\xi_{tT}]&=(k_1-k_0)^2\rho_1 \rho_0 \nonumber
	\\&=-(k_0-k_0\rho_0-k_1\rho_1)(k_1-k_0\rho_0-k_1\rho_1) \nonumber
	\\ &=-(k_0-X_{tT})(k_1-X_{tT}).
\end{align}
Thus, recalling (\ref{eq:SDE}), we see that the SDE of $\{X_{tT}\}$ is
\begin{equation}
	\dd X_{tT}=r_t X_{tT} \d t-\frac{P_{tT}(k_0-X_{tT})(k_1-X_{tT})}{T-t} \d W_t,
\end{equation}
with the initial condition $X_{0T}=P_{0T}(k_0p+k_1(1-p))$.
For $K\in(P_{tT}k_0,P_{tT}k_1)$, we are able to solve the equation $\Lambda(t,x)=K$ for $x$.
We have
\begin{align}
	\Lambda(t,x)&=P_{tT} \left(k_0\, \Q[X_T=k_0\,|\,\xi_{tT}=x]+k_1\, \Q[X_T=k_1\,|\,\xi_{tT}=x]\right) \nonumber
		\\ &=P_{tT} \left(k_1-(k_1-k_0)\, \Q[X_T=k_0\,|\,\xi_{tT}=x]\right),
\end{align}
so the solution to $\Lambda(t,x)=K$ is
\begin{equation}
	\xi_t^*=\frac{t}{2T}(k_0+k_1)-\frac{T-t}{k_1-k_0} \, \log\left[\frac{p}{1-p}\frac{K-P_{tT}k_0}{P_{tT}k_1-K} \right].
\end{equation}
The price of a call option on $X_{tT}$ is
\begin{equation}
	C_{st}=P_{st}\sum_{i=0}^1 (P_{tT}k-K)\, \Phi\left[\frac{M(k_i)-\xi^*_t}{\sqrt{V}} \right] \Q[X_T=k_i\,|\,\xi_{sT}].
\end{equation}

%% file: chap05.tex
%
%

\chapter{Variance-gamma information} \label{chap:VG}
We presented some of the properties of the VG process and the VG bridge in Section \ref{sec:VG}.
We now consider the (standard) variance gamma random bridge (VGRB).

In a similar way to a VG bridge, we have two terminal value decompositions of a VGRB.
The first decomposition writes a VGRB in terms of its terminal value, a Brownian bridge, a gamma bridge, and a random volatility factor.
The second decomposition writes a VGRB in terms of its terminal value, two gamma bridges, and a random volatility factor.
These lead to two efficient algorithms for the simulation of VGRBs.

We demonstrate the flexibility of VG information models by way of two simple examples.
The first example recovers the equity model of \citet{MCC1998}.
In this case we consider a single $X$-factor $X_T$ which, after a linear transformation, represents the log-return on a non-dividend-paying stock over the time period $[0,T]$.
We assume that, \emph{a priori}, $X_T$ is an asymmetric VG random variable, and that the market filtration is generated by a standard VGRB whose terminal value is pinned to the value of $X_T$.
Under these conditions, we show that the stock price process is an exponentiated asymmetric VG process.

The second example is an application to credit risk.
We price a binary bond when the market filtration is generated by a VGRB.
Following \citet{BHM1}, we include a rate parameter $\s$ which controls the speed at which information is released.
Through simulations, we explore the effect of $\s$ on the sample paths of bond price processes in the case of default.
When $\s$ is high, information about the impending default enters the market quickly, and the price of the bond becomes small well before the maturity date.
When $\s$ is low, information about the default is slow to reach the market, and there is a sudden collapse in the bond price close to the maturity date.
Further simulations are performed that demonstrate the effect of the VG parameter $m$ on bond price trajectories,
where $m^{-1}$ was the variance of the gamma time change at $t=1$.
When $m$ is small, bond price processes exhibit more large jumps.
When $m$ is large, bond price processes are smoother, and look more like the prices generated in \citep{BHM1} using Brownian information.

\section{Variance-gamma random bridge} \label{sec:VGRB}
Let $\{V_{tT}\}$ be a standard VGRB, so in the transition law (\ref{eq:LRBtranslaw}) we take $f_t(x)=f^{(m)}_t(x)$ which we defined in (\ref{eq:VGden}) to be
\begin{equation}
		f_t^{(m)}(x)=\sqrt{\frac{2}{\pi}}\frac{m^{mt}}{\G[mt]}\left(\frac{x^2}{2m} \right)^{\frac{mt}{2}-\frac{1}{4}} 
			K_{mt-\half}\left[ \sqrt{2mx^2} \right],
\end{equation}
for $m>0$.
We assume that $V_{TT}$ has marginal law $\nu$.
To ensure that $0<f_T(x)<\infty$ $\nu$-a.s., we require that $\Q[V_{TT}=0]=\nu(\{0\})=0$ or $T>(2m)^{-1}$.
From (\ref{eq:VGdecompI}), we can decompose $\{V_{tT}\}$ as
\begin{equation}
	\label{eq:VGRBI}
	V_{tT} = V_{TT} \gb{t} +\sqrt{\Sigma_T} \, \b(\gb{t}) \qquad (0\leq t \leq T),
\end{equation}
where 
(i) $\{\b(t)\}$ is a Brownian bridge to the value 0 at time 1, 
(ii) $\{\gb{t}\}$ is a gamma bridge to the value 1 at time $T$ (with parameter $m>0$), and 
(iii) given $V_{TT}$, $\Sigma_T$ is a GIG random variable with parameter set $\{mT-1/2,|V_{TT}|,\sqrt{2m}\}$.
Also, from (\ref{eq:VGdecompII}), we can decompose $\{V_{tT}\}$ as
\begin{equation}
	\label{eq:VGRBII}
	V_{tT}=  V_{TT}\gbb{t} +  Y_T \mu\left(\gbb{t}-\gb{t} \right) \qquad (0\leq t \leq T),
\end{equation}
where 
(i) $\mu=(m/2)^{-1/2}$, 
(ii) $\{\gb{t}\}$ and $\{\gbb{t}\}$ are independent gamma bridges to the value 1 at time $T$, and 
(iii) given $V_{TT}$, $Y_T$ is a random variable with density
\begin{equation}
	\label{eq:condY_X}
 y\mapsto \1_{\{y>(-V_{TT})^+\}}\frac{m^{2mt}}{\G[mT]^2 f_T^{(m)}(V_{TT})} (yV_{TT}+y^2)^{mT-1} \e^{-m(V_{TT}+2y)}.
\end{equation}

\subsection{Example}
Recall that in (\ref{eq:VG_k}) we defined $k_{(m,\th,\rho)}$ as
\begin{equation}
	k_{(m,\th,\s)}=\sqrt{1+\frac{\th^2}{2m\s^2}}.
\end{equation}
Define $\rho>0$ by
\begin{equation}
	\label{eq:VGrho}
	\rho^2=\half \left(1+\sqrt{1+\frac{2\th^2}{m}}\right),
\end{equation}
then  $\rho$ satisfies $\rho=k_{(m,\th,\rho)}$.
Now suppose that, \emph{a priori}, $V_{TT}$ is a VG random variable with density $f_T^{(m,\th,\rho)}(x)$ as given by (\ref{eq:AVGden}).
We then have
\begin{align}
	\psi_t(\dd z; \xi)&=\frac{f^{(m)}_{T-t}(z-\xi)}{f_T^{(m)}(z)}f^{(m,\th,\rho)}_{T}(z) \d z  \nonumber
	\\ &= \e^{\th\xi/\rho^2}\left(k_{(m,\th,\rho)}\right)^{-2mt}f_{T-t}^{(m,\th,\rho)}(z-\xi) \d z,
\end{align}
for $t\in[0,T)$.
The transition law of $\{V_{tT}\}$ is
\begin{align}
	\Q\left[V_{tT} \in \dd y \left|\, V_{sT}=x\right.\right]&=\frac{\psi_t(\R;y)}{\psi_s(\R;x)} f_{t-s}^{(m)}(y-x)\d y \nonumber
	\\ &=\e^{\th(y-x)/\rho^2}\left(k_{(m,\th,\rho)}\right)^{-2m(t-s)}f_{t-s}^{(m)}(y-x) \d y \nonumber
	\\ &=f_{t-s}^{(m,\th,\rho)}(y-x) \d y,
\end{align}
and similarly
\begin{equation}
	\Q\left[V_{TT} \in \dd y \left|\, V_{sT}=x\right.\right]=f_{T-s}^{(m,\th,\rho)}(y-x) \d y.
\end{equation}
Over the time period $[0,T]$, $\{V_{tT}\}$ is then a two-parameter VG process (since $\rho$ is a function of $\th$ and $m$).
The scaled LRB $\{\s V_{tT} / \rho\}$ is a three parameter VG process with parameters $m$, $\th$, and $\s$.
We have shown that an asymmetric VG process can be constructed as a random bridge of a symmetric VG process.

\section{Simulation}
In this section we will assume that we can generate sample paths for Brownian and gamma bridges (see \citet{ALT2003}, and \citet{RW2002}),
gamma and GIG random variates (see \citet{D1986}), and random variates from the law of $V_{TT}$.

From the constructions (\ref{eq:VGRBI}) and (\ref{eq:VGRBII}), we have two natural ways of simulating paths of VGRBs.
Indeed, all we need to do is be able to sample from the distributions of $Y_T$ and $\Sigma_T$, given the value of $V_{TT}$.
Since the conditional distribution of $\Sigma_T$ is GIG, this causes no problems.
To sample from the conditional distribution of $Y_T$, we note the following:
If $V_{TT}\leq 0$ then, from (\ref{eq:condY_X}), the density of $Y_T$ is proportional to
\begin{align}
	\1_{\{y>-V_{TT}\}}(y(V_{TT}+y))^{mT-1}\e^{-2my} &\leq \1_{\{y>-V_{TT}\}}y^{2mT-2}\e^{-2my} \nonumber
	\\ &\propto\1_{\{y>-V_{TT}\}} g_{T-\frac{1}{2m}}^{(m)}(y),
\end{align}
where $g^{(m)}_t(x)$ is the gamma density given in (\ref{eq:gamma_m_den}).
If $V_{TT}>0$ then the density of $Y_T$ is proportional to
\begin{align}
	\1_{\{y>0\}}(y(V_{TT}+y))^{mT-1}\e^{-(V_{TT}+2y)} &\leq \1_{\{y>-V_{TT}\}}(V_{TT}+y)^{2mT-2}\e^{-2m(V_{TT}+y)} \nonumber
	\\ &\propto\1_{\{y>0\}} g_{T-\frac{1}{2m}}^{(m)}(V_{TT}+y).
\end{align}
So, given $V_{TT}$, if $T>(2m)^{-1}$ we can simulate $Y_T$ with the following acceptance-rejection algorithm:

\noindent If $V_{TT}\leq 0$
\begin{enumerate}
	\item
		Generate $G$ from the gamma distribution with density $g_{T-\frac{1}{2m}}^{(m)}(V_{TT}+y)$.
	\item
		If $G\leq -V_{TT}$ go to step 1.
	\item
		Generate $U$ from the standard uniform distribution.
	\item
		If $(1+V_{TT}/G)^{mT-1}<U$ go to step 1.
	\item
		Return $G$.
\end{enumerate} 

\noindent If $V_{TT}> 0$
\begin{enumerate}
	\item
		Generate $G$ from the gamma distribution with density $g_{T-\frac{1}{2m}}^{(m)}(V_{TT}+y)$.
	\item
		If $G\leq V_{TT}$ go to step 1.
	\item
		Generate $U$ from the standard uniform distribution.
	\item
		If $(1+V_{TT}/G)^{1-mT}<U$ go to step 1.
	\item
		Return $G-V_{TT}$.
\end{enumerate}

\section{VG equity model} \label{sec:VG_Equity}
\citet{MCC1998} used an asymmetric VG process to model the log-returns of a single stock, and priced European options on the stock.
It follows from the asymptotic behaviour of the Bessel function $K_{\nu}[z]$ given in (\ref{eq:K_inf}) that the exponential moment of the VG distribution exists only if $\th+\s^2/2<m$, which we now assume is true.
We model the stock price as
\begin{equation}
	S_t=S_0 \exp\left[{rt+L_t+w t}\right] \qquad (t\geq 0),
\end{equation}
where, under the risk-neutral measure, $\{L_t\}$ is a three-parameter VG process with parameters $(m,\th,\s)$, $r>0$ is the constant rate of interest, and
\begin{equation}
	w=m \log\left[1-\frac{\th}{m}-\frac{\s^2}{2m} \right].
\end{equation}
The characteristic function of $\{L_t\}$ was given in (\ref{eq:VGCF}), from which it becomes apparent that the drift term $wt$ ensures that
the discounted stock price process $\{\e^{-rt} S_t\}$ is a martingale under the risk-neutral measure.

We will show that this asymmetric VG model can be recovered using the information-based approach when the information process is a symmetric VGRB.
First we fix some suitably distant future date $T$ (for example, this could be the maturity date of the longest-dated, liquidly-traded, European option on the stock).
We then let $X_T$ be a VG random variable with density $f^{(m,\th,\rho)}_T(x)$, where $\rho$ is given by (\ref{eq:VGrho}), and set
\begin{equation}
	h(x)=S_0\exp(rT+\s x /\rho+wT ).
\end{equation}
Let the information process $\{\xi_{tT}\}$ be a standard VGRB with parameter $m$ such that $\xi_{TT}=X_T$. 
From Section \ref{sec:VGRB} we have that $\{\s \xi_{tT} / \rho\}$ is a VG process with parameters $\{m,\th,\s\}$.
We then have
\begin{align}
	H_{tT}&=\exp(-r(T-t)) \E[h(X_T) \,|\, \xi_{tT}] \nonumber
	\\ &=S_0\exp(rt+\s\xi_{tT}/\rho+wT) \, \E[\e^{\s(\xi_{TT}-\xi_{tT})/\rho}] \nonumber
	\\ &=S_0\exp(rt+\s\xi_{tT}/\rho+wt),
\end{align}
for $t\in[0,T]$.
Hence we have $\{H_{tT}\}_{0\leq t \leq T}\law\{S_t\}_{0\leq t \leq T}$.
European options on the price process $\{H_{tT}\}$ can then be calculated using the techniques in \citep{MCC1998}.

\section{VG binary bond} \label{sec:VG_BB}
In the example of Section \ref{sec:Binary_Bond} we derived the price of a binary bond in the Brownian information model.
\citet{BHM1} derived a similar price process, only they included a rate parameter $\s>0$.
This rate parameter behaved like the volatility parameter in the Black-Scholes model.
We can construct a similar model using a VGRB as the information process.

\begin{figure}[ht]
	\begin{center}
		\subfigure[Information process when $m=10$.]{\includegraphics[scale=.9]{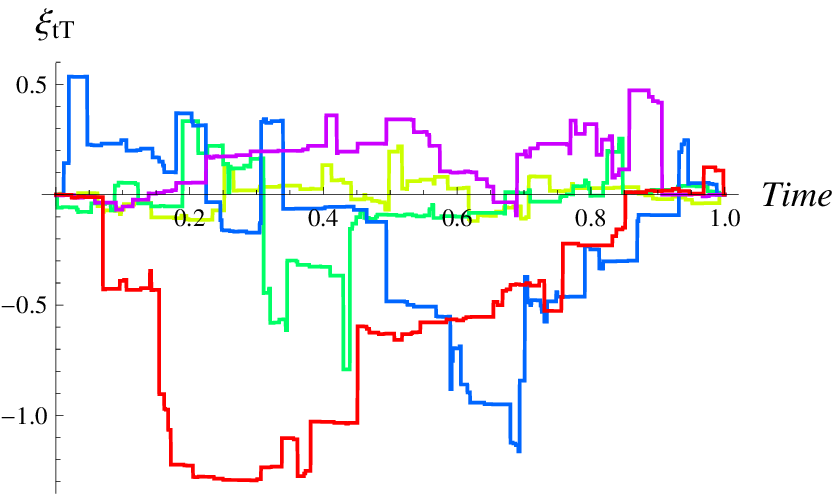}}
		\subfigure[Bond price with $m=10$.]{\includegraphics[scale=.9]{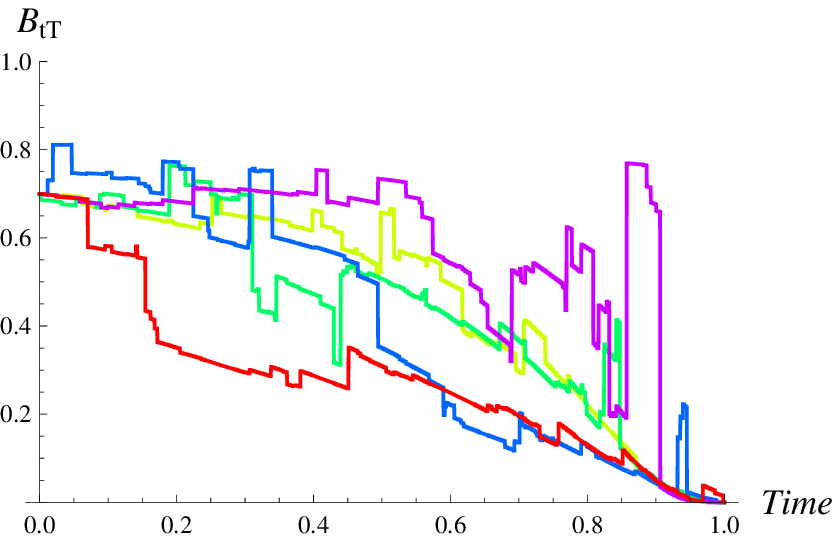}}
		\\
		\subfigure[Information process when $m=25$.]{\includegraphics[scale=.9]{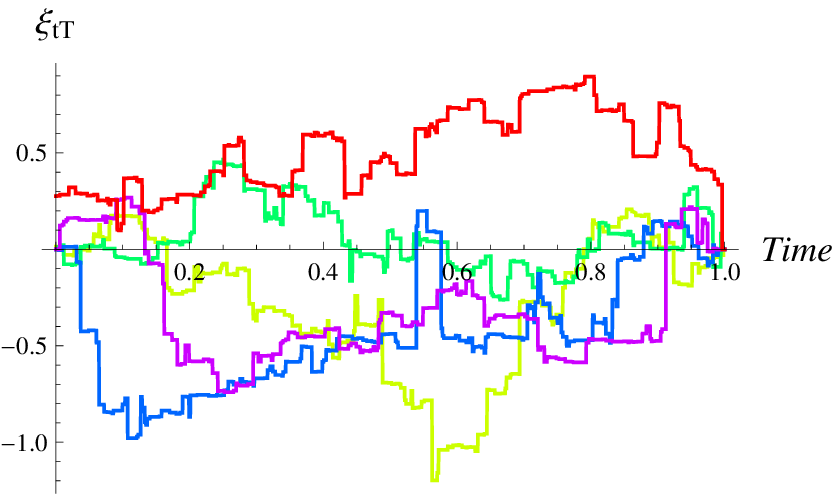}}
		\subfigure[Bond price with $m=25$.]{\includegraphics[scale=.9]{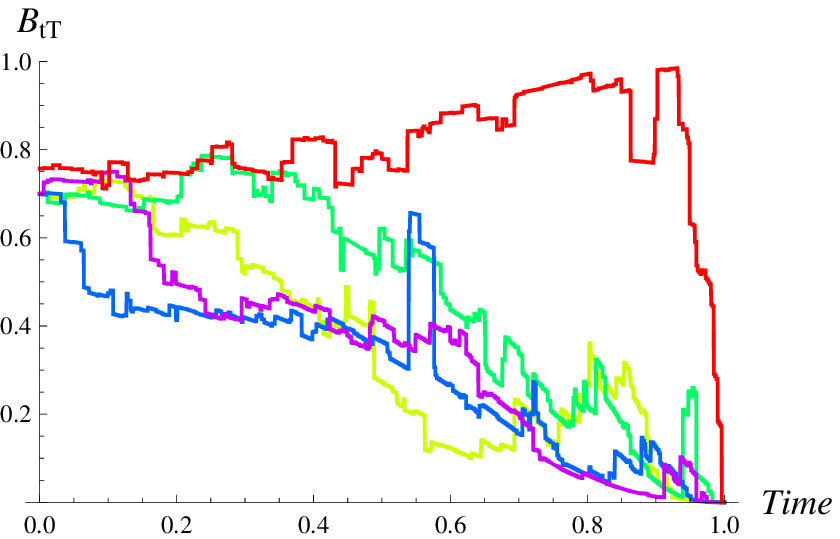}}
		\\
		\subfigure[Information process when $m=100$.]{\includegraphics[scale=.9]{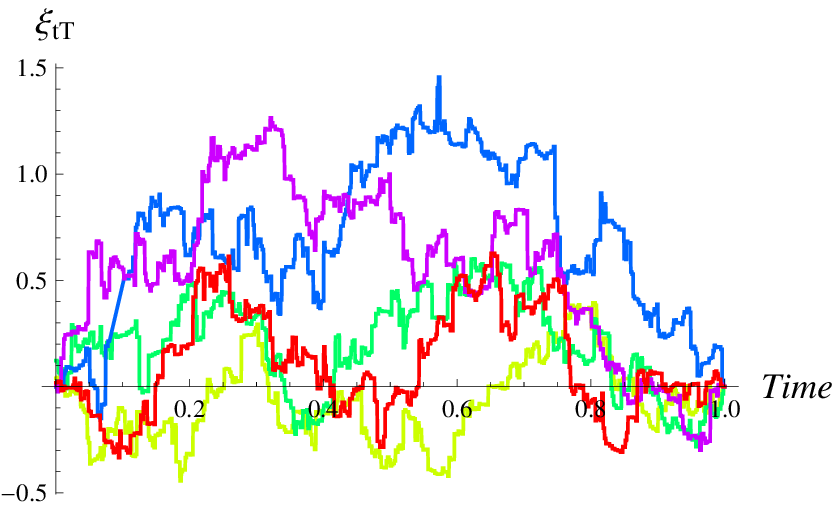}}
		\subfigure[Bond price with $m=100$.]{\includegraphics[scale=.9]{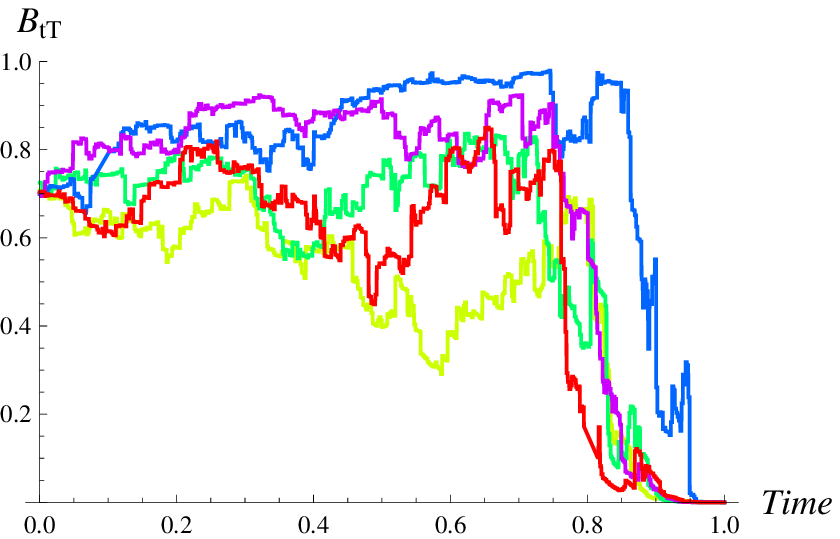}}
	\end{center}
	\caption[VG binary bond simulations: defaulting bonds]{%
						Simulations of VG information processes and bond price processes in cases where the bond defaults.
						Various values of the parameter $m$ used.
						The other parameter values are fixed as $r_t=0$, $p=0.3$, $T=1$, and $\s=1$.}
		\label{fig:VG1}
\end{figure}

\begin{figure}[ht]
	\begin{center}
		\subfigure[Information process when $m=10$.]{\includegraphics[scale=.9]{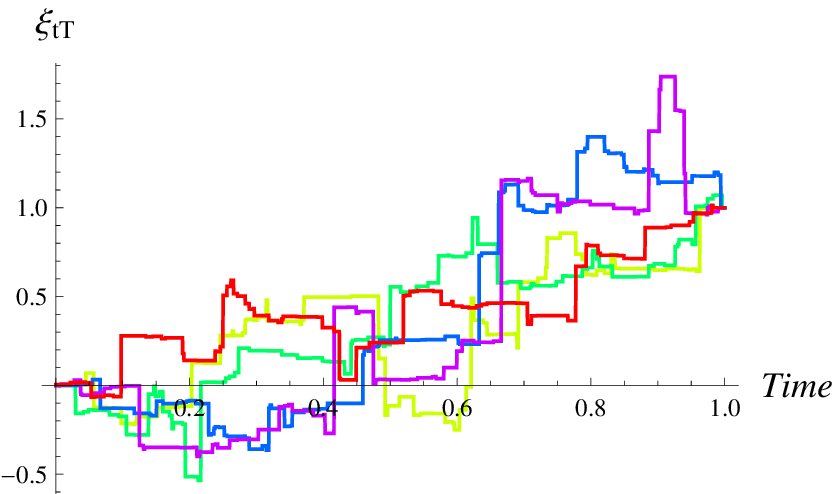}}
		\subfigure[Bond price with $m=10$.]{\includegraphics[scale=.9]{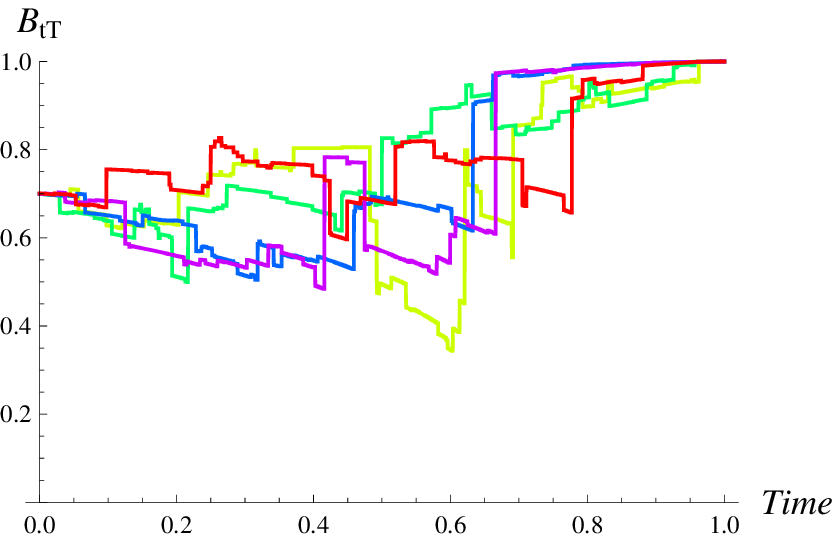}}
		\\
		\subfigure[Information process when $m=25$.]{\includegraphics[scale=.9]{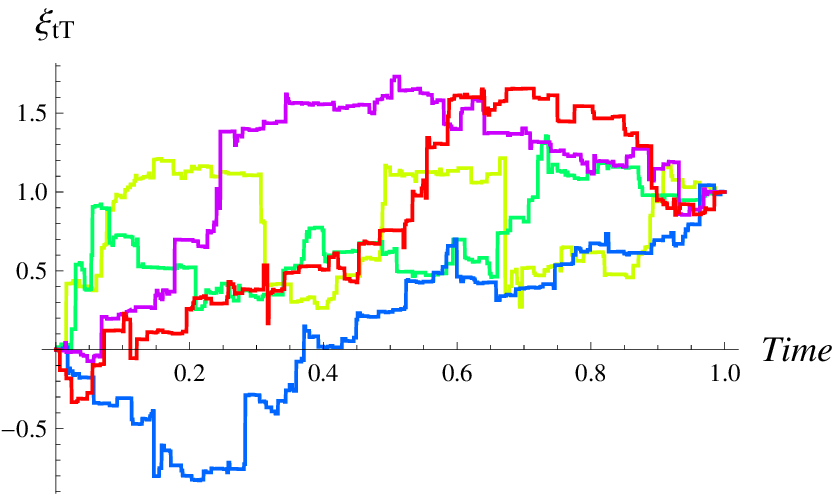}}
		\subfigure[Bond price with $m=25$.]{\includegraphics[scale=.9]{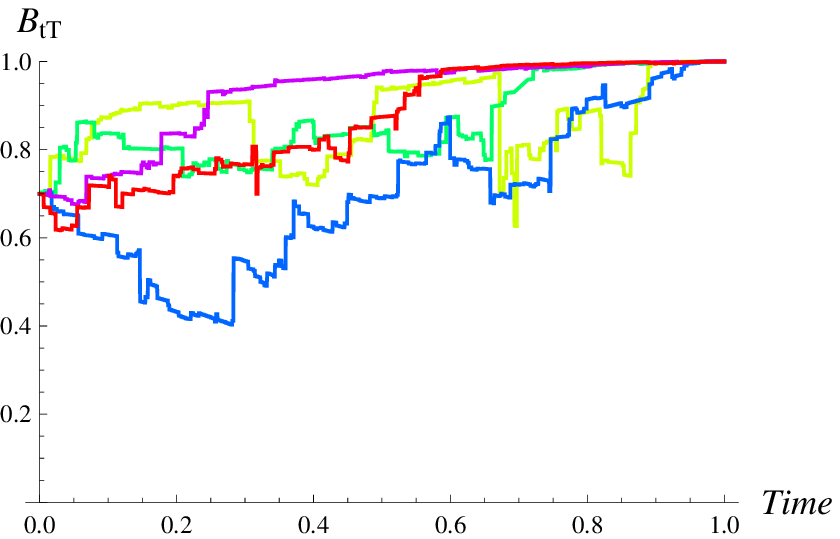}}
		\\
		\subfigure[Information process when $m=100$.]{\includegraphics[scale=.9]{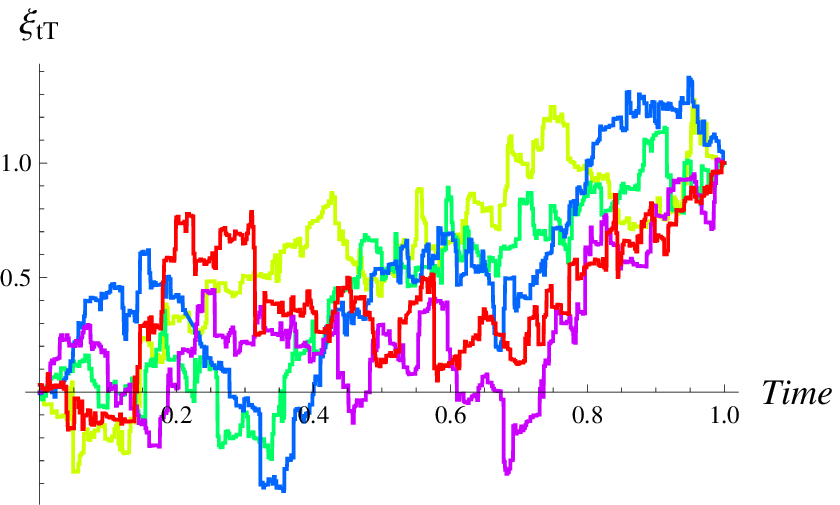}}
		\subfigure[Bond price with $m=100$.]{\includegraphics[scale=.9]{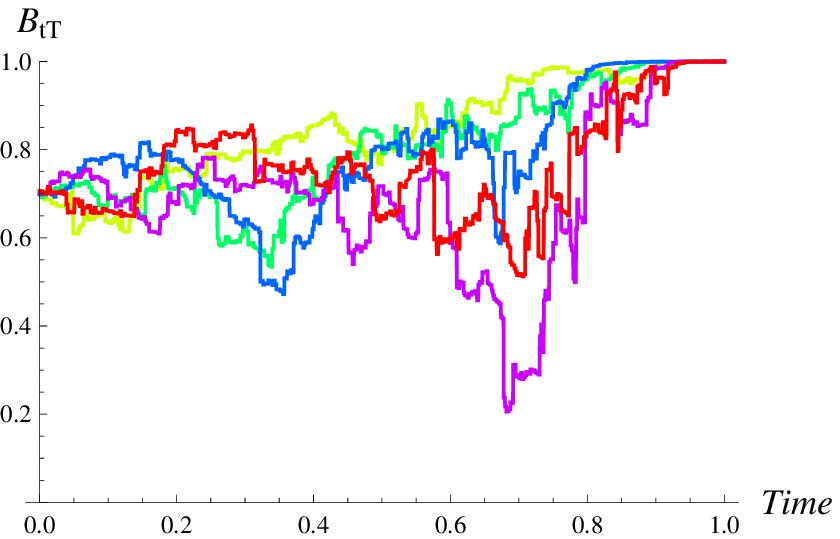}}
	\end{center}
	\caption[VG binary bond simulations: non-defaulting bonds]{%
						Simulations of VG information processes and bond price processes in cases where the bond does not default.
						Various values of the parameter $m$ used.
						The other parameter values are fixed as $r_t=0$, $p=0.3$, $T=1$, and $\s=1$.}
		\label{fig:VG2}
\end{figure}

\begin{figure}[ht]
	\begin{center}
		\subfigure[Information process when $\s=0.1$.]{\includegraphics[scale=.9]{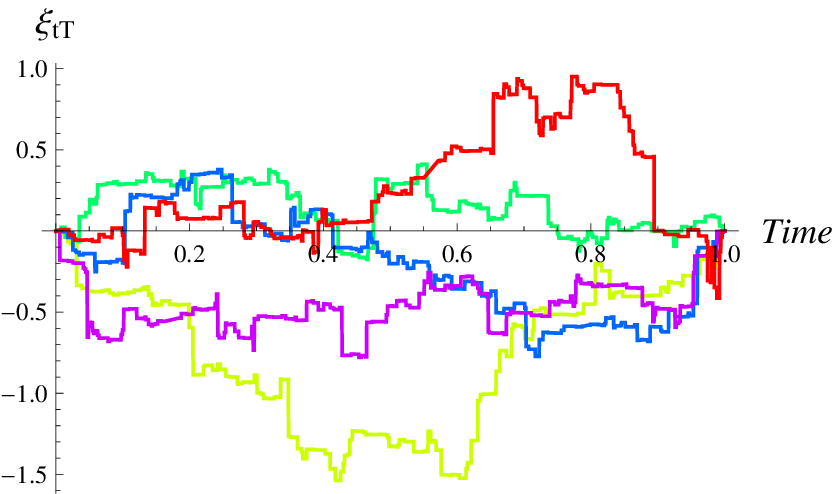}}
		\subfigure[Bond price with $\s=0.1$.]{\includegraphics[scale=.9]{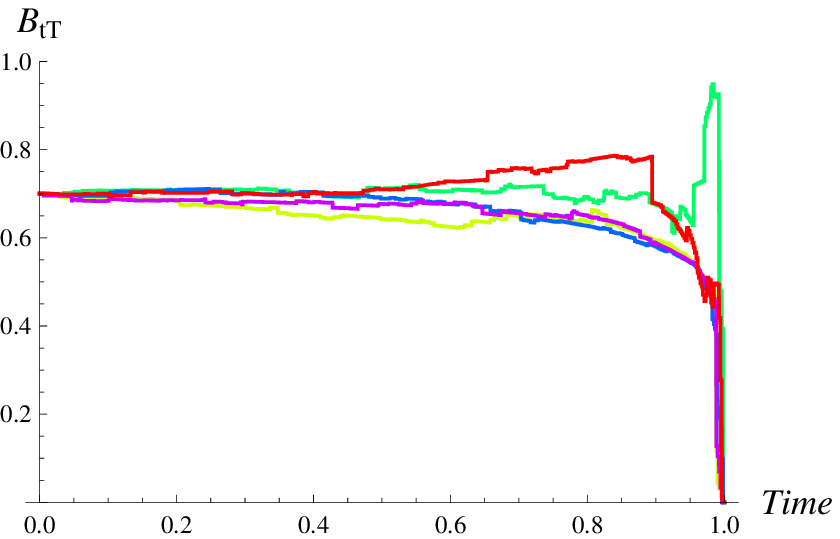}}
		\\
		\subfigure[Information process when $\s=1$.]{\includegraphics[scale=.9]{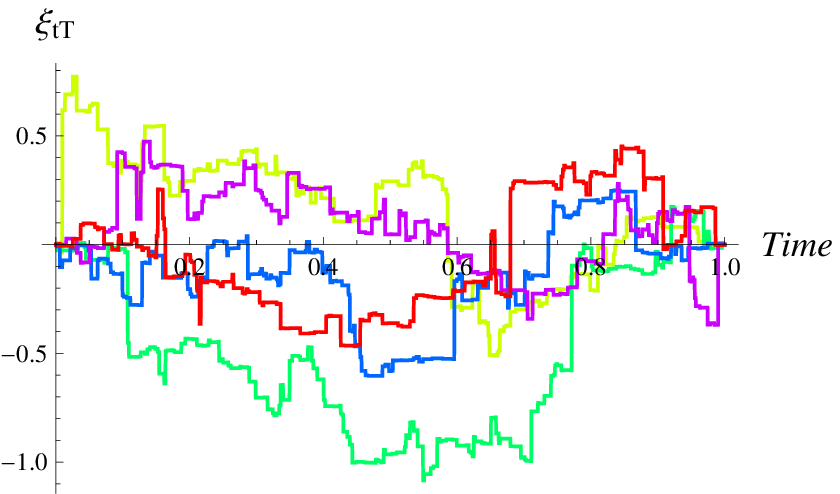}}
		\subfigure[Bond price with $\s=1$.]{\includegraphics[scale=.9]{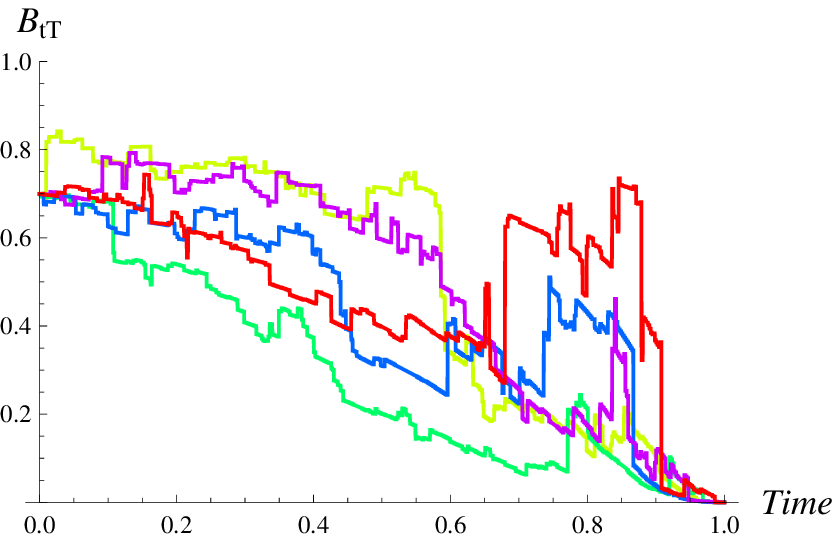}}
		\\
		\subfigure[Information process when $\s=10$.]{\includegraphics[scale=.9]{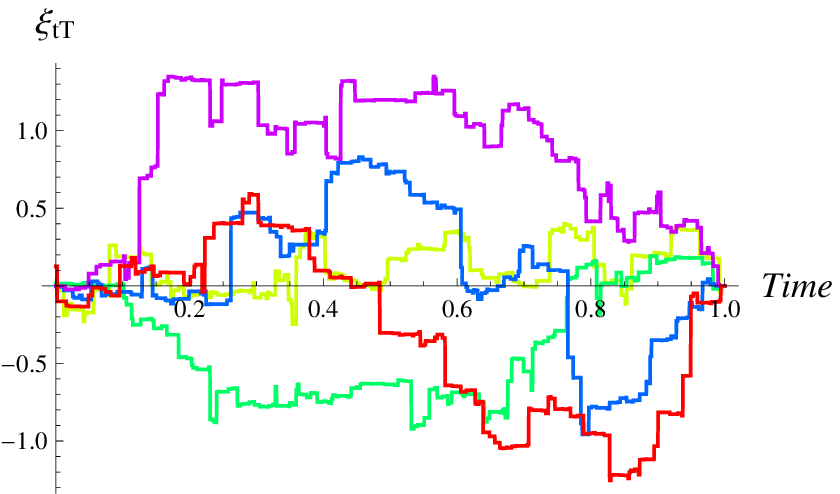}}
		\subfigure[Bond price with $\s=10$.]{\includegraphics[scale=.9]{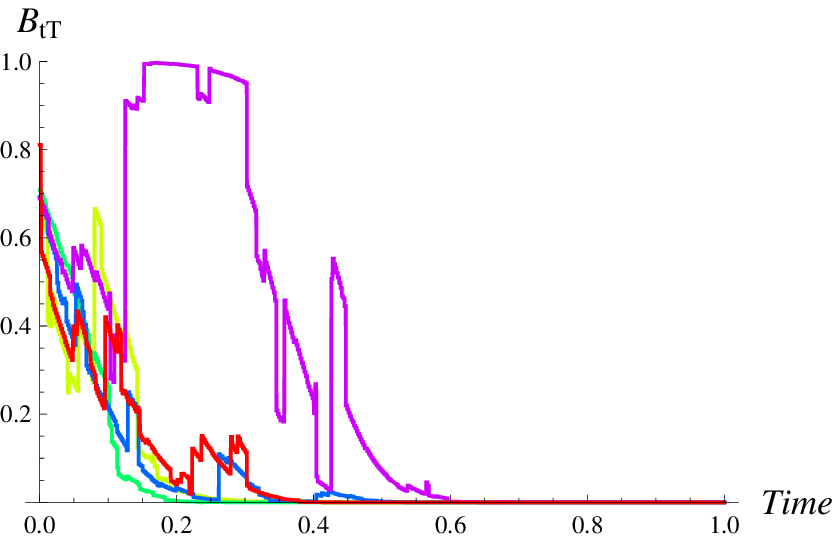}}
	\end{center}
	\caption[VG binary bond simulations: varying the rate parameter]{%
						Simulations of VG information processes and bond price processes in cases where the bond defaults.
						Various values of the information rate parameter $\s$ are used.
						The other parameter values are fixed as $r_t=0$, $p=0.3$, $T=1$, and $m=25$.}
		\label{fig:VG3}
\end{figure}

Let $X_T$ be an $X$-factor such that $\Q[X_T=0]=p$ and $\Q[X_T=\s]=1-p$.
Set $h(x)=x/\s$, and let $H_T=h(X_T)$ be the redemption amount of a credit-risky, zero-coupon bond.
We assume that the market filtration is generated by a standard VGRB $\{\xi_{tT}\}$ such that $\xi_{TT}=X_T$.
For $\{\xi_{tT}\}$ to be well defined, its parameter $m$ must satisfy $T>(2m)^{-1}$.
The time-$t$ price of the bond is then given by
\begin{equation}
	B_{tT}=P_{tT} \, \E[h(X_T)\,|\,\xi_{tT}]=P_{tT} \, \Q[X_T=\s \,|\, \xi_{tT}].
\end{equation}
From Section \ref{sec:Binary_Bond}, we have
\begin{align}
	B_{tT}&=P_{tT} \left(1+{\frac{f_T^{(m)}(\s)}{f_T^{(m)}(0)}
							\frac{f_{T-t}^{(m)}(-\xi_{tT})}{f_{T-t}^{(m)}(\s-\xi_{tT})}\frac{p}{1-p}} \right)^{-1} \nonumber
	\\ &=P_{tT} \left(1+c \left| \frac{\xi_{tT}}{\s-\xi_{tT}} \right|^{m(T-t)-\half}
							\frac{K_{m(T-t)-1/2}\left[\sqrt{2m\xi_{tT}^2}\right]}{K_{m(T-t)-1/2}\left[\sqrt{2m(\s-\xi_{tT})^2}\right]} \right)^{-1},
\end{align}
where
\begin{equation}
	c=2p\left(\frac{m\s^2}{2}\right)^{\frac{mT}{2}-\frac{1}{4}}\frac{K_{mT-1/2}\left[ \sqrt{2m\s^2}\right]}
		{(1-p) \G[mT-1/2]}.
\end{equation}
See Figures \ref{fig:VG1}, \ref{fig:VG2}, and \ref{fig:VG3} for example simulations.

%% file: chap06.tex
%
%

\chapter{Stable-1/2 information} \label{chap:insurance}
The term `stable process' refers here to a strictly stable process with index $\a\in(0,2)$;
thus, we are excluding the case of Brownian motion ($\a=2$).
The use of stable processes for the modelling of prices in financial markets was proposed by \citet{BBM1963} in connection with his analysis of cotton futures.
In the tails, the \levy densities of stable processes exhibit power-law decay.
As a result, the behaviour of stable processes is wild, and their trajectories exhibit frequent large jumps.
The variance of a stable random variable is infinite.
If $\a\leq 1$, the expectation either does not exist or is infinite. 
This heavy-tailed behaviour makes stable processes ill-suited to some applications in finance, such as forecasting and option pricing.

To overcome some of the drawbacks of the stable processes, so-called tempered stable processes have been introduced (see \citet{CT2004}, for example, for details).
A tempered stable process is a pure-jump \levy process, and its \levy density is the exponentially dampened \levy density of a stable process.
The exponential dampening of the \levy density improves the integrability of the process to the extent that all the moments of a tempered stable process exist.
Tempered stable processes do not possess the time-scaling property of stable processes.

In this chapter we apply stable-1/2 random bridges to the modelling of cumulative losses.
The techniques presented can equally be applied to cumulative gains.
The integrability of a stable-1/2 random bridge depends on the integrability of its terminal distribution.
At some fixed future time, the $n$th moment of the process is finite if and only if the $n$th moment of its terminal value is finite.
Thus a stable-1/2 random bridge with an integrable terminal distribution can be considered to be a dampened stable-1/2 subordinator.
In fact, the stable-1/2 random bridge is a generalisation of the tempered stable-1/2 subordinator.
If the \levy density of a stable-1/2 subordinator is exponentially dampened, the resulting process is an IG process.
We shall see that the IG process is a special case of a stable-1/2 random bridge.

We look in detail at the non-life reserving problem.
An insurance company will incur losses when certain events occur.
An event may be, for example, a period of high wind, a river flooding, or a motor accident. 
The losses are the costs associated with recompensing policy-holders who have been disadvantaged by an event. 
These costs might, for example, cover repairs to property, replacement of damaged items, loss of business, medical care, and so on.
Although a loss is incurred by the insurance company on the date of an event (the `loss date' or `accident date'), payment is generally not made immediately.
Delays will occur because loss is not always immediately reported to the company, the full extent of the costs takes time to emerge, the insurance company's obligation to pay takes time to establish, and so forth.

In return for covering policy-holder risk, the insurance company receives premiums.
The premiums received over a fixed period of time should, typically, be sufficient to cover the losses the company incurs over that period.
Since losses can take years to pay in full, the company sets aside some of the premiums to cover future payments; these are called `reserves'.
If the reserves are set too low, the company may struggle to cover its liabilities, leading to insolvency.
Large upward moves in the reserves due to a worsening in the expected future development of liabilities can cause similar problems.
If the reserves are set too high, the company may be accused by shareholders or regulators of inappropriately withholding profits.
Hence, it is the interest of the company to forecast its ultimate liability as accurately as possible when deciding the level of reserves to set.

We use a stable-1/2 random bridge to model the paid-claims process (i.e.~cumulative amount paid to-date) of an insurance company. 
The losses contributing to the paid-claims process are assumed to have occurred in a fixed interval of time.
Sometimes claims-handling information about individual losses is known, such as that contained in police or loss-adjuster reports.
In the model presented here, such information is disregarded, and the paid-claims process is regarded as providing all relevant information.
We derive the conditional distribution of the company's total liability given the paid-claims process.
We then estimate recoveries from reinsurance treaties on the total liability.
The expressions arising in such estimates are similar to the expectations encountered in the pricing of call spreads on stock prices.

We shall examine the upper-tail of the conditional distribution of the ultimate liability,
and find that it is as heavy as the \emph{a priori} tail.
This has an interesting interpretation in the case when the insurer is exposed to a catastrophic loss.
At time $t<T$, the probability of a catastrophic loss occurring in the interval $[t,T]$ decreases as $t$ approaches $T$.
However, in some sense, the \emph{size} of a catastrophic loss does not decrease as $t$ approaches $T$, since the tail of the conditional distribution of the cumulative loss does not thin.

When the \emph{a priori} total loss distribution is a generalized inverse-Gaussian distribution, we find that the model is particularly tractable.
We present a family of special cases where the expected total loss can be expressed as a rational function of the current value of the paid-claims process.
That is, each member of the family is a martingale that can be written as a rational function of an increasing process. 

The model can be extended to include more than one paid-claims process.
We consider the case where there are two processes that are not independent, and which have different activity parameters.
We then have two ultimate losses to estimate.
We provide expressions for the expected values of the ultimate losses given both paid-claims processes.
The numerical computations required to evaluate these expressions are no more difficult than those of the one-dimensional case.
We demonstrate how to calculate the \emph{a priori} correlation between the ultimate liabilities.
The correlation can be used as a calibration tool when modelling cumulative losses arising from related lines of business (e.g.~personal motor and commercial motor business).

We also describe how to simulate sample paths of the stable-1/2 random bridge, and how to use a deterministic time-change to adjust the model when the paid-claims process is expected to develop non-linearly.

\section{Stable-1/2 random bridge}
In this chapter we take $f_t(x)$ to be the increment density of the stable-1/2 subordinator as given in (\ref{eq:StableDen}); that is
\begin{equation} 
	\label{eq:stable_den_2}
	f_t(x)=\1_{\{x>0\}} \,\frac{ct}{\sqrt{2\pi}\,x^{3/2}} \, \exp\left(-\half \frac{c^2t^2}{x}  \right).
\end{equation}
Let $\{\xi_{tT}\}$ be a process with law $\lrbc([0,T],\{f_t\},\nu)$.
We say that $\{\xi_{tT}\}$ is a stable-1/2 random bridge.
Recall that $\nu$ must concentrate mass where $f_T(x)$ is positive and finite. 
Hence $\nu$ must be a probability law with support on the positive half-line.

\section{Insurance model}

We approach the non-life insurance claims reserving problem by modelling a paid-claims process by a stable-1/2 random bridge.
The stable-1/2 random bridge is a suitable candidate model for a paid-claims process because it is (i) increasing, and (ii) tractable (in particular, its density has a simple form).
We shall look at the problem of calculating the reserves required to cover the losses arising from a single line of business when we observe the paid-claims process.
\citet{EA1989} and \citeauthor{RN1} \citep{RN1,RN2} provide general descriptions of the problem.
\citet{EV2002} survey some of the existing actuarial models.
\citet{HB1970} and \citet{TK2004} contain related topics.
The present work ties in with that of \citet{BHM3} who use a gamma random bridge process to model a cumulative loss or gain.

The method we use has a flavour of the Bornhuetter-Ferguson model from actuarial science \citep{BF1972} (see also \citep{EV2002}).
In implementing the Bornhuetter-Ferguson model, one begins with an \emph{a priori} estimate for the \emph{ultimate loss} 
(the total cumulative loss arising from the underwritten risks).
Periodically, this estimate is revised using a chain-ladder technique to take into account both the \emph{a priori} estimate
and the development of the total paid (or reported) claims to date.

In the proposed model, we assume an \emph{a priori} distribution for the ultimate loss.
By conditioning on the development of the paid-claims process, we revise the ultimate loss distribution using the Bayesian methods described in Chapter \ref{chap:info_based}.
In this way, we continuously update the conditional distribution for the total loss.
This is as opposed to the deterministic Bornhuetter-Ferguson model in which only a point estimate is updated.
Knowledge of the conditional distribution allows one to calculate confidence intervals around the expected loss, and to calculate expected reinsurance recoveries.

The main assumptions of the model are the following:
\begin{enumerate}
	\item
		The claims arising from the line of business have \emph{run-off} at time $T$.
		That is, at time $T$ all claims have been settled, and the ultimate loss $U_T$ is known.
	\item
		$U_T$ has \emph{a priori} law $\nu$ such that $U_T>0$ and $\E[U_T^2]<\infty$.
	\item
		The paid-claims process $\{\xi_{tT}\}$ is a stable-1/2 random bridge, and $\xi_{TT}=U_T$.
	\item \label{A4}
		The \emph{best estimate} of the ultimate loss is $U_{tT}=\E\left[U_T \left| \F^{\xi}_t \right.\right]$,
		where $\{\F^{\xi}_t\}$ is the natural filtration of $\{\xi_{tT}\}$.
\end{enumerate}
A few remarks should be made about assumption \ref{A4}.
First, using the natural filtration of $\{\xi_{tT}\}$ as the reserving filtration means that the paid-claims process is the \emph{only} source of information
about the ultimate loss once the measure $\nu$ is set.
We do not consider the situation where we have access to information about claims that have been reported but not yet paid in full (such as case estimates).
Second, the expectation is taken with respect to $\Q$, which may or may not be the `real-world' measure.
Let us call $\Q$ the \emph{actuarial measure}.
When reserving, practitioners routinely discount data before modelling.
Discounting may adjust the data for the time-value of money or for the effects of claims inflation.
Claims inflation, and interest rates, though understood to be stochastic, usually only provide a small amount of uncertainty to the forecasting of the ultimate loss, relative to the uncertainty surrounding the frequency and (discounted) sizes of insurance claims.
Furthermore, it is often for practical purposes reasonable to assume that claims inflation and interest rates are independent of claim frequency and size.
Hence, a stochastic reserving model may lose little from the assumption that interest rates and inflation rates are deterministic.
We make this assumption, and further assume that the paid-claims process has been discounted for the effects of interest and inflation.

\section{Estimating the ultimate loss}
The time-$t$ conditional law of $U_T$ is
\begin{align}
	\nu_t(\dd z)&=\frac{\psi_t(\dd z;\xi_{tT})}{\psi_t(\R;\xi_{tT})} \nonumber
	\\&=\frac{\1_{\{z>\xi_{tT}\}}\left( \frac{z}{z-\xi_{tT}} \right)^{3/2}
				\exp\left( -\frac{c^2}{2}\left(\frac{(T-t)^2}{z-\xi_{tT}}-\frac{T^2}{z} \right) \right) \nu(\dd z)}
					{\int_{\xi_{tT}}^{\infty} \left( \frac{u}{u-\xi_{tT}} \right)^{3/2}
				\exp\left( -\frac{c^2}{2}\left(\frac{(T-t)^2}{u-\xi_{tT}}-\frac{T^2}{u} \right) \right) \nu(\dd u)}.
\end{align}
Given this law, confidence intervals and quantiles for the ultimate loss are readily calculated.
The best-estimate ultimate loss is
\begin{equation}
	U_{tT}=\int_{\xi_{tT}}^{\infty} z \, \nu_t(\dd z).
\end{equation}
At time $t<T$, the total amount of claims yet to be paid is $U_T-\xi_{tT}$.
The amount that the insurance company sets aside to cover this unknown amount is called the \emph{reserve}.
The expected value of the total future payments is called the \emph{best-estimate reserve}, and can be expressed by
\begin{equation}
	R_{tT}=U_{tT}-\xi_{tT}.
\end{equation}
For prudence, the reserve may be greater than the best-estimate reserve.
However, for regulatory reasons it is sometimes required that the best-estimate reserve is reported.
The variance of the total future payments is the variance of the ultimate loss,
which is
\begin{equation}
	\var\left[U_T-\xi_{tT}\left| \F^{\xi}_t \right.\right]=\var\left[U_T\left| \F^{\xi}_t \right.\right]=\int_{\xi_{tT}}^{\infty} (z-U_{tT})^2\, \nu_t(\dd z).
\end{equation}

\section{The paid-claims process}
We give the first two conditional moments of the paid-claims process.
From Corollary \ref{coro:ExpectLRB}, we have
\begin{equation}
	\label{eq:expectation}
	\E\left[\xi_{tT} \left|\, \F^{\xi}_s \right.\right]=\frac{T-t}{T-s}\xi_{sT}+\frac{t-s}{T-s} U_{sT}.
\end{equation}
Equation (\ref{eq:expectation}) implies that the paid-claims development is expected to be linear.
We return to this point later.
Using Proposition \ref{prop:mom} and a straightforward conditioning argument, we have
\begin{align}
	\E\!\left[\xi_{tT}^2\right]&=\frac{t}{T} \int_0^{\infty}
	z^2\left\{1-c(T-t)\e^{\frac{c^2T^2}{2z}}\sqrt{\frac{2\pi}{z}}\,\Phi\left[-cTz^{-1/2} \right]  \right\} \nu(\dd z) \nonumber
	\\ &=\frac{t}{T} \E\left[U_T^2\right]-c(T-t)\sqrt{2\pi}\int_0^{\infty}
	z^{3/2}\,\e^{\frac{c^2T^2}{2z}}\,\Phi\left[-cTz^{-1/2}\right] \nu(\dd z). \label{eq:var}
\end{align}
Fix $s<T$ and define the relocated process $\{\eta_{tT}\}_{s\leq t \leq T}$ by
\begin{equation}
	\eta_{tT}=\xi_{tT}-\xi_{sT}.
\end{equation}
The dynamic consistency property implies that, given $\xi_{sT}$, $\{\eta_{tT}\}$ is
a stable-1/2 random bridge with the marginal law of $\eta_{TT}$ being $\nu^*(A)=\nu_s(A+\xi_{sT})$.
Then we have
\begin{align}
	\E\!\left[\xi_{tT}^2\left|\, \F^{\xi}_s \right.\right] \nonumber
	&=\E\!\left[\eta_{tT}^2\left|\, \xi_{sT} \right.\right]+2\xi_{sT}\,\E\left[\eta_{tT} \left|\, \xi_{sT} \right.\right]+\xi_{sT}^2
	\\&=\frac{T-t}{T-s}\xi_{sT}^2+\frac{t-s}{T-s}\E\left[U^2_T\left|\,\xi_{sT}\right.\right] \nonumber
	\\& \quad-c(T-t)\sqrt{2\pi}\int_{\xi_{sT}}^{\infty}
					(z-\xi_{sT})^{\frac{3}{2}}\,\e^{\frac{c^2(T-s)^2}{2(z-\xi_{sT})}}\,\Phi\!\!\left[-\frac{c(T-s)}{\sqrt{z-\xi_{sT}}} \right]\! \nu_s(\dd z).
\end{align}

\section{Reinsurance}
An insurance company may buy reinsurance to protect against adverse claim development.
The \emph{stop-loss} and \emph{aggregate excess-of-loss} treaties are two types of reinsurance that cover some or all of the total amount of claims paid over a fixed threshold.
Under a stop-loss treaty, the reinsurance covers all the losses above a prespecified level.
If this level is $K$, then the reinsurance provider pays a total amount $(U_T-K)^+$ to the insurance company.
The `aggregate $L$ excess of $K$' treaty is a capped stop-loss, and covers the layer $[K,K+L]$.
In this case the reinsurance provider pays an amount $(U_T-K)^+-(U_T-K-L)^+$.

The insurance company typically receives money from the reinsurance provider periodically. 
The amount received depends on the amount the company has paid on claims to-date.
If the insurer has the paid-claims process $\{\xi_{tT}\}$, and receives payments from a stop-loss treaty (at level $K$) on the fixed dates $t_1<t_2<\cdots <t_n=T$, then the amount received on date $t_i$ is
\begin{equation}
	\label{eq:RI_payment}
	(\xi_{t_i,T}-K)^+-(\xi_{t_{i-1},T}-K)^+.
\end{equation}
Recall the following expressions from Section \ref{sec:stable_half}: the stable-1/2 bridge marginal density function
\begin{align}
	f_{tT}(y;z)&=\frac{f_t(y)f_{T-t}(z-y)}{f_T(z)}
	\\ &=\1_{\{ 0<y\leq z\}}\frac{1}{\sqrt{{2\pi}}}  \nonumber
		\frac{ct(T-t)}{T}\frac{\exp\left( -\frac{1}{2} \frac{c^2(Ty-tz)^2}{yz(z-y)}\right)}{\left(y-y^2/z\right)^{3/2}},
		\label{eq:stable_br_den_2}
\end{align}
the stable-1/2 bridge marginal distribution function
\begin{align}
		F_{tT}(y;z)&=\int_0^{y} f_{tT}(x;z) \d x \nonumber
	\\		&=	\Phi\left[\frac{c(Ty-tz)}{\sqrt{yz(z-y)}}\right]+\left(1-\frac{2t}{T} \right)\e^{2c^2t(T-t)/z}\,
				\Phi\left[\frac{c((2t-T)y-tz)}{\sqrt{yz(z-y)}}\right],
\end{align}
and the stable-1/2 bridge partial first moment
\begin{align}
		M_{tT}(y;z)&=\int_0^{y}z\, f_{tT}(x;z) \d x \nonumber
	\\		&=\frac{t}{T}z\left\{ \Phi\left[\frac{c(Ty-tz)}{\sqrt{yz(z-y)}}\right]-\e^{2c^2t(T-t)/z}\,
					\Phi\left[\frac{c((2t-T)y-tz)}{\sqrt{yz(z-y)}}\right] \right\}.
\end{align}
The expected value of reinsurance payments such as (\ref{eq:RI_payment}) can be calculated using the following:

\begin{prop}
	Fix $t\in(0,T)$.
	At time $s<t$, the expected exceedence of $\xi_{tT}$ over some fixed $K>0$ is
		\begin{align}
			D_{st}&=\E\left[ (\xi_{tT}-K)^+\left|\, \F^{\xi}_s \right.\right] \nonumber
			\\&=\frac{T-t}{T-s}\xi_{sT}+\frac{t-s}{T-s}U_{sT}-K \nonumber
			\\&\qquad+\1_{\{K>\xi_{sT}\}}(K-\xi_{sT})\int_{K}^{\infty}
					F_{t-s,T-s}(K-\xi_{sT};z-\xi_{sT})\, \nu_s(\dd z) \nonumber
			\\&\qquad-\1_{\{K>\xi_{sT}\}}\int_{K}^{\infty}M_{t-s,T-s}(K-\xi_{sT};z-\xi_{sT}) \, \nu_s(\dd z).
		\end{align}
\end{prop}
\begin{proof}
	If $K\leq \xi_{sT}$ then
	\begin{align}
		\E\left[ (\xi_{tT}-K)^+\left|\, \F^{\xi}_s \right.\right]&=\E\left[ \xi_{tT}\left|\, \F^{\xi}_s \right.\right]-K \nonumber
		\\ &=\frac{T-t}{T-s}\xi_{sT}+\frac{t-s}{T-s}U_{sT}-K.
	\end{align}
	Thus we need only consider the case when $K>\xi_{sT}$.
	
	The $\F^{\xi}_s$-conditional law of $\xi_{tT}$ is
	\begin{equation}
		\Q[\xi_{tT}\in \dd y \,|\, \F^{\xi}_s]=\frac{\psi_t(\R;\xi_{tT})}{\psi_s(\R;\xi_{sT})}f_{t-s}(y-\xi_{sT}) \d y.
	\end{equation}
	Hence we have
	\begin{align}
		D_{st}&= \frac{1}{\psi_s(\R;\xi_{sT})}\int_K^{\infty}(y-K)\psi_t(\R;y)f_{t-s}(y-\xi_{sT})\d y \nonumber
		\\ &= \frac{1}{\psi_s(\R;\xi_{sT})} \nonumber
				\int_K^{\infty}(y-K)\int_K^{\infty}\frac{f_{T-t}(z-y)}{f_T(z)}\,\nu(\dd z) \, f_{t-s}(y-\xi_{sT})\d y
		\\ &= \frac{1}{\psi_s(\R;\xi_{sT})}\int_{K}^{\infty}\int_{K}^{z}(y-K)\frac{f_{T-t}(z-y)f_{t-s}(y-\xi_{sT})}{f_T(z)} \d y \,\nu(\dd z) \nonumber
		\\ &= \int_{K}^{\infty}\int_{K}^{z}(y-K)f_{t-s,T-s}(y-\xi_{sT};z-\xi_{sT}) \d y \, \nu_s(\dd z).
	\end{align}
	Making the change of variable $x=y-\xi_{sT}$ yields
	\begin{align}
		D_{st} &= \int_{K}^{\infty}\int_{K-\xi_{sT}}^{z-\xi_{sT}}(x+\xi_{sT}-K)f_{t-s,T-s}(x;z-\xi_{sT}) \d x \, \nu_s(\dd z) \nonumber
		\\ &= \int_{K}^{\infty}\left\{\frac{t-s}{T-s}(z-\xi_{sT})-M_{t-s,T-s}(K-\xi_{sT};z-\xi_{sT})\right\} \nu_s(\dd z) \nonumber
		\\ &\qquad\qquad + (\xi_{sT}-K)\int_{K}^{\infty}\left\{1-F_{t-s,T-s}(K-\xi_{sT};z-\xi_{sT})\right\} \nu_s(\dd z) \nonumber
		\\ &= \frac{T-t}{T-s}\xi_{sT}+\frac{t-s}{T-s}U_{sT}-K \nonumber
	+\int_{K}^{\infty}(K-\xi_{sT})F_{t-s,T-s}(K-\xi_{sT};z-\xi_{sT})\, \nu_s(\dd z) \nonumber
		\\&\qquad\qquad-\int_{K}^{\infty}M_{t-s,T-s}(K-\xi_{sT};z-\xi_{sT}) \, \nu_s(\dd z).
	\end{align}
\end{proof}

Suppose that the insurance company has limited its liability by entering into a stop-loss reinsurance contract.
At time $s\in[0,T)$, the expected reinsurance recovery between times $t$ and $u$ is
\begin{equation}
	\label{eq:ReRec}
	\E\left[ (\xi_{uT}-K)^+- (\xi_{tT}-K)^+ \left|\, \F^{\xi}_s \right.\right]=D_{su}-D_{st}, 
\end{equation}
for $s<t<u\leq T$.

Using a similar method to the calculation of $D_{st}$, we can calculate the expectation of $\xi_{tT}$ conditional on it exceeding a threshold.
For a threshold $\th>\xi_{sT}$, we find
\begin{equation}
	\E[\xi_{tT} \,|\, \xi_{sT}, \xi_{tT}>\th]=
		 \frac{\frac{T-t}{T-s}\xi_{sT}+\frac{t-s}{T-s} U_{sT}-\int_{\xi_{sT}}^{\infty}M_{t-s,T-s}(\th-\xi_{sT};z-\xi_{sT}) \,\nu_s(\dd z)}
				{1-\int_{\xi_{sT}}^{\infty}F_{t-s,T-s}(\th-\xi_{sT};z-\xi_{sT}) \,\nu_s(\dd z)}.
\end{equation}
Sometimes called the conditional value-at-risk ($\mathrm{CVaR}$), this expected value is a coherent risk measure, and is a useful tool for risk management (see \citet{MFE2005}).
Note that $\mathrm{CVaR}$ is normally defined as an expected value conditional on a shortfall in profit.
Since we are modelling loss, and not profit, the risk we most wish to manage is on the upside.
Hence, conditioning on an exceedence is of greater interest.

\section{Tail behaviour}
In this section we consider how the probability of extreme events is affected by the paid-claims development.
Suppose that the line of business we are modelling is exposed to rare but `catastrophic' large loss events.
In this case we assume that the \emph{a priori} distribution of the ultimate loss has a heavy right-tail.
If a catastrophic loss could hit the insurance company at any time before run-off, then it is important that any conditional distributions for the ultimate loss
retain the heavy-tail property.
We shall see that in the stable-1/2 random bridge model, the conditional distributions are as heavy tailed as the \emph{a priori} distribution.

Assume that $U_T$ has a continuous density $p(z)$ which is positive for all $z$ above some threshold.
Then the value of $U_T$ is unbounded in the sense that
\begin{equation}
	\Q[U_T>x]>0, \quad\text{for all $x\in\R$.}
\end{equation}
Define
\begin{equation}
	\mathrm{Tail}_t=
		\lim_{L\rightarrow\infty} \frac{\Q\left[ \xi_{TT}>L \right]}{\Q\left[ \xi_{TT}-\xi_{tT}>L \left|\, \xi_{tT} \right.\right]}.
\end{equation}
If $\mathrm{Tail}_t=\infty$ then the tail of the future-payments distribution at time $t>0$ is not as heavy as the \emph{a priori} tail.
That is, a catastrophic loss at time $t$ is `smaller' than a catastrophic loss at time 0.
If $\mathrm{Tail}_t=0$ then the tail of the future-payments distribution is greater at time $t$ than \emph{a priori}.
If $0<\mathrm{Tail}_t<\infty$ then the tail is as heavy at time $t$ as \emph{a priori}.
Using L'H\^opital's rule, we have
\begin{align}
	\mathrm{Tail}_t
			&= \lim_{L\rightarrow\infty}
			\frac{\psi_t(\R;\xi_{tT})\int_{L}^{\infty} p(z) \d z} 
			{\int_{L+\xi_{tT}}^{\infty} \left( \frac{z}{z-\xi_{tT}} \right)^{3/2}
				\exp\left( -\frac{c^2}{2}\left(\frac{(T-t)^2}{z-\xi_{tT}}-\frac{T^2}{z} \right) \right) p(z) \d z} \nonumber
			\\&=  \lim_{L\rightarrow\infty}
			\frac{\psi_t(\R;\xi_{tT})\,p(L)}
			{\left( \frac{L+\xi_{tT}}{L} \right)^{3/2}
				\exp\left( -\frac{c^2}{2}\left(\frac{(T-t)^2}{L}-\frac{T^2}{L+\xi_{tT}} \right) \right)p(L+\xi_{tT})} \nonumber
			\\&={\psi_t(\R;\xi_{tT})}\,\lim_{L\rightarrow\infty}\frac{p(L)}{p(L+\xi_{tT})},
\end{align}
for $t\in(0,T)$.
Some examples follow:
\begin{itemize}
	\item
		If $p(z)\propto \1_{\{z>0\}}\e^{-z}$ (exponential) then $\mathrm{Tail}_t=\psi_t(\R;\xi_{tT})\,\e^{\xi_{tT}}$.
	\item
		If $p(z)\propto \1_{\{z>0\}}\e^{-z^2}$ (half-normal) then $\mathrm{Tail}_t=\psi_t(\R;\xi_{tT})\,\e^{\xi_{tT}^2}$.
	\item
		If $p(z)\propto \1_{\{z>0\}}z^{-3/2} \e^{-1/z}$ (L\'evy) then $\mathrm{Tail}_t=\psi_t(\R;\xi_{tT})$.
\end{itemize}
This property has an interesting parallel with the \emph{subexponential} distributions.
By definition, $X$ has a subexponential distribution if
\begin{equation}
	\lim_{L\rightarrow\infty} \frac{\Q\left[ \sum_{i=1}^n X_i>L \right]}{\Q\left[ X>L \right]} = n,
\end{equation}
where $\{X_i\}_{i=1}^n$ are independent copies of $X$ (see \citet{EKM1997}).
We note that
\begin{equation}
\lim_{L\rightarrow\infty} \frac{\Q\left[ Z_{T}>L \right]}{\Q\left[ Z_T-Z_t>L \left|\, Z_t \right.\right]}=\infty,
\end{equation}
for $\{Z_t\}$ a Brownian motion, a geometric Brownian motion or a gamma process.
If $\{Z_t\}$ is a stable-1/2 subordinator, so the increments of $\{Z_t\}$ are subexponential, then
\begin{equation}
\lim_{L\rightarrow\infty} \frac{\Q\left[ Z_{T}>L \right]}{\Q\left[ Z_T-Z_t>L \left|\, Z_t \right.\right]}=\frac{T}{T-t}.
\end{equation}

\section{Generalized inverse-Gaussian prior}
The generalized inverse-Gaussian (GIG) distribution is a three-parameter family of distributions on the positive half-line
(see \citet{Jorg1982} or \citet{EH2004} for further details).
In the present work it has so far appeared only tangentially.
We now provide some of its properties.
The density of the GIG distribution is
\begin{equation}
	\label{eq:GIGdensity}
	 f_{\textit{GIG}}(x;\l,\delta,\g)= 
	 \1_{\{x>0\}} \left( \frac{\g}{\delta}\right)^{\l} \frac{1}{2\,K_{\l}[\g\delta]}x^{\l-1}\exp\left(-\tfrac{1}{2}(\delta^2x^{-1}+\g^2x) \right).
\end{equation}
Here $K_{\nu}[z]$ is the modified Bessel function seen in Section \ref{sec:VG}.
The permitted parameter values are
\begin{align}
	&\delta\geq 0, && \g>0, 		&&\text{if $\l>0$,}
	\\ &\delta> 0, && \g>0, 		&&\text{if $\l=0$,}
	\\ &\delta> 0, && \g\geq 0, &&\text{if $\l<0$.}
\end{align}
For $\l>0$, taking the limit $\delta\rightarrow 0^+$ yields the gamma distribution.
For $\l<0$, taking the limit $\g\rightarrow 0^+$ yields the reciprocal-gamma distribution---this includes the \levy distribution for $\l=-1/2$
(recall that the \levy distribution is the increment distribution of \SH subordinators).
The case $\l=-1/2$ and $\g>0$ corresponds to the inverse-Gaussian (IG) distribution.
If $X$ has density (\ref{eq:GIGdensity}) then the moment $\mu_k=\E[X^k]$ is given by
\begin{align}
	\mu_k&=\frac{K_{\l+k}[\g\delta]}{K_{\l}[\g\delta]}\left(\frac{\delta}{\g} \right)^k \qquad \text{for $\l\in\R$, $\delta>0$, $\g>0$},
	\\\mu_k&=\left\{
					\begin{aligned}
						&\frac{\G[\l+k]}{\G[\l]} \left(\frac{2}{\g^2} \right)^k && k>-\l
						\\ &\infty && k\leq-\l
					\end{aligned}
				\right. \text{and $\l>0$, $\delta=0$, $\g>0$,}
	\\\mu_k&=\left\{
					\begin{aligned}
						&\frac{\G[-\l-k]}{\G[-\l]} \left(\frac{\delta^2}{2} \right)^k && k<-\l
						\\ &\infty && k\geq-\l
					\end{aligned}
				\right. \text{and $\l<0$, $\delta>0$, $\g=0$.}
\end{align}

For convenience, we recall some facts about the IG process that appeared in Section \ref{subsec:IG}.
The IG process is a \levy process with increment density
\begin{equation}
		q_t(x)=\1_{\{x>0\}} \frac{c t}{\sqrt{2\pi}}\frac{1}{x^{3/2}}\exp\left(-\frac{\g^2}{2 x}\left(x-\tfrac{c}{\g}t\right)^2 \right).
\end{equation}
We see that $q_t(x)=f_{\textit{GIG}}(x;-\tfrac{1}{2},ct,\g)$.
The $k$th moment of $q_t(x)$ is
\begin{equation}
	m_t^{(k)}=\sqrt{\frac{2}{\pi}}\,\g\e^{\g c t}\left(\frac{c t}{\g} \right)^{k+\frac{1}{2}} K_{k-1/2}[\g c t],
\end{equation}
for $k>0$.

\subsection{GIG terminal distribution}
We shall see that the GIG distributions constitute a natural class of \emph{a priori} distributions for the ultimate loss.
With $\g>0$ and $c>0$ fixed, we examine some properties of a paid-claims process $\{\xi_{tT}\}$ with time-$T$ density $f_{\textit{GIG}}(z;\l,cT,\g)$.
The transition law is
\begin{align}
		\Q[\xi_{tT} \in \dd y \,|\, \xi_{sT}=x]&=\frac{\psi_t(\R;y)}{\psi_s(\R;x)} f_{t-s}(y-x)\d y,
		\\\Q[\xi_{TT} \in \dd y \,|\, \xi_{sT}=x]&=\frac{\psi_s (\dd y;x)}{\psi_s(\R;x)},
\end{align}
where
\begin{align}
	\psi_0(\dd z;\xi)&=f_{\textit{GIG}}(z;\l, c T,\g) \d z,
	\\ \psi_t(\dd z;\xi)&=(1-\tfrac{t}{T}) \1_{\{z>\xi\}} 
				\frac{\exp\left({-\frac{c^2}{2}\left(\frac{(T-t)^2}{z-\xi}-\frac{T^2}{z} \right)}\right)}{({1-\xi/z})^{3/2}} 
				\,f_{\textit{GIG}}(z;\l,c T,\g) \d z.
\end{align}
Writing
\begin{equation}
	\kappa=\left(\frac{\g}{c T}\right)^{\l} \frac{1}{2\, K_{\l}[\g\sqrt T ]},
\end{equation}
we have
\begin{align}
	\psi_t(\R;y)&=\kappa(1-\tfrac{t}{T})\e^{-\frac{1}{2}\g^2 y} \int_{y}^{\infty}z^{\l+\frac{1}{2}}
			\frac{\e^{-\frac{c^2}{2}\frac{(T-t)^2}{z-y}-\frac{1}{2}\g^2(z-y)}}{(z-y)^{3/2}} \d z \nonumber
	\\ &=\kappa(1-\tfrac{t}{T})\e^{-\frac{1}{2}\g^2 y} \int_{0}^{\infty}(z+y)^{\l+\frac{1}{2}}
			\frac{\e^{-\frac{c^2}{2}\frac{(T-t)^2}{z}-\frac{1}{2}\g^2z}}{z^{3/2}} \d z \nonumber
	\\ &=\frac{\kappa\sqrt{2\pi}}{cT}\e^{-\frac{1}{2}\g^2 y- \g c(T-t)} \int_{0}^{\infty}(z+y)^{\l+\frac{1}{2}}
			q_{T-t}(z) \d z.
\end{align}
Given $\xi_{tT}=y$, the best-estimate ultimate loss is given by
\begin{align}
	U_{tT}=\psi_t(\R;y)^{-1}\int_{y}^{\infty} z \, \psi_t(\dd z;y)
	 =\frac{\int_0^{\infty}(z+y)^{\l+\frac{3}{2}} q_{T-t}(z) \d z}{\int_0^{\infty}(z+y)^{\l+\half} q_{T-t}(z) \d z}.
\end{align}

\subsection{Case $\l=-1/2$}
When $\l=-1/2$ we have
\begin{align}
	\frac{\psi_t(\R;y)}{\psi_s(\R;x)} f_{t-s}(y-x) 
		&=\1_{\{y-x>0\}}\frac{1}{\sqrt{2\pi}}\frac{c(t-s)}{(y-x)^{3/2}}
		\exp\left(-\frac{\g^2}{2}\frac{\left((y-x)- c(t-s)/\g \right)^2}{y-x} \right) \nonumber
	\\	& =q_{t-s}(y-x).
\end{align}
Hence $\{\xi_{tT}\}$ is an IG process.
Note that in this case $\{\xi_{tT}\}$ has independent increments.

\subsection{Case $\l=n-\frac{1}{2}$}
We now consider the case where $\l=n-\half$, for $n\in\mathbb{N_+}$.
For convenience we write
\begin{equation}
	q^{(k)}_t(x)=f_{\textit{GIG}}(x;k-1/2,ct,\g).
\end{equation}
Hence we have $q^{(0)}_t(x)=q_t(x)$.
The transition density of $\{\xi_{tT}\}$ is then
\begin{align}
	\frac{\psi_t(\R;y)}{\psi_s(\R;x)} f_{t-s}(y-x) 
	 &=q_{t-s}(y-x)\frac{\int_{0}^{\infty}(z+y)^n q_{T-t}(z) \d z}{\int_{0}^{\infty}(z+x)^n q_{T-s}(z) \d z} \nonumber
	\\	& =q_{t-s}(y-x)\frac{\sum_{k=0}^n\binom{n}{k} m_{T-t}^{(n-k)} \,y^{k}}{\sum_{k=0}^n\binom{n}{k} m_{T-s}^{(n-k)} \,x^{k}}.
\end{align}
When $n=1$ this is
\begin{align}
	\frac{\psi_t(\R;y)}{\psi_s(\R;x)} f_{t-s}(y-x)&=q_{t-s}(y-x)\frac{y+\frac{c}{\g} (T-t)}{x+\frac{c}{\g} (T-s)} \nonumber
				\\&= \left(1- \frac{c (t-s)}{\g x+ c (T-s) }\right) q_{t-s}^{(0)}(y-x) \nonumber
				\\&\qquad\qquad+\left( \frac{ c (t-s)}{\g x+ c (T-s)} \right) q_{t-s}^{(1)}(y-x).
\end{align}
Thus the increment density is a weighted sum of GIG densities.
We shall now derive a weighted sum representation for general $n$.
We can write
\begin{align}
	\int_{0}^{\infty}(z+y)^n \,q_{T-t}(z) \d z
	&=\int_{0}^{\infty}((z+x)+(y-x))^n\, q_{T-t}(z) \d z \nonumber
	\\&=\sum_{k=0}^n \binom{n}{k} (y-x)^{n-k} \int_{0}^{\infty}(z+x)^{k} \,q_{T-t}(z) \d z \nonumber
	\\&=\sum_{k=0}^n \binom{n}{k} (y-x)^{n-k} \sum_{j=0}^k \binom{k}{j}m^{(k-j)}_{T-t}\, x^{k}.
\end{align}
Then we have
\begin{align}
	\frac{\psi_t(\R;y)}{\psi_s(\R;x)} f_{t-s}(y-x) 
	 &=q_{t-s}(y-x)\frac{\int_{0}^{\infty}(z+y)^n q_{T-t}(z) \d z}{\int_{0}^{\infty}(z+x)^n q_{T-s}(z) \d z} \nonumber
	\\  &=q_{t-s}(y-x)\frac{\sum_{k=0}^n \binom{n}{k} (y-x)^{n-k} 
					\sum_{j=0}^k \binom{k}{j}m^{(k-j)}_{T-t}\, x^{j}}{\sum_{k=0}^n\binom{n}{k} m^{(n-k)}_{T-s} \,x^{k}}.\label{eq:IGSB_inc_den}
\end{align}
However, when $k\in\mathbb{N}_0$,
\begin{align}
	\frac{z^{k}\,q_{t-s}(z)}{q_{t-s}^{(k)}(z)}&=\frac{z^{k}\,f_{\textit{GIG}}(z;-1/2, c (t-s),\g)}{f_{\textit{GIG}}(z;k-1/2,c (t-s),\g)} \nonumber
		\\ &=\left( \frac{ c (t-s)}{\g} \right)^{k} \frac{K_{k-1/2}[\g c (t-s)]}{K_{1/2}[\g c (t-s)]} \nonumber
		\\ &=m^{(k)}_{t-s}. 
\end{align}
Hence we have
\begin{equation}
	\label{eq:den_id}
	(y-x)^{n-k}\, q_{t-s}(y-x)=m^{(n-k)}_{t-s} \, q_{t-s}^{(n-k)}(y-x).
\end{equation}
Using the identity (\ref{eq:den_id}), (\ref{eq:IGSB_inc_den}) can be expanded to obtain
\begin{equation}
	\frac{\psi_t(\R;y)}{\psi_s(\R;x)} f_{t-s}(y-x) =\sum_{k=0}^n w^{(k)}_{st}(x) \, q_{t-s}^{(k)}(y-x),
\end{equation}
where
\begin{equation}
	w^{(k)}_{st}(x)=\frac{\binom{n}{k} m^{(n-k)}_{t-s} \sum_{j=0}^k \binom{k}{j}m^{(k-j)}_{T-t}\, x^{j}}
		{\sum_{j=0}^n\binom{n}{j} m^{(n-j)}_{T-s} \,x^{j}}.
\end{equation}
Notice that $w^{(k)}_{st}(x)$ is a rational function whose denominator is a polynomial of order $n$, and whose numerator is a polynomial of order $k\leq n$.
Thus the transition probabilities of $\{\xi_{tT}\}$ depend on the first $n$ integer powers of the current value.
The conditional law of the ultimate loss is
\begin{equation}
	\frac{\psi_s(\dd y;\xi_{sT})}{\psi_s(\R;\xi_{sT})}=\frac{y^n q^{(0)}_{T-s}(y-\xi_{sT})}{\sum_{k=0}^n \xi_{sT}^k \, m_{T-s}^{(n-k)}} \d y.
\end{equation}

We can verify that $\sum_{k=0}^n w^{(k)}_{st}(x)=1$ using the fact that IG densities are closed under convolution.
We have
\begin{equation}
	q_{T-s}(z)=\int_{0}^{z} q_{T-t}(y) q_{t-s}(z-y) \d y, \quad \text{for $0\leq s<t<T$}.
\end{equation}
For fixed $n\in\mathbb{N}_+$, we then have
\begin{align}
	\sum_{k=0}^n\binom{n}{k} m^{(n-k)}_{T-s} \,x^{k}&=\int_0^{\infty} (z+x)^n q_{T-s}(z) \d z \nonumber
	\\ &= \int_0^{\infty} (z+x)^n \int_{0}^{z} q_{T-t}(y) q_{t-s}(z-y) \d y \d z \nonumber
	\\ &= \int_{0}^{\infty} q_{T-t}(y) \int_y^{\infty} (z+x)^n \, q_{t-s}(z-y)  \d z \d y \nonumber
	\\ &= \int_{0}^{\infty} q_{T-t}(y) \int_0^{\infty} (z+y+x)^n \, q_{t-s}(z)  \d z \d y \nonumber
	\\ &= \int_{0}^{\infty} q_{t-s}(y) \left[ \sum_{k=0}^n \binom{n}{k} m^{(n-k)}_{t-s} (y+x)^{k} \right] \d y \nonumber
	\\ &= \sum_{k=0}^n \binom{n}{k} m^{(n-k)}_{t-s} \int_{0}^{\infty} (y+x)^{k} \,q_{t-s}(y) \d y \nonumber
	\\ &= \sum_{k=0}^n \binom{n}{k} m^{(n-k)}_{t-s} \sum_{j=0}^{k} \binom{k}{j} m^{(k-j)}_{T-t} x^{j},
\end{align}
which gives
\begin{equation}
	\sum_{k=0}^n\frac{ \binom{n}{k} m^{(n-k)}_{t-s} \sum_{j=0}^{k} \binom{k}{j} m^{(k-j)}_{T-t} x^{j}}
		{\sum_{j=0}^n\binom{n}{j} m^{(n-j)}_{T-s} \,x^{j}}=1.
\end{equation}

\subsection{Moments of the paid-claims process}
The best-estimate ultimate loss simplifies to
\begin{equation}
	U_{tT}=\frac{\sum_{k=0}^{n+1}\binom{n+1}{k}m^{(n+1-k)}_{T-t} \xi_{tT}^k}{\sum_{k=0}^{n}\binom{n}{k}m^{(n-k)}_{T-t} \xi_{tT}^k}.
\end{equation}
For example, when $n=1$ we obtain
\begin{equation}
	U_{tT}=\frac{c(T-t)(1+\g c(T-t))+2\g^2 c(T-t)\xi_{tT}+\g^3 \xi_{tT}^2}{\g^2c(T-t)+\g^3\xi_{tT}}.
\end{equation}
By similar calculations, we have
\begin{equation}
	\E[U_{T}^m\,|\,\xi_{tT}]=\frac{\sum_{k=0}^{n+m}\binom{n+m}{k}m^{(n+m-k)}_{T-t} \xi_{tT}^k}{\sum_{k=0}^{n}\binom{n}{k}m^{(n-k)}_{T-t} \xi_{tT}^k}
	\quad \text{for $m\in\mathbb{N_+}$,}
\end{equation}
and
\begin{equation}
	\E\left[\left.\e^{\half\a^2U_{T}} \,\right| \xi_{tT}\right]=
	\\	\frac{\sum_{k=0}^{n}\binom{n}{k}\bar{m}^{(n-k)}_{T-t} \xi_{tT}^k}{\sum_{k=0}^{n}\binom{n}{k}m^{(n-k)}_{T-t} \xi_{tT}^k}
					\, \exp\left(\tfrac{1}{2} \a^2\xi_{tT}-(T-t)(\bar{\g}-\g)\right),
\end{equation}
for $0<\a<\g$, where $\bar{\g}=\sqrt{\g^2-\a^2}$, and $\bar{m}^{(k)}_{t}$ is the $k$th moment of the IG distribution with parameters $\delta=ct$ and $\g=\bar{\g}$.

\section{Exposure adjustment}
We have seen that
\begin{equation}
	\E[\xi_{tT}]=\frac{t}{T}\E[U_T];
\end{equation}
thus in the model so far the development of the paid-claims process is expected to be linear.
This is not always the case in practice.
In some cases the marginal exposure is (strictly) decreasing as the development approaches run-off.
This manifests itself as
\begin{equation}
	\frac{\partial^2}{\partial t^2} \E[\xi_{tT}]<0,
\end{equation}
for $t$ close to $T$.
A straightforward method to adjust the development pattern is through a time change.
We describe the marginal exposure of the insurer through time by a deterministic function $\varepsilon:[0,T]\rightarrow\R_+$.
The total exposure of the insurer is
\begin{equation}
	\int_{0}^T \varepsilon(s) \d s.
\end{equation}
We define the increasing function $\tau(t)$ by
\begin{equation}
	\tau(t)=T\frac{\int_0^t \varepsilon(s) \d s}{\int_0^T \varepsilon(s) \d s}.
\end{equation}
By construction $\tau(0)=0$ and $\tau(T)=T$.

Now let $\tau(t)$ determine the \emph{operational time} in the model.
We define the time-changed paid-claims process $\{\xi^{\t}_{tT}\}$ by
\begin{equation}
	\xi^{\t}_{tT}=\xi(\t(t),T),
\end{equation}
and set the reserving filtration to be natural filtration of $\{\xi_{tT}^\t\}$.
Then we have
\begin{align}
	\E[\xi^{\t}_{tT}]=\frac{\int_0^t \varepsilon(s)\d s}{\int_0^T \varepsilon(s)\d s} \E[U_{T}]
	\\\intertext{and}
	\frac{\partial^2}{\partial t^2} \E[\xi_{tT}^{\t}]=\frac{\E[U_T]}{\int_0^T\varepsilon(s) \d s}\varepsilon'(t).
\end{align}

\subsection{Craighead curve}
\begin{figure}[ht]
	\begin{center}
		\includegraphics[scale=1]{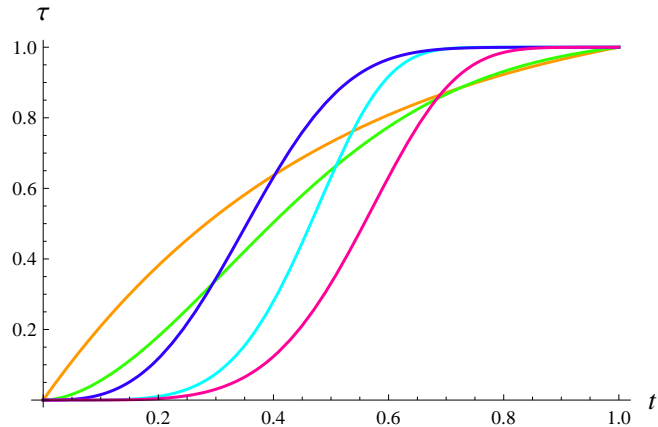}
	\end{center}
	\caption[Craighead curves: truncated Weibull time-changes]{%
		Plots of the truncated Weibull time change for various parameters, and with $T=1$.
		The expected paid-claims development of the model will have the same profile as $\tau(t)$ (scaled by $\E[U_T]$).
		Hence, under any one of the above time changes, when $t$ is close to $T$ the marginal exposure falls 
		(i.e.~$\frac{\partial^2}{\partial t^2} \E[\xi_{tT}^{\tau}]<0$).
		\label{fig:Weibull}
}
\end{figure}

\citet{Craig1979} proposed fitting a Weibull distribution function to the development pattern of paid claims for forecasting the ultimate loss (see also \citet{BE1997}).
In actuarial work, the Weibull distribution function is sometimes referred to as the `Craighead curve'.
To achieve a similar development pattern we can use the Weibull density as the marginal exposure:
\begin{equation}
	\varepsilon(t)=\frac{b}{a}\left(x/a\right)^{b-1} \e^{(x/a)^b},
\end{equation}
for $a,b>0$.
Then the time change $\tau(t)$ is the renormalised, truncated Weibull distribution function
\begin{equation}
	\tau(t)=T \frac{1-\e^{(t/a)^b}}{1-\e^{(T/a)^b}}.
\end{equation}
See Figure \ref{fig:Weibull} for plots of this function.
When $b\leq 1$, $\t'(t)$ is decreasing.
Under such a time change, the marginal exposure is decreasing for all $t\in[0,T]$.
When $b>1$, $\t'(t)$ achieves its maximum at
\begin{equation}
	t^*=a\left(\frac{b-1}{b}\right)^{1/b},
\end{equation}
and $\t'(t)$ is decreasing for $t\geq t^*$.
Thus if $T>t^*$ then the marginal exposure is decreasing for $t\in[t^*,T]$.
If $T\leq t^*$ then the marginal exposure is increasing for $t\in[0,T]$.

\section{Simulation} \label{sec:Stable_Sim}
\begin{figure}[ht]
	\begin{center}
		\subfigure[$c=3$]{\includegraphics[scale=.9]{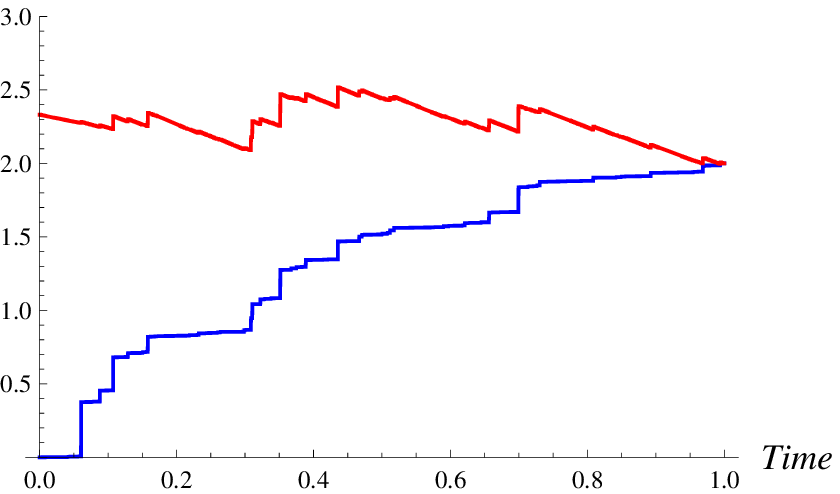}}
		\subfigure[$c=5$]{\includegraphics[scale=.9]{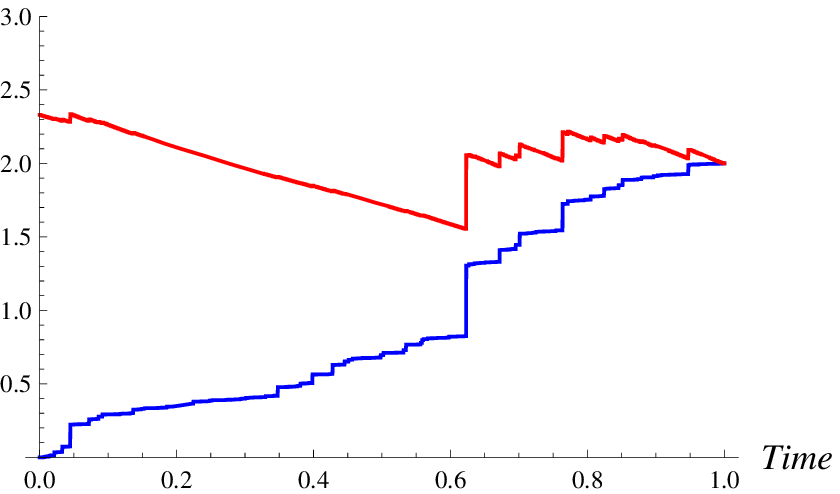}}
		\\
		\subfigure[$c=7$]{\includegraphics[scale=.9]{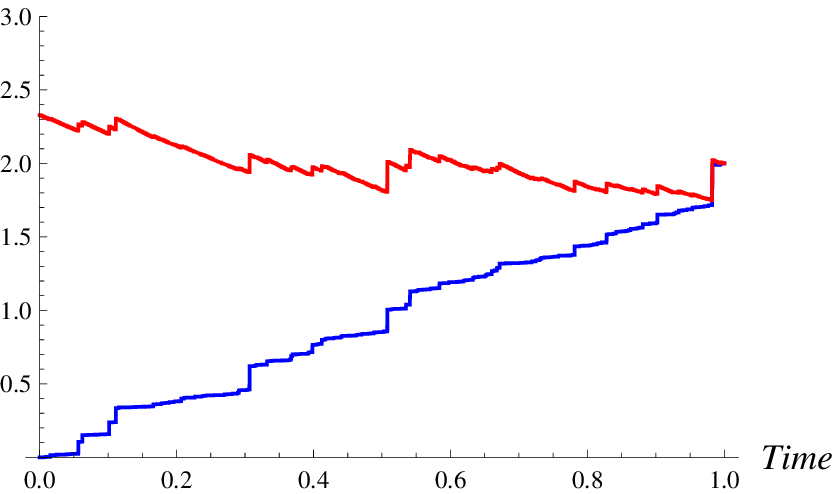}}
		\subfigure[$c=10$]{\includegraphics[scale=.9]{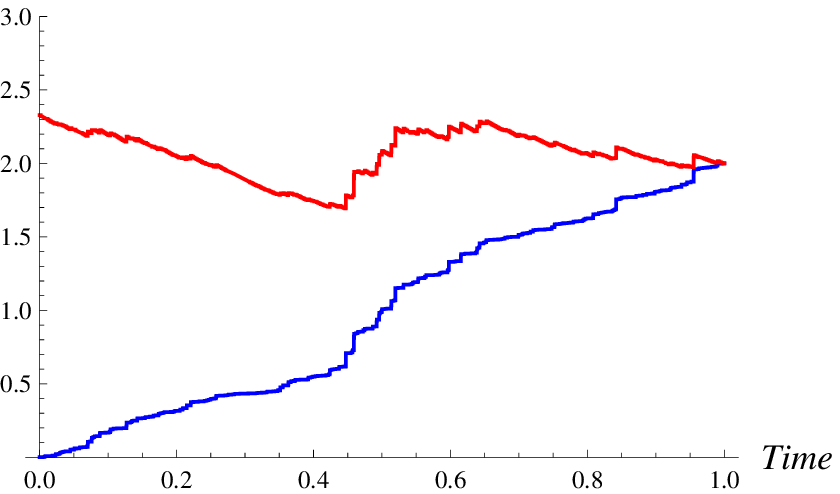}}
	\end{center}
	\caption[Simulations from the stable-1/2 random bridge reserving model]{%
					Simulations of the paid-claims process $\{\xi_{tT}\}$ (bottom line) and the best-estimate process $\{U_{tT}\}$ (top line). 
					Various values of the activity parameter $c$ are used.
					\emph{A priori}, the ultimate loss $U_T$ has a generalized Pareto distribution (GPD) with density 
					\mbox{$f_{\textit{GPD}}(x)=\1_{\{x>1\}}\left(1+\frac{x-1}{4} \right)^{-5}.$}
					(This is the GPD with scale parameter $\s=1$, location parameter $\mu=1$, and shape parameter $\xi=1/4$.)}
		\label{fig:srb}
\end{figure}

We consider the simulation of sample paths of a stable-1/2 random bridge.
First, we can generalise (\ref{eq:GT2}) to
\begin{multline}
	\label{eq:simalgo}
	\left[\left. \xi(\tfrac{s+t}{2},T)\,\right| \xi(s,T)=y, \xi(t,T)=z \right]\law 
	 \\y+\half (z-y)\left(1+\frac{Z}{\sqrt{c^2(t-s)^2/(z-y)+Z^2}} \right),
\end{multline}
where $0< s<t\leq T$, and $Z$ is a standard Gaussian.
We can then generate a discretised path $\{\hat{\xi}(t_i,T)\}_{i=0}^{2^n}$, where $t_i=iT2^{-n}$, by the following recursive algorithm:
\begin{enumerate}
	\item Generate the variate $\hat{\xi}(T,T)$ with law $\nu$, and set $\hat{\xi}(0,T)=0$.
	\item Generate $\hat{\xi}(\tfrac{T}{2},T)$ from $\hat{\xi}(0,T)$ and $\hat{\xi}(T,T)$ using identity (\ref{eq:simalgo}).
	\item Generate $\hat{\xi}(\tfrac{T}{4},T)$ from $\hat{\xi}(0,T)$ and $\hat{\xi}(\tfrac{T}{2},T)$,
				and then generate $\hat{\xi}(\tfrac{3T}{4},T)$ from $\hat{\xi}(\tfrac{T}{2},T)$ and $\hat{\xi}(T,T)$.
	\item Generate $\hat{\xi}(\tfrac{T}{8},T)$, $\hat{\xi}(\tfrac{3T}{8},T)$, $\hat{\xi}(\tfrac{5T}{8},T)$, $\hat{\xi}(\tfrac{7T}{8},T)$.
	\item Then iterate.
\end{enumerate}
See Figure \ref{fig:srb} for example simulations.

\section{Multiple lines of business}
We shall generalise the paid-claims model to achieve two goals.
The first is to allow more than one paid-claims process, and allow dependence between the processes.
The second is to keep the dimensionality of the calculations low with a view to practicality.
The following results can be applied to the modelling of multiple lines of business or multiple origin years when there is dependence between loss processes.

\subsection{Two paid-claims processes}
We consider a case with two paid-claims processes, but the results can be extended to higher dimensions.
In what follows, we set $f_t^c(x)=f_t(x)$ as given by (\ref{eq:stable_den_2}), and $f^c_{tT}(x)=f_{tT}(x)$ as given by (\ref{eq:stable_br_den_2}).
Here we have introduced the superscript to emphasise the dependence on $c$.
Let $\{S(t,T^*)\}$ be a stable-1/2 random bridge with terminal density $p(z)=\nu(\dd z)/\dd z$, and with activity parameter $c$.
Fix a time $T<T^*$, then define two paid-claims processes by
\begin{align}
	\xi^{(1)}_{tT}&=S(t,T^*) && (0\leq t\leq T),
	\\\xi^{(2)}_{tT}&=k^{2}S(\l t+T,T^*)-k^2 S(T,T^*) && (0\leq t\leq T),
\end{align}
where $\l=T^*/T-1$, and $k=c_2/(c\l)$ for some $c_2>0$.
The density of $\xi^{(1)}_{TT}$ is given by
\begin{align}
	p^{(1)}(x)&=f_T^c(x)\int_{0}^{\infty}\frac{f_{T^*-T}^c(z-x)}{f_{T^*}^c(z)}p(z) \d z \nonumber
				\\ &=\int_{0}^{\infty}f_{T,T^*}^c(x;z) \, p(z) \d z,
	\\\intertext{and the density of $\xi^{(2)}_{TT}$ is}
	p^{(2)}(x)&=k^{-2}f_{T^*-T}^c(k^{-2}x)\int_{0}^{\infty}\frac{f_{T}^c(z-k^{-2}x)}{f_{T^*}^c(z)}p(z) \d z \nonumber
				\\ &=k^{-4}f_{T^*-T}^c(k^{-2}x)\int_{0}^{\infty}\frac{f_{T}^c(k^{-2}z-k^{-2}x)}{f_{T^*}^c(k^{-2}z)}p(k^{-2}z) \d z 
				\label{eq:denI}
				\\ &=k^{-4}\int_{0}^{\infty}f_{T^*-T,T^*}^c(k^{-2}x;k^{-2}z)\,p(k^{-2}z) \d z 
				\label{eq:denII}
				\\ &=k^{-2}\int_{0}^{\infty}f_{T,\l^{-1}T^*}^{c_2}(x;z)\,p(k^{-2}z) \d z.
				\label{eq:denIV}
\end{align}
Here (\ref{eq:denI}) follows after a change of variable, (\ref{eq:denII}) follows from the definition of $f_{tT}(y;z)$ given in (\ref{eq:def_ftT}), and (\ref{eq:denIV}) follows from the functional form of $f_{tT}(y;z)$ for stable-1/2 bridges given in (\ref{eq:kernel}).
It follows from Corollary \ref{coro:inc} that $\{\xi^{(1)}_{tT}\}$ is a stable-1/2 random bridge with law 
$\lrbc([0,T],\{f^c_t\},p^{(1)}(z)\dd z)$, and (combined with the scaling property of stable-1/2 bridges) $\{\xi^{(2)}_{tT}\}$ has law $\lrbc([0,T],\{f^{c_2}_t\},p^{(2)}(z)\dd z)$.
The conditional, joint density of $(\xi^{(1)}_{tT},k^{-2}\xi^{(2)}_{tT})$ is
\begin{multline}
	\Q\left[ \xi^{(1)}_{tT}\in\dd y_1, k^{-2} \xi^{(2)}_{tT}\in\dd y_2 \left|\,\xi^{(1)}_{sT}=x_1, k^{-2}\xi^{(2)}_{sT}=x_2 \right.\right]=
	\\ \left\{
	\int_{z=x_1+x_2}^{\infty} \frac{f_{T^*-(1+\l)t}^c(z-(y_1+y_2))}{f^c_{T^*-(1+\l)s}(z-(x_1+x_2))} p(z) \d z \right\}
	f_{t-s}^c(y_1-x_1) \d y_1\, f_{\l (t-s)}^c(y_2-x_2) \d y_2,
\end{multline}
for $0\leq s<t\leq T$.
Then we have
\begin{align}
	&\Q\left[ \xi^{(1)}_{tT}+k^{-2} \xi^{(2)}_{tT}\in\dd y \left|\,\xi^{(1)}_{sT}=x_1, k^{-2}\xi^{(2)}_{sT}=x_2 \right.\right] \nonumber
	\\ &\quad=\left\{ 
		\int_{z=x_1+x_2}^{\infty} \frac{f_{T^*-(1+\l)t}^c(z-y)}{f^c_{T^*-(1+\l)s}(z-(x_1+x_2))} p(z) \d z \right\} 
		f_{(1+\l) (t-s)}^c(y-(x_1+x_2)) \d y \nonumber
	\\ &\quad=\left\{ \int_{z=x_1+x_2}^{\infty} f_{(1+\l)(t-s),T^*-(1+\l)s}^c(z-(x_1+x_2);y-(x_1+x_2)) \, p(z) \d z \right\} \dd y;
\end{align}
and, given $\xi^{(1)}_{sT}=x_1$ and $k^{-2}\xi^{(2)}_{sT}=x_2$, the marginal density of $\xi^{(1)}_{tT}$ is
\begin{align}
	y_1&\mapsto\int_{z=x_1+x_2}^{\infty}f^c_{t-s,T^*-(1+\l)s}(y_1-x_1;z-(x_1+x_2)) \, p(z) \d z,
	\\\intertext{and the marginal density of $k^{-2}\xi^{(2)}_{tT}$ is}
	y_2&\mapsto\int_{z=x_1+x_2}^{\infty}f^c_{\l(t-s),T^*-(1+\l)s}(y_2-x_2;z-(x_1+x_2)) \, p(z) \d z.
\end{align}

\subsection{Correlation}
The \emph{a priori} correlation between the terminal values is well defined when the second moment of $\nu$ is finite.
The correlation can be used as a tool in the calibration of the model.
Assuming that $\E[S(T^*,T^*)^2]<\infty$, the correlation is defined as
\begin{equation}
	\frac{\E\left[\xi_{TT}^{(1)} \, \xi^{(2)}_{TT} \right]-\E\left[\xi^{(1)}_{TT}\right]\E\left[\xi^{(2)}_{TT}\right]}
	{\sqrt{\left(\E\left[ \left(\xi^{(1)}_{TT}\right)^2 \right]-\E\left[\xi^{(1)}_{TT}\right]^2\right)
	\left(\E\left[ \left(\xi^{(2)}_{TT}\right)^2 \right]-\E\left[\xi^{(2)}_{TT}\right]^2\right)}}.
	\label{eq:correlation}
\end{equation}
We shall calculate each of the components of (\ref{eq:correlation}) separately.
First, we have
\begin{equation}
	\label{eq:corr_E1}
	\E\left[\xi^{(1)}_{TT}\right]=\E[S(T,T^*)]=\frac{T}{T^*}\E[S(T^*,T^*)].
\end{equation}
Noting that 
\begin{align}
	\xi^{(2)}_{TT}&=k^2(S(T^*,T^*)-S(T,T^*)) \nonumber
	\\&\law k^2 S(T^*-T,T^*),
\end{align}
we have
\begin{align}
	\E\left[\xi^{(2)}_{TT}\right]&=k^2 \E[S(T^*-T,T^*)] \nonumber
	\\&=k^2\left(1-\frac{T}{T^*}\right)\E[S(T^*,T^*)]. \label{eq:corr_E2}
\end{align}
The second moments of $\xi^{(1)}_{TT}$ and $\xi^{(2)}_{TT}$ follow from (\ref{eq:var}), and are given by
\begin{align}
	\E\left[\left(\xi^{(1)}_{TT}\right)^2\right]&=	\frac{T}{T^*} \E\left[S(T^*,T^*)^2\right]-(T^*-T)C_{T^*}, \label{eq:mom1}
	\\\intertext{and}
	\E\left[\left(\xi^{(2)}_{TT}\right)^2\right]&=k^4 \left(1-\frac{T}{T^*}\right) \E\left[S(T^*,T^*)^2\right]-k^4TC_{T^*},\label{eq:mom2}
\end{align}
where
\begin{equation}
	C_{T^*}=c\sqrt{2\pi}\int_0^{\infty}
						z^{3/2}\,\e^{\frac{c^2T^{*2}}{2z}}\,\Phi\left[-cT^*z^{-1/2}\right] \, p(z) \d z.
\end{equation}
The final term required for working out the correlation is the cross moment.
This is
\begin{align}
	\E\left[\xi^{(1)}_{TT} \, \xi^{(2)}_{TT} \right]
		&= k^2\E\left[S(T,T^*)\left(S(T^*,T^*)-S(T,T^*) \right) \right] \nonumber
		\\ &=k^2\E\left[S(T,T^*)S(T^*,T^*)\right]-k^2\E\left[S(T,T^*)^2 \right]. \label{eq:corr_eq}
\end{align}
The first term on the right of (\ref{eq:corr_eq}) is 
\begin{align}
	k^2\E\left[S(T,T^*)S(T^*,T^*)\right]
		&=k^2\int_0^{\infty}\int_0^{\infty}x\, y \, \frac{f_T^c(x)f_{T^*-T}^c(y-x)}{f_{T^*}^c(y)} \d x \, p(y) \d y \nonumber
		\\ &=k^2\int_0^{\infty}\int_0^{\infty}x \, y\, f_{T,T^*}^c(x;y) \d x \, p(y) \d y \nonumber
		\\ &=k^2\frac{T}{T^*}\int_0^{\infty}y^2 \, p(y) \d y \nonumber
		\\ &=k^2\frac{T}{T^*}\E[S(T^*,T^*)^2].
\end{align}
The second term on the right of (\ref{eq:corr_eq}) is given by (\ref{eq:mom1}).
Hence we have
\begin{equation}
	\label{eq:corr_x}
	\E\left[\xi^{(1)}_{TT} \, \xi^{(2)}_{TT} \right]=k^2(T^*-T)C_{T^*}.
\end{equation}
The expression for the correlation follows by putting (\ref{eq:corr_E1}), (\ref{eq:corr_E2}), (\ref{eq:mom1}), (\ref{eq:mom2}), and (\ref{eq:corr_x}) together.

\subsection{Ultimate loss estimation}
We now estimate the terminal values of the paid-claims processes.
At time $t<T$, the best-estimate ultimate loss of $\{\xi_{tT}^{(1)}\}$ (or, indeed, $\{\xi_{tT}^{(2)}\}$) depends on the two values $\xi_{tT}^{(1)}$ and $\xi_{tT}^{(2)}$.
The best-estimate ultimate loss of $\{\xi_{tT}^{(1)}\}$ is
\begin{align}
	U_{tT}^{(1)}&=\E\left[\xi^{(1)}_{TT} \left|\, \xi^{(1)}_{tT}=x_1, \xi^{(2)}_{tT}=x_2 \right.\right] \nonumber
	\\ &=\E\left[S(T,T^*) \left|\, S(t,T^*)=x_1, S(T+\l t,T^*)-S(T,T^*)=k^{-2}x_2 \right.\right] \nonumber
	\\ &=\E\left[S(T+\l t,T^*)\left|\, S(t,T^*)=x_1, S(T+\l t,T^*)-S(T,T^*)=k^{-2}x_2 \right.\right]-k^{-2}x_2 \nonumber
	\\ &=\E\left[S(T+\l t,T^*) \left|\, S(t,T^*)=x_1, S((1+\l)t,T^*)-S(t,T^*)=k^{-2}x_2 \right.\right]-k^{-2}x_2 \label{eq:usereorder}
	\\ &=\E\left[S(T+\l t,T^*) \left|\, S((1+\l)t,T^*)=x_1+k^{-2}x_2 \right.\right]-k^{-2}x_2 \label{eq:useMarkov}
	\\ &=\frac{T-t}{T^*-(1+\l)t} \left( \E\left[S(T^*,T^*) \left|\, S((1+\l)t,T^*)=x_1+k^{-2}x_2\right.\right]-k^{-2}x_2\right) \nonumber
	\\ &\qquad\qquad +  \frac{T^*-(T-t)}{T^*-(1+\l)t}x_1. \label{eq:useexp}
\end{align}
Equation (\ref{eq:usereorder}) holds since reordering the increments of an LRB yields an LRB with same law,
 (\ref{eq:useMarkov}) follows from the Markov property of LRBs,
and (\ref{eq:useexp}) follows from (\ref{eq:expectation}).
We also have
\begin{equation}
	\E\left[S(T^*,T^*) \left|\, S((1+\l)t,T^*)=x_1+k^{-2}x_2\right.\right]=\int_{0}^{\infty} z \, p_t(z) \d z,
	\label{eq:key_int}
\end{equation}
where
\begin{multline}
	p_t(z)=\1_{\{z>x_1+k^{-2}x_2\}}K^{-1} \left( \frac{z}{z-(x_1+k^{-2}x_2)} \right)^{3/2}
			\\ \times \,	\exp\left( -\frac{c^2}{2}\left(\frac{(T^*-(1+\l)t)^2}{z-(x_1+k^{-2}x_2)}-\frac{T^{*2}}{z} \right) \right) p(z),
\end{multline}
for $K$ a constant chosen to normalise the density $p_t(z)$.
Similarly, the best-estimate ultimate loss of $\{\xi_{tT}^{(2)}\}$ is
\begin{multline}
	U_{tT}^{(2)}= k^2\frac{T^*-(T-t)}{T^*-(1+\l)t}\left( \E\left[S(T^*,T^*) \left|\, S((1+\l)t,T^*)=x_1+k^{-2}x_2\right.\right]-x_1\right) 
	\\ +  \frac{T-t}{T^*-(1+\l)t}x_2.
\end{multline}
To compute both $U_{tT}^{(1)}$ and $U_{tT}^{(2)}$ we need to perform at most two one-dimensional integrals 
(the integral we need is (\ref{eq:key_int}), but we note that $p_t(x)$ includes a normalising constant $K$---which is found be evaluating a second integral).
We are saved the complication of performing double integrals. 

To extend these results to higher dimensions we can split the `master' process $\{S_{tT}\}$ into more than two subprocesses.
Regardless of the number of subprocesses (i.e.~paid-claims processes), all of the best-estimate ultimate losses can be computed by performing at most two one-dimensional integrals.
This makes such a multivariate model highly computationally efficient.

It should be noted that the case where all the subprocesses are identical in law (in the example above, this is the case when $k=1$), many of the results can be extended to an arbitrary master LRB using Corollary \ref{coro:inc}.
Importantly, the computational efficiency can be achieved for an arbitrary LRB.

%% file: chap07.tex
%
%

\chapter{Cauchy information} \label{chap:Cauchy}
In this chapter, as in the last, we examine the random bridges of a stable process.
We look here at LRBs based on a driftless Cauchy process, which is strictly stable with index $\a=1$.
We call these LRBs Cauchy random bridges (CRBs).
The third moment of the (one-dimensional) marginal distribution of a CRB does not exist for $t\in(0,T)$, regardless of the choice of terminal distribution, since the corresponding moments of a Cauchy bridge do not exist.
Integrability is obtainable for moments of order less than three.
Hence, conditioning a Cauchy process on its time-$T$ distribution can moderate its wild behaviour.
This moderation cannot be as severe as exponentially-dampening the process's \levy density, which would produce a tempered-stable process.
We provide expressions for the first two moments of a CRB (when they exist).
These follow from the analysis of Cauchy bridges in Section \ref{sec:Cauchy}.
We then outline an algorithm for the simulation of CRBs.
The algorithm uses the representation of a Cauchy process as a Brownian motion time-changed by a stable-1/2 subordinator.
Returning to information-based pricing, we consider the valuation of a binary bond when the information process is a CRB.
We price a call option on the binary bond price from first principles.
This proves tractable, and a closed-form expression is derived for the call price.
Plots of simulations of bond price processes are provided at the end, which show the influence of the activity parameter $c$.

\section{Cauchy random bridges}
As in (\ref{eq:CauchyDensity}), we define
\begin{equation}
	f_t(y)=\frac{ct}{\pi(y^2+c^2t^2)},
\end{equation}
which is the density of a driftless Cauchy process.
Let $\{Z_{tT}\}$ be a random bridge of a driftless Cauchy process such that the marginal law of $Z_{TT}$ is $\nu$.
That is, $\{Z_{tT}\}$ is a CRB with activity parameter $c$.
Since $f_T(x)$ is bounded and positive on the whole real line, $Z_{TT}$ can be restricted to take values in any Borel set of $\R$.
For fixed $k>0$, let $\{\bar{Z}_{t,kT}\}$ be a CRB with activity parameter $c$ such that $\bar{Z}_{kT,kT} \law k\,Z_{TT}$, i.e.~%
\begin{equation}
	\Q[\bar{Z}_{kT,kT}\leq z]=\Q[k \, Z_{kT}\leq z]=\nu((-\infty,z/k]).
\end{equation}
Then it follows from Proposition \ref{prop:StableBridge} that
\begin{equation}
	\left\{ k^{-1} \bar{Z}_{kt,kT} \right\}_{0\leq t \leq T}\law \left\{Z_{tT} \right\}_{0\leq t \leq T}.
\end{equation}
If $\E[Z_{TT}]<\infty$ then we have
\begin{equation}
		\label{eq:cm1}
		\E\left[Z_{tT}\left|\, Z_{sT}=x \right.\right]=\frac{T-t}{T-s} x+\frac{t-s}{T-s}\E[Z_{TT} \,|\, Z_{sT}=x],
\end{equation}
and if $\E[Z_{TT}^2]<\infty$ then, using (\ref{eq:CBexp2}),
\begin{equation}
	\label{eq:cm2}
	\E\left[\left.Z_{tT}^2 \,\right| Z_{sT}=x \right]=\frac{T-t}{T-s} x^2 +\frac{t-s}{T-s}(\E[Z_{TT}^2 \,|\, Z_{sT}=x]+c^2(T-s)(T-t)),
\end{equation}
for $0\leq s <t <T$.
From (\ref{eq:cm1}) and (\ref{eq:cm2}) it is straightforward to show that
\begin{multline}
	\var[Z_{tT} \,|\, Z_{sT}]=\\ \frac{(T-t)(t-s)}{(T-s)^2}\left(Z_{sT}-\E[Z_{TT}\,|\,Z_{sT}]\right)^2+\frac{t-s}{T-s} \var[Z_{TT}\,|\,Z_{sT}]
										+c^2(t-s)(T-t).
\end{multline}
The $Z_{sT}$-conditional marginal law of $Z_{TT}$ is
\begin{align}
	\nu_s(\dd z)&=\frac{\psi_s(\dd z;Z_{sT})}{\psi_s(\R;Z_{sT})} \nonumber
	\\&=
		\frac{\frac{z^2+c^2T^2}{(z-Z_{sT})^2+c^2(T-s)^2}\, \nu (\dd z)}{\int_{-\infty}^{\infty}\frac{z^2+c^2T^2}{(z-Z_{sT})^2+c^2(T-s)^2}\, \nu (\dd z)}.
\end{align}
Then the $Z_{sT}$-conditional marginal density and distribution functions of $Z_{tT}$ can be written
\begin{align}
	y &\mapsto \int_{-\infty}^{\infty} f_{t-s,T-s}(y-Z_{sT};z-Z_{sT}) \, \nu_s(\dd z),
	\\\intertext{and}
	y &\mapsto \int_{-\infty}^{\infty} F_{t-s,T-s}(y-Z_{sT};z-Z_{sT}) \, \nu_s(\dd z), \label{eq:CauchyCondDist}
\end{align}
respectively. 
Here $f_{tT}(y;z)$ is given by (\ref{eq:CBden}), and $F_{tT}(y;z)$ is given in the statement of Proposition \ref{prop:CBDF}.

\subsection{Example}
Consider the case where $Z_{TT}$ can only take values in the countable set $\{a_i\}_{i=-\infty}^{\infty}$.
Writing $q(i)=\nu(\{a_i\})$, the transition law of $\{Z_{tT}\}$ is given by
\begin{equation}
	\Q[Z_{tT}\in\dd y \,|\, Z_{sT}=y]=\frac{c(T-t)(t-s)}{\pi(T-s)}\frac{\sum_i w_{tT}(i)\, q(i)}{\sum_i w_{sT}(i)\, q(i)} ((y-x)^2+c^2(t-s)^2)^{-1},
\end{equation}
and
\begin{equation}
	\Q[Z_{TT}=a_j \,|\, Z_{sT}=y]=\frac{w_{sT}(j) \, q(j)}{\sum_i w_{sT}(i)\, q(i)},
\end{equation}
where
\begin{equation}
	w_{tT}(i)=\frac{a_i^2+c^2T^2}{(a_i-y)^2+c^2(T-t)^2}.
\end{equation}

\section{Simulation}
The Cauchy process is a Brownian motion time-changed by an independent stable-1/2 subordinator.
For $\{W(t)\}$ a Brownian motion and $\{S_t\}$ a stable-1/2 subordinator, the joint density of
the pair $(S_T,W(S_T))$ is
\begin{equation}
	(y,z)\mapsto \1_{\{y>0\}}\frac{\e^{-z^2/(2y)}}{\sqrt{2\pi y}} \frac{cT}{\sqrt{2\pi}}\e^{-c^2T^2/(2y)}.
\end{equation}
Then, conditional on $W(S_T)=z$, the density of $S_T$ is
\begin{equation}
	\label{eq:con_den_cauchy}
	y\mapsto \1_{\{y>0\}} \half (z^2+c^2T^2)\frac{\e^{-(z^2+c^2T^2)/(2y)}}{y^2}.
\end{equation}
Integrating (\ref{eq:con_den_cauchy}) yields
\begin{equation}
	\Q[S_T> y \,|\, W(S_T)=z]=\exp\left( -\frac{z^2+c^2T^2}{2y} \right) \qquad \text{for $y>0$}.
\end{equation}
Given $W(S_T)$, $S_T$ is the reciprocal of an exponential random variable, and we have
\begin{equation}
	S_T  \law -\frac{z^2+c^2T^2}{\log U},
\end{equation}
where $U$ is a standard uniform random variable.
Conditioned on $W(S_T)$, the time-change process $\{S_t\}_{0\leq t\leq T}$ is stable-1/2 random bridge.

An algorithm for the simulation of stable-1/2 random bridges appeared in Section \ref{sec:Stable_Sim}.
If we can generate a value of $Z_{TT}$ from its \emph{a priori} distribution then we can construct a
CRB using the following algorithm:
\begin{enumerate}
	\item
		Generate a standard uniform random variate $U$, and then set $S_T=-(Z_{TT}^2+c^2T^2)/(\log U)$.
	\item
		Generate a path $\{S_t\}_{0\leq t \leq T}$ of a stable-1/2 random bridge to $S_T$.
	\item
		Generate a path $\{\b(t)\}_{0\leq t \leq 1}$ of a Brownian bridge with the terminal value 0.
	\item
		Return $\{ (S_t/S_T) Z_{TT} + \sqrt{S_T}\, \b(S_t/S_T) \}_{0\leq t \leq T}$.
\end{enumerate}

\citet{SM2010} describe a general algorithm for the simulation of a symmetric stable bridge using Monte Carlo methods.
This method could be adapted for the simulation of Cauchy random bridge paths.

\section{Cauchy binary bond}
We revisit the information-based binary bond model, this time taking the information process $\{\xi_{tT}\}$ to be a CRB.
The $X$-factor $X_T$ is the redemption amount of a credit-risky, zero-coupon, binary bond, and we set $\xi_{TT}=X_T$.
\emph{A priori}, we have $\Q[X_T=k_0]=p$ and $\Q[X_T=k_1]=1-p$, where $k_0<k_1$.
The $\xi_{tT}$-conditional distribution of $\xi_{TT}=X_T$ is given by
\begin{equation}
	\Q[\xi_{TT}=k_0 \,|\, \xi_{tT}=\xi ]=
	\left(1+\frac{(1-p)(c^2T^2+k_1^2)(c^2(T-t)^2+(k_0-\xi)^2)}{p(c^2T^2+k_0^2)(c^2(T-t)^2+(k_1-\xi)^2)}  \right)^{-1}.
\end{equation}
For convenience, we define
\begin{align}
	q_t(0)&=\Q[\xi_{TT}=k_0 \,|\, \xi_{tT}],
	\\\intertext{and} q_t(1)&=\Q[\xi_{TT}=k_1 \,|\, \xi_{tT}]=1-q_t(0).
\end{align}
We define the function $\Lambda(t,x):\R_+\times\R\rightarrow\R$ by
\begin{equation}
	\Lambda(t,x)=P_{tT}\,\E[X_T \,|\, \xi_{tT}=x].
\end{equation}
Then the process $\{P_{0t}\,\Lambda(t,\xi_{tT})\}_{0\leq t \leq T}$ is a martingale, and the bond price process $\{B_{tT}\}$ is given by
\begin{equation}
	B_{tT}=\Lambda(t,\xi_{tT}).
\end{equation}
After a straightforward calculation, we find
\begin{equation}
	\Lambda(t,x)=P_{tT} \frac{\a_0+\a_1 \, x+\a_2\, x^2}{\b_0+\b_1 \,x+\b_2\, x^2},
\end{equation}
where
\begin{equation}
\begin{aligned}
	 		\a_0&=p k_0 (c^2T^2+k_0^2)(c^2(T-t)^2+k_1^2)
	 				+(1-p) k_1 (c^2T^2+k_1^2)(c^2(T-t)^2+k_0^2),
	\\	\a_1&=-2k_0k_1\left(p(c^2T^2+k_0^2)+(1-p)(c^2T^2+k_1^2)\right),
	\\	\a_2&=p k_0(c^2T^2+k_0^2)+(1-p)k_1(c^2T^2+k_1^2),
\end{aligned}
\end{equation}
and
\begin{equation}
\begin{aligned}
	  \b_0&=p (c^2T^2+k_0^2)(c^2(T-t)^2+k_1^2)
	 			+(1-p)(c^2T^2+k_1^2)(c^2(T-t)^2+k_0^2),
	\\	\b_1&=-2\left(k_1p(c^2T^2+k_0^2)+k_1(1-p)(c^2T^2+k_1^2)\right),
	\\	\b_2&=p (c^2T^2+k_0^2)+(1-p)(c^2T^2+k_1^2).
\end{aligned}
\end{equation}

\subsection{Call option price}
We shall calculate the price of a call option on the bond price.
We approach the problem from first principles, as opposed to referring to the general formula derived in Section \ref{sec:Call_Option}.

At time $s<t$, the price of a call option on the bond price $B_{tT}$ is
\begin{align}
	C_{st}&=P_{st} \, \E\left[\left. (B_{tT}-K)^+ \,\right| \xi_{sT}  \right] \nonumber
	\\ &=P_{st} \, \E\left[\left. (\Lambda(t,\xi_{tT})-K)^+ \,\right| \xi_{sT}  \right],
\end{align}
where $K$ is the strike price.
Defining the set $A_t$ by
\begin{equation}
	\label{eq:A_t}
	A_t=\left\{ x: \Lambda(t,x)>K \right\},
\end{equation}
the call option price is
\begin{align}
	C_{st}&=P_{st} \, \E\left[\left.\1_{\{\xi_{tT}\in A_t\}} (\Lambda(t,\xi_{tT})-K) \,\right| \xi_{sT}  \right] \nonumber
		\\&=P_{st} \, \E\left[\left.\1_{\{\xi_{tT}\in A_t\}} \Lambda(t,\xi_{tT}) \,\right| \xi_{sT}  \right]
		- P_{st} K \, \Q[\xi_{tT}\in A_t \,|\, \xi_{sT}].
\end{align}
Note that when discussing the price of an option on an asset in Section \ref{sec:Call_Option}, we used the notation $B_t$ instead of $A_t$. 
We have changed to $A_t$ here to avoid confusion with the bond price $B_{tT}$.
We proceed by finding an explicit expression for the set $A_t$.

\begin{figure}[ht]
	\begin{center}
		\includegraphics{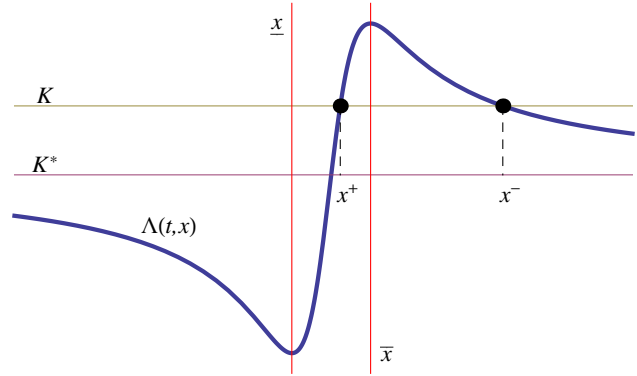}
	\end{center}
	\caption[The Cauchy binary bond price plotted as a function of the information process]{%
						An example plot of $\Lambda(t,x)$ as a function of $x$.}
	\label{fig:Cauchy}
\end{figure}

For fixed $t$, classical calculus reveals that $\Lambda(t,x)$ is maximised and minimised at the points
\begin{align}
	\overline{x}&=\half\left(k_0+k_1+\sqrt{(k_0-k_1)^2+4c^2(T-t)^2}\right),
	\\\intertext{and} 
	\underline{x}&=\half\left(k_0+k_1-\sqrt{(k_0-k_1)^2+4c^2(T-t)^2}\right),
\end{align}
respectively.
As a function of $x$, $\Lambda(t,x)$ is decreasing on the intervals $(-\infty,\underline{x}]$ and $[\overline{x},\infty)$,
and is increasing on the interval $[\underline{x},\overline{x}]$.
If $K\geq \Lambda(t,\overline{x})$ then $C_{sT}=0$, and if $K\leq \Lambda(t,\underline{x})$ then $C_{sT}=B_{sT}$.
See Figure \ref{fig:Cauchy} for an example plot of $x\mapsto\Lambda(t,x)$.

\begin{figure}[ht]
	\begin{center}
		\subfigure[Information process when $c=10^{-3}$.]{\includegraphics[scale=.9]{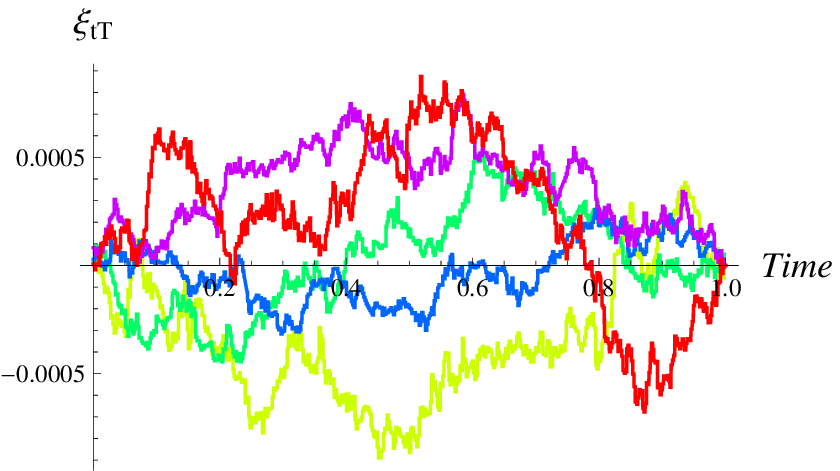}}
		\subfigure[Bond price with $c=10^{-3}$.]{\includegraphics[scale=.9]{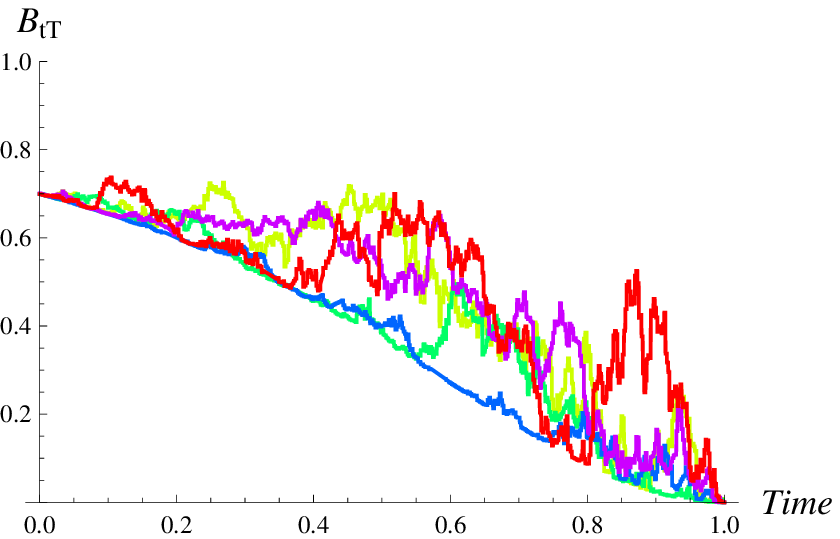}}
		\\
		\subfigure[Information process when $c=1$.]{\includegraphics[scale=.9]{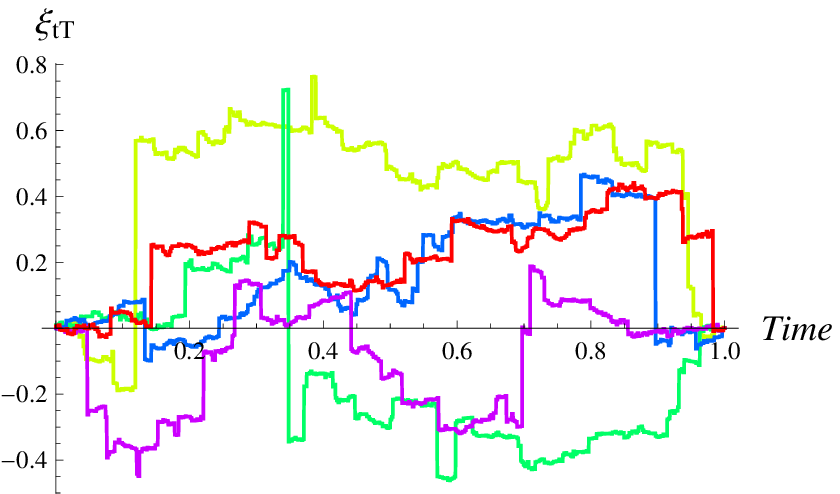}}
		\subfigure[Bond price with $c=1$.]{\includegraphics[scale=.9]{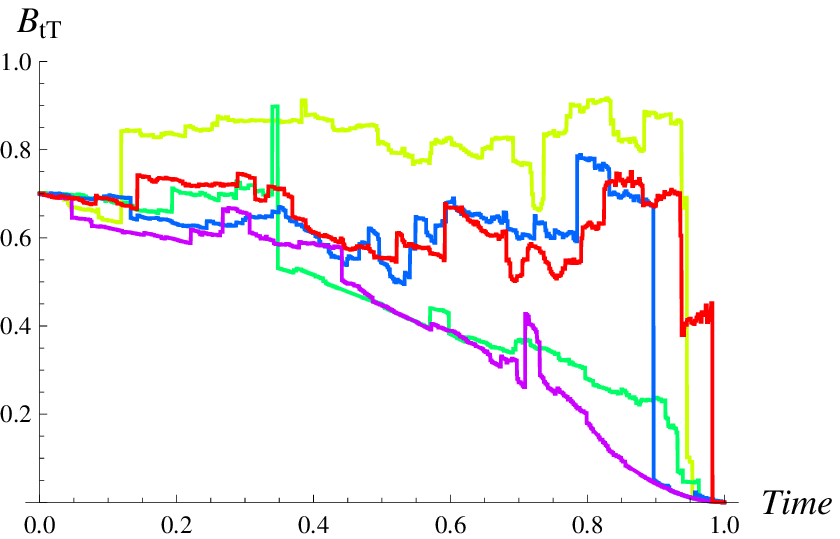}}
		\\
		\subfigure[Information process when $c=10$.]{\includegraphics[scale=.9]{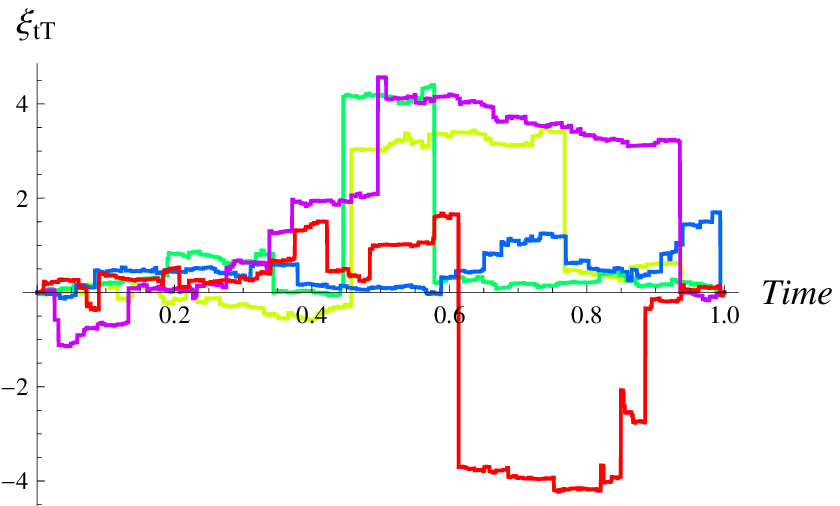}}
		\subfigure[Bond price with $c=10$.]{\includegraphics[scale=.9]{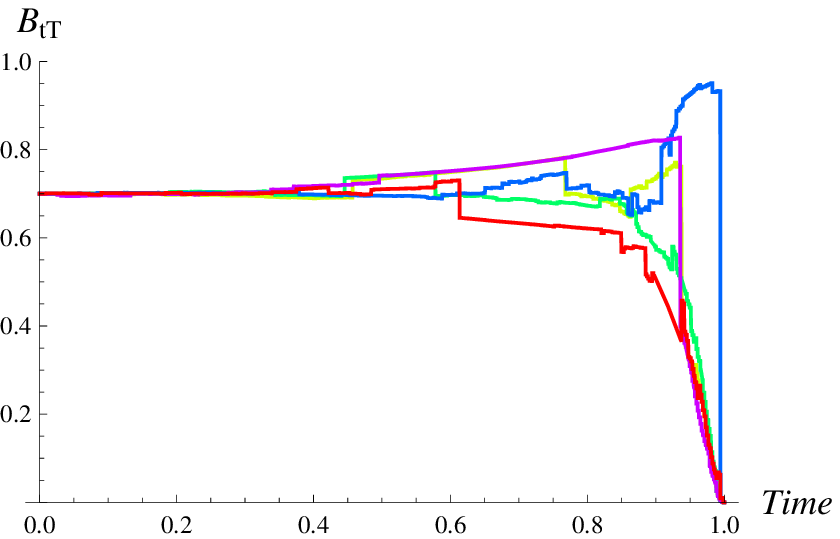}}
	\end{center}
	\caption[Cauchy binary bond simulations: defaulting bonds]{%
						Simulations of Cauchy information processes and bond price processes in cases where the bond defaults.
						Various values of the activity parameter $c$ are used.
						The other parameter values are fixed as $r_t=0$, $p=0.3$, and $T=1$.}
		\label{fig:C1}
\end{figure}

\begin{figure}[ht]
	\begin{center}
		\subfigure[Information process when $c=10^{-3}$.]{\includegraphics[scale=.9]{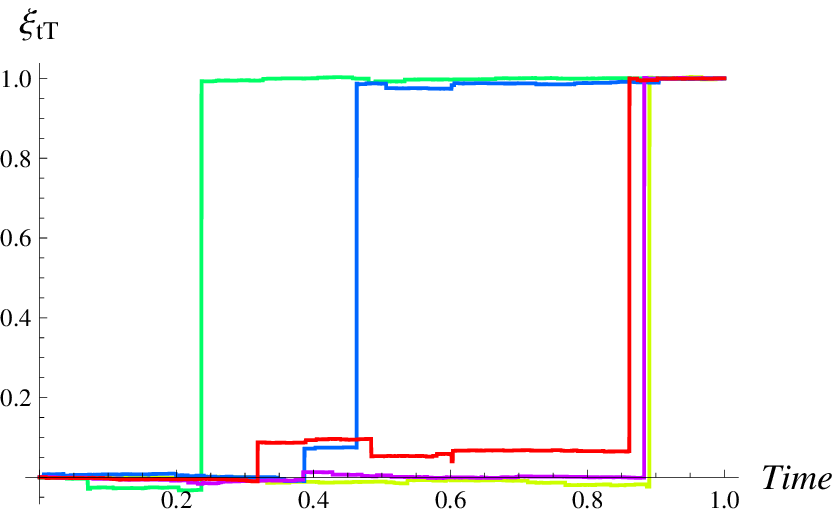}}
		\subfigure[Bond price with $c=10^{-3}$.]{\includegraphics[scale=.9]{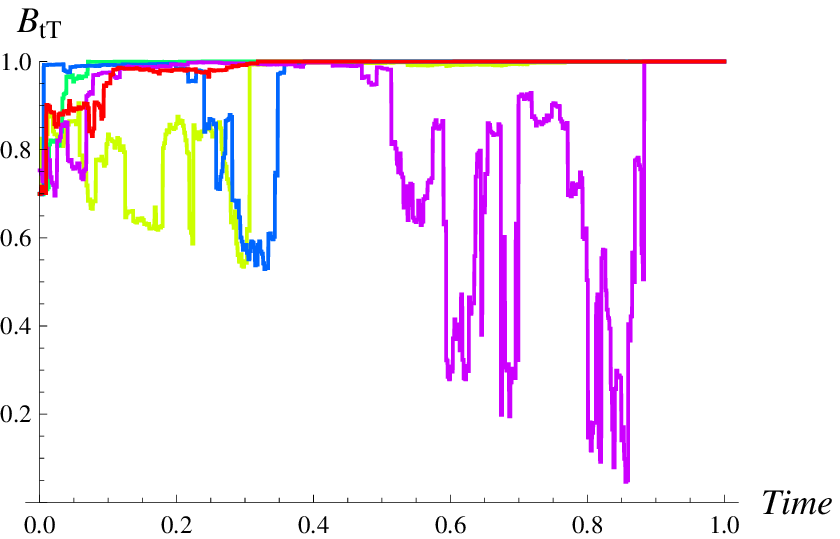}}
		\\
		\subfigure[Information process when $c=1$.]{\includegraphics[scale=.9]{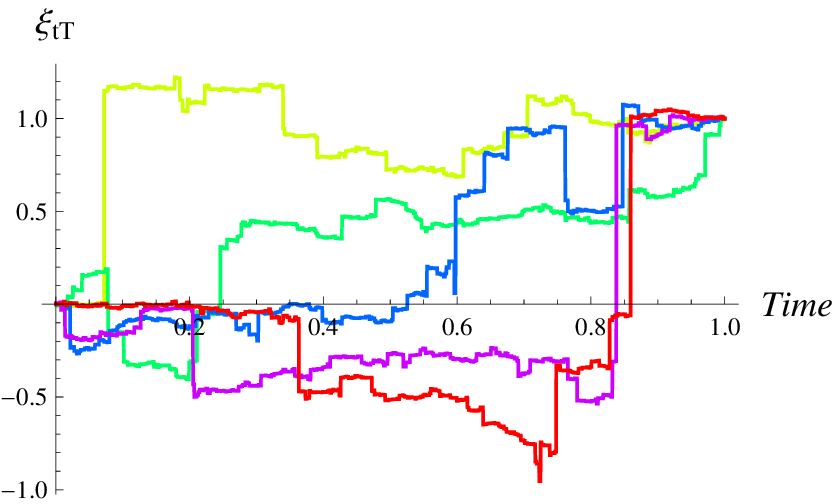}}
		\subfigure[Bond price with $c=1$.]{\includegraphics[scale=.9]{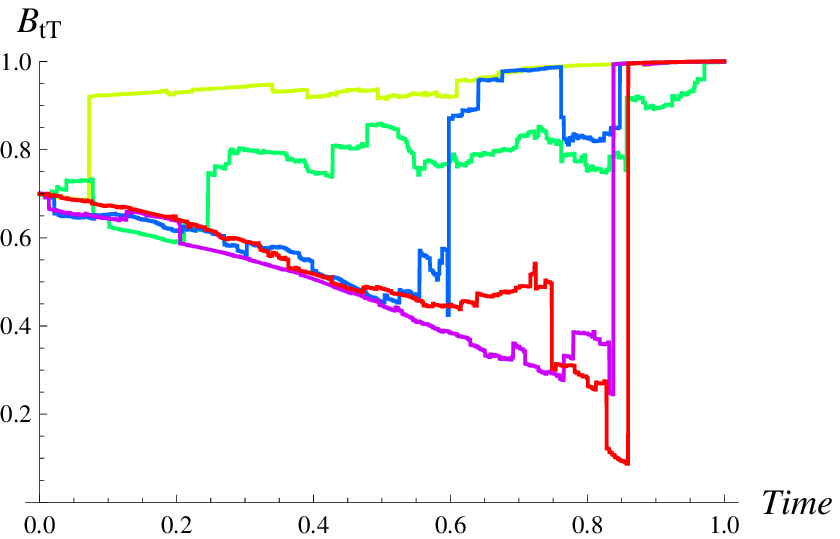}}
		\\
		\subfigure[Information process when $c=10$.]{\includegraphics[scale=.9]{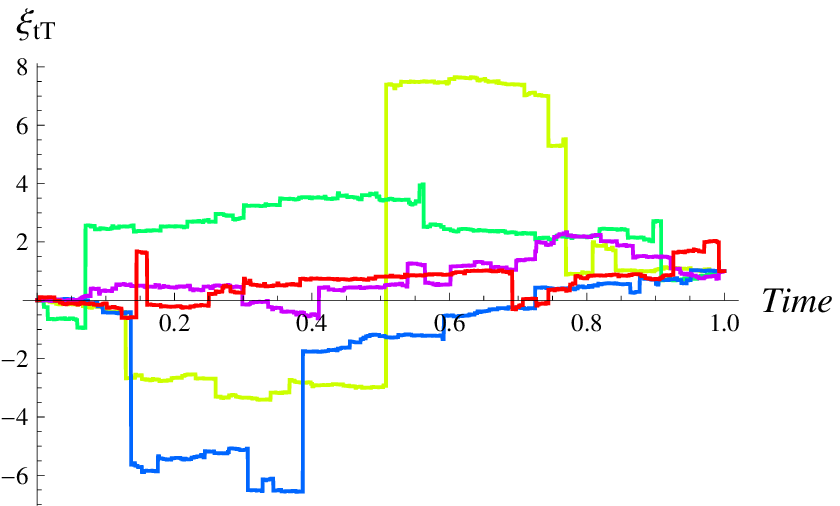}}
		\subfigure[Bond price with $c=10$.]{\includegraphics[scale=.9]{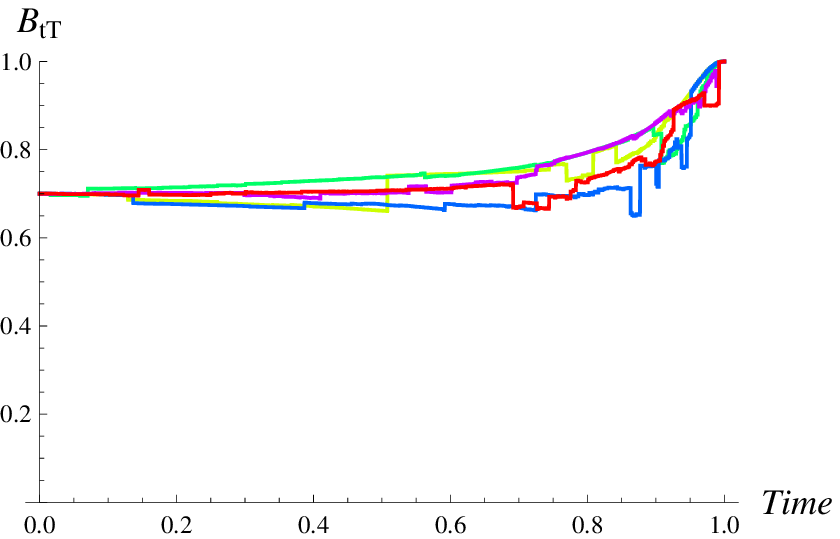}}
	\end{center}
	\caption[Cauchy binary bond simulations: non-defaulting bonds]{%
						Simulations of Cauchy information processes and bond price processes in cases where the bond does not default.
						Various values of the activity parameter $c$ are used.
						The other parameter values are fixed as $r_t=0$, $p=0.3$, and $T=1$.}
		\label{fig:C2}
\end{figure}

We define
\begin{equation}
	K^*=\lim_{x\rightarrow \pm \infty} \Lambda(t,x)=P_{tT} \frac{\a_2}{\b_2}.
\end{equation}
When $K\in(\underline{x},\overline{x})\backslash \{K^*\}$, the equation
\begin{equation}
	\label{eq:criticalK}
	\Lambda(t,x)=K
\end{equation}
is quadratic in $x$, and has the two real solutions
\begin{equation}
	x^{\pm}=\frac{-(P_{tT}\a_1-K\b_1)\pm \sqrt{(P_{tT}\a_1-K\b_1)^2-4(P_{tT}\a_2-K\b_2)(P_{tT}\a_0-K\b_0)}}{2(P_{tT}\a_2-K\b_2)}.
\end{equation}
When $K>K^*$ we have $x^+<x^-$, and when $K<K^*$ we have $x^+>x^-$.
When $K=K^*$ we set
\begin{align}
	x^+&=-\frac{\a_1\b_2-\a_2\b_0}{\b_2\a_1-\a_2\b_1},
	\\ x^-&=+\infty;
\end{align}
in this way $x^+$ is the only solution of equation (\ref{eq:criticalK}).
Then we can write
\begin{equation}
	A_t=\left\{
		\begin{aligned}
			&(x^+,x^-) && \text{for $K\geq K^*$,}
			\\ &\R\backslash [x^-,x^+] && \text{for $K<K^*$.}
		\end{aligned}
		\right.
\end{equation}
It follows from (\ref{eq:CauchyCondDist}) that
\begin{equation}
	\Q[\xi_{tT}\in A_t \,|\, \xi_{sT}]=
	 \1_{\{K< K^*\}}+\sum_{i=0}^1 q_t(i) \left( F_{st}^{-}(i)-F_{st}^{+}(i)\right),
\end{equation}
where 
\begin{equation}
	F_{st}^{\pm}(i)=F_{t-s,T-s}(x^{\pm}-\xi_{sT};k_i-\xi_{sT}).
\end{equation}
We also have
\begin{align}
	\E\left[\left.\1_{\{\xi_{tT}\in A_t\}} \Lambda(t,\xi_{tT}) \,\right| \xi_{sT}  \right]
	 &=\int_{y\in A_t} \Lambda(t,y) \, \Q[\xi_{tT}\in \dd y \,|\, \xi_{sT}] \nonumber
	\\ &=P_{tT}\int_{y\in A_t} \int_{z=-\infty}^{\infty} z \, \frac{\psi_t(\dd z;y)}{\psi_t(\R;y)}
					\frac{\psi_t(\R;y)}{\psi_s(\R;\xi_{sT})} f_{t-s}(y-\xi_{sT}) \d y \nonumber
	\\ &=P_{tT}\int_{z=-\infty}^{\infty}\int_{y\in A_t} z\, f_{t-s,T-s}(y-\xi_{sT};z-\xi_{sT}) \d y \, \nu_s(\dd z).
	\label{eq:C_bin_opt}
\end{align}
Noting that
\begin{multline}
	\int_{A_t} f_{t-s,T-s}(y-\xi_{sT};z-\xi_{sT}) \d y=
	\\\1_{\{K< K^*\}}+ F_{t-s,T-s}(x^- -\xi_{sT};z-\xi_{sT})-F_{t-s,T-s}(x^+ -\xi_{sT};z-\xi_{sT}),
\end{multline}
equation (\ref{eq:C_bin_opt}) simplifies to
\begin{multline}
	\E\left[\left.\1_{\{\xi_{tT}\in A_t\}} \Lambda(t,\xi_{tT}) \,\right| \xi_{sT}  \right]=
	\\ \1_{\{K< K^*\}}P_{st}^{-1}\Lambda(s,\xi_{sT})+P_{tT}\sum_{i=0}^1 k_i\, q_t(i) \left( F_{st}^{-}(i)-F_{st}^{+}(i)\right).
\end{multline}
The call price is then
\begin{equation}
	C_{st}=\1_{\{K<K^*\}} (B_{sT}-P_{st}K)+ 
	\sum_{i=0}^1  q_t(i)(P_{sT} k_i-P_{st}K) \left( F_{st}^{-}(i)-F_{st}^{+}(i)\right).
\end{equation}
Simulations of binary bond prices in this model are given in Figures \ref{fig:C1} and \ref{fig:C2}.

%% file: chap08.tex
%
%

\chapter{Normal inverse-Gaussian information} \label{chap:NIG}
This chapter is a mirror of Chapter \ref{chap:VG}, only we now consider the NIG process instead of the VG process.
Since the VG process and the NIG process are similar, we find that the NIG random bridge (NIGRB) is very similar to the VG random bridge.
To avoid too much repetition, we briefly list the contents of this chapter:
\begin{itemize}
	\item It is shown that an asymmetric NIGRB is a scaled standard NIGRB.
	\item An algorithm is provided for the simulation of NIGRB sample paths.
	\item An exponential NIG equity model is shown to be a special case of a single-factor information-based model when the information process is an NIGRB.
	\item The price of binary bond is derived in a single-factor information-based model.
	\item Simulations of binary bond price processes are plotted.
\end{itemize}

\section{Normal inverse-Gaussian random bridge} \label{sec:NIGRB}
Let $\{Y_{tT}\}$ be a standard NIGRB, so in the transition law (\ref{eq:LRBtranslaw}), we take $f_t(x)=f^{(\a)}_t(x)$ which we defined in (\ref{eq:NIG_den}) to be
\begin{equation}
	f^{(\a)}_t(x)=\frac{\a^2t\,\e^{\a^2 t}}{\pi\sqrt{\a^2t^2+x^2}}
		K_1\left[ \a\sqrt{\a^2t^2+x^2} \right].
\end{equation}
We assume that $Y_{TT}$ has marginal law $\nu$.
Since $0<f_T^{(\a)}(x)<\infty$ for all $x\in\R$, $Y_{TT}$ can be restricted to take values in any Borel set of $\R$.

\subsection{Example}
Fix $c>0$ and $\th\in\R$, and define
\begin{equation}
	\label{eq:rep_NIG_k_a}
	k^2=c^{-1}\sqrt{\th^2+c^2}, \qquad \a^2=c\sqrt{\th^2+c^2}.
\end{equation}
Suppose that $\{Y_{tT}\}$ is a standard NIGRB with parameter $\a$ and terminal density
\begin{equation}
	\label{eq:NIG_den1}
	f^{(c,\th,k)}_T(z)=\frac{cT}{k\pi}\e^{c^2 T+\th z/k^2} \sqrt{\frac{c^2k^2+\th^2}{c^2k^2T^2+z^2}}
		K_1\left[ k^{-2}\sqrt{(\th^2+c^2k^2)(c^2k^2 T^2+z^2)} \right].
\end{equation}
Equation (\ref{eq:NIG_den1}) is the time-$T$ density of an NIG process with parameters $c$, $\th$, and $k$.
From (\ref{eq:NIG_den_simple}) and (\ref{eq:NIG_k_a}), we can write
\begin{equation}
	\nu(\dd z)=\e^{(c^2-\a^2)T+\th z/k^2}  f^{(\a)}_T(z) \d z.
\end{equation}
We then have
\begin{align}
	\psi_t(\dd z; \xi)&=\frac{f^{(\a)}_{T-t}(z-\xi)}{f_T^{(\a)}(z)}f^{(c,\th,k)}_{T}(z) \d z  \nonumber
	\\ &= \e^{(c^2-\a^2)t+\th \xi/k^2} f_{T-t}^{(c,\th,k)}(z-\xi) \d z,
\end{align}
for $t\in[0,T)$.
For $0\leq s <t <T$, the transition law of $\{Y_{tT}\}$ is
\begin{align}
	\Q\left[Y_{tT} \in \dd y \left|\, Y_{sT}=x\right.\right]&=\frac{\psi_t(\R;y)}{\psi_s(\R;x)} f_{t-s}^{(\a)}(y-x)\d y \nonumber
	\\ &=\e^{(c^2-\a^2)(t-s)+\th (y-x)/k^2} f_{t-s}^{(\a)}(y-x) \d y \nonumber
	\\ &=f_{t-s}^{(c,\th,k)}(y-x) \d y.
\end{align}
Similarly
\begin{equation}
	\Q\left[Y_{TT} \in \dd y \left|\, Y_{sT}=x\right.\right]=f_{T-s}^{(c,\th,k)}(y-x) \d y.
\end{equation}
Over the time period $[0,T]$, $\{Y_{tT}\}$ is a two-parameter NIG process (since $k$ is a function of $\th$ and $c$).
The scaled LRB $\{\s Y_{tT} / k\}$ is a three parameter NIG process with parameters $c$, $\th$, and $\s$.

\section{Simulation}
Recall that a standard NIG process can be written as $\{W(X_t)\}$, where $\{W_t\}$ is a standard Brownian motion, and $\{X_t\}$ is an independent IG process.
The joint density of $(X_T,W(X_T))$ is
\begin{equation}
	(y,z)\mapsto \1_{\{y>0\}} \frac{c T\e^{c^2 T}}{\sqrt{2\pi}}\frac{1}{y^{3/2}}\exp\left(-\tfrac{c^2}{2}(T^2y^{-1}+y) \right)
									\frac{\exp\left( -\frac{z^2}{2y} \right)}{\sqrt{2\pi y}}.
\end{equation}
Given $W(X_T)=z$, up to a normalisation the density of $X_T$ is
\begin{equation}
	y\mapsto \1_{\{y>0\}} y^{-2} \exp\left(-\frac{1}{2}\left(\frac{c^2T^2+z^2}{y}+c^2y\right) \right).
\end{equation}
Hence the conditional distribution of $X_T$ is generalized inverse-Gaussian with density $f_{\textit{GIG}}(y;-1,\sqrt{c^2T^2+z^2},c)$, where
\begin{equation}
	 f_{\textit{GIG}}(x;\l,\delta,\g)= 
	 \1_{\{x>0\}} \left( \frac{\g}{\delta}\right)^{\l} \frac{1}{2\,K_{\l}[\g\delta]}x^{\l-1}\exp\left(-\tfrac{1}{2}(\delta^2x^{-1}+\g^2x) \right).
\end{equation}
Furthermore, given $W(X_T)$ the time-change $\{X_t\}_{0\leq t \leq T}$ is a stable-1/2 random bridge (since an IG random bridge is a stable-1/2 random bridge).
We can simulate a sample path of $\{Y_{tT}\}$ by the following algorithm:
\begin{enumerate}
	\item
		Generate a variate $Y_{TT}$ from the probability law $\nu$.
	\item
		Generate a variate $X_{TT}$ from the GIG distribution with parameters $\l=-1$, $\delta=\sqrt{c^2T^2+Y_{TT}^2}$, and $\g=c$.
	\item
		Simulate a path $\{X_{tT}\}$ of a stable-1/2 random bridge with parameter $c$, and terminating at value $X_{TT}$ at time $T$.
	\item
		Simulate a standard Brownian bridge $\{\b(t)\}_{0\leq t \leq 1}$, terminating at value 0 at time 1.
	\item
		Return $\{ X_{tT} Y_{TT}/X_{TT}+\sqrt{X_{TT}} \b(X_{tT}/X_{TT}) \}_{0\leq t \leq T}$.
\end{enumerate} 
We note that \citet{RW2003} apply related bridge-based constructions to the simulation of the NIG process.

\section{NIG equity model}
In a similar way to the VG equity model of Section \ref{sec:VG_Equity}, we can use a three-parameter NIG process to model the log-returns of a single stock (see \citet{RW2003}).
It follows from the asymptotic behaviour of the Bessel function $K_{\nu}[z]$ (see (\ref{eq:K_inf})) that the exponential moment of the NIG distribution exists only if $c^2>\s^2+2\th$, which we assume holds.
We model the stock price as
\begin{equation}
	S_t=S_0 \exp\left[{rt+L_t+w t}\right] \qquad (t\geq 0),
\end{equation}
where, under the risk-neutral measure, $\{L_t\}$ is an NIG process with parameter set $\{c,\th,\s\}$, $r>0$ is the constant rate of interest, and
\begin{equation}
	w=c\sqrt{c^2-\s^2-2\th}-c^2.
\end{equation}
Since $\E[\e^{L_t}]=\exp(-wt)$, the drift term $wt$ ensures that the discounted stock price process $\{\e^{-rt} S_t\}$ is a martingale under the risk-neutral measure.

We now show how to recover this model using the information-based approach when the information process is a symmetric NIGRB.
We let $X_T$ be a NIG random variable with density $f^{(c,\th,k)}_T(x)$, where $k=k(c,\th)$ is given by (\ref{eq:rep_NIG_k_a}), and set
\begin{equation}
	h(x)=\exp(rT+\s x /k+wT ).
\end{equation}
Let the information process $\{\xi_{tT}\}$ be a standard NIGRB with parameter $\a$ given by (\ref{eq:rep_NIG_k_a}), such that $\xi_{TT}=X_T$. 
Then $\{\s \xi_{tT} / k\}$ is an NIG process with parameter set $\{c,\th,\s\}$.
Furthermore
\begin{align}
	H_{tT}&=\exp(-r(T-t)) \E[h(X_T) \,|\, \xi_{tT}] \nonumber
	\\ &=\exp(rt+\s\xi_{tT}/k+wT) \, \E[\e^{\s(\xi_{TT}-\xi_{tT})/k}] \nonumber
	\\ &=\exp(rt+\s\xi_{tT}/k+wt),
\end{align}
for $t\in[0,T]$.
Hence we have $\{H_{tT}\}_{0\leq t \leq T}\law\{S_t\}_{0\leq t \leq T}$.

\section{NIG binary bond}
\begin{figure}[ht]
	\begin{center}
		\subfigure[Information process when $c=10$.]{\includegraphics[scale=.9]{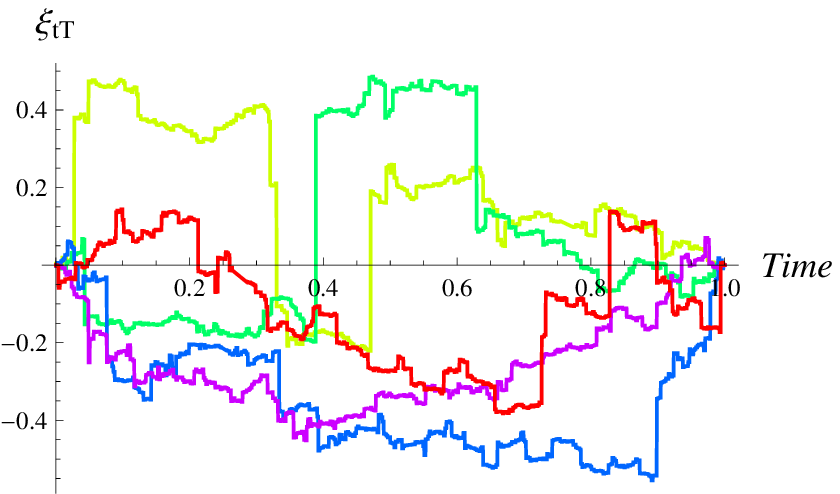}}
		\subfigure[Bond price with $c=10$.]{\includegraphics[scale=.9]{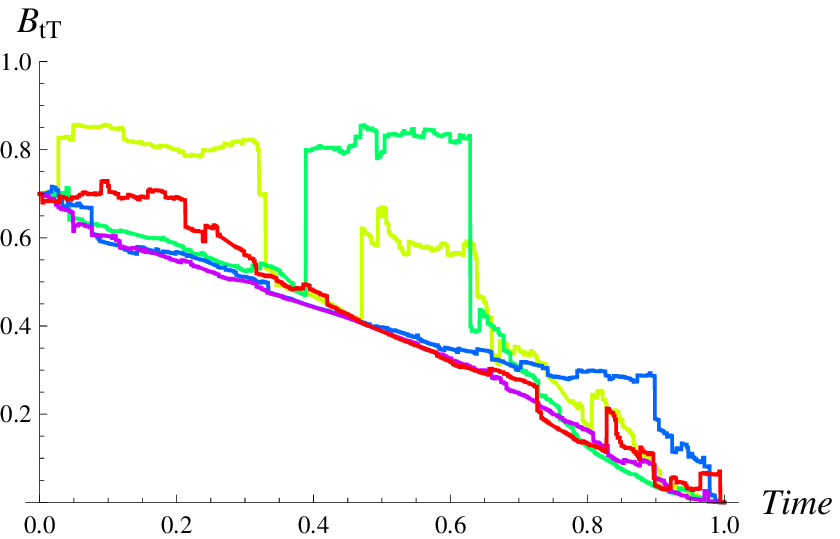}}
		\\
		\subfigure[Information process when $c=25$.]{\includegraphics[scale=.9]{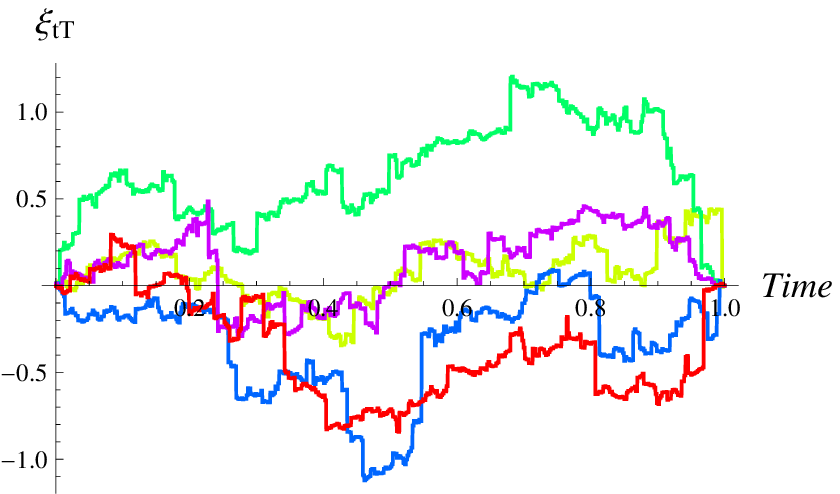}}
		\subfigure[Bond price with $c=25$.]{\includegraphics[scale=.9]{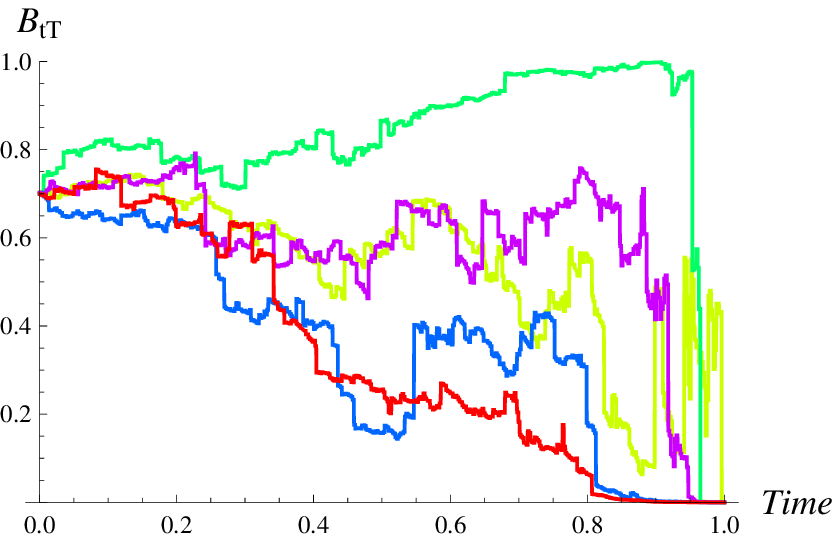}}
		\\
		\subfigure[Information process when $c=100$.]{\includegraphics[scale=.9]{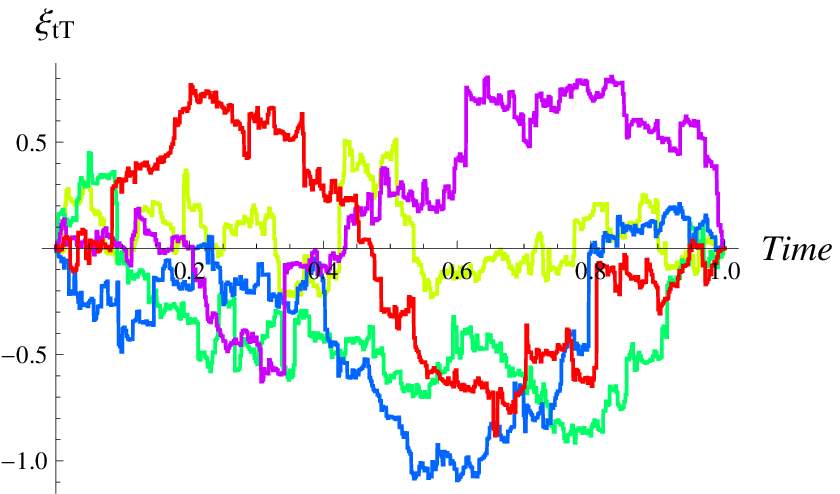}}
		\subfigure[Bond price with $c=100$.]{\includegraphics[scale=.9]{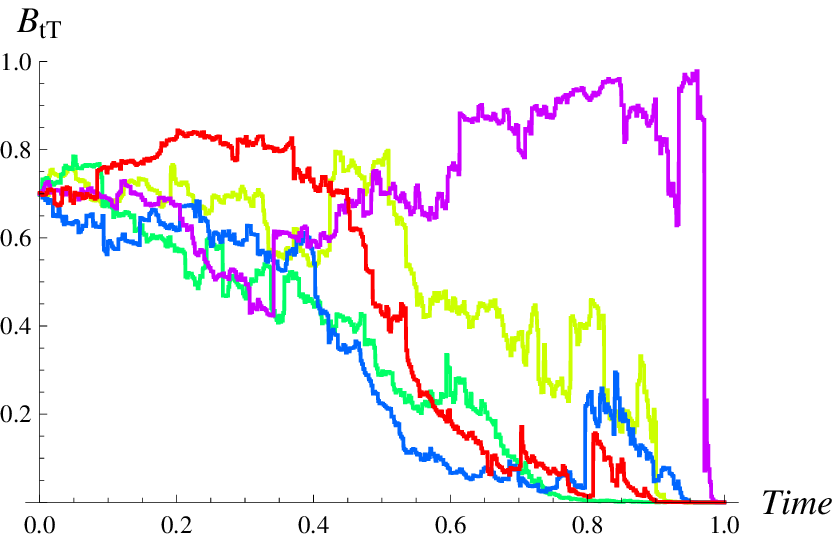}}
	\end{center}
	\caption[NIG binary bond simulations: defaulting bonds]{%
						Simulations of NIG information processes and bond price processes in cases where the bond defaults.
						Various values of the parameter $c$ used.
						The other parameter values are fixed as $r_t=0$, $p=0.3$, $T=1$, and $\s=1$.}
		\label{fig:NIG1}
\end{figure}

\begin{figure}[ht]
	\begin{center}
		\subfigure[Information process when $c=1$.]{\includegraphics[scale=.9]{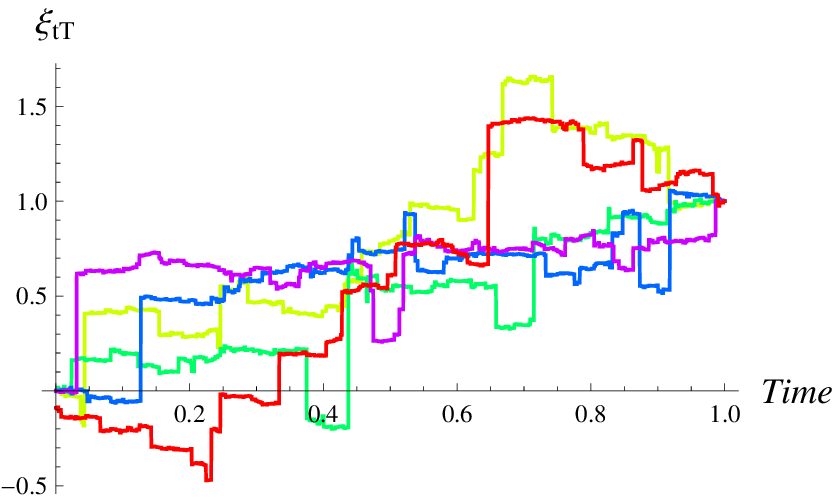}}
		\subfigure[Bond price with $c=1$.]{\includegraphics[scale=.9]{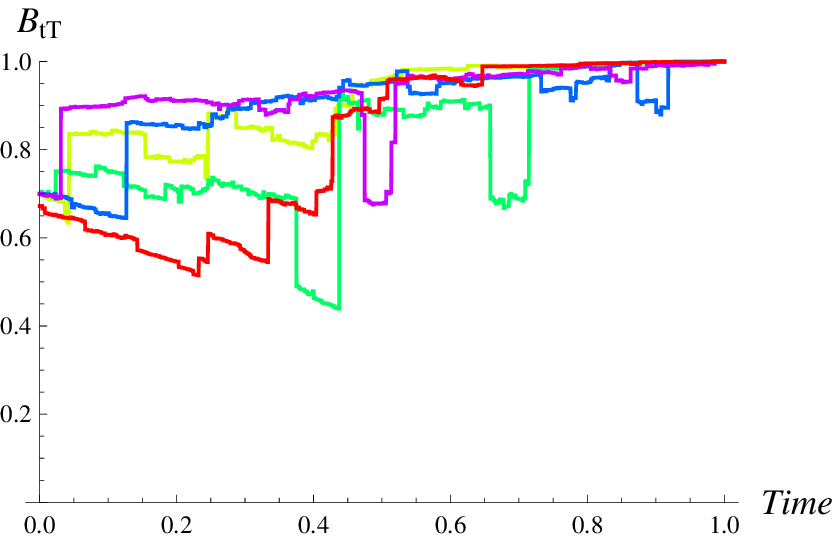}}
		\\
		\subfigure[Information process when $c=10$.]{\includegraphics[scale=.9]{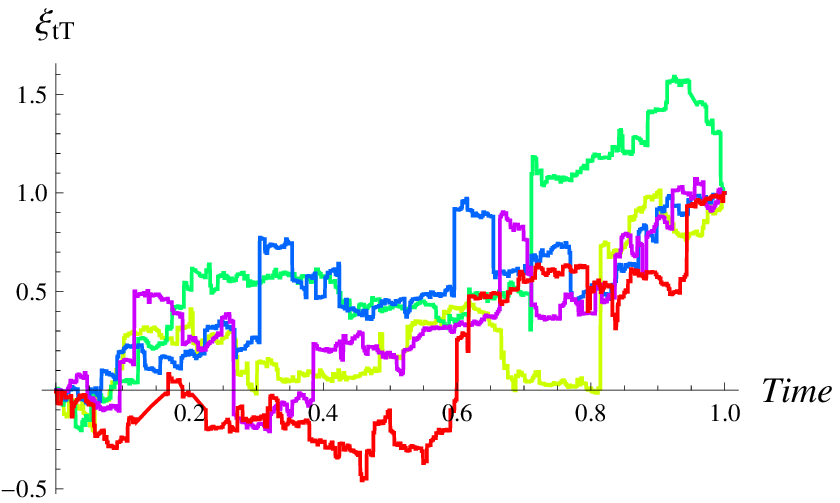}}
		\subfigure[Bond price with $c=10$.]{\includegraphics[scale=.9]{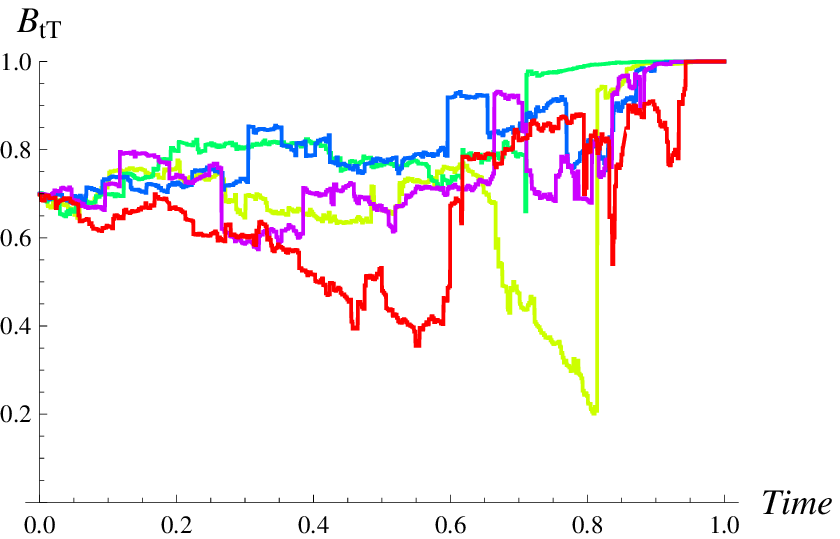}}
		\\
		\subfigure[Information process when $c=50$.]{\includegraphics[scale=.9]{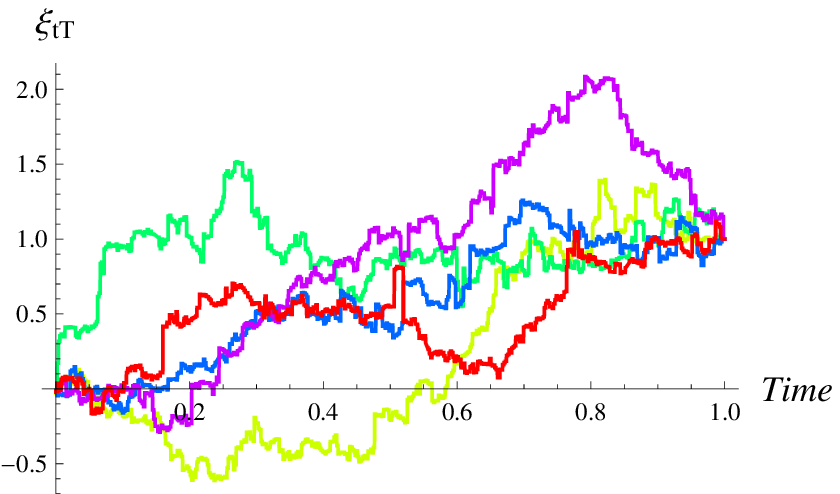}}
		\subfigure[Bond price with $c=50$.]{\includegraphics[scale=.9]{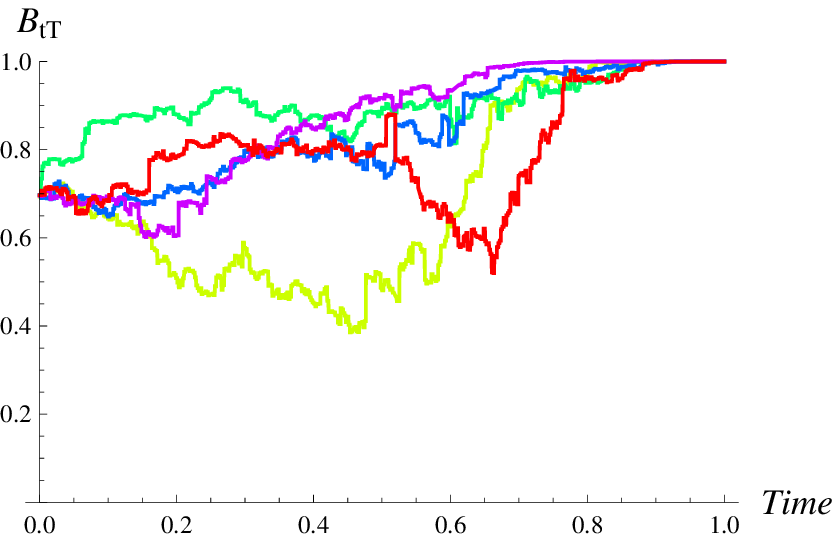}}
	\end{center}
	\caption[NIG binary bond simulations: non-defaulting bonds]{%
						Simulations of NIG information processes and bond price processes in cases where the bond does not default.
						Various values of the parameter $c$ used.
						The other parameter values are fixed as $r_t=0$, $p=0.3$, $T=1$, and $\s=1$.}
			\label{fig:NIG2}
\end{figure}

\begin{figure}[ht]
	\begin{center}
		\subfigure[Information process when $\s=0.1$.]{\includegraphics[scale=.9]{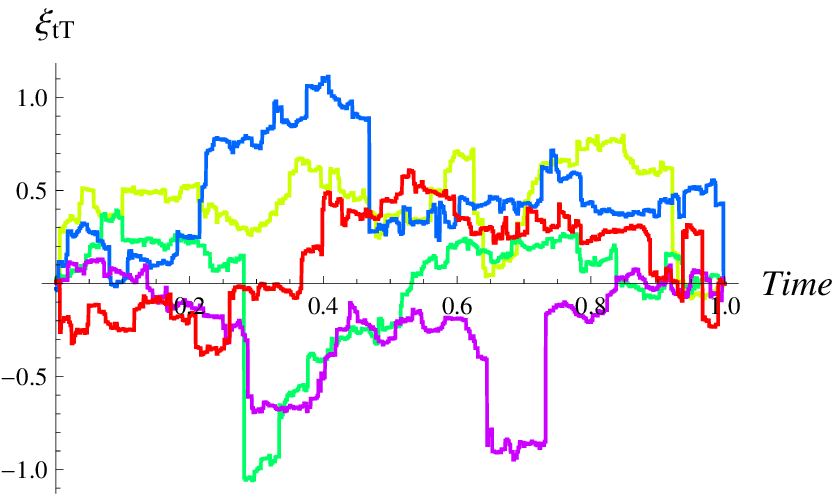}}
		\subfigure[Bond price with $\s=0.1$.]{\includegraphics[scale=.9]{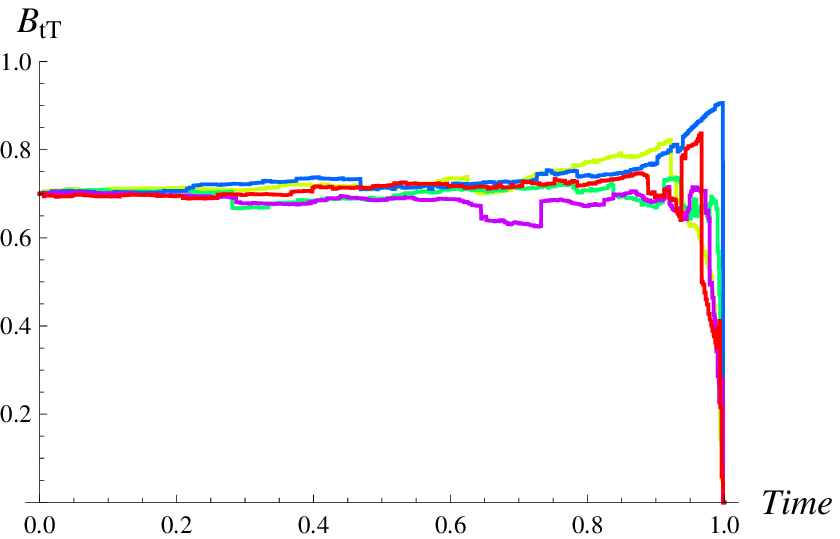}}
		\\
		\subfigure[Information process when $\s=1.0$.]{\includegraphics[scale=.9]{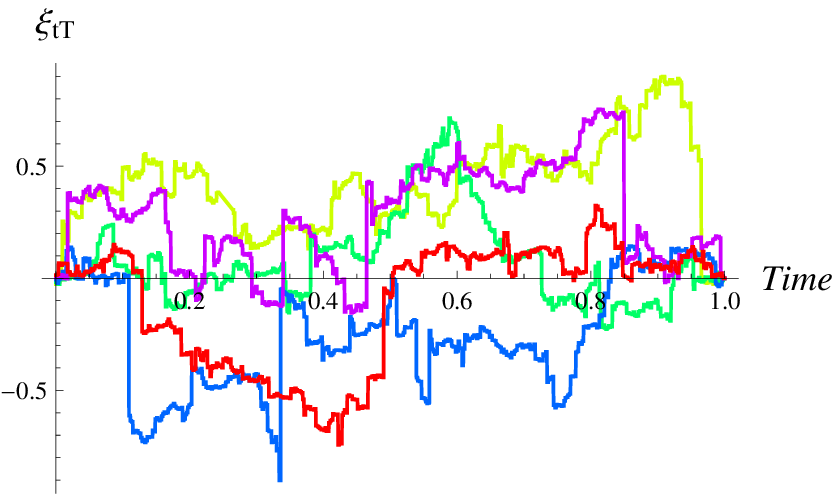}}
		\subfigure[Bond price with $\s=1$.]{\includegraphics[scale=.9]{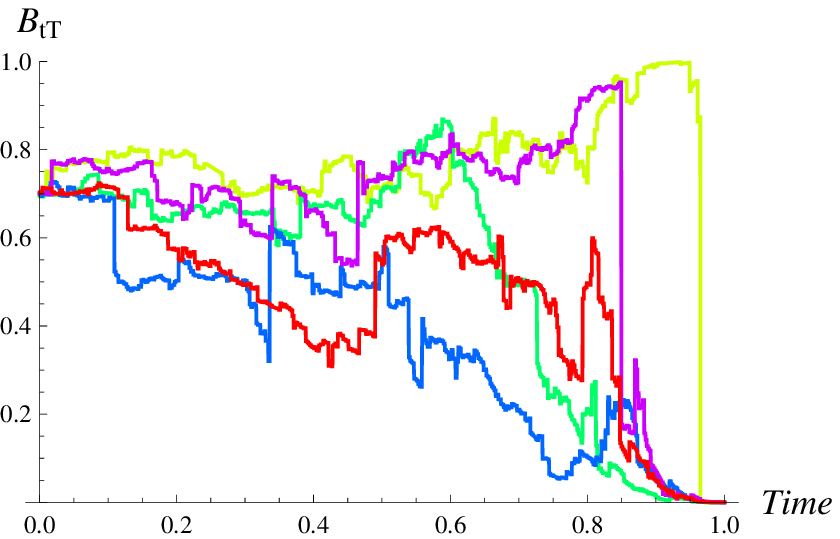}}
		\\
		\subfigure[Information process when $\s=10$.]{\includegraphics[scale=.9]{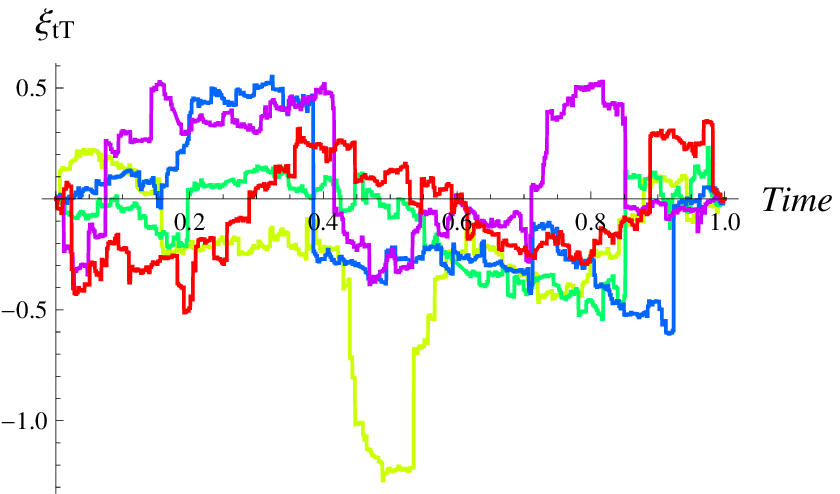}}
		\subfigure[Bond price with $\s=10$.]{\includegraphics[scale=.9]{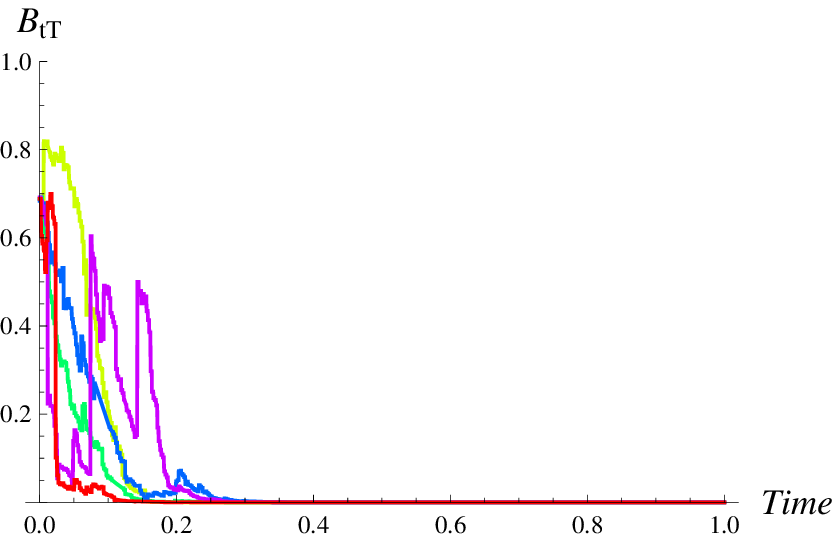}}
	\end{center}
	\caption[NIG binary bond simulations: varying the rate parameter]{%
						Simulations of NIG information processes and bond price processes in cases where the bond defaults.
						Various values of the information rate parameter $\s$ are used.
						The other parameter values are fixed as $r_t=0$, $p=0.3$, $T=1$, and $c=10$.}
		\label{fig:NIG3}
\end{figure}

We price a binary bond in an NIG information model including a rate parameter $\s>0$.
We let $X_T$ be an $X$-factor such that $\Q[X_T=0]=p$ and $\Q[X_T=\s]=1-p$.
Setting $h(x)=x/\s$, we suppose that $H_T=h(X_T)$ is the redemption amount of a credit-risky, zero-coupon bond.
We assume that the market filtration is generated by a standard NIGRB $\{\xi_{tT}\}$ such that $\xi_{TT}=X_T$.
The time-$t$ price of the bond is then given by
\begin{align}
	B_{tT}&=P_{tT} \, \E[h(X_T)\,|\,\xi_{tT}]
	\\&=P_{tT} \, \Q[X_T=\s \,|\, \xi_{tT}].
\end{align}
From Section \ref{sec:Binary_Bond}, we have
\begin{align}
	B_{tT}&=P_{tT} \left(1+{\frac{f_T^{(\a)}(\s)}{f_T^{(\a)}(0)}
							\frac{f_{T-t}^{(\a)}(-\xi_{tT})}{f_{T-t}^{(\a)}(\s-\xi_{tT})}\frac{p}{1-p}} \right)^{-1} \nonumber
	\\ &=P_{tT} \left(1+\g \sqrt{\frac{\a^2(T-t)^2+(\s-\xi_{tT})^2}{\a^2(T-t)^2+\xi_{tT}^2}}
				\frac{K_{1}\left[\a\sqrt{\a^2(T-t)^2+\xi_{tT}^2}\right]}{K_{1}\left[\a\sqrt{\a^2(T-t)^2+(\s-\xi_{tT})^2}\right]} \right)^{-1},
\end{align}
where
\begin{equation}
	\g=\frac{p}{1-p}\sqrt{\frac{\a^2T^2}{\a^2T^2+\s^2}} 
		\frac{K_1\left[\a\sqrt{\a^2T^2+\s^2}  \right]}{K_1\left[\a^2T\right]}.
\end{equation}
See Figures \ref{fig:NIG1}, \ref{fig:NIG2}, and \ref{fig:NIG3} for example simulations.

%% file: chap09.tex
%
%

\chapter{Poisson information}\label{chap:Poisson}
In this chapter we present some properties of a discrete random bridge, namely, the Poisson random bridge (PRB).
PRBs are counting processes, and we shall see that they are Poisson processes with a state-dependent intensity.

For a PRB, the distribution function of the waiting time to the next jump takes an interesting form.
It can be written as an elementary transformation of the probability generating function of the PRB's final value.
By differentiating this distribution function, we find that that the intensity of the PRB is a linear function of the PRB's expected final value.

We consider the case when the terminal value of a PRB has a negative binomial distribution.
We find that all of the increments of this PRB have a negative binomial distribution.
The intensity of this process is a linear function of the current value.
When a jump occurs, this causes the intensity to jump, and in the periods between jumps the intensity decreases smoothly.
We see therefore that this process exhibits a kind of contagion in the sense that every jump instantaneously increases the probability of another jump happening in some fixed future time interval.
A PRB whose terminal value has a log-series distribution also has negative binomial transition probabilities.
It can be considered to be a limiting case of a PRB with a negative binomial terminal distribution.

A mixed Poisson distribution is a Poisson distribution with a mixed mean.
Thus, the mixed Poisson distribution is a distribution on the non-negative integers.
When the terminal distribution of a PRB is mixed Poisson, then every increment of the PRB has a mixed Poisson distribution.
Such a process is called a mixed Poisson process.
If the gamma distribution is used to mix the Poisson mean, the negative binomial PRB is recovered.
Since the log-series distribution is a distribution on the positive integers, it is not a mixed Poisson distribution.
In fact, if the terminal distribution of a PRB takes values in a strict subset of $\p{N}_0$, then it is not a mixed Poisson process.
Hence, the class of PRBs does not coincide with the class of mixed Poisson processes.

The simulation of sample paths of a PRB is straightforward using the order statistics property of the jump times of the process.
We shall provide example simulations of a PRB with its conditional expected terminal value and intensity.

Using a construction similar to that of the compound Poisson process, the compound Poisson random bridge is a PRB with random jump sizes.
We shall derive an explicit expression for the marginal characteristic function of this process.
Although not explored in the present work, these compound processes could be applied to problems in finance and insurance (as an alternative to the compound Poisson process).

At the end of this chapter, we provide an application of PRBs to finance.
We consider the problem of pricing an $n$th-to-default credit swap.
The buyer of such a contract pays a continuous premium in exchange for a lump-sum payment on the date of the $n$th default from a basket of credit risky assets.
Modelling the default times of the credit risky assets as the jump times of a PRB we are able to provide an explicit expression for the value of the swap, with the additional simplifying assumptions that (a) the interest rate is constant, and (b) premiums are paid continuously.

\section{Poisson random bridge}
Let $\{N_{tT}\}$ be an LRB with law $\lrbd([0,T],\{Q_t\},P)$, where
\begin{equation}
	\label{eq:Po_PMF}
	Q_t(k)=\frac{\e^{-\l t}(\l t)^{k}}{k!}, \qquad k\in \p{N}_0,
\end{equation}
for some constant $\l>0$.
We call $\{N_{tT}\}$ a \emph{Poisson random bridge}.
One can consider $\{N_{tT}\}$ to be a Poisson process conditioned to have the probability mass function \mbox{$P:\p{N}_0\rightarrow\R_+$} at time $T$.
We assume that
\begin{equation}
	\E[N_{TT}]=\sum_k k P(k)<\infty.
\end{equation}

We can use the results from Section \ref{sec:Poisson} and Chapter \ref{chap:LRB} to derive various properties of $\{N_{tT}\}$.
It follows from (\ref{eq:D_terminal}) that the $N_{sT}$-conditional probability mass function of $N_{TT}$ is
\begin{equation}
	\label{eq:Ps}
	P_s(k)=\frac{\frac{k!}{(k-N_{sT})!}\left(1-\frac{t}{T}\right)^k P(k)}
					{\sum_{k=N_{sT}}^{\infty}\frac{k!}{(k-N_{sT})!} \left(1-\frac{t}{T}\right)^k P(k)}.
\end{equation}
Corollary \ref{coro:ExpectLRB} gives
\begin{equation}
	\E\left[N_{tT}\left|\,N_{sT}\right.\right]=\frac{T-t}{T-s}N_{sT}+\frac{t-s}{T-s}\E\left[N_{TT}\left|\,N_{sT}\right.\right].
\end{equation}
If $\E[N_{TT}^2]<\infty$, then conditioning on the terminal value $N_{TT}$ and recalling equation (\ref{eq:PBmom2}) gives
\begin{align}
	\E\left[N_{tT}^2\right]&=\E\left[\E\left[\left. N_{tT}^2\,\right| N_{TT} \right]\right] \nonumber
	\\ &=\frac{t}{T}\left(1-\frac{t}{T}\right) \E[N_{TT}]+\left(\frac{t}{T}\right)^2 \E\left[ N_{TT}^2\right].
\end{align}
This result can be generalised to the following:
\begin{prop}
	If $\sum_k k^2 P(k)<\infty$, then
	\begin{multline}
		\E\left[N_{tT}^2\left|\,N_{sT}\right.\right]
			=\left(\frac{T-t}{T-s}\right)^2N_{sT}^2+\frac{t-s}{T-s}\frac{T-t}{T-s}
					\left(2\,\E\left[N_{TT}\left|\,N_{sT}\right.\right]-1 \right)N_{sT}
			\\  +\frac{t-s}{T-s}\frac{T-t}{T-s}\,\E\left[N_{TT}\left|\,N_{sT}\right.\right]
							+\left(\frac{t-s}{T-s}\right)^2\E\left[\left.N_{TT}^2\,\right|N_{sT}\right].
	\end{multline}
\end{prop}
\begin{proof}
	Fix $s<T$ and define the process $\{\eta_{tT}\}_{s \leq t \leq T}$ by $\eta_{tT}=N_{tT}-N_{sT}$.
	It follows from the dynamic consistency property that, given $N_{sT}$, $\{\eta_{tT}\}$ is an LRB with terminal probability mass function $P^*(i)=P_s(i+N_{sT})$.
	With the understanding that $\E_s[A]=\E[A\,|\,N_{sT}]$, we have
	\begin{align}
		\E_s\left[N_{tT}^2\right]&=\E_s\left[(N_{sT}+\eta_{tT})^2 \right] \nonumber
		\\ &=N_{sT}^2+2 N_{sT} \E_s[\eta_{tT}]+\E_s[\eta_{tT}^2] \nonumber
		\\ &=N_{sT}^2+2 N_{sT} \tfrac{T-t}{T-s}\E_s[\eta_{TT}]+\tfrac{t-s}{T-s}\tfrac{T-t}{T-s}\E_s[\eta_{TT}]+\left(\tfrac{T-t}{T-s}\right)^2\E_s[\eta_{TT}^2] \nonumber
		\\ &=N_{sT}^2+2 N_{sT} \tfrac{T-t}{T-s}\left(\E_s[N_{TT}]-N_{sT}\right)+\tfrac{t-s}{T-s}\tfrac{T-t}{T-s}\left(\E_s[N_{TT}]-N_{sT}\right) \nonumber
				\\&\qquad +\left(\tfrac{T-t}{T-s}\right)^2\left(\E_s[N_{TT}^2]-2N_{sT}\E_s[N_{TT}]+N_{sT}^2\right) \nonumber
		\\ &=\left(\tfrac{T-t}{T-s}\right)^2N_{sT}^2+\tfrac{t-s}{T-s}\tfrac{T-t}{T-s}\left(2\,\E_s[N_{TT}]-1 \right)N_{sT} \nonumber
			\\  &\qquad+\tfrac{t-s}{T-s}\tfrac{T-t}{T-s}\,\E_s[N_{TT}]+\left(\tfrac{t-s}{T-s}\right)^2\E_s[N_{TT}^2].
	\end{align}
\end{proof}

Counting processes are characterised by the distribution of their jump times.
We shall see that the distribution function of the $i$th jump of the PRB is an infinite sum of weighted beta probabilities.
This is a consequence of the order statistics property of the jump times of Poisson processes.
From this result it becomes apparent that the distribution of the next-jump time of a PRB can be expressed in terms of the conditional probability generating function of its terminal value.
We shall then use the next-jump time distribution to write the intensity of the PRB in a form that highlights the impact of new information.

Denote the probability generating function of $P$ by $G_{P}(z)$, i.e.~
\begin{equation}
	G_{P}(z)=\E[z^{N_{TT}}]=\sum_{k=0}^{\infty} z^k P(k).
\end{equation}
Note that $G_{P}(z)<\infty$ for $0\leq z \leq 1$, so the definition of $G_{P}$ is non-degenerate.
Indeed, we have
\begin{equation}
	P(k)=\frac{G^{(k)}_P(0)}{k!}.
\end{equation}
Furthermore, since we have assumed that $\E[N_{TT}]<\infty$, it follows that $G'_{P}(z)$ exists for $0\leq z\leq 1$, with
\begin{equation}
	G'_{P}(1)=\lim_{z\rightarrow 1^-} G'_{P}(z)=\sum_{k=0}^{\infty}k \,P(k)=\E[N_{TT}].
\end{equation}
The probability generating function of $N_{tT}$ is
\begin{align}
	\E\left[z^{N_{tT}}\right]&=\E\left[\E\left[\left.z^{N_{tT}}\,\right| N_{TT}\right]\right] \nonumber
	\\ &=\E\left[\left(1-\frac{t}{T}+\frac{t}{T}z \right)^{N_{TT}} \right] \nonumber
	\\ &=G_{P}\!\left(1-\frac{t}{T}+\frac{t}{T}z \right),
\end{align}
for $z\leq 1$.
Define the $i$th jump time of $\{N_{tT}\}$ by
\begin{equation}
	T_i=\inf\{t\in [0,T]:N_{tT}=i\}.
\end{equation}
We adopt the convention that $\inf \emptyset=\infty$; if $N_{TT}<i$ then $T_i=\infty$.
From equation (\ref{eq:PB_JumpTime}), the distribution function of the $i$th jump time $T_i$ can be written in terms of the regularized incomplete beta function $I_z[a,b]$ in the form
\begin{align}
	\Q\left[T_i\leq t  \right]&=\sum_{k=0}^{\infty} \Q\left[\left. T_i\leq x \,\right| N_{TT}=k  \right] P(k) \nonumber
	\\ &=\sum_{k=0}^{\infty} \1_{\{1\leq i\leq k\}}I_{t/T}[i,k-i+1] P(k),
\end{align}
for $0\leq t \leq T$.
We also have
\begin{equation}
	\Q[T_i=\infty]=\Q[N_{TT}<i]=\sum_{k=1}^iP(k). \label{eq:pointmass}
\end{equation}
The distribution of $T_i$ is \emph{mixed} in the sense that it has a point mass at $\infty$ given by (\ref{eq:pointmass}), and a
continuous part with density
\begin{equation}
	\label{eq:Tdensity}
	t\mapsto \1_{\{0\leq t\leq T\}} \sum_{k=0}^{\infty} \1_{\{1\leq i\leq k\}}
		\frac{\left(\tfrac{t}{T}\right)^{i-1}\left(1-\tfrac{t}{T} \right)^{k-i}}{T\,\mathrm{B}[i,k-i+1]} P(k).
\end{equation}
Note that
\begin{equation}
	I_z[1,k]=1-(1-z)^k,
\end{equation}
from which it follows that
\begin{align}
	\Q[T_1\leq t]&=\sum_{k=1}^{\infty} \left[1-\left(1-\frac{t}{T}\right)^k \right] P(k) \nonumber
	\\&= 1-\sum_{k=0}^{\infty} \left(1-\frac{t}{T}\right)^k P(k) \nonumber
	\\&=1-G_{P}\!\left(1-\frac{t}{T} \right),
\end{align}
for $t\in [0,T]$.
The distribution function of $T_1$ is given by
\begin{equation}
	F_{T_1}(t)=\left\{
		\begin{aligned}
			&0 && t<0,
			\\ &1-G_{P}(1-\tfrac{t}{T}) && 0\leq t \leq T,
			\\ &1-P(0) && T<t<\infty,
			\\ &1 && t=\infty.
		\end{aligned}
			\right.
\end{equation}
The density of the continuous part of the distribution is
\begin{equation}
	t\mapsto \1_{\{0\leq t\leq T\}} \tfrac{1}{T} \, G'_{P}\!\left(1-\frac{t}{T} \right),
\end{equation}
and there is a point mass $P(0)$ at $t=\infty$ corresponding to the case where there are no jumps.
For $t\in[0,T]$, we have
\begin{equation}
	\Q[N_{tT}=0]=1-\Q[T_1\leq t]=G_{P}\!\left(1-\frac{t}{T} \right).
\end{equation}
Since $\{N_{tT}\}$ has stationary increments, we also have
\begin{equation}
	\Q[N_{u+t , T}-N_{u,T}=0]=G_{P}\!\left(1-\frac{t}{T} \right),
\end{equation}
where $t,u$ satisfy $0\leq u<u+t <T$.
We can use the dynamic consistency property to find the conditional distribution of the waiting time until the next jump.
We define the process $\{\eta_{tT}\}_{s\leq t\leq T}$ by $\eta_{tT}=N_{tT}-N_{sT}$.
Then, given $N_{sT}$, $\{\eta_{tT}\}$ is a PRB with terminal probability mass function 
\begin{equation}
	P^*(k)=P_s(k+N_{sT}).
\end{equation}
The probability generating function of $P^*$ is
\begin{align}
	G_{P^*}(z)&= \sum_{k=0}^{\infty} z^k \, P^*(k) \nonumber
	\\ &= \sum_{k=0}^{\infty} z^k \, P_s(k+N_{sT}) \nonumber
	\\ &= \sum_{n=N_{sT}}^{\infty} z^{n-N_{sT}} \, P_s(n) \nonumber
	\\ &= z^{-N_{sT}} \, G_{P_s}(z).
\end{align}
Denoting the $i$th jump time of $\{\eta_{tT}\}$ by $T_i^{(\eta)}$, we have
\begin{align}
	\Q[W_{i+1}\leq t \,|\, N_{sT}=i]&=\Q[T_1^{(\eta)}\leq t] \nonumber
	\\ &=1-G_{P^*}\!\!\left(\tfrac{T-t}{T-s} \right) \nonumber
	\\ &=1-\left(\tfrac{T-t}{T-s} \right)^{-N_{sT}}G_{P_s}\!\!\left(\tfrac{T-t}{T-s} \right),
\end{align}
for $t\in[s,T]$.
This is equivalent to
\begin{equation}
	\Q[N_{u+t,T}-N_{uT}>0 \,|\, N_{sT}]=1- \left(\tfrac{T-(s+t)}{T-s} \right)^{-N_{sT}}G_{P_s}\!\!\left(\tfrac{T-(s+t)}{T-s} \right),
\end{equation}
for $0\leq s \leq u<u+t <T$.

\begin{prop}
The intensity of the PRB is
\begin{equation}
	\l_s=\frac{\E[N_{TT} \,|\, N_{sT}]-N_{sT}}{T-s}.
\end{equation}
\end{prop}
\begin{proof}
	\begin{align}
		\lim_{t\rightarrow 0^+} \frac{\Q[N_{s+t,T}-N_{sT}=1 \,|\, N_{sT}]}{t} &= 
		\lim_{t\rightarrow 0^+} \frac{1- \left(\tfrac{T-(s+t)}{T-s} \right)^{-N_{sT}}G_{P_s}\!\!\left(\tfrac{T-(s+t)}{T-s} \right)}{t} \nonumber
		\\ &= \frac{\dd}{\dd u} \left[1- \left(\tfrac{T-u}{T-s} \right)^{-N_{sT}}G_{P_s}\!\!\left(\tfrac{T-u}{T-s} \right)   \right]_{u=s}	 \nonumber
		\\ &=\frac{G'_{P_s}(1)-N_{sT} \, G_{P_s}(1)}{T-s} \nonumber
		\\ &= \frac{\E[N_{TT} \,|\, N_{sT}]-N_{sT}}{T-s}.
	\end{align}
\end{proof}

\begin{rem}
	The intensity of a counting process is related to survival probabilities of jump times by
	\begin{equation}
		\Q[T_{1+N_{sT}}>t \,|\, N_{sT}]=\E\left[\left. \exp\left(-\int_{s}^t \l_{u_-} \d u \right)\,\right| N_{sT} \right],
	\end{equation}
	where $0\leq s< t \leq T$.
\end{rem}

\subsection{Example: Negative binomial prior}
Consider the case where $P$ is a negative binomial probability mass function, i.e.~
\begin{equation}
	\label{eq:NBlaw}
	P(k)=\binom{k+m-1}{k}p^m(1-p)^k, \quad \text{for $k\in\mathbb{N}_0$,}
\end{equation}
where $m>0$ and $0<p<1$ are constants.
For $m\in \p{N}_+$, the negative binomial distribution gives probabilities relating to the number of failures of independent Bernoulli trials.
If $\{B_i\}_{i=1}^{\infty}$ is a sequence of outcomes from independent Bernoulli trials, each with probability of success $p$, then
\begin{align}
	P(k)&=\Q[\text{$B_{k+m}$ is the $m$th failure}]
		\\&=\Q\left[\#\left\{B_i: B_i=0, 1\leq i \leq k+m \right\}=m\right].
\end{align}
The probability generating function for the negative binomial distribution is
\begin{equation}
	G_{P}(z)=\left[ \frac{p}{1-(1-p)z} \right]^m.
\end{equation}
For the remainder of this chapter, if $x!$ is written and $x\notin\p{N}_0$, then we adopt the convention that $x!=\G[x+1]$.

Let $\{N_{tT}\}$ be a PRB where $N_{TT}$ has the probability mass function $P$ given in (\ref{eq:NBlaw}).
Then we have
\begin{align}
	\phi_t(k;j)&=\frac{Q_{T-t}(k-j)}{Q_T(k)} P(k) \nonumber
	\\&=\1_{\{0\leq j\leq k\}}\frac{\e^{t\l}k!}{\l^jT^j(k-j)!}\left(1-\tfrac{t}{T} \right)^{k-j}\binom{k+m-1}{k}p^m(1-p)^k,
\end{align}
and
\begin{align}
	\sum_{k=-\infty}^{\infty}\phi_t(k,j) 
	&=\frac{\e^{t\l}}{\l^jT^j}\frac{(j+m-1)!}{(m-1)!}p^m(1-p)^j
						\sum_{k=j}^{\infty}\left[\left(1-\tfrac{t}{T}\right)(1-p) \right]^{k-j}\binom{k+m-1}{k-j} \nonumber
	\\&=\frac{\e^{t\l}j!}{\l^jT^j}\left[\frac{p}{p+\frac{t}{T}(1-p)} \right]^m\left[\frac{1-p}{p+\frac{t}{T}(1-p)} \right]^j\binom{j+m-1}{j}* \nonumber
	\\&\qquad*\sum_{n=0}^{\infty}\left[\left(1-\tfrac{t}{T}\right)(1-p) \right]^{n}\left[p+\tfrac{t}{T}(1-p)\right]^{j+m}\binom{n+j+m-1}{n} \nonumber
	\\&=\frac{\e^{t\l}j!}{\l^jT^j}\left[\frac{p}{p+\frac{t}{T}(1-p)} \right]^m\left[\frac{1-p}{p+\frac{t}{T}(1-p)} \right]^j\binom{j+m-1}{j}.
\end{align}
Hence the transition probabilities of $\{N_{tT}\}$ are given by
\begin{multline}
	\label{eq:com11}
	\Q[N_{tT}=j \,|\, N_{sT}=i] =\frac{\sum_k\phi_t(k,j)}{\sum_k\phi_s(k,i)}Q_{t-s}(j-i)
	\\=\1_{\{0\leq i\leq j\}} \binom{j+m-1}{j-i}\left[ \frac{p+\tfrac{s}{T}(1-p)}{p+\tfrac{t}{T}(1-p)} \right]^{i+m}
												\left[ \frac{\tfrac{t-s}{T}(1-p)}{p+\tfrac{t}{T}(1-p)} \right]^{j-i},
\end{multline}
and
\begin{multline}
	\label{eq:com12}
	\Q[N_{TT}=j \,|\, N_{sT}=i]=\frac{\phi_s(j,i)}{\sum_k\phi_s(k,i)}
	\\ =\1_{\{0\leq i\leq j\}} \binom{j+m-1}{j-i}\left[ p+\tfrac{s}{T}(1-p)\right]^{i+m}
												\left[ \left(1-\tfrac{s}{T} \right)(1-p)\right]^{j-i}.
\end{multline}
Remarkably, all the increments of the process $\{N_{tT}\}_{0\leq t \leq T}$ have negative binomial distributions.
Note that
\begin{equation}
	\Q[N_{tT}=0]=\left[ \frac{p}{p+\tfrac{t}{T}(1-p)} \right]^{m}=G_{P}\!\left(1-\tfrac{t}{T}  \right),
\end{equation}
as expected.
By definition, we have
\begin{equation}
	P_s(j)=\Q[N_{tT}=j \,|\, N_{sT}],
\end{equation}
and so
\begin{align}
	G_{P_s}(z)&=\sum_{k=N_{sT}}^{\infty} z^k P_s(k) \nonumber
	\\ &=z^{N_{sT}} \sum_{k=0}^{\infty} z^k P_s(k+N_{sT}) \nonumber
	\\ &=z^{N_{sT}}\left[ \frac{p+\tfrac{s}{T}(1-p)}{1-(1-p)(1-\tfrac{s}{T})z}\right]^{m+N_{sT}}.
\end{align}
It is then straightforward to verify that
\begin{align}
	\Q[N_{tT}-N_{sT}=0\,|\,N_{sT}]&=\left[ \frac{p+\tfrac{s}{T}(1-p)}{p+\tfrac{t}{T}(1-p)} \right]^{m+N_{sT}} \nonumber
		\\&=\left(\frac{T-t}{T-s}  \right)^{-N_{sT}}G_{P_s}\!\!\left(\frac{T-t}{T-s}  \right),
\end{align}
as expected.
At time $s$, the conditional expectation of $N_{TT}$ is
\begin{align}
	\E[N_{TT} \,|\, N_{sT}]&=N_{sT}+(N_{sT}+m)\frac{(1-\frac{s}{T})(1-p)}{p+\frac{s}{T}(1-p)} \nonumber
	\\ &=\frac{N_{sT}+m(1-\frac{s}{T})(1-p)}{p+\frac{s}{T}(1-p)} ,
\end{align}
and the intensity of the process $\{N_{tT}\}$ is
\begin{equation}
	\frac{\E[N_{TT} \,|\, N_{sT}]-N_{sT}}{T-s}=\frac{(N_{sT}+m)(1-p)}{pT+s(1-p)}.
\end{equation}
We see that a jump in the process $\{N_{tT}\}$ causes a jump in its intensity process.
\citet[2.2.4]{HB1970} proposes a \emph{positive contagion} model where the intensity of a Poisson process is a linear function of its current level, which bears some similarity with the negative binomial PRB.
One stark difference is that \citeauthor{HB1970} assumes the linear form for the intensity process in an \emph{ad hoc} manner, whereas the intensity of the PRB is a consequence of Bayesian updating.

\subsection{Example: Log-series prior}
We shall see that a PRB with log-series terminal distribution has similar dynamics to the a PRB with a negative binomial terminal distribution.
Notably, after the first jump (which occurs with probability 1) the increment distributions of the PRB are negative binomial.

For $p\in(0,1)$, log-series distribution has probability mass function
\begin{equation}
	\label{eq:log_series}
	P(k)=\frac{-1}{\log p} \frac{(1-p)^k}{k} \qquad k\in\p{N}_+.
\end{equation}
Assume that $\{N_{tT}\}$ is a PRB with the terminal probability mass function given in (\ref{eq:log_series}).
Note that $N_{TT}\geq 1$. 
When $N_{sT}=0$, the transition probabilities of $\{N_{tT}\}$ are given by
\begin{align}
	\Q[N_{tT}=j \,|\, N_{sT}=0]&=\left\{
		\begin{aligned}
			&\frac{\log[p+\tfrac{t}{T}(1-p)]}{\log[p+\tfrac{s}{T}(1-p)]} && (j=0),
			\\&\frac{1}{j \log[p+\tfrac{s}{T}(1-p)]}\left[\frac{\frac{t-s}{T}(1-p)}{p+\frac{t}{T}(1-p)}\right]^j && (j\in\p{N}_+),
		\end{aligned}
		\right.
	\\\intertext{and}	
	\Q[N_{TT}=j \,|\, N_{sT}=0]&=
						\frac{-(1-p)^j(1-\tfrac{s}{T})^j}{j \log[p+\tfrac{s}{T}(1-p)]} \qquad (j\in\p{N}_+).
\end{align}
When $N_{sT}>0$ the transition probabilities are similar to negative binomial case:
\begin{align}
	\label{eq:com21}
	\Q[N_{tT}=j \,|\, N_{sT}=i]&=\binom{j-1}{j-i}\left[ \frac{p+\tfrac{s}{T}(1-p)}{p+\tfrac{t}{T}(1-p)} \right]^{i}
												\left[ \frac{\tfrac{t-s}{T}(1-p)}{p+\tfrac{t}{T}(1-p)} \right]^{j-i},
\\	\label{eq:com22}
	\Q[N_{TT}=j \,|\, N_{sT}=i]&=\binom{j-1}{j-i}\left[ p+\tfrac{s}{T}(1-p)\right]^{i}
												\left[ \left(1-\tfrac{s}{T} \right)(1-p)\right]^{j-i},
\end{align}
for $i\in\p{N}_+$ and $j\geq i$.
Comparing (\ref{eq:com21}) and (\ref{eq:com22}) to (\ref{eq:com11}) and (\ref{eq:com12}) we see that $\{N_{tT}\}$ can be considered a PRB with negative binomial terminal distribution in the limiting case $m\rightarrow 0$.

The probability generating function of the log-series distribution is
\begin{equation}
	G_P(z)=\frac{\log[1-(1-p)z]}{\log p}.
\end{equation}
The distribution function of the first jump time of $\{N_{tT}\}$ is
\begin{equation}
	\Q[T_1\leq t]=\1_{\{t\geq 0\}}-\1_{\{0\leq t\leq T\}} \frac{\log[p+\frac{t}{T}(1-p)]}{\log p}.
\end{equation}
Then $T_1$ has a continuous distribution with density
\begin{equation}
	t\mapsto \1_{\{0\leq t\leq T\}}\frac{-(1-p)}{(pT+t(1-p))\log p}.
\end{equation}
Given $N_{sT}=0$, the distribution function of $T_1$ updates to
\begin{equation}
	\Q[T_1\leq t \,|\,N_{sT}=0]=\1_{\{t\geq s\}}-
		\1_{\{s\leq t\leq T\}} \frac{\log[1-\tfrac{t-s}{T-s}(1-p)(1-\tfrac{s}{T})]}{\log[p+\tfrac{s}{T}(1-p)]}.
\end{equation}
If $N_{sT}>0$ is given, then the jump time distributions are recovered from the negative binomial case by setting $m=0$.

\section{Mixed Poisson processes}
A wide and tractable class of PRBs consists of the \emph{mixed Poisson processes}.
\citet{Grendell1997} provides a thorough exposition of these processes (see also \citet{OL1940}).
If every increment of a process has a mixed Poisson distribution (i.e.,~a Poisson distribution with a mixed mean) then it is a mixed Poisson process.
We shall see that when the terminal distribution of a PRB is a mixed Poisson distribution, then the PRB is a mixed Poisson process.

Let $\{N_{TT}\}$ be a PRB with terminal probability mass function $P(k)$ given by
\begin{equation}
	P(k)= \frac{T^k}{k!}\int_{0}^{\infty} \th^k \e^{-\th T} \pi(\th) \d \th,
\end{equation}
for some probability density $\pi:\R_+\rightarrow\R_+$:
We say that $N_{TT}$ has a \emph{mixed Poisson distribution}.
If $\Theta$ is a random variable with density $\pi(\th)$, then $N_{TT}$ can be interpreted as a Poisson random variable with the unknown parameter $\Theta T$.
Without loss of generality, we my assume that $\l=1$ in (\ref{eq:Po_PMF}), so
\begin{equation}
	Q_t(k)=\frac{t^k\e^{-t}}{k!}.
\end{equation}
Then we have
\begin{align}
	\phi_t(k;j)&=\frac{Q_{T-t}(k-j)P(k)}{Q_T(k)}
	\\ &=\frac{(T-t)^{k-j}\e^t}{(k-j)!} \int_0^{\infty} \th^k \e^{-\th T} \pi(\th) \d \th,
\end{align}
for $j,k\in\p{N}_0$ satisfying $j\leq k$.
It is straightforward to show
\begin{equation}
	\sum_{k=j}^{\infty} \phi_t(k;j)=\e^t\int_0^{\infty} \th^j \e^{-\th t} \pi(\th) \d \th.
\end{equation}
Then the transition probabilities of $\{N_{tT}\}$ are given by
\begin{align}
	\Q[N_{tT}=j \,|\, N_{sT}=i]&=\frac{\sum_{k=j}^{\infty} \phi_t(k;j)}{\sum_{k=i}^{\infty} \phi_s(k;i)}Q_{t-s}(j-i) \nonumber
		\\&=\frac{\int_0^{\infty} \th^j \e^{-\th t} \pi(\th) \d \th}
					{\int_0^{\infty} \th^i \e^{-\th s} \pi(\th) \d \th}\,\frac{(t-s)^{j-i}\e^{-(t-s)}}{(j-i)!},
\end{align}
where $i,j\in\p{N}_0$ satisfy $i\leq j$, and $s,t$ satisfy $0\leq s<t<T$; and
\begin{align}
	\Q[N_{TT}=k \,|\, N_{sT}=i]&=\frac{\phi_s(k;i)}{\sum_{k=i}^{\infty} \phi_s(k;i)}
		\\&=\frac{\int_0^{\infty} \th^k \e^{-\th T} \pi(\th) \d \th}
					{\int_0^{\infty} \th^i \e^{-\th s} \pi(\th) \d \th}\,\frac{(T-s)^{k-j}}{(k-j)!},
\end{align}
where $i,k\in\p{N}_0$ satisfy $i \leq k$, and $0\leq s <T$.
When $\pi(\th)=\delta_{\mu}(\th)$ for some $\mu>0$, the transition probabilities reduce to those of a Poisson process with parameter $\mu$.
So, given the value of $\Theta>0$, we have
\begin{equation}
	\label{cond_MPP}
	\Q[N_{tT}=j\,|\, N_{sT}=i, \Theta]=\frac{(\Theta(t-s))^{j-i} \e^{-\Theta(t-s)}}{(j-i)!},
\end{equation}
for $i\leq j$ and $s\leq t \leq T$.

We shall show that the transition distributions of $\{N_{tT}\}$ are mixed Poisson distributions.
First, we find the conditional distribution of $\Theta$ given the information to date.
For $0< t_1<t_2<\cdots<t_n<T$, the joint law of $(\{N_{t_i,T}\}_i,N_{TT},\Theta)$ is given by
\begin{multline}
	\Q\left[N_{t_1,T}=k_1,N_{t_2,T}=k_2,\ldots,N_{t_n,T}=k_n,N_{T}=K,\Theta \in \dd \th\right]=
	\\ \frac{Q_{T-t_n}(K-k_n) \prod_{i=1}^n Q_{t_i-t_{i-1}}(k_i-k_{i-1})}{Q_T(K)} \frac{(\th T)^K \e^{-\th T}}{K!} \pi(\th) \d \th,
\end{multline}
where $t_0=0$ and $k_0=0$.
Using the Bayes theorem, we have
\begin{align}
	&\Q\left[\left.N_{TT}=K,\Theta \in \dd \th \,\right|  N_{t_1,T}=k_1, N_{t_2,T}=k_2,\ldots,N_{t_n,T}=k_n\right] \nonumber
	\\ &\qquad\qquad\qquad\qquad= \frac{\frac{Q_{T-t_n}(K-k_n)}{Q_T(K)} \frac{(\th T)^K \e^{-\th T}}{K!} \pi(\th) \d \th}
				{\int_{0}^{\infty}\sum_{K=k_n}^{\infty}\frac{Q_{T-t_n}(K-k_n)}{Q_T(K)} \frac{(\th T)^K \e^{-\th T}}{K!} \pi(\th) \d \th} \nonumber
	\\ &\qquad\qquad\qquad\qquad= \frac{Q_{T-t_n}(K-k_n) \th^K \e^{-\th T} \pi(\th) \d \th}
				{\int_{0}^{\infty}\sum_{K=k_n}^{\infty}Q_{T-t_n}(K-k_n) \th^K \e^{-\th T} \pi(\th) \d \th}.
\end{align}
This implies that
\begin{align}
	\Q\left[\Theta \in \dd \th \left|\,  \F^{N}_t \right.\right] 
		&=\frac{\sum_{k=N_{tT}}^{\infty} Q_{T-t}(k-N_{tT}) \th^k \e^{-\th T} \pi(\th) \d \th}
					{\int_{0}^{\infty}\sum_{k=N_{tT}}^{\infty}Q_{T-t_n}(k-N_{tT}) \th^k \e^{-\th T} \pi(\th) \d \th} \nonumber
			\\&=\frac{ \th^{N_{tT}} \e^{-\th t} \pi(\th) \d \th}
					{\int_{0}^{\infty}\th^{N_{tT}} \e^{-\th t} \pi(\th) \d \th},
\end{align}
where $\{\F^N_t\}$ is the filtration generated by $\{N_{tT}\}$.
We write
\begin{equation}
	\pi_t(\th)=\frac{ \th^{N_{tT}} \e^{-\th t} \pi(\th)}
					{\int_{0}^{\infty}\th^{N_{tT}} \e^{-\th t} \pi(\th) \d \th};
\end{equation}
thus $\pi_t(\th)$ is the $\F^N_t$-conditional density of $\Theta$.
From equation (\ref{cond_MPP}), we then have
\begin{align}
	\Q[N_{tT}=j \,|\, N_{sT}]&=\int_{\th=0}^{\infty} \Q[N_{tT}=j \,|\, N_{sT},\Theta=\th] \, \Q[\Theta \in \dd \th \,|\, N_{sT}] \nonumber
	\\ &= \frac{ (t-s)^{j-N_{sT}}}{(j-N_{sT})!} \int_0^{\infty}\th^{j-N_{sT}} \e^{-\th(t-s)}\pi_s(\th) \d \th.
\end{align}
Hence, all the increments of $\{N_{tT}\}$ have a mixed Poisson distribution.

\begin{rem}
	The negative binomial PRB encountered in the previous section can be recovered by taking
	\begin{equation}
		\pi(\th)=\frac{1}{\G[m]}\left(\frac{pT}{1-p}\right)^m \th^{m-1} \exp\left(-\frac{pT}{1-p} \th \right).
	\end{equation}
	In this case $\Theta$ is a gamma random variable.
\end{rem}

\begin{rem}
	The log-series distribution is not a mixed Poisson distribution (note that $P(0)=0$ for the log-series distribution).
	Hence, if the terminal distribution of a PRB is log-series then the PRB is not a mixed Poisson process.
	Indeed, if $P(k)=0$ for any $k\in\N_0$, then a PRB with terminal mass function $P$ is not a mixed Poisson process.
\end{rem}

The $N_{sT}$-conditional distribution of $N_{TT}$ is
\begin{align}
	\E[N_{TT}\,|\,N_{sT}]&=
	\sum_{k=N_{sT}}^{\infty} k	\frac{ (T-s)^{k-N_{sT}}}{(k-N_{sT})!} \int_0^{\infty}\th^{k-N_{sT}} \e^{-\th(T-s)}\pi_s(\th) \d \th \nonumber
	\\ &=\int_0^{\infty}\th \, \pi_s(\th) \d \th \nonumber
	\\ &=\E[\Theta \,|\, N_{sT}].
\end{align}
Then the intensity of $\{N_{tT}\}$ is given by
\begin{equation}
	\frac{\E[\Theta \,|\, N_{sT}]-N_{sT}}{T-s}=\frac{1}{T-s}\left(\frac{ \int_{0}^{\infty}\th^{1+N_{sT}} \e^{-\th s} \pi(\th) \d \th}
																										{\int_{0}^{\infty}\th^{N_{sT}} \e^{-\th s} \pi(\th) \d \th}-N_{sT} \right).
\end{equation}

\section{Simulation}
\begin{figure}[ht]
	\begin{center}
		\subfigure{\includegraphics[scale=.9]{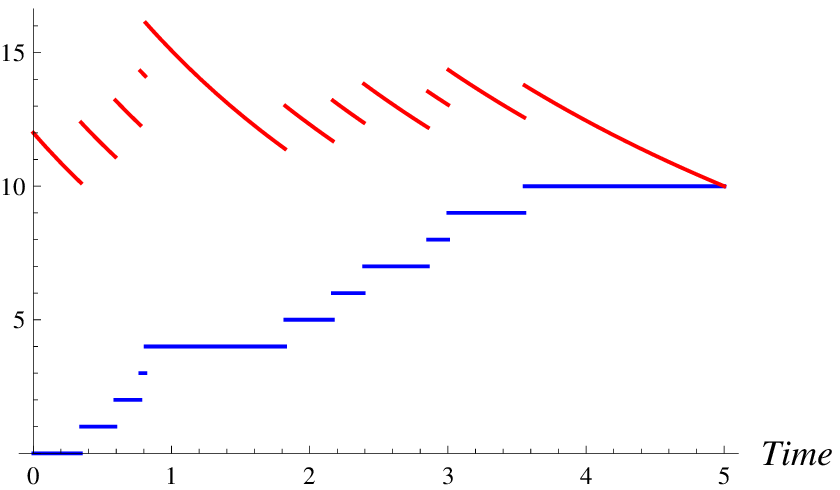}}
		\subfigure{\includegraphics[scale=.9]{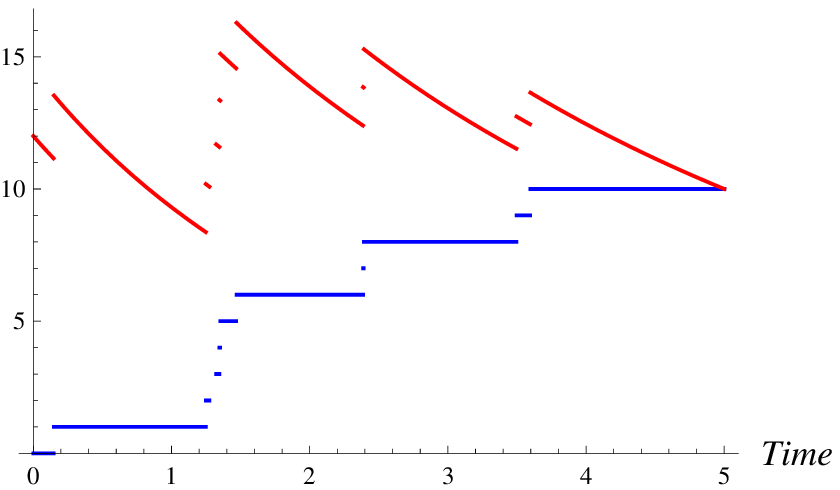}}
		\\
		\subfigure{\includegraphics[scale=.9]{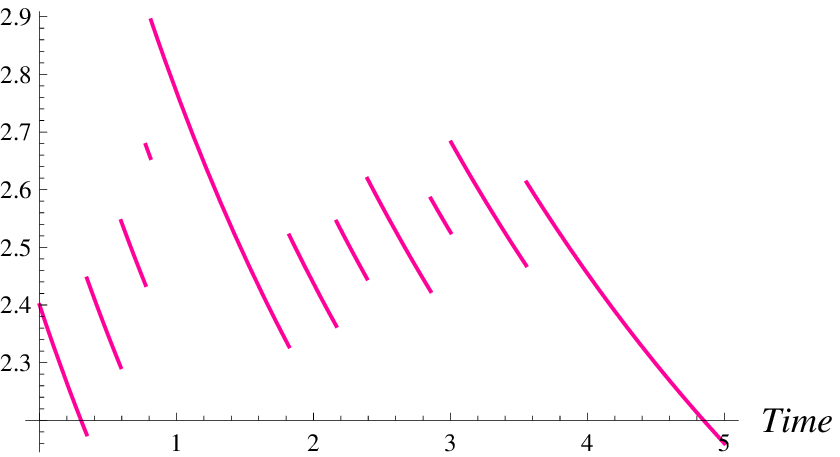}}
		\subfigure{\includegraphics[scale=.9]{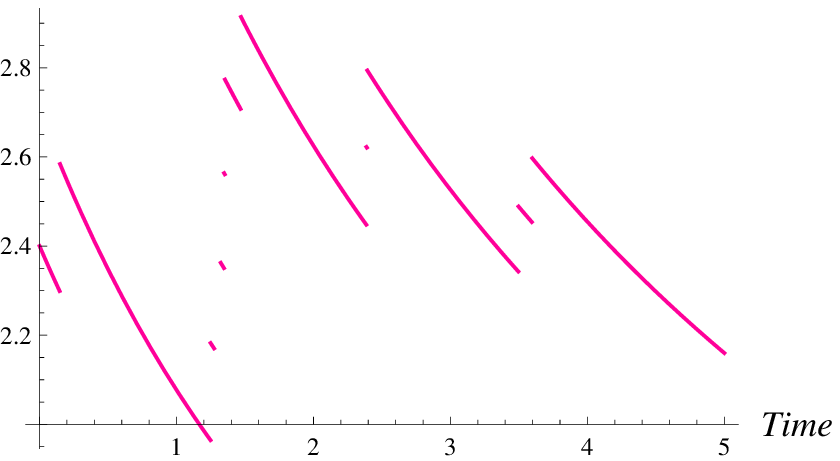}}
	\end{center}
	\caption[Poisson random bridge simulations]{%
					Top row: Two simulations of a Poisson random bridge with a negative binomial terminal distribution are plotted (bottom line) 
						with the conditional expectation of the terminal value (top line).
					Bottom row: The intensity process of the Poisson random bridge is plotted.
						The negative-binomial parameters are $m=8$, $p=0.4$.}
		\label{fig:Po}
\end{figure}

An efficient method for the simulation of a PRB is to generate the jump times.
The case $N_{TT}=0$ is trivial since we have $N_{tT}=0$ for all $t\in[0,T]$.
When $N_{TT}>0$ we set the jump times to be given by
\begin{equation}
	T_i=\frac{\sum_{j=1}^{i}E_j}{\sum_{j=1}^{1+N_{TT}} E_j} \qquad \text{for $1\leq i \leq N_{TT}$,}
\end{equation}
where the $E_j$'s are independent, identically distributed exponential random variables (with arbitrary parameter).
See Figure \ref{fig:Po} for example simulations.

\section{Compound Poisson random bridge}
Compound Poisson processes are \levy processes.
If the jump size distribution of a compound Poisson process is discrete then associated discrete LRBs can be defined,
and if the jump size distribution of a compound Poisson process is sufficiently smooth then associated continuous LRBs can be defined.
We call such LRBs compound-Poisson random bridges (C-PRBs).
Hence, a C-PRB can be considered to be a compound Poisson process conditioned to have a particular marginal distribution at time $T$.
The \emph{a priori} distribution of the number of jumps and the distribution of the jump sizes depend on the choice of terminal distribution.
The jumps of the C-PRBs occur at the jump times of a PRB that is adapted to the filtration of the C-PRB, and in general the jump sizes of a C-PRB are not independent.
Progress can be made in the analysis of C-PRBs, but dependence between the jump-size and jump-time distributions makes the path tricky.

A simpler but more tractable class of processes can be constructed by allowing the jump sizes of a PRB to be random and \emph{independent}.
We call such processes compound Poisson random bridges (CPRBs), and trust that the absence of a hyphen is enough to avoid confusion with C-PRBs.
It is CPRBs that we consider for the rest of this section.

Let the process $\{N_{tT}\}_{0\leq t\leq T}$ be a PRB, and let $N_{TT}$ have arbitrary marginal probability mass function $P\ll Q_T$.
Let $\{X_i\}$ be a sequence of independent, identically distributed random variables with characteristic function $\chi_X$.
We assume that the $X_i$'s are independent of $\{N_{tT}\}$.
We then define the CPRB $\{Y_{tT}\}$ by
\begin{equation}
	Y_{tT}=\sum_{i=1}^{N_{tT}} X_i.
\end{equation}
If $X_i\neq 0$ then $\{Y_{tT}\}$ jumps at the same times as $\{N_{tT}\}$, but the jump sizes of $\{Y_{tT}\}$ are random.
The characteristic function of $Y_{tT}$ is
\begin{align}
	\E\left[ \e^{\i\a Y_{tT}} \right]&=\E\left[ \exp\left(\i\a\sum_{i=1}^{N_{tT}}X_i\right)\right] \nonumber
	\\&=\E\left[\prod_{i=1}^{N_{tT}}\E\left[\left. \exp\left(\i\a X_i\right) \right| N_{tT}\right]\right] \nonumber
	\\&=\E\left[\chi_X(\a)^{N_{tT}}\right] \nonumber
	\\&=G_{P}\!\left(1-\frac{t}{T}+\frac{t}{T}\,\chi_X(\a) \right).
\end{align}
Using the dynamic consistency property, this can be generalised to
\begin{align}
	\E\left[\left. \e^{\i\a Y_{tT}} \,\right| Y_{sT}, N_{sT} \right]
			&=\e^{\i\a Y_{sT}}\, \E\left[\left. \exp\left(\i\a\sum_{i=N_{sT}+1}^{N_{tT}}X_i\right) \,\right| N_{sT}\right] \nonumber
	\\&=\e^{\i\a Y_{sT}}\, \E\left[\prod_{i=N_{sT}+1}^{N_{tT}}\E\left[\left. \exp\left(\i\a X_i\right) \right| N_{sT},N_{tT}\right]\right] \nonumber
	\\&=\e^{\i\a Y_{sT}}\, \E\left[\left. \chi_X(\a)^{N_{tT}-N_{sT}}\,\right| N_{sT}\right] \nonumber
	\\&=\e^{\i\a Y_{sT}}\, G_{P^*}\!\left(1-\frac{t-s}{T-s}+\frac{t-s}{T-s}\,\chi_X(\a) \right) \nonumber
	\\&=\e^{\i\a Y_{sT}}\,\frac{ G_{P_s}\!\left(\frac{T-t}{T-s}+\frac{t-s}{T-s}\,\chi_X(\a) \right)}
		{\left(\frac{T-t}{T-s}+\frac{t-s}{T-s}\chi_X(\a) \right)^{N_{sT}}}.
\end{align}
Counting processes compounded by random jump sizes are common in the insurance literature, and we refer the reader to \citet{HB1970}, \citet{OL1940}, and \citet{TK2004} for further details and references.

\section{$n$th-to-default baskets}
We now consider the pricing of an $n$th-to-default credit swap.
The swap provides protection against defaults occuring in a basket of $K$ homogeneous credit risks.
In return for a continuously paid premium, the buyer receives an amount $1-R$ on the date of the $n$th default, if this default occurs before date $T$.
The buyer pays the premium until the earlier time of the $n$th default and $T$.
Here $R$ represents the recovery rate on the credit risks, and is for simplicity assumed constant.
We also assume that interest rates are constant, i.e.~$r_t=r$, for all $t$.

We assume that the defaults occur on the jump times of the PRB $\{N_{tT}\}$ with terminal probability mass function $P:\{1,2,\ldots,K\}\rightarrow\R_+$.
Denoting the $n$th jump of $\{N_{tT}\}$ by $T_n$, the value of the swap to the buyer is
\begin{equation}
	V_0=\E\left[\int_{0}^{T_n \wedge T}\e^{-rt}\e^{q  t} \d t+(1-R)\1_{\{T_n\leq T\}}\e^{-rT_n} \right],
\end{equation}
where $q $ is the premium rate.
This can be rewritten as
\begin{equation}
	V_0=\frac{1}{q -r} \left( \E\left[\1_{\{T_n\leq T\}}\e^{(q -r)T_n}\right]
		+\e^{(q -r)T}\, \Q\left[T_n>T\right]-1\right) 
			 +(1-R)\,\E\left[\1_{\{T_n\leq T\}}\e^{-rT_n} \right].
\end{equation}
Note that
\begin{equation}
	\Q\left[T_n>T\right]=\Q\left[N_{TT}<n\right]=\sum_{k=1}^{n-1}P(k).
\end{equation}
Recalling the expression for density of $T_n$ given in (\ref{eq:Tdensity}), we have
\begin{align}
	\E\left[\1_{\{T_n\leq T\}}\e^{-rT_n} \right]
			&=\sum_{k=n}^{K} P(k)\int_{0}^{T} 
				\frac{\left(\tfrac{t}{T}\right)^{n-1}\left(1-\tfrac{t}{T} \right)^{k-n}}{T\,\mathrm{B}[n,k-n+1]}\e^{-rt} \d t \nonumber
			\\&=\sum_{k=n}^{K} M[n,k+1,-rT] P(k),
\end{align}
where Kummer's function $M[\a,\b,z]$ was given in (\ref{eq:Kummer}).
The initial value of the contract is then
\begin{multline}
	V_0=\frac{1}{q -r} \left( \sum_{k=n}^{K} M[n,k+1,(q -r)T] P(k)
		+\e^{(q -r)T}\, \sum_{k=0}^{n-1}P(k)-1\right) 
			\\ \qquad +(1-R)\,\sum_{k=n}^{K} M[n,k+1,-rT] P(k).
\end{multline}
Typically, the premium rate $q$ would be set to ensure that $V_0=0$, so the swap has zero initial value.
At time $s<T$, if the $n$th default has not yet occurred, the value of the swap to the buyer is
\begin{equation}
	V_s=\E\left[\left.\int_{s}^{T_n \wedge T}\e^{-rt}\e^{q  t} \d t+(1-R)\1_{\{T_n\leq T\}}\e^{-rT_n}\,\right| N_{sT} \right].
\end{equation}
By the dynamic consistency property, if we know $N_{sT}$ we can define an LRB by
\begin{equation}
	\eta_{tT}=N_{tT}-N_{sT} \qquad (s\leq t \leq T).
\end{equation}
Then we can write
\begin{align}
	V_s&=\frac{1}{q -r}
		\E\left[\left.\1_{\{T^{(\eta)}_{n-N_{sT}}\leq T\}}\e^{(q -r)T^{(\eta)}_{n-N_{sT}}}\,\right| N_{sT} \right]\nonumber
		\\&\qquad+\frac{1}{q -r}\e^{(q -r)T}\, \Q\left[T_n>T\,|\,N_{sT}\right]-\frac{1}{q -r}\e^{(q -r)s} \nonumber
		\\&\qquad\qquad+(1-R)\,\E\left[\left.\1_{\{T^{(\eta)}_{n-N_{sT}}\leq T\}}\e^{-rT^{(\eta)}_{n-N_{sT}}}\,\right| N_{sT}  \right].
\end{align}
In a similar way to the calculation of $V_0$, we find
\begin{align}
	V_s&=\frac{1}{q -r}  \sum_{k=n}^{K} M[n-N_{sT},k-N_{sT}+1,(q -r)(T-s)] P_s(k) \nonumber
		\\&\qquad +\frac{1}{q -r}\e^{(q -r)T}\, \sum_{k=N_{sT}}^{n-1}P_s(k)-\frac{1}{q -r}\e^{(q -r)s} \nonumber
		\\&\qquad\qquad +(1-R)\,\sum_{k=n}^{K} M[n-N_{sT},k-N_{sT}+1,-r(T-s)] P_s(k),
\end{align}
where $P_s$ is the $N_{sT}$-conditional probability mass function of $N_{TT}$, and is given by (\ref{eq:Ps}).

%% file: chap10.tex
\chapter{Some integrals}

\section{Proof of Proposition \ref{prop:DFIGLRB}}
\label{apdx:SBDF}

\begin{prop*}
	For $y\in[0,z]$, the distribution function of the random variable $S^{(z)}_{tT}$ is given by
	\begin{equation}
		F_{tT}(y;z)=
			\Phi\left[\frac{c(Ty-tz)}{\sqrt{yz(z-y)}}\right]+\left(1-\frac{2t}{T} \right)\e^{2c^2t(T-t)/z}\,
				\Phi\left[\frac{c((2t-T)y-tz)}{\sqrt{yz(z-y)}}\right].
	\end{equation}
\end{prop*}

\begin{proof}
	We need to show that 
	\begin{multline}
	\label{eq:main_int}
		\frac{1}{\sqrt{2\pi}} \frac{ct(T-t)}{T}\int_0^y \frac{\exp\left( -\frac{1}{2} \frac{c^2(Tu-tz)^2}{uz(z-u)}\right)}{\left(u-u^2/z\right)^{3/2}} \d u
		\\=\Phi\left[\frac{c(Ty-tz)}{\sqrt{yz(z-y)}}\right]+\left(1-\frac{2t}{T} \right)\e^{2c^2t(T-t)/z}\,
				\Phi\left[\frac{c((2t-T)y-tz)}{\sqrt{yz(z-y)}}\right].
	\end{multline}
	First we define
	\begin{equation}
		\label{eq:apdx_var_change}
		x(u)=\frac{c(Tu-tz)}{\sqrt{uz(z-u)}}.
	\end{equation}
	Inverting this function is a matter of finding the positive root of a quadratic equation, and it gives
	\begin{equation}
		u(x)=\frac{z(2c^2tT+x^2z)+xz\sqrt{4c^2t(T-t)z+x^2z^2}}{2(c^2T^2+x^2z)}.
	\end{equation}
	Differentiating $x(u)$ yields
	\begin{equation}
		\label{eq:dxdu}
		x'(u)=\left(\frac{(T-2t)u+tz}{2z}\right) \frac{c}{(u-u^2/z)^{3/2}}.
	\end{equation}
	Writing
	\begin{equation}
		\label{eq:alphadef}
		\a(x)=\sqrt{4c^2t(T-t)z+x^2z^2},
	\end{equation}
	we have
	\begin{align}
		\frac{t(T-t)}{T}&\frac{2z}{(T-2t)u(x)+tz} \nonumber
		\\ &=\frac{4t(T-t)(c^2T^2+x^2z)}{4c^2tT^2(T-t)+T^2x^2z+T(T-2t)x\a(x)} \nonumber
		\\ &=\frac{T^2 \a(x)^2-(T-2t)^2x^2z^2}{T^2\a(x)^2+T(T-2t)xz\a(x)} \nonumber
		\\ &=\frac{(T\a(x)+(T-2t)xz)(T\a(x)-(T-2t)xz)}{T\a(x)(T\a(x)+(T-2t)xz)} \nonumber
		\\ &=1-\frac{(T-2t)xz}{T\a(x)}.	\label{eq:hardslog}
	\end{align}
	Then (\ref{eq:dxdu}) and (\ref{eq:hardslog}) give
	\begin{equation}
		\left(1-\frac{(T-2t)x(u)z}{T\sqrt{4c^2t(T-t)z+x(u)^2z^2}}\right) x'(u)=\frac{ct(T-t)}{T(u-u^2/z)^{3/2}}.
	\end{equation}
	So making the change of variable $x=x(u)$ on the left hand side of (\ref{eq:main_int}) gives
	\begin{align}
		&\int_{-\infty}^{x(y)} \frac{\e^{-x^2/2}}{\sqrt{2 \pi}} \left(1-\frac{(T-2t)xz}{T\sqrt{4c^2t(T-t)z+x^2z^2}}\right) \d x \nonumber
		\\& \quad= \Phi[x(y)]-\left(1-\frac{2t}{T}\right) \int_{-\infty}^{x(y)} \frac{x}{\sqrt{4c^2t(T-t)/z+x^2}}\frac{\e^{-x^2/2}}{\sqrt{2 \pi}} \d x \nonumber
		\\& \quad= \Phi[x(y)]-\left(1-\frac{2t}{T}\right) \e^{2c^2t(T-t)/z} \left(\Phi\left[\sqrt{4c^2t(T-t)/z+x(y)^2}\right]-1\right) \nonumber
		\\& \quad= \Phi[x(y)]+\left(1-\frac{2t}{T}\right) \e^{2c^2t(T-t)/z} \,\Phi\left[-\sqrt{4c^2t(T-t)/z+x(y)^2}\right] \nonumber
		\\& \quad= \Phi\left[\frac{c(Ty-tz)}{\sqrt{yz(z-y)}}\right]+\left(1-\frac{2t}{T} \right)\e^{2c^2t(T-t)/z}\,
				\Phi\left[\frac{c((2t-T)y-tz)}{\sqrt{yz(z-y)}}\right].
	\end{align}
\end{proof}

\section{Proof of Proposition \ref{prop:mom}}
\label{apdx:SBmom}

\begin{prop*}
	Define the incomplete first moment of $S^{(z)}_{tT}$ by
	\begin{equation}
		M_{tT}(y;z)=\int_{0}^y u \, f_{tT}(u;z) \d u \qquad (0\leq y\leq z).
	\end{equation}
	Then we have
	\begin{equation}
		M_{tT}(y;z)=\frac{t}{T}z \left\{ \Phi\left[\frac{c(Ty-tz)}{\sqrt{yz(z-y)}}\right]-\e^{2c^2t(T-t)/z}\,
					\Phi\left[\frac{c((2t-T)y-tz)}{\sqrt{yz(z-y)}}\right] \right\},	
	\end{equation}
	and the second moment of $S^{(z)}_{tT}$ is given by
	\begin{equation}
		\E\left[ \left(S^{(z)}_{tT}\right)^2 \right]=
		\frac{t}{T}z^2\left\{1-c(T-t)\e^{\frac{c^2T^2}{2z}}\sqrt{\frac{2\pi}{z}}\,\Phi\left[-cTz^{-1/2} \right]  \right\}.
	\end{equation}
\end{prop*}

\begin{proof}
	Making the change of variable (\ref{eq:apdx_var_change}), we have
	\begin{equation}
		\int_{0}^y u \, f_{tT}(u;z) \d u=\frac{t(T-t)}{T} \int_{-\infty}^{x(y)} \frac{2zu(x)}{(T-2t)u(x)+tz} \frac{\e^{-x^2/2}}{\sqrt{2\pi}} \d x.
	\end{equation}
	Recalling equation (\ref{eq:hardslog}), we have
	\begin{align}
		\frac{t(T-t)}{T} \frac{2zu(x)}{(T-2t)u(x)+tz}&=\frac{t(T-t)z}{T(T-2t)}\left[ 2-\frac{2tz}{(T-2t)u(x)+tz} \right] \nonumber
		\\ &=\frac{tz}{T(T-2t)}\left[ T-2t+\frac{(T-2t)xz}{\a(x)} \right] \nonumber
		\\ &=\frac{t}{T}z\left[1+\frac{xz}{\a(x)}\right]. \label{eq:meanint}
	\end{align}
	Then we have
	\begin{align}
		&\int_{0}^y u \, f_{tT}(u;z) \d u \nonumber
		\\&\quad=\frac{t(T-t)}{T} \int_{-\infty}^{x(y)} \frac{2zu(x)}{(T-2t)u(x)+tz} \frac{\e^{-x^2/2}}{\sqrt{2\pi}} \d x \nonumber
		\\&\quad=\frac{t}{T}z \left\{ \Phi[x(y)]+\int_{-\infty}^{x(y)} \frac{x}{\sqrt{4c^2t(T-t)/z+x^2}}\frac{\e^{-x^2/2}}{\sqrt{2 \pi}} \d x\right\} \nonumber
		\\&\quad=\frac{t}{T}z \left\{ \Phi[x(y)]-\e^{2c^2t(T-t)/z} \,\Phi\left[-\sqrt{4c^2t(T-t)/z+x(y)^2}\right] \right\} \nonumber
		\\&\quad=\frac{t}{T}z \left\{ \Phi\left[\frac{c(Ty-tz)}{\sqrt{yz(z-y)}}\right]-\e^{2c^2t(T-t)/z}\,
				\Phi\left[\frac{c((2t-T)y-tz)}{\sqrt{yz(z-y)}}\right] \right\},
	\end{align}
	as required.
		
	Changing the variable in the second integral of the proposition gives
	\begin{equation}
		\int_{0}^z u^2 \, f_{tT}(u;z) \d u=\frac{t(T-t)}{T} \int_{-\infty}^{\infty} \frac{2zu(x)^2}{(T-2t)u(x)+tz} \frac{\e^{-x^2/2}}{\sqrt{2\pi}} \d x.
	\end{equation}
	Now, using (\ref{eq:alphadef}) and (\ref{eq:meanint}), we have
	\begin{align}
		&\frac{t(T-t)}{T} \frac{2zu(x)^2}{(T-2t)u(x)+tz} \nonumber
		\\ &\quad=\frac{t}{T}zu(x)\left[1+\frac{xz}{\a(x)}\right] \nonumber
		\\ &\quad=\frac{t}{T}z^2\left[\frac{c^2tT+x^2z}{c^2T^2+x^2z}+\frac{x\a(x)^2+xz(2c^2tT+x^2z)}{2(c^2T^2+x^2z)\a(x)}\right] \nonumber
		\\ &\quad=\frac{t}{T}z^2\left[\frac{c^2tT+x^2z}{c^2T^2+x^2z}+\frac{xz(c^2t(3T-2t)+x^2z)}{2(c^2T^2+x^2z)\a(x)}\right] \nonumber
		\\ &\quad=\frac{t}{T}z^2\left[\frac{c^2tT+x^2z}{c^2T^2+x^2z}+\frac{xz(c^2t(3T-2t)+x^2z)}{2(c^2T^2+x^2z)\sqrt{4c^2t(T-t)+x^2z^2}}\right].
	\end{align}
	Note that the second term above is odd in $x$.
	Hence
	\begin{align}
		&\int_{0}^y u^2 \, f_{tT}(u;z) \d u \nonumber
		\\ &\quad=\frac{t}{T}z^2 \int_{-\infty}^{\infty}\frac{c^2tT+x^2z}{c^2T^2+x^2z}\frac{\e^{-x^2/2}}{\sqrt{2\pi}} \d x \nonumber
		\\ &\quad=\frac{t}{T}z^2\left\{1-\frac{2c^2T(T-t)}{z}\int_0^{\infty}\frac{1}{c^2T^2/z+x^2} \frac{\e^{-x^2/2}}{\sqrt{2\pi}} \d x \right\} \nonumber
		\\ &\quad=\frac{t}{T}z^2\left\{1-\frac{2c^2T(T-t)}{z}\left( \sqrt{\frac{\pi}{2}}\sqrt{\frac{z}{c^2T^2}} \, \e^{\frac{c^2T^2}{2z}}
						\, \Phi\left[ -\sqrt{c^2T^2/z}\right] \right) \right\} \label{eq:erfint}
		\\ &\quad= \frac{t}{T}z^2\left\{1-c(T-t)\e^{\frac{c^2T^2}{2z}}\sqrt{\frac{2\pi }{z}}\,\Phi\left[-cTz^{-1/2} \right]  \right\}.
	\end{align}
	The equality (\ref{eq:erfint}) uses the identity \citep[7.4.11]{AS1964}.
\end{proof}

\section{Proof of Proposition \ref{prop:CBDF}}
\label{apdx:CBDF}

\begin{prop*}
	The distribution function of $Z_{tT}^{(z)}$ is
	\begin{align}
		F_{tT}(y;z)=\half&+\frac{(T-t)(c^2T(T-2t)+z^2)}{\pi T(c^2(T-2t)^2+z^2)} \arctan\left[ \frac{y}{ct} \right]\nonumber
		\\ &+\frac{t(c^2T(T-2t)-z^2)}{\pi T(c^2(T-2t)^2+z^2)} \arctan \left[\frac{z-y}{c(T-t)} \right]\nonumber
		\\ &+\frac{ct(T-t)z}{\pi T(c^2(T-2t)^2+z^2)} \log\left[\frac{y^2+c^2t^2}{(z-y)^2+c^2(T-t)^2}  \right],
	\end{align}
	where $0<t<T$, $c>0$.
\end{prop*}

\begin{proof}
	Writing $a=ct$ and $b=c(T-t)$, we need to calculate the integral
	\begin{equation}
		F_{tT}(y;z)=\frac{ab}{\pi(a+b)} (z^2+(a-b)^2) \int_{-\infty}^y \frac{\dd x}{(x^2+a^2)((z-x)^2+b^2)}.
	\end{equation}
	First we note that
	\begin{equation}
		\frac{1}{(x^2+a^2)((z-x)^2+b^2)}=\frac{u_1x +v_1}{x^2+a^2}+\frac{u_2x+v_2}{(z-x)^2+b^2},
	\end{equation}
	where
	\begin{equation}
			\begin{aligned}
				u_1&=\frac{2z}{(z^2+(a+b)^2)(z^2+(a-b)^2)},
				\\ v_1&=\frac{z^2+b^2-a^2}{(z^2+(a+b)^2)(z^2+(a-b)^2)},
				\\ u_2&=-u_1,
				\\ v_2&=2u_1z-v_1.
			\end{aligned}
	\end{equation}
	For $-R<y$, we define
	\begin{equation}
		I_R(y)= \int_{-R}^y \frac{\dd x}{(x^2+a^2)((z-x)^2+b^2)}.
	\end{equation}
	Then we can write
	\begin{align}
		I_R(y)&=\int_{-R}^y \left[\frac{u_1x +v_1}{x^2+a^2}+\frac{u_2x+v_2}{(z-x)^2+b^2}\right] \d x \nonumber
		\\&= \frac{u_1}{2} \log\left[\frac{y^2+a^2}{R^2+a^2} \right] \nonumber
							+\frac{u_2}{2} \log\left[\frac{(y-z)^2+b^2}{(R+z)^2+b^2} \right] \nonumber
		\\&\quad+\frac{v_1}{a} \left\{\arctan\left[ \frac{y}{a}\right]+\arctan\left[ \frac{R}{a}\right]\right\} \nonumber
		\\&\quad+\left(\frac{v_2+u_2z}{b}\right) \left\{\arctan\left[ \frac{y-z}{b}\right]+\arctan\left[ \frac{R+z}{b}\right]\right\} \nonumber
		\\&= \frac{u_1}{2} \log\left[\frac{y^2+a^2}{R^2+a^2}\times\frac{(R+z)^2+b^2}{(y-z)^2+b^2} \right] \nonumber
		\\&\quad+\frac{v_1}{a} \left\{\arctan\left[ \frac{y}{a}\right]+\arctan\left[ \frac{R}{a}\right]\right\} \nonumber
		\\&\quad+\left(\frac{u_1z-v_1}{b}\right) \left\{\arctan\left[ \frac{y-z}{b}\right]+\arctan\left[ \frac{R+z}{b}\right]\right\}.
	\end{align}
	Then we have
	\begin{multline}
		\lim_{R\rightarrow\infty} I_R(y)=\left(\frac{v_1}{a}+\frac{u_1z-v_1}{b}\right)\frac{\pi}{2}+\frac{u_1}{2} \log\left[\frac{y^2+a^2}{(y-z)^2+b^2} \right] 
		\\+\frac{v_1}{a} \arctan\left[ \frac{y}{a}\right]+\left(\frac{u_1z-v_1}{b}\right) \arctan\left[ \frac{y-z}{b}\right].
	\end{multline}
	The distribution function is then
	\begin{align}
		F_{tT}(y;z)&=\frac{ab}{\pi(a+b)}(z^2+(a-b)^2) \lim_{R\rightarrow\infty} I_R(y) \nonumber
		\\ &=\half+\frac{ab}{\pi(a+b)}\frac{z}{z^2+(a+b)^2} \log\left[\frac{y^2+a^2}{(z-y)^2+b^2}  \right]\nonumber
		\\ &\qquad+\frac{b}{\pi(a+b)}\frac{b^2-a^2+z^2}{z^2+(a+b)^2} \arctan\left[ \frac{y}{a} \right]\nonumber
		\\ &\qquad+\frac{a}{\pi(a+b)}\frac{b^2-a^2-z^2}{z^2+(a+b)^2} \arctan \left[\frac{z-y}{b} \right].
	\end{align}
	Substituting for $a$ and $b$ yields the required result.
\end{proof}

\section{Proof of Proposition \ref{prop:CBmom}}
\label{apdx:CBmom}

\begin{prop*}
	The first two moments of $Z_{tT}^{(z)}$ exist, and are given by
	\begin{align}
		\E\left[Z_{tT}^{(z)} \right]&=\frac{t}{T}z,
		\\\intertext{and} \E\left[\left(Z_{tT}^{(z)}\right)^2 \right] &=\frac{t}{T}(z^2+c^2T(T-t)).
	\end{align}
\end{prop*}

\begin{proof}
	Note the following two integrals:
	\begin{align}
		I_1&=\int_{-\infty}^{\infty} (y^2+c^2t^2) f_{tT}(y;z) \d y \nonumber
		\\ &=\frac{ct(T-t)}{\pi T} \int_{-\infty}^{\infty} \frac{z^2+c^2T^2}{(z-y)^2+c^2(T-t)^2} \d y \nonumber
		\\ &=\frac{ct(T-t)}{\pi T} \int_{-\infty}^{\infty} \frac{z^2+c^2T^2}{y^2+c^2(T-t)^2} \d y \nonumber
		\\ &=\frac{t}{T} (z^2+c^2T^2),
		\\\intertext{and}
		I_2&=\int_{-\infty}^{\infty} ((z-y)^2+c^2(T-t)^2) f_{tT}(y;z) \d y \nonumber
		\\ &=\frac{ct(T-t)}{\pi T} \int_{-\infty}^{\infty} \frac{z^2+c^2T^2}{y^2+c^2t^2} \d y \nonumber
		\\ &=\left(1-\frac{t}{T}\right) (z^2+c^2T^2).
	\end{align}
	Then we have
	\begin{align}
		\int_{-\infty}^{\infty}y\,f_{tT}(y;z)\d y&=\frac{I_1-I_2-c^2t^2+c^2(T-t)^2+z^2}{2z} \nonumber
		\\&=\frac{t}{T}z,
	\\\intertext{and}
			\int_{-\infty}^{\infty}y^2\,f_{tT}(y;z)\d y&=I_1-c^2t^2 \nonumber
		\\&=\frac{t}{T}(z^2+c^2T(T-t)),
	\end{align}
as required.
\end{proof}

%% file: Literature.tex
\addcontentsline{toc}{chapter}{\protect\numberline{}{Bibliography}}

\fancyhead{}
\fancyfoot{}
\pagestyle{fancy} 
\renewcommand{\chaptermark}[1]{\markboth{\thechapter \ #1}{}}
\fancyhead[RO,LE]{\sffamily\small \thepage}
\fancyhead[LO,RE]{\sffamily\small \nouppercase{\rightmark}}
\pagestyle{fancy}
\renewcommand{\headrulewidth}{0.4pt}
\renewcommand{\footrulewidth}{0.0pt}

\nocite{*}
\bibliography{THESIS_REF}